\documentclass{article}

\usepackage{xcolor}
\usepackage{arxiv}
\usepackage{graphicx}
\usepackage{subfig}

\usepackage{lmodern}
\usepackage{siunitx}
\usepackage{booktabs}
\usepackage{etoolbox}
\usepackage{soul}

\usepackage[utf8]{inputenc} 
\usepackage[T1]{fontenc}    
\usepackage{hyperref}       
\usepackage{url}            
\usepackage{booktabs}       
\usepackage{amsfonts}       
\usepackage{nicefrac}       
\usepackage{microtype}      
\usepackage[toc,page,titletoc]{appendix}
\usepackage{lineno}
\usepackage{amsmath,bm}
\usepackage{amsfonts,amssymb}
\usepackage{amsmath}
\DeclareMathOperator{\Tr}{Tr}

\usepackage{soul}

\title{COVID-19 Forecasts Using Internet Search Information in the United States}

\author{Simin Ma\thanks{H. Milton Stewart School of Industrial and Systems Engineering, Georgia Institute of Technology, USA} \\ \And Shihao Yang \footnotemark[1]}

\begin{document}
\maketitle

\begin{abstract}
As the COVID-19 ravaging through the globe, accurate forecasts of the disease spread is crucial for situational awareness, resource allocation, and public health decision-making. Alternative to the traditional disease surveillance data collected by the United States (US) Centers for Disease Control and Prevention (CDC), big data from Internet such as online search volumes has been previously shown to contain valuable information for tracking infectious disease dynamics such as influenza epidemic. In this study, we evaluate the feasibility of using Internet search volume of relevant queries to track and predict COVID-19 pandemic. We find strong association between COVID-19 death trend and the search volume of symptom-related queries such as ``loss of taste''. Then we further develop a previously-proposed influenza-tracking ARGO model (AutoRegression with GOogle search data) to predict future 4-week COVID-19 deaths on the US national level, by combining search volume information with COVID-19 time series information. Encouraged by the 20\% average error reduction of ARGO on national level comparing to the baseline time series model, we additionally build state-level COVID-19 deaths models. We introduce variants of ARGOX (Augmented Regression with GOogle data CROSS space), leveraging the cross-state cross-resolution spatial temporal framework that pools information from search volume and COVID-19 reports across states, regions and the nation. These variants of ARGOX are then aggregated in a winner-takes-all ensemble fashion to produce the final state-level 4-week forecasts. Numerical experiments demonstrate that our method steadily outperforms time series baseline models, and achieves the state-of-the-art performance among the publicly available benchmark models. Overall, we show that disease dynamics and relevant public search behaviors co-evolve during the COVID-19 pandemic, and capturing their dependencies while leveraging historical cases/deaths as well as spatial-temporal cross-region information will enable stable and accurate US national and state-level forecasts.
\end{abstract}


\section*{Author Summary}
Big data from the Internet has great potential to track infectious diseases at multiple geographical levels, such as estimating influenza activity from online search volume data. With the current COVID-19 pandemic, accurate forecasts of the disease dynamics can help resource allocation and public health decision-making. In this work, we further develop a previously-proposed influenza-tracking model for the short-term COVID-19 deaths forecasting in the United State at both national and state levels. Our model efficiently combines publicly available Internet search data at multiple resolutions (national, regional, and state-level) with the surveillance data from the Centers for Disease Control and Prevention (CDC), accounting for the spatial-temporal structure in the disease spread and online search pattern. Our method, across all states, performs competitively with the current state-of-arts benchmark models, demonstrating great potential in assisting CDC's current ensemble predictions. Our model is robust and easy to implement, with the flexibility to incorporate additional information from other sources and resolutions, making it generally applicable to tracking other social, economic or public health events at the state or local level.

\section{Introduction}
COVID-19, an acute respiratory syndrome disease caused by novel coronavirus SARS-CoV-2, has spread to more than 200 countries worldwide, leading to more than 224 million confirmed cases and 4.98 million deaths as of Oct 25, 2021 \cite{COVID-19-GlobalPan-Treatment}. Understanding how the disease spread dynamics progress over time is much needed, given the fluid situation and the potential rapid growth of COVID-19 infections. The implementation of efficient intervention policy and the allocation of emergency resources all depend on the accurate forecasts of the disease situation \cite{shinde2020forecasting}. Currently, machine learning methods \cite{DeepCovid_GT, UCSB_attention, Delphi_KF} and compartmental models \cite{COVID19Simulator, UCLA_sueir, epiforecasts_MIT, arik2020interpretable, yang2021estimating} are the most popular and prevailing approaches for the publicly-available COVID-19 spread forecasts, according to the weekly forecast reports compiled by the Centers for Disease Control and Prevention (CDC) \cite{CDC_Ensemble}. On the other hand, statistical models utilizing internet search behaviors for COVID-19 predictions have not attracted much attention.

In the last decade, numerous studies have shown that Internet-based big data could be a valuable complementary data source to monitor the prevalence of infectious diseases and provide near real-time disease estimations \cite{ARGO, Santillana_2015, Lu_2019, GFT_2008}, alternative to the traditional surveillance approach. For example, Yang et al. \cite{ARGO} provided a robust way to estimate real-time influenza situation in the United States using Google search data; Ning et al. \cite{ARGO2_Regional} and Yang et al. \cite{ARGOX} further extended the method for the regional-level and the state-level influenza estimation using the Google search data at the finer resolution; Yang et al. \cite{yang2017advances} demonstrated the success of search data to track dengue fever in five tropical countries. Other types of internet-based data include cloud-based electronic health records \cite{yang2017using} and social media messages \cite{santillana2015combining}. Currently, during the ongoing COVID-19 pandemic, a large amount of pandemic-related online searches are generated, indicating actual infections and general concerns, which might contain useful information to estimate and predict the disease spread.

However, building COVID-19 prediction model with online search data is undoubtedly challenging. First of all, COVID-19 pandemic is a novel disease outbreak with rapid development, which creates difficulties in identifying relevant keyword queries of the search data. Even with the expert-curated queries, the online search frequency data can contain a high level of noise with many unusual spikes, due to the general searches driven by non-disease factors such as media coverage or public concern. Besides, the ground truth data compiled by CDC could also be noisy with frequent retrospective revisions of daily/weekly cases accounting for the mistakes in data collection and reporting.

\subsection*{Related literature}
The correlation between search engine data (such as Google Trends \cite{GoogleTrends}, Baidu \cite{ARGONet_Baidu}, Twitter, and Youtube searches \cite{rufai2020world}) and the COVID-19 situation has been well documented for multiple countries \cite{GT_Analysis_World, Lampos_Media, prasanth2021forecasting}, including specific studies in China \cite{li2020retrospective, husnayain2020applications}, Europe \cite{mavragani2020tracking, walker2020use}, India \cite{GT_India}, Iran \cite{walker2020use, ayyoubzadeh2020predicting}, U.S. \cite{walker2020use, hong2020population, GT_analysis_US, GT_Analysis_USStates}, and Spain \cite{GTStudy_Spain}. However, these articles mostly focus on the pure correlation exercise, including correlation computation \cite{prasanth2021forecasting, GTStudy_Spain, GT_Analysis_USStates, GT_Analysis_World}, rank analysis \cite{GT_analysis_US}, and cross correlation for time delay between search peaks and COVID-19 cases/deaths \cite{li2020retrospective, husnayain2020applications, GT_India}. None of them examines the importance among the search queries, considers the spatial-temporal structure of the data, or attempts to make weeks-ahead predictions.

The relevant search queries in existing literature include general COVID-19 terms such as ``coronaviours'' or ``COVID'' \cite{li2020retrospective}, public safety precautions such as ``handwashing'' \cite{husnayain2020applications}, symptom-related queries such as ``loss of smell'' \cite{GT_Analysis_USStates, walker2020use}. However, most of the existing articles focus on a handful of query terms under a specific class, without a large-scale data-driven query identification process. 

While most of existing articles argue for the potential of online search data for COVID-19 forecasts \cite{rufai2020world, li2020retrospective, husnayain2020applications, mavragani2020tracking, walker2020use, hong2020population, GT_India, GT_Analysis_World, GT_Analysis_USStates}, only a few actually build the prediction models \cite{GT_analysis_US, prasanth2021forecasting, ayyoubzadeh2020predicting, Lampos_Media}. Specifically, Mavragani and Gkillas\cite{GT_analysis_US} demonstrates the search data predictive power via quantile regression in U.S. states. Prasanth et al. \cite{prasanth2021forecasting} uses the search data from selected queries on a long-short term memory (LSTM) framework for U.S, U.K and India. Ayyoubzadeh et al. \cite{ayyoubzadeh2020predicting} takes it further by combining linear regression with LSTM model to provide short-term COVID-19 cases forecast in Iran. Lampos et al. \cite{Lampos_Media} conducts a prediction study in several European countries, using transfer-learning and Gaussian processes.

However, none of the articles above fully utilizse the predictive power of internet search data by accounting for the spatial-temporal structure, including the time series information of COVID-19 or internet searches in near-by regions/areas. Other qualitative analysis, ad-hoc correlation exercise, or off-the-shelf model application are even further away from a real impact. The only internet-search-based model, to the best of our knowledge, that accounts for spatial structure is ARGONet \cite{ARGONet_Baidu}, which uses a clustering and $L_1$-penalized data augmentation technique for 2-day ahead COVID-19 cases forecast in China. So far, none of the existing Internet-search-based methods provides robust weeks-ahead forecasts for different geographical areas in the United States that account for spatial correlations.

\subsection*{Our contribution}
In this paper, we propose a simple framework with Google search queries for United States national and state level 4-week-ahead COVID-19 deaths forecasts. In particular, we identify relevant queries through a large-scale cross-correlation exercise, and de-noise the search frequency data from unusual spikes using inter-quantile-range method. We detect strong correlation between lagged  Google search data and COVID-19 deaths with symptom-related terms, and select important Google search queries among the large-batch of related terms to increase the robustness and interpretability of our forecasts. We then utilize the detected predictive power and combine the selected Google search data and lagged COVID-19 time series information to produce national level forecasts. We further incorporate a multi-resolution spatial temporal framework for state-level forecasts, leveraging cross-state, cross-region COVID-19 and Google search information, accounting for the geographical proximity and correlations in infections. Unlike the clustering technique in ARGONet \cite{ARGONet_Baidu}, our state level method exhibits stronger internal spatial structure and better model interpretability. Lastly, we incorporate a winner-takes-all mechanism to generate more coherent final United States national and state level predictions. Numerical comparisons show that our method performs competitively with other publicly available COVID-19 forecasts. The success of our method demonstrates that previously-developed models for influenza predictions \cite{ARGO, ARGOX} using online search data can be re-purposed for accurate and robust forecasts of COVID-19, further emphasizing the general applicability of our method and the power of big data disease detection.

\section{Data Acquisition and Pre-processing}
This paper focuses on the 50 states of the United States, plus Washington D.C. We use confirmed cases, confirmed death and Google search query frequencies as inputs. 

\subsection{COVID-19 Reporting Data}\label{sec:COVID-Data}
We use reported COVID-19 confirmed cases and death of United States from New York Times (NYT) \cite{NYT_COVID} as features in our model. This data is collected from January 21, 2020 to Oct 9, 2021. 

When comparing against other Centers for Disease Control and Prevention (CDC) official predictions, we use COVID-19 confirmed cases and death from JHU CSSE COVID-19 dataset \cite{JHU_Data} as the groundtruth. This data is curated dataset used by the CDC at their official website, collected from January 22, 2020 to Oct 10, 2021.

We do not use JHU COVID-19 dataset as input features in our model because JHU COVID-19 dataset retrospectively corrects past confirmed cases and death due to reporting error, and federal and state policy changes, while NYT dataset does not revise past data, which gives more realistic forecasts.

\subsection{Google Search Data}
The online search data used in this paper is obtained from Google Trends \cite{GoogleTrends}, where one can obtain the search frequencies of a term of interest in a specific region and time frame by typing in the search query on the website. With Google Trends API, we are able to obtain a daily time series of the search frequencies for the term of interest, including all searches that contain all of its words (un-normalized). The search term's frequencies time series from Google Trends is based on a sampling approach, which looks at the search query representative of all raw Google searches frequencies \cite{GoogleTrends}.

This paper uses 256 top searched COVID-19 related Google search queries, including common searched COVID related terms, COVID related symptoms, COVID pandemic policies implemented, COVID related resource allocated, and etc. Example terms include ``Coronavirus'', ``COVID 19'', ``COVID Vaccine'', ``loss of taste'', ``loss of smell'', ``cough'', and ``fever''. We obtain all the Google search queries' daily frequencies for national and state level. Regional level Google search queries are obtained by simply summing up the state level Google search query volumes that are in the region. Note that some of the search terms might seem identical, e.g. "coronavirus vaccine" and "covid 19 vaccine.", but we treat them separately in our model due to  linguistic heterogeneity, as terms with similar meaning but differently phrased are embedded with different search frequencies by the public due to different linguistic preferences.

Google Trends also truncates data to 0 if the search volume for the query is too low. Consequently, for a given query and state, the zeros in Google Trends data indicate missing data due to low volume of searches for the the specified query and state, which is very common in practice. We account for the high level of sparsity in the state level data by borrowing information from regional level to ``enrich'' state-level sparsity through a weighted average of state-level search frequency (2/3 weight) and regional-level search frequency (1/3 weight) \cite{ARGOX}.

\subsubsection{Inter-Quantile Range (IQR) Filter for Google Search Data}
Instability and sudden spikes/drops in Google search volume data can due to natural noises in Google Trends' sampling approach. Meanwhile, sparsity might still exist in some states' search queries after regional-enrichment of state-level Google search data. Such instability and sparsity severely reduce the prediction accuracy at the national, regional and state level. Therefore, we introduce an Inter-Quantile Range (IQR) based data filtering mechanism to reduce the noise in the data.

We first drop the Google search terms that have frequencies lower than the median number of all other Google search queries frequencies. Then, we identify the outliers of a Google search query. The large-valued outlier are those above 99.9 percent quantile and also three standard deviations above past-week rolling average. The small-value outliers aer those below 1 percent quantile. We overwrite the outlier values to be the past three-day average. 

The reason behind different data processing approaches for large and small valued outliers lies in the hypothesis that an sudden increase of search volume is more probable to be true than a drop in search frequency of a Google search query, which is possibly due to inadequate search intensity. For instance, sudden increases in search frequencies of COVID-19 related terms occur when COVID-19 first hit the U.S. in mid March 2020, while decreases/sparsities in search query volumes are resulted from low search volume and missing data, especially in state-level search queries. Therefore, the IQR filter is ``looser'' on large valued outliers and ``more strict'' on small valued outliers. As an result, we believe that a large valued outlier is indeed an ``unreasonable'' spike if they are significantly larger than search frequencies from the other days in the same week.

This IQR inspired filtering mechanism is able to further account for the sparsity in Google search queries as well as removing unusual spikes, which improves our model's forecast accuracy.

\subsubsection{Optimal Lag}\label{sec:Opt_lag}
\begin{figure}[htbp]
\centering
\includegraphics[width=0.7\textwidth]{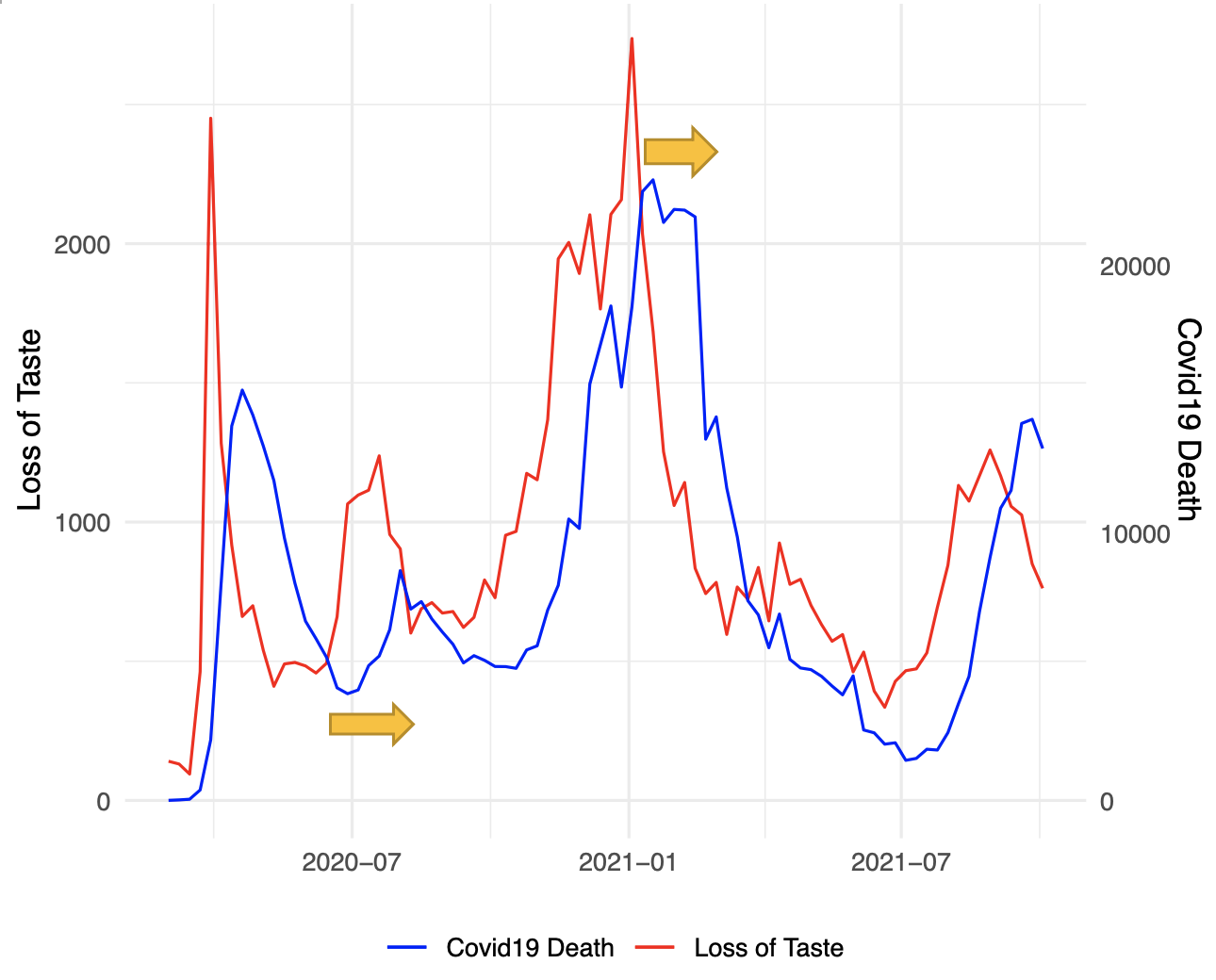}
\caption{{\bf Google search query ``loss of taste'' and COVID-19 weekly incremental death}
Illustration of delay in peak between Google search query search frequencies (Loss of Taste in red) and COVID-19 national level weekly incremental death (blue). Y-axis are adjusted accordingly.}
\label{fig1}
\end{figure}

It is typical to see the peak of COVID-19 search volume ahead of the peak in reported cases or deaths, see Fig \ref{fig1} for an illustration for query ``loss of taste''. One hypothesis is that the early-stage infected people could search for COVID-19 related information online before their arrival at a clinic or tested positive.

As such, using delayed Google search frequencies for forecast is essential for our predictions. One simple way is to enumerate all possible delaying lags of Google search frequencies as exogenous feature variables. Yet, this will significantly increase the number of exogenous variables and impact prediction accuracy which could result in over-fitting. Thus, we derive the optimal lag for each Google search query and only consider those optimal lags in forecasting model. 

We use the period from April 1st 2020 to June 30th 2020 to find the optimal lags. We will use the period after July 1st 2020 for comparing forecast accuracy. Because media-driven or information-seeking searches are very common at the beginning of the pandemic  \cite{MediaFear_Ebola,Media_H1N1}, we exclude period prior to April 1st 2020 in the analysis, so that the query terms identified are more likely to be driven by actual infection. For each query, we fit a linear regression of COVID-19 daily death against lagged Google search frequency, considering a range of lags (4 to 35 days). We select the lagged Google search frequency that has the lowest mean square error (MSE) as the optimal lag for that query. Table \ref{tab:23query_optimallag} in supplementary materials shows all the optimal lags for the selected important Google search terms, ranked by their optimal lags. For national, regional and state level Google search information, we consider the same optimal lag displayed in table \ref{tab:23query_optimallag} throughout this study.

\subsubsection{Highly Correlated 23 Terms}\label{sec:23Terms}
Though we removed low frequency queries and sparsity in the remaining queries through the IQR filter, some remaining queries might still exhibit high variability and do not obtain a clear trend comparing to COVID-19 death. To further obtain most useful terms for predicting COVID-19 death and eventually reduce our model complexity, we computed Pearson correlation coefficient between each of the optimal lagged search term and COVID-19 daily death during the period from April 1st 2020 to June 30th 2020, where the detail derivation of optimal delay is shown in section \ref{sec:Opt_lag}. Table \ref{tab:23query_PC} in supplementary materials lists all the Google search query terms and their Pearson correlation coefficient against COVID-19 daily death, in which only positive Pearson correlation terms are displayed.

We select the 23 terms that have Pearson correlations above $0.5$ as ``important terms'' and only use them in our forecast model, shown in table \ref{tab:23query_optimallag}.

\section{Methods}
\subsection{ARGO Inspired Prediction}\label{sec:ARGO_Nat}
Let $X_{i,t,m}$ be the Google Trends data of search term $i$ day $t$ of area $m$; $y_{t,m}$ be the New York Times COVID-19 death increment at day $t$ of area $m$; $c_{t,m}$ be the New York Times COVID-19 confirmed case increment at day $t$ of area $m$, where the area $m$ can refer to the entire nation, one specific HHS region (such as New England), or one specific state (such as Georgia). Let $O_k$ be the optimal lag for the $k$th Google search term, which is the same for all area $m$. Let $\mathbb{I}_{\{t, r\}}$ be the weekday $r$ indicator for $t$ (i.e., $\mathbb{I}_{\{t, 1\}}$ indicates day $t$ being Monday, and $\mathbb{I}_{\{t, 6\}}$ indicates day $t$ being Saturday), which accounts for the weekday seasonality in COVID-19 incremental death time series.

Inspired by ARGO method \cite{ARGO}, with information available as of time $T$, to estimate $y_{T+l,m}$ for $l>0$, the incremental COVID-19 death on day $T+l$ of area $m$, an $L_1$ regularized linear estimator is used:
\begin{equation}\label{eqn:argo_yhat}
\begin{aligned}
    \hat{y}_{T+l,m} = \hat{\mu}_{y,m} + &\sum^{I}_{i=0}\hat{\alpha}_{i,m}y_{T-i,m} + \sum_{j\in \mathcal{J}}\hat{\beta}_{j,m}c_{T+l-j,m} + \sum^{K}_{k=1}\hat{\delta}_{k,m}X_{k,T+l-\hat{O}_k,m} \\
    \;\;\;& + \sum^6_{r=1}\hat{\gamma}_{r,m}\mathbb{I}_{\{T+l, r\}}
\end{aligned}
\end{equation}
where we use lagged death, lagged confirmed cases and optimal lagged Google search terms for death prediction.  For $l^{\text{th}}$ day ahead prediction at area $m$, the coefficients $\{\mu_{y,m},\bm{\alpha}=(\alpha_{1,m},\ldots,\alpha_{I,m}), \bm{\beta}=(\beta_{1,m},\ldots,\beta_{|\mathcal{J}|,m}), \bm{\delta}=(\delta_{1,m},\ldots,\delta_{K,m}), \bm{\gamma}=(\gamma_{1,m}, \ldots, \gamma_{6,m})\}$ are obtained via
\begin{equation}\label{eqn:argo_obj}
    \begin{aligned}
    \underset{\mu_{y,m},\bm{\alpha},\bm{\beta},\bm{\delta},\gamma_m, \bm{\lambda}}{\mathrm{argmin}} 
    \sum_{t=T-M-l+1}^{T-l} &\Bigg( y_{t+l,m}-\mu_{y,m} - \sum^{6}_{i=0}{\alpha}_{i,m}y_{t-i,m} - \sum_{j\in \mathcal{J}}{\beta}_{j,m}c_{t+l-j,m} \\
    \;\;\;&-  \sum^{23}_{k=1}{\delta}_{k,m}X_{k,t+l-\hat{O}_k,m} -\sum^6_{r=1}\gamma_{r,m}\mathbb{I}_{\{t+l,r\}} \Bigg)^2\\
    \;\;\;&
    + \lambda_\alpha\|\bm{\alpha}\|_1+\lambda_\beta\|\bm{\beta}\|_1+\lambda_\delta\|\bm{\delta}\|_1 + \lambda_\gamma \|\bm{\gamma}\|_1
    \end{aligned}
\end{equation}

We set $M=56$, i.e. 56 days as training period; $I=6$ considering consecutive 1 week lagged death; $\mathcal{J}=\max\left(\{7,14,21,28\},l\right)$ considering weekly lagged confirmed cases; $K=23$ highly correlated Google search terms; $\hat{O}_k=\max\left(O_k,l\right)$ be the adjusted optimal lag of $k$th Google search term subject to $l^{\text{th}}$ day ahead prediction. We set hyperparameters $\bm{\lambda}=(\lambda_\alpha,\lambda_\beta,\lambda_\delta, \lambda_\gamma)$ through cross-validation. For simplicity, we constrain $\lambda_\alpha=\lambda_\beta=\lambda_\delta=\lambda_\gamma$.

To further impose smoothness into our predictions, we use the three-day moving average of the coefficients for predicting day $T+l$, which slightly boosts our prediction accuracy.

Using the above formulation, we forecast future 4 weeks of daily incremental COVID-19 death of area $m$, i.e. $\{\hat{y}_{T+1,m},\ldots,\hat{y}_{T+28,m}\}$, and aggregate them into weekly prediction. In other words, $\hat{y}_{T+1:T+7,m}=\sum^7_{i=1}\hat{y}_{T+i,m}$ is first week, $\hat{y}_{T+8:T+14,m}= \sum^{14}_{i=8}\hat{y}_{T+i,m}$ is the second week, $\hat{y}_{T+15:T+21,m}= \sum^{21}_{i=15}\hat{y}_{T+i,m}$, is the third week, and $\hat{y}_{T+22:T+28,m}= \sum^{28}_{i=22}\hat{y}_{T+i,m}$ is the fourth week ARGO incremental death prediction. We denote this method as ``ARGO Inspired Prediction''.

\subsection{ARGOX Inspired State Level Prediction}
\label{sec:ARGOX_Joint_Alone}
In the previous section, our prediction is based on Google search terms and past COVID-19 cases and death information. The predictions perform competitively on national level, but fall short in regional and state level due to the spatial correlation induced by geographical proximity, transportation connectivity and region-state-wise related spread. Thus, we incorporate ARGOX \cite{ARGOX}, a unified spatial-temporal statistical framework that combines multi-resolution, multi-source information while maintaining coherency with regional and national COVID-19 death.
ARGOX \cite{ARGOX} operates in two steps: the first step extracts state level internet search information via LASSO, and the second step enhances the estimates using cross-region, cross-resolution spatial temporal framework. The second step takes a dichotomous approach for the joint states and alone states, identified through geographical separations and multiple correlations on COVID-19 growth trends. In particular, it incorporates the cross-state, cross-source correlations for the joint model of connected states, while leaving the freedom for stand-alone modeling for a handful of isolated states. For the connected states, it gathers raw estimates for state/regional/national-level weekly COVID-19 incremental deaths and past time step's COVID-19 incremental deaths, and applies a penalized best linear predictor to get the final state level estimates. See SI for detailed formulation in our particular setting of joint model and stand-alone model.

\subsection{State Level Forecasting with National Constraint}\label{sec:ARGOX_Nat_Constraint}
Although ARGOX is a consistent framework in cascading fashion, the exact predictions from ARGOX are not consistent through aggregation from bottom up. There exists inconsistency between the aggregation of state-level ARGOX prediction and national level prediction ARGO prediction, illustrated in fig \ref{fig:Nat_ConstraintProof} and \ref{fig:Nat_ConstraintProof_3_4} in supplementary materials, which shows that the sum of ARGOX 51 states' COVID-19 incremental death predictions does not equal the national level ARGO prediction. Furthermore, our ARGO inspired national level predictions (section \ref{sec:ARGO_Nat}) are more accurate for the national level groundtruth than the sum of ARGOX 51 states' predictions. Therefore, we propose a constrained second step to ARGOX inspired state level prediction, by restricting the sum of state-level ARGOX predictions to be close to the ARGO inspired national prediction.
 
After the first step in section \ref{sec:ARGOX_Joint_Alone}, instead of separating the 51 US states into ``joint'' and ``alone'' states and estimating state-level COVID-19 deaths separately, we treat them as a whole (except HI and VT) as we are constraining the sum of all states' death estimations. We separate out HI and VT, since these two states have the lowest incremental COVID-19 deaths, and such sparsity causes instability when deriving the covariance matrices. We estimate HI and VI COVID-19 weekly incremental death using the ARGO inspired state-level estimates through equation (\ref{eqn:argo_obj}).

For the 49 states (except HI and VT), our raw estimates for the state-level weekly COVID-19 incremental death $\bm{y}_\tau=(y_{\tau,1},\ldots,y_{\tau,49})$ are $\hat{\bm{y}}^{GT}_{\tau} = (\hat{y}^{GT}_{\tau,1},\dots, \hat{y}^{GT}_{\tau,49})^\intercal$, $\hat{\bm{y}}_\tau^{reg}=(\hat{y}_{\tau, r_1}^{reg}, \dots,\hat{y}_{\tau, r_{49}}^{reg})^\intercal$ and $\hat{\bm{y}}^{nat}_{\tau} = (\hat{y}^{nat}_{\tau},\dots, \hat{y}^{nat}_{\tau})^\intercal$, where $r_m$ is the region number for state $m$. Here, we denote $state$ and $reg$ to be state/regional estimates with internet search information only, where the detailed derivations can be found in SI, and $nat$ to be national estimates from equation (\ref{eqn:argo_yhat}). Similar to the second step in section \ref{sec:ARGOX_Joint_Alone} and in ARGOX \cite{ARGOX}, we denote the state-level death increment at week $\tau$ as $\bm{Z}_{\tau}=\Delta\bm{y}_{\tau}=\bm{y}_{\tau}-\bm{y}_{\tau-1}$ and it has four predictors: (i) $\bm{Z}_{\tau-1}=\Delta\bm{y}_{\tau-1}$, (ii) $\hat{\bm{y}}_{\tau}^{GT} - \bm{y}_{\tau-1}$, (iii) $\hat{\bm{y}}_{\tau}^{reg} - \bm{y}_{\tau-1}$, and (iv) $\hat{\bm{y}}_{\tau}^{nat} - \bm{y}_{\tau-1}$. Let $\bm{W}_{\tau}$ denote the collection of these four vectors $\bm{W}_{\tau}=(\bm{Z}_{\tau-1}^\intercal, (\hat{\bm{y}}_{\tau}^{GT} - \bm{y}_{\tau-1})^\intercal, (\hat{\bm{y}}_{\tau}^{reg} - \bm{y}_{\tau-1})^\intercal, (\hat{\bm{y}}_{\tau}^{nat} - \bm{y}_{\tau-1})^\intercal)^\intercal$.

We denote $\hat y^{*\text{ARGO}}_{\tau,nat}$ as the week $\tau$ national level ARGO-inspired COVID-19 incremental death estimation excluding HI and VT. To predict the week $\tau$ state-level COVID-19 death increment, we solve the constrained optimization problem below that minimizes the variance between groundtruth and our predictor, subject to the constraint that the sum of our predictor ought to be close to national level COVID-19 incremental death estimations. 
\begin{equation}\label{eqn:ARGOX_Nat_Constraint_Problem}
    \begin{aligned}
    &\min_{\bm{A}}\, \Tr(\mathrm{Var}(\bm{Z}_{\tau}-\bm{A}\bm{W}_\tau))\\
    &\text{s.t.}\;\; \mathbf{1}^\intercal(\mu_Z+\bm{A}\bm{W}_\tau)=\hat y^{*\text{ARGO}}_{\tau,nat}-\mathbf{1}^\intercal \bm{y}_{\tau-1}
\end{aligned}
\end{equation}
where $\mathbf{1}=(1,\ldots,1)^\intercal$ is a length 49 vector of 1s, and $\mu_Z$ is the mean of $\bm{Z}_\tau$. In particular if the above optimization problem is unconstrained, the solution will be the best-linear predictor (no ridge-penalty) in ARGOX \cite{ARGOX}. The detail derivation of the optimization problem in equation (\ref{eqn:ARGOX_Nat_Constraint_Problem}) as well as the final closed-form solution can be found in the supplementary materials.

\subsection{Winner-takes-all State Level Ensemble Forecast}
To further boost the state-level COVID-19 death prediction accuracy, we incorporate an ensemble framework that combines our previous estimations and selects the best predictor for each week. For all 51 U.S. states, we denote the ARGO-inspired state-level prediction for week $\tau$ as ``ARGO'' (section \ref{sec:ARGO_Nat}), ARGOX-inspired joint-alone state prediction as  ``ARGOX-2step'' (section \ref{sec:ARGOX_Joint_Alone}), and ARGOX-inspired national constrained prediction as ``ARGOX-NatConstraint'' (section \ref{sec:ARGOX_Nat_Constraint}). For a training period of 15 weeks, we evaluate each predictor with mean squared error (MSE) and select the one with lowest MSE as the ensemble predictor for week $\tau+1$, $\tau+2$, $\tau+3$, and $\tau+4$. Such winner-takes-all approach has been previously shown to be effective for influenza estimation \cite{Lu_2019}.

\section{Retrospective Evaluation}
\subsection{Evaluation Metrics}\label{sec:Error_Metric}
We use three metrics to evaluate the accuracy of an estimate of COVID-19 death against the actual COVID-19 Death published by John Hopkins University (JHU): the root mean squared error (RMSE), the mean absolute error (MAE), and the Pearson correlation (Correlation). RMSE between an estimate $\hat{y}_t$ and the true value $y_t$ over period $t=1,\ldots, T$ is $\sqrt{\frac{1}{T}\sum_{t=1}^T \left(\hat{y}_t - y_t\right)^2}$. MAE between an estimate $\hat{y}_t$ and the true value $y_t$ over period $t=1,\ldots, T$ is $\frac{1}{T}\sum_{t=1}^T \left|\hat{y}_t - y_t\right|$. 
Correlation is the Pearson correlation coefficient between $\hat{\bm{y}}=(\hat{y}_1, \dots, \hat{y}_T)$ and $\bm{y}=(y_1,\dots, y_T)$.

\subsection{Comparison among our Methods}
\subsubsection{National Level}
We focus on United State's 51 states/districts (including Washington DC) for all comparisons in this section. In this section, we conduct sensitivity analysis by comparing our own methods on national level and state level for 1 to 4 weeks ahead COVID-19 incremental death prediction.

For national level, we compare with four other simplified models (i) persistence (Naive), (ii) ARGO prediction (section \ref{sec:ARGO_Nat}), (iii) AR-only prediction, and (iv) GT-only prediction. The Naive (persistence) predictions use current week's incremental death counts from New York Times (NYT) as next 1 to 4 weeks' estimation, since NYT does not retrospectively correct past data. AR-only prediction uses the model setup in section \ref{sec:ARGO_Nat} but only with lagged death and cases information, 
whereas GT-only prediction is using Google search information only. 
For fair comparisons, both AR-only and GT-only predictions include weekday indicators, and use 56 days training period. Daily COVID-19 incremental deaths are estimated using the three methods above, and aggregated into weekly incremental deaths for the time period of July 4, 2020 to Oct 9th, 2021. The period before July 4, 2020 is excluded for comparison analysis, as the optimal Google search terms' lags are selected using that period.  The groundtruth is COVID-19 weekly incremental death from JHU COVID-19 dataset (section \ref{sec:COVID-Data}). 

\begin{table}[ht]
\sisetup{detect-weight,mode=text}
\renewrobustcmd{\bfseries}{\fontseries{b}\selectfont}
\renewrobustcmd{\boldmath}{}
\newrobustcmd{\B}{\bfseries}
\addtolength{\tabcolsep}{-4.1pt}
\footnotesize
\centering
\caption{National Level Comparison Error Metrics}\label{tab:Nat_Our}
\begin{tabular}{lrrrr}
  \hline
& 1 Week Ahead & 2 Weeks Ahead & 3 Weeks Ahead & 4 Weeks Ahead\\ 
    \hline \multicolumn{1}{l}{RMSE} \\
\hspace{1em} AR Only & 1991.510 & 2967.008  & 3765.499  & 4908.492\\
\hspace{1em} ARGO & \B1751.555& \B 2226.426  & \B2889.028 & \B4127.143\\
\hspace{1em} GT Only & 2228.040& 2936.837   &3167.419 &4398.494 \\
\hspace{1em} Naive &  2001.722  & 2955.116    &3829.775  &4694.051 \\
\multicolumn{1}{l}{MAE} \\
\hspace{1em} AR Only &1193.806    & 1812.925   & 2607.313  &3559.030 \\
\hspace{1em} ARGO & \B1187.866 & \B 1583.627   &\B2106.448 &\B2907.970  \\
\hspace{1em} GT Only & 1341.418 & 2002.000 &2279.597 &3118.358  \\
\hspace{1em} Naive &1374.672   & 2110.254  &2888.000   & 3616.149  \\
\multicolumn{1}{l}{Correlation} \\
\hspace{1em} AR Only &   0.948& 0.893 &0.866 & 0.824\\
\hspace{1em} ARGO  &\B0.959 &\B0.938   &\B0.903 & \B0.843\\
\hspace{1em} GT Only & 0.937  & 0.876  & 0.859  & 0.738\\
\hspace{1em} Naive &  0.944 & 0.876 & 0.791 &0.684\\
\hline
\end{tabular}

\vspace{1ex}
{\raggedright National level 1 to 4 weeks ahead COVID-19 incremental prediction comparisons in 3 error metrics. Boldface highlights the best performance for each metric in each study period. All comparisons are based on the original scale of COVID-19 national incremental death. On average, ARGO is able to achieve around 20\% RMSE, 15\% MAE reduction, and around 5\% correlation improvement, compared to the Naive or AR Only predictions. \par}
\end{table}

Table \ref{tab:Nat_Our} summarizes the accuracy metrics for all estimation methods for the period from July 4, 2020 to Oct 9th, 2021 on national level, which shows that ARGO’s estimates outperform all other
simplified models, in every accuracy metric for the whole time period. Fig. 1 displays the estimates against the observed COVID-19 weekly incremental death. 

\begin{figure}[htbp]
\centering
\subfloat[1 Week Ahead National Level Predictions]
{\includegraphics[width=0.49\textwidth]{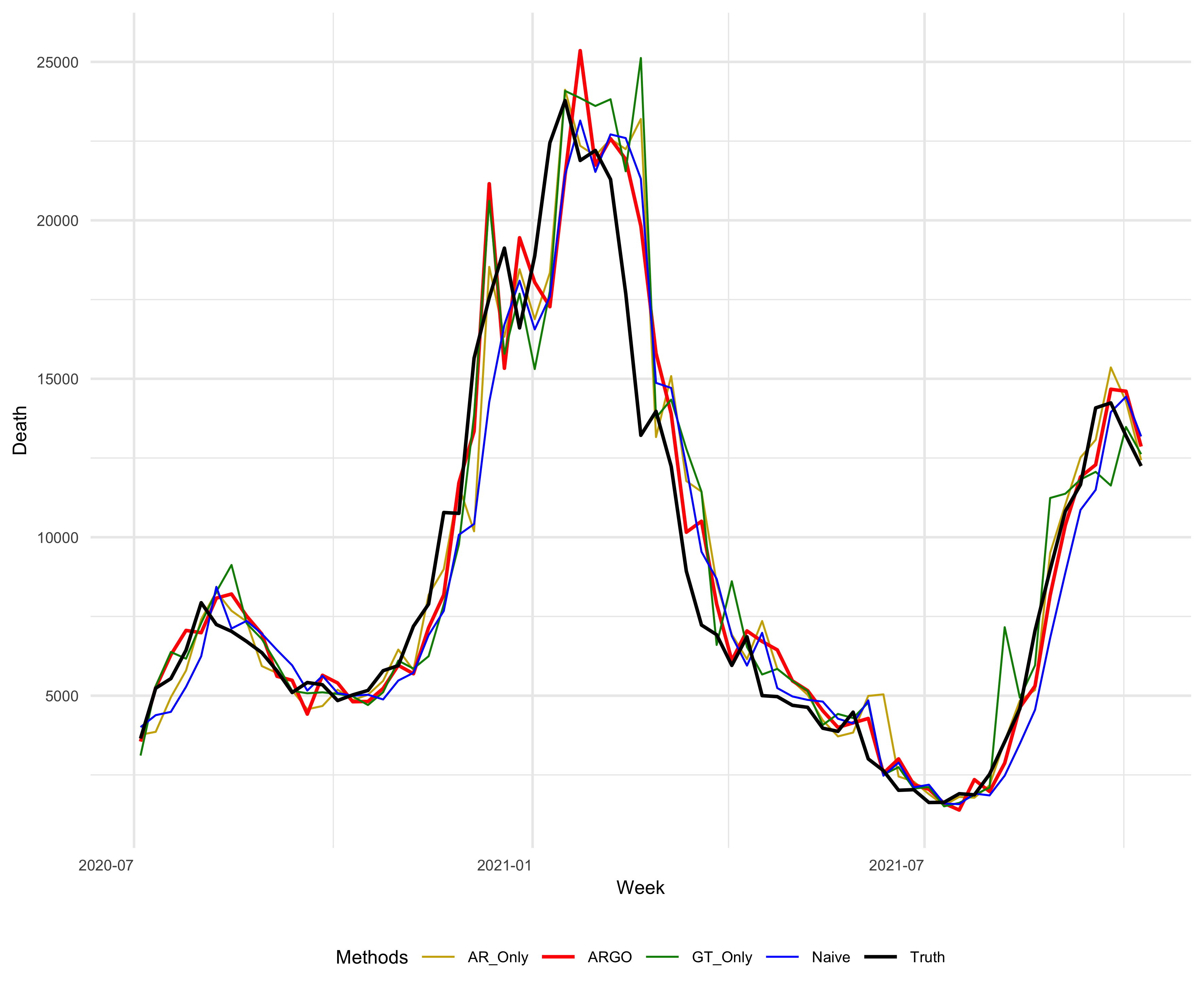}\label{fig:Nat_Compare_Our_1Week}}
\hfill
\subfloat[2 Weeks Ahead National Level Predictions]
{\includegraphics[width=0.49\textwidth]{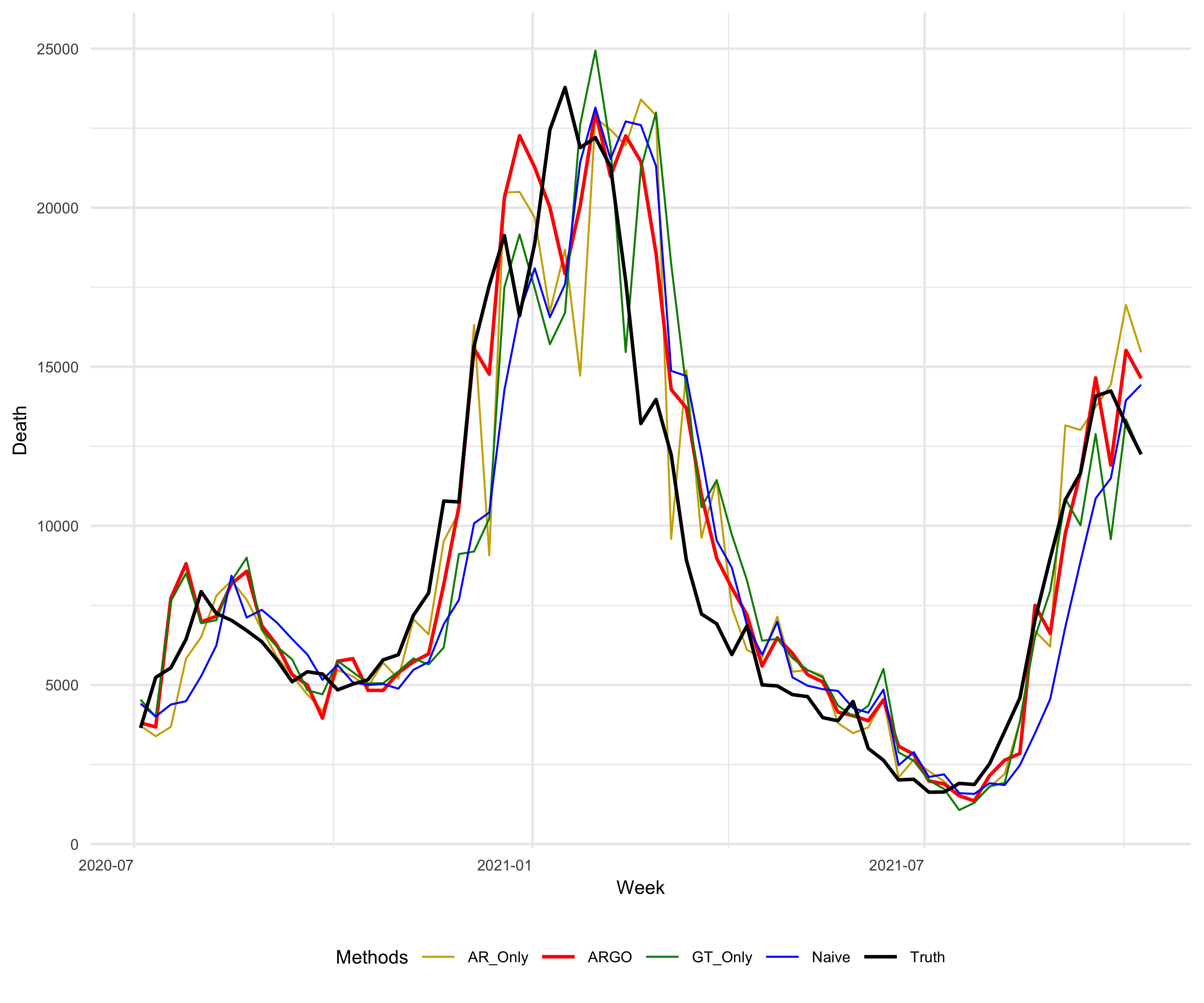}\label{fig:Nat_Compare_Our_2Weeks}}
\hfill
\subfloat[3 Weeks Ahead National Level Predictions]
{\includegraphics[width=0.49\textwidth]{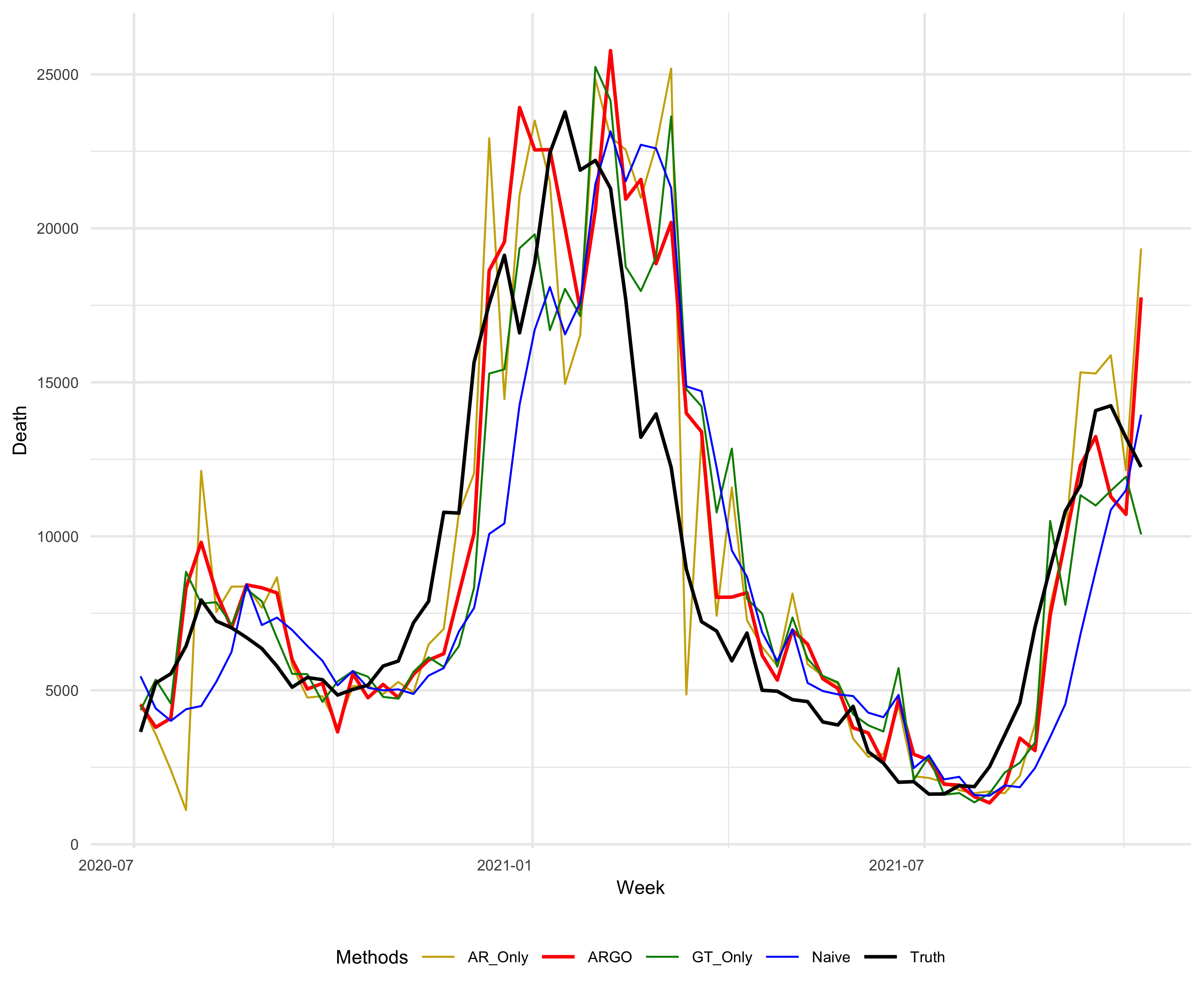}\label{fig:Nat_Compare_Our_3Weeks}}
\hfill
\subfloat[4 Weeks Ahead National Level Predictions]
{\includegraphics[width=0.49\textwidth]{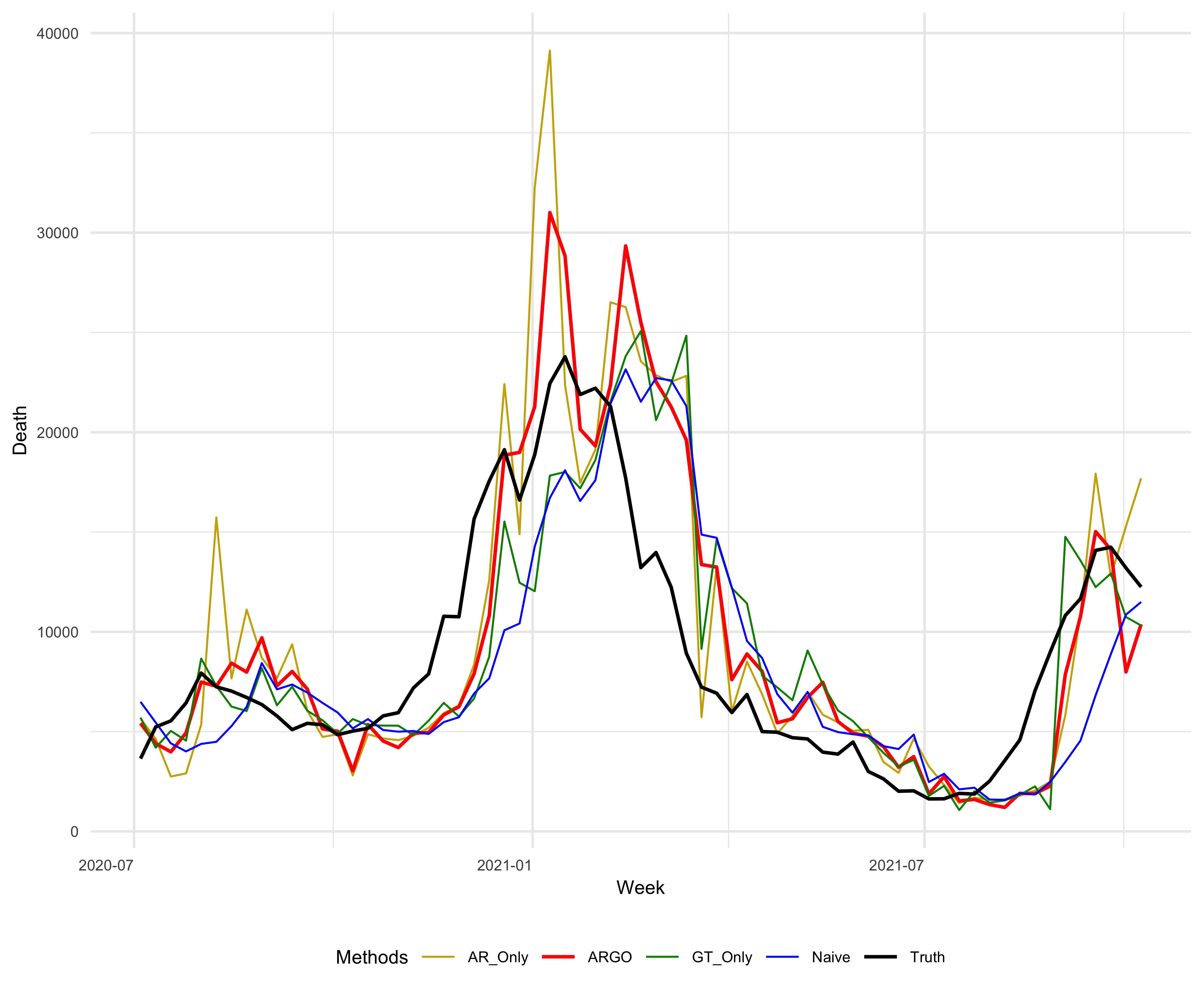}\label{fig:Nat_Compare_Our_4Weeks}}
\caption{1 to 4 weeks ahead national level COVID-19 weekly incremental death predictions' comparisons weekly from 2020-07-04 to 2021-10-09. The method included are AR-Only, ARGO, GT-Only, Naive (persistence), truth.
Estimation results for COVID-19 1 (top left), 2 (top right), 3 (bottom left), and 4 (bottom right) weeks ahead weekly incremental death. ARGO estimations (thick red), contrasting with the true COVID-19 death from JHU dataset (thick black) as well as the estimates from AR-Only (gold), GT-Only (Green), and Naive (blue).}
\label{fig2}
\end{figure}

ARGO outperforms simplified models in terms of RMSE, MAE, and Pearson correlation throughout 1 to 4 weeks ahead predictions. AR-Only predictions, on the other hand, have higher MSE comparing against Naive predictions for 2 and 4 weeks ahead, suggesting the importance of Google search terms in the method. COVID-19 death's trend is highly correlated with the optimal lagged important terms we selected, which boosts the model's accuracy. However, using only Google search information isn't good enough as well, as GT-only predictions are barely beating Naive predictions for 1 to 4 weeks ahead considering MAE and fall behind if considering RMSE, indicating autoregressive information (lagged COVID-19 cases and death) can help predictions using solely Google search terms by correcting its trend to not overshoot or under estimate, as shown in the period between December 2020 to March 2021 in Fig \ref{fig2}. Delaying behavior exists in all methods for 1 to 4 weeks ahead predictions, due to the lagged latest information to train for predictions, especially when forecast horizon extends to 3 and 4 weeks ahead. Yet, utilizing people's search behavior to foresee future trends, ARGO is able to overcome such delay effect in almost all the weeks for 1 week ahead predictions and majority of the weeks in 2020 for 2 to 4 weeks ahead predictions. Also, ARGO is the only method that captures the COVID-19 death peak in the week of January 16, 2021, for all 1 to 4 weeks ahead predictions. The integration of time series information and Google search terms leads to a trend-capturing estimation curve without undesired spikes in 1 to 3 weeks ahead forecasts, and robust recovering of spikes in 4 weeks ahead forecasts comparing to other benchmark methods.

The results presented above demonstrate ARGO's accuracy and robustness. The optimal lagged 23 important Google search terms appear to be key factors in the enhanced accuracy of ARGO, as well as the past week's COVID-19 deaths, as shown in Fig \ref{fig:Nat_Coef} and \ref{fig:Nat_Coef_3_4} in supplementary materials, which reflect a strong temporal autocorrelation after the feature selection of $L_1$ penalty.

\subsubsection{State Level}
For state level, we compare persistence (Naive) predictions, ARGO prediction, ARGOX-2step prediction, ARGOX-NatConstraint and Winner-takes-all Ensemble. ARGO predictions for each state are obtained from a $L_1$ penalized regression, utilizing the lagged state-level COVID-19 cases, death and optimal lagged Google search terms (section \ref{sec:ARGO_Nat}). ARGOX-2step combines cross-state cross-region information to obtain model outputs as predictions (section \ref{sec:ARGOX_Joint_Alone}). ARGOX-NatConstraint, revised upon ARGOX-2step, constrains the sum of state-level prediction to align with national-level prediction while pooling cross-resolution cross-temporal information (section \ref{sec:ARGOX_Nat_Constraint}). ARGO daily predictions use 56 days as training period, and are aggregated into 1 to 4 weeks ahead weekly incremental death predictions. ARGOX-2step, ARGOX-NatConstraint, and winner-takes-all ensemble method all use 30 consecutive and overlapping weeks as training period.

We conduct retrospective estimation of the 1 to 4 weeks ahead 51 U.S. state level COVID-19 incremental deaths, for the period of July 4th 2020 to Oct 9th 2021. To evaluate the accuracy of our estimation, we conduct sensitivity analysis among our methods by comparing their estimations with the actual COVID-19 weekly incremental death released by JHU dataset in multiple error metrics, including RMSE, MAE, and the Pearson correlation (section \ref{sec:Error_Metric}).

\begin{table}[ht]
\sisetup{detect-weight,mode=text}
\renewrobustcmd{\bfseries}{\fontseries{b}\selectfont}
\renewrobustcmd{\boldmath}{}
\newrobustcmd{\B}{\bfseries}
\addtolength{\tabcolsep}{-4.1pt}
\footnotesize
\centering
\begin{tabular}{lrrrr}
  \hline
& 1 Week Ahead & 2 Weeks Ahead & 3 Weeks Ahead & 4 Weeks Ahead \\ 
    \hline \multicolumn{1}{l}{RMSE} \\
 \hspace{1em} ARGO &  86.62 & 108.51 & 129.48 & 168.49 \\ 
   \hspace{1em} ARGOX 2Step & 86.26 & 122.76 & 157.56 & 239.85 \\ 
   \hspace{1em} ARGOX NatConstraint & 86.44 & 111.33 & 142.80 & 200.79 \\ 
   \hspace{1em} Ensemble & \B75.88 & \B88.37 & \B104.04 & \B119.26 \\ 
   \hspace{1em} Naive& 83.12 & 108.01 & 131.06 & 155.28 \\ 
   \multicolumn{1}{l}{MAE} \\
  \hspace{1em} ARGO &49.92 & 64.16 & 81.10 & 105.30 \\ 
   \hspace{1em} ARGOX 2Step & 49.59 & 73.50 & 98.36 & 148.06 \\ 
   \hspace{1em} ARGOX NatConstraint & 52.33 & 71.89 & 95.68 & 133.17 \\ 
   \hspace{1em} Ensemble & \B42.29 & \B50.65 & \B62.72 & \B80.10 \\ 
   \hspace{1em} Naive& 46.10 & 62.72 & 81.36 & 101.86 \\ 
   \multicolumn{1}{l}{Correlation} \\
 \hspace{1em} ARGO &0.77 & 0.71 & 0.67 & 0.58 \\ 
   \hspace{1em} ARGOX 2Step & 0.81 & 0.76 & 0.72 & 0.63 \\ 
   \hspace{1em} ARGOX NatConstraint & 0.70 & 0.60 & 0.50 & 0.39 \\ 
   \hspace{1em} Ensemble & \B0.82 & \B0.80 & \B0.77 & \B0.71 \\ 
   \hspace{1em} Naive& 0.79 & 0.71 & 0.63 & 0.53 \\ 
   \hline
\end{tabular}
\caption{Comparison of different methods for state-level COVID-19 1 to 4 weeks ahead incremental death in 51 U.S. states. The averaged RMSE, MAE, and correlation are reported and best performed method is highlighted in boldface.} 
\label{tab:State_Our}
\end{table}

\begin{figure}[htbp]
\centering
\includegraphics[width=0.8\textwidth]{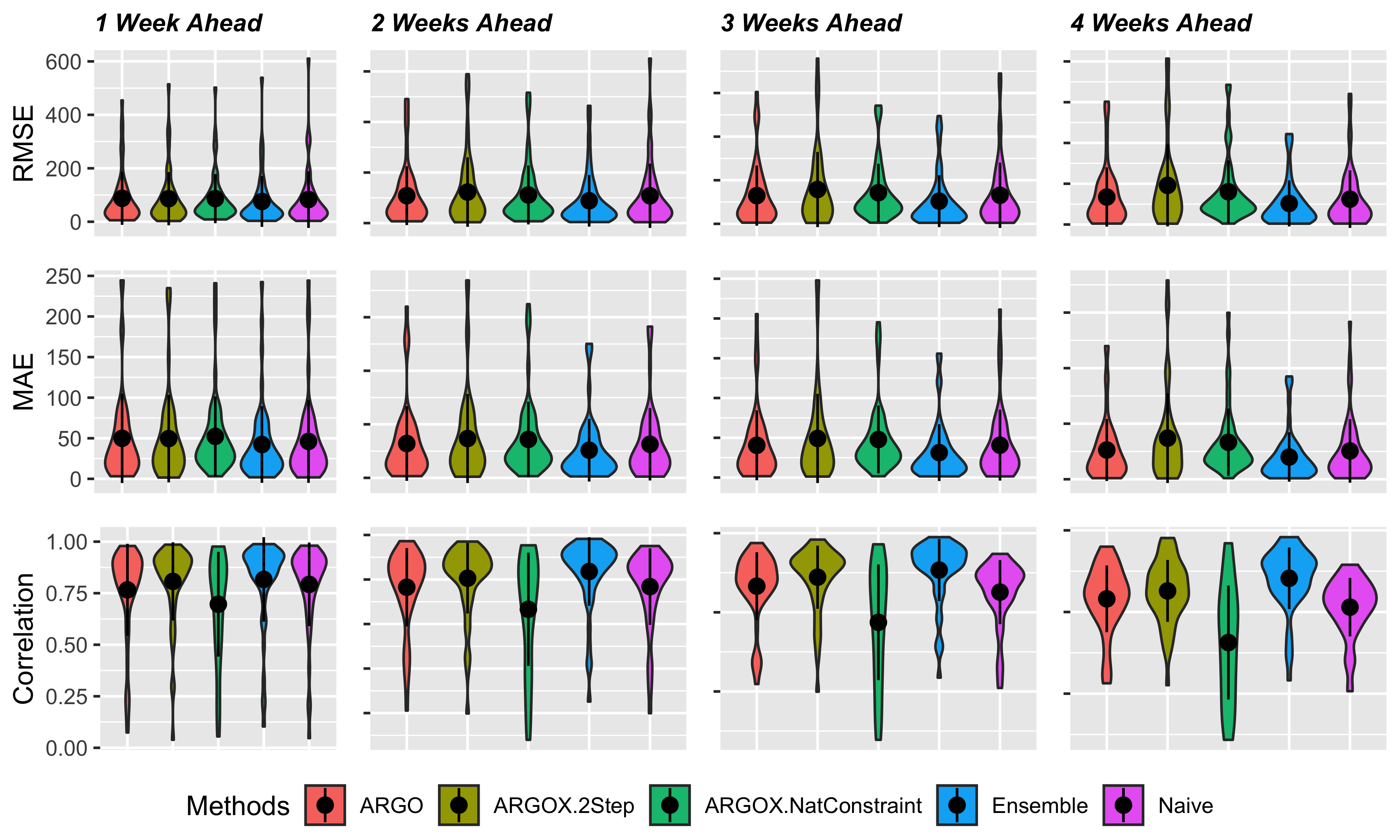}
\caption{The distribution of values for each metric for each model, over the 51 states for 1 to 4 weeks (from left to right) ahead predictions during the period from 2020-07-04 to 2021-10-09. The embedded black dot and vertical line indicate mean and 1 standard deviation range. Average of each error metric across 51 states are reported in table \ref{tab:State_Our}.}
\label{fig3}
\end{figure}

Table \ref{tab:State_Our} summarizes the overall results of ARGO prediction, ARGOX-2step prediction, ARGOX-NatConstraint, Winner-takes-all Ensemble and Naive, averaging over the 51 states for the whole period of July 4th 2020 to Oct 9th 2021. Our winner-takes-all ensemble method gives the leading performance uniformly in all metrics as shown in table \ref{tab:State_Our}, which achieves around 18\% error reduction in RMSE, around 20\% error reduction in MAE and around 8\% increase in Pearson correlation compared to the best alternative in the whole period on average across 1 to 4 weeks ahead predictions. Among all the methods we compare in this section, winner-takes-all ensemble is the only method that uniformly outperforms the naive predictions. The robustness and accuracy are further illustrated in Fig \ref{fig3}, which shows the 51 state's RMSE, MSE and Pearson correlation in the violin charts for 1 to 4 weeks ahead predictions. The winner-takes-all ensemble approach outperforms all other approaches in the three metrics, in terms of mean and standard deviation range over all 51 states. 

Detailed numerical results for each state are reported in tables \ref{tab:State_Ours_AK}–\ref{tab:State_Ours_WY} and fig \ref{fig:State_Ours_AK}-\ref{fig:State_Ours_WY} in supplementary material, where the ensemble approach demonstrates its accuracy in majority of the states. Notice that JHU ground-truth data exists jumps and spikes in some states due to retrospective edition. Such noisy ground-truth data is a key challenge to forecasting tasks. Our ensemble framework, on the other hand, reveals its robustness over geographical variability and extracts a strong combination from all other ARGO, ARGOX approaches.

To show detail break down of methods contributing to the ensemble approach, we display the ensemble approach's selection proportion among ARGO, ARGOX-2step and ARGOX-NatConstraint 1 to 4 weeks ahead predictions for all 51 U.S. states from 2020-07-04 to 2021-10-09, in table \ref{tab:Ensemble_Selection}. Throughout the period for all 51 states, the ensemble approach selects state-level ARGO predictions the most (around 40\%), and ARGOX-NatConstraint the least (around 25\%) throughout 1 to 4 weeks ahead estimations. This is evident as table \ref{tab:State_Our} indicates state-level ARGO performs competitively against naive estimations for all 1 to 4 weeks ahead predictions on average across all the states for almost all three error metrics. Yet, state-level ARGO cannot beat naive predictions uniformly across all the states,  as shown in supplementary materials tables \ref{tab:State_Ours_AK}–\ref{tab:State_Ours_WY} and fig \ref{fig:State_Ours_AK}-\ref{fig:State_Ours_WY}. On the other hand, though ARGOX-NatConstraint and ARGOX-2Step seem to perform poorly against naive predictions as shown in table \ref{tab:State_Our}, they contribute to the ensemble approach and boost its accuracy drastically. Detailed ensemble approach state-level selections for 1 to 4 weeks ahead estimations are displayed in fig \ref{fig:EnsembleSelect_1Week} and \ref{fig:EnsembleSelect_2Weeks} in supplementary materials. It seems that state-level ARGO, ARGOX-NatConstraint and ARGOX-2Step all have different strength and weakness depending on the particular state and time. Through adaptive selections of the best performer for each state, the winner-takes-all ensemble approach takes the best part of each of the three models, and thus achieves most robust prediction performance.

\begin{table}[ht]
\sisetup{detect-weight,mode=text}
\renewrobustcmd{\bfseries}{\fontseries{b}\selectfont}
\renewrobustcmd{\boldmath}{}
\newrobustcmd{\B}{\bfseries}
\addtolength{\tabcolsep}{-4.1pt}
\footnotesize
\centering
\centering
\begin{tabular}{lrrrr}
\hline
& 1 Week Ahead & 2 Weeks Ahead & 3 Weeks Ahead & 4 Weeks Ahead\\
\hline
ARGO & 0.3916 & 0.4011 & 0.4204 & 0.4398 \\
ARGOX-2Step & 0.3570 & 0.3494 & 0.3185 & 0.3071 \\
ARGOX-NatConstraint & 0.2513 & 0.2494 & 0.2611 & 0.2530\\
\hline
\end{tabular}
\caption{Ensemble approach total selection across all 51 U.S. states for 1 to 4 weeks ahead predictions.}
\label{tab:Ensemble_Selection}
\end{table}
In addition to the weekly estimates, ARGOX Ensemble also gives confidence intervals, by simply taking the confidence intervals of selected method. Table \ref{tab:CI_Coverage} (in supplementary materials) shows the coverage of the confidence intervals for all 51 states. The nominal 95\% confidence interval has an actual 88.2\% coverage on average for 1 week ahead predictions, suggesting that our confidence intervals reasonably measure
the accuracy of our weekly estimates, albeit with over-confidence.

\subsection{Comparison with Other Publicly Available Methods}

We obtain all available CDC published forecasts from teams making projections of COVID-19 cumulative and incident deaths for comparison \cite{UMASS,CDC_Ensemble,MOBS_GLEAM,LANL_GrowthRate,UA_EpiGro,epiforecast}. CDC official predictions are a compilation of predictions from all teams that submit their weekly predictions every Monday since January 15th 2020, contributed by different research groups or individuals. Note that all prediction models have their strengths in different date ranges and are providing robust and consistent predictions. They are the current state-of-art methods top performances among 30 number of methods submitted to CDC, considering comparison period from July 4th 2020 to Oct 9th 2021. We use both ARGOX Ensemble method and persistence (Naive) method for this comparison period (29 prediction submission records). We only consider top 6 CDC published teams for prediction comparison, after filtering out teams having missing values in their reporting over the period and states we are considering. Again, the groundtruth is COVID-19 weekly incremental death from JHU COVID-19 dataset, shown in section \ref{sec:COVID-Data}, and naive approach uses this week's death published by NYT as 1 to 4 weeks ahead predictions. We summarize the national level comparison results in table \ref{tab:Nat_Other} and the state level comparison results in table \ref{tab:State_Other}, where we compared RMSE, MAE and Correlation of 1 to 4 weeks ahead national and state level COVID-19 death predictions. We rank the teams according to the average of each error metric we used. We further show the distribution of state level comparison results across all three error metric in Fig \ref{fig4} violin charts (in supplementary material), where mean and standard deviations of each methods are displayed. Detailed state-by-state error metric are shown in fig \ref{fig:State_Compare_Other_RMSE_Heatmap_12} to \ref{fig:State_Compare_Other_PC_Heatmap_34}. 
 
\begin{table}[ht]
\sisetup{detect-weight,mode=text}
\renewrobustcmd{\bfseries}{\fontseries{b}\selectfont}
\renewrobustcmd{\boldmath}{}
\newrobustcmd{\B}{\bfseries}
\addtolength{\tabcolsep}{-4.1pt}
\footnotesize
\centering
\centering
\begin{tabular}{lrrrrr}
  \hline
  & 1 Week Ahead & 2 Weeks Ahead & 3 Weeks Ahead  & 4 Weeks Ahead & Average \\ 
    \hline \multicolumn{1}{l}{RMSE} \\
\hspace{1em}UMass-MechBayes\cite{UMASS} & 1320.17 & 1741.42 & 2127.58 & 2242.40 & 1857.89 \\ 
 \hspace{1em} COVIDhub-ensemble\cite{CDC_Ensemble} & 1389.41 & 1706.14 & 2015.08 & 2431.96 & 1885.65 \\ 
\hspace{1em}  MOBS-GLEAM\_COVID \cite{MOBS_GLEAM} & 1403.75 & 1800.84 & 2241.63 & 2684.33 & 2032.63 \\ 
\hspace{1em}  \underline{ARGOX\_Ensemble} & (\#5) 1749.21 & (4) 2224.44 & (4) 2889.18 & (7) 4123.99 & (\#4) 2746.71 \\ 
\hspace{1em}  LANL-GrowthRate\cite{LANL_GrowthRate} & 1614.51 & 2279.68 & 2893.12 & 4234.14 & 2755.36 \\ 
\hspace{1em}  UA-EpiCovDA \cite{UA_EpiGro} & 2051.75 & 2794.84 & 3346.70 & 3348.33 & 2885.40 \\ 
\hspace{1em}  epiforecasts-ensemble1 \cite{epiforecast}& 2236.67 & 2545.07 & 3169.34 & 4203.65 & 3038.68 \\ 
\hspace{1em}  Naive & 1977.97 & 2954.15 & 3814.30 & 4671.89 & 3354.58 \\ 

\hline \multicolumn{1}{l}{MAE} \\
\hspace{1em} UMass-MechBayes\cite{UMASS} & 925.67 & 1115.24 & 1408.36 & 1630.82 & 1270.02 \\ 
\hspace{1em}  COVIDhub-ensemble \cite{CDC_Ensemble} & 965.55 & 1168.32 & 1436.95 & 1779.86 & 1337.67 \\ 
\hspace{1em}  MOBS-GLEAM\_COVID \cite{MOBS_GLEAM}& 1064.64 & 1362.83 & 1752.92 & 2033.36 & 1553.44 \\ 
\hspace{1em} \underline{ARGOX\_Ensemble} & (\#4) 1183.57 & (\#4) 1582.22 & (\#4) 2103.16 & (\#6) 2906.57 & (\#4) 1943.88 \\ 
\hspace{1em}  UA-EpiCovDA\cite{UA_EpiGro}  & 1383.43 & 1770.43 & 2203.41 & 2601.15 & 1989.61 \\ 
 \hspace{1em} epiforecasts-ensemble1 \cite{epiforecast}& 1390.14 & 1656.43 & 2124.52 & 2832.15 & 2000.81 \\ 
\hspace{1em}  LANL-GrowthRate\cite{LANL_GrowthRate} & 1245.88 & 1808.40 & 2275.33 & 3096.98 & 2106.65 \\ 
\hspace{1em}  Naive & 1369.52 & 2110.81 & 2875.33 & 3575.81 & 2482.87 \\

\hline \multicolumn{1}{l}{Correlation} \\
\hspace{1em} UMass-MechBayes\cite{UMASS}  & 0.98 & 0.96 & 0.94 & 0.94 & 0.95 \\ 
 \hspace{1em} COVIDhub-ensemble \cite{CDC_Ensemble} & 0.97 & 0.96 & 0.94 & 0.92 & 0.95 \\ 
\hspace{1em}  MOBS-GLEAM\_COVID \cite{MOBS_GLEAM} & 0.97 & 0.96 & 0.94 & 0.91 & 0.94 \\ 
\hspace{1em}  LANL-GrowthRate\cite{LANL_GrowthRate}  & 0.97 & 0.94 & 0.92 & 0.86 & 0.92 \\ 
\hspace{1em}  \underline{ARGOX\_Ensemble} & (\#4) 0.96 & (\#4) 0.94 & (\#4) 0.90 & (\#5) 0.84 & (\#4) 0.91 \\ 
\hspace{1em}  UA-EpiCovDA\cite{UA_EpiGro} & 0.96 & 0.91 & 0.87 & 0.88 & 0.91 \\ 
\hspace{1em}  epiforecasts-ensemble1\cite{epiforecast} & 0.93 & 0.91 & 0.86 & 0.80 & 0.87 \\ 
 \hspace{1em} Naive & 0.94 & 0.87 & 0.79 & 0.68 & 0.82 \\ 
  \hline
\end{tabular}
\caption{Comparison among different models' 1 to 4 weeks ahead U.S. national level weekly incremental death predictions (from 2020-07-04 to 2021-10-10). The RMSE, MAE, Pearson correlation and their averages are reported. Methods are sorted based on their average. Our ARGOX-Ensemble's ranking for each error metric are included in parenthesis.}
\label{tab:Nat_Other}
\end{table}

\begin{table}[ht]
\sisetup{detect-weight,mode=text}
\renewrobustcmd{\bfseries}{\fontseries{b}\selectfont}
\renewrobustcmd{\boldmath}{}
\newrobustcmd{\B}{\bfseries}
\addtolength{\tabcolsep}{-4.1pt}
\footnotesize
\centering
\begin{tabular}{lrrrrr}
  \hline
  & 1 Week Ahead & 2 Weeks Ahead & 3 Weeks Ahead  & 4 Weeks Ahead & Average \\ 
    \hline \multicolumn{1}{l}{RMSE} \\
\hspace{1em} COVIDhub-ensemble\cite{CDC_Ensemble}  & 62.19 & 71.39 & 80.59 & 92.81 & 76.74 \\ 
\hspace{1em} UMass-MechBayes\cite{UMASS} & 63.02 & 74.29 & 87.13 & 107.05 & 82.87 \\ 
\hspace{1em} MOBS-GLEAM\_COVID \cite{MOBS_GLEAM}  & 70.13 & 81.04 & 95.61 & 112.56 & 89.83 \\ 
\hspace{1em} \underline{ARGOX\_Ensemble} & (\#4) 75.92 & (\#4) 87.09 & (\#4) 102.05 & (\#5) 123.91 & (\#4) 97.24 \\ 
\hspace{1em} UA-EpiCovDA\cite{UA_EpiGro}  & 80.13 & 100.85 & 115.18 & 121.61 & 104.44 \\ 
\hspace{1em} LANL-GrowthRate \cite{LANL_GrowthRate} & 77.83 & 97.79 & 114.21 & 131.67 & 105.37 \\ 
\hspace{1em}   Naive & 82.90 & 106.00 & 125.90 & 144.99 & 114.95 \\ 
\hspace{1em}   epiforecasts-ensemble1 \cite{epiforecast} & 111.41 & 157.79 & 167.75 & 197.94 & 158.72 \\

\hline \multicolumn{1}{l}{MAE} \\
\hspace{1em} COVIDhub-ensemble\cite{CDC_Ensemble}  & 36.04 & 42.14 & 49.24 & 57.20 & 46.16 \\ 
 \hspace{1em} UMass-MechBayes\cite{UMASS} & 36.09 & 43.74 & 51.61 & 62.15 & 48.40 \\ 
\hspace{1em}  MOBS-GLEAM\_COVID \cite{MOBS_GLEAM} & 43.01 & 50.70 & 60.32 & 70.73 & 56.19 \\ 
\hspace{1em}  \underline{ARGOX\_Ensemble} & (\#3) 42.11 & (\#4) 51.04 & (\#4) 62.11 & (\#4) 76.93 & (\#4) 58.05 \\ 
\hspace{1em}  LANL-GrowthRate  \cite{LANL_GrowthRate} & 46.81 & 58.62 & 70.43 & 83.33 & 64.80 \\ 
\hspace{1em}  UA-EpiCovDA \cite{UA_EpiGro} & 48.01 & 62.61 & 71.97 & 78.51 & 65.28 \\ 
\hspace{1em}  Naive & 45.77 & 61.58 & 77.65 & 93.40 & 69.60 \\ 
\hspace{1em}  epiforecasts-ensemble1 & 51.84 & 65.12 & 76.77 & 92.41 & 71.53 \\

\hline \multicolumn{1}{l}{Correlation} \\
\hspace{1em} COVIDhub-ensemble\cite{CDC_Ensemble}  & 0.86 & 0.83 & 0.81 & 0.77 & 0.82 \\ 
\hspace{1em}  UMass-MechBayes\cite{UMASS} & 0.86 & 0.82 & 0.79 & 0.76 & 0.81 \\ 
\hspace{1em}  MOBS-GLEAM\_COVID\cite{MOBS_GLEAM} & 0.84 & 0.80 & 0.76 & 0.72 & 0.78 \\ 
\hspace{1em}  \underline{ARGOX\_Ensemble} & (\#4) 0.82 & (\#4) 0.80 & (\#3) 0.77 & (\#4) 0.71 & (\#4) 0.77 \\ 
\hspace{1em}  LANL-GrowthRate\cite{LANL_GrowthRate} & 0.82 & 0.75 & 0.71 & 0.65 & 0.73 \\ 
\hspace{1em}  UA-EpiCovDA\cite{UA_EpiGro} & 0.78 & 0.70 & 0.64 & 0.63 & 0.69 \\ 
 \hspace{1em} epiforecasts-ensemble1 & 0.78 & 0.71 & 0.65 & 0.58 & 0.68 \\ 
 \hspace{1em} Naive & 0.79 & 0.71 & 0.64 & 0.54 & 0.67 \\ 
  \hline
\end{tabular}
\caption{Comparison among different models' 1 to 4 weeks ahead U.S. states level weekly incremental death predictions (from 2020-07-04 to 2021-10-09). The RMSE, MAE, Pearson correlation and their averages are reported. Methods are sorted based on their average. Our ARGOX-Ensemble's ranking for each error metric are included in parenthesis.}
\label{tab:State_Other}
\end{table}

From table \ref{tab:Nat_Other} and table \ref{tab:State_Other}, we can observe that our ARGOX-Ensemble model produces competitive accuracy for COVID-19 1 to 4 weeks ahead predictions, and is among the top 8 models in term of all three error metrics. Fig \ref{fig4} (in supplementary materials) further displays ARGOX-Ensemble's accuracy and robustness in terms of the metrics' mean and standard deviation range over all 51 states. Overall, all the error metric comparisons demonstrate ARGOX-Ensemble's competitiveness during the comparison time periods.

\section{Discussion}

While our ARGOX-Ensemble approach shows strong results, its accuracy and robustness depends on the reliability of its inputs. One limitation of our method is that the Google search query volumes are sensitive to media coverage, and such instability could propagate into our COVID-19 death predictions. Fortunately, media driven searches die down as pandemic progresses. In addition, our model also mitigate such instability via adaptive training. 

We use the summer period to identify the optimal lag and the 23 highly correlated queries. Such idea of optimal lag captures the intuition that people tend to search before clinic visits. It is interesting to observe different indications of epidemiological plausibility and infections from the optimal delays of the queries (table \ref{tab:23query_optimallag}).  COVID-19 incremental death trend has the longest delay from COVID-19 cases and tests related queries (more than 4 weeks), and follows by mild (3-4 weeks) and severe symptoms (2-3 weeks) related queries, indicating symptom to death is in moderate horizon and symptom's severity increases while the optimal lag decreases. On the other hand, there are still some Google search queries affected by media coverage and general public fear. Namely, vaccination related queries have the shortest delays (table \ref{tab:23query_optimallag}), which is a short term signal as well as general concern, intensified by news media coverage and spikes in cases or death trends, since the vaccination is not yet available in summer 2020. Nonetheless, our model is able to robustly capture the COVID-19 death trend, by determining the optimal lags during the summer period.

Information in Google search data deteriorates as forecast horizons expands, which could potentially impact the robustness and accuracy of our 4 weeks ahead predictions (Fig \ref{fig2} and Tab \ref{tab:Nat_Other}). Nevertheless, the L1 penalty and the dynamic training are able to capture the most relevant search terms and time series information for COVID-19 national level death estimation, and our model is still better than GT-only or AR-only models (Table \ref{tab:Nat_Our}) in all forecasting horizons. Meanwhile, ARGOX-Ensemble is able to robustly select accurate 1-4 weeks ahead state-level predictions from the three ARGOX alternative methods, despite that JHU dataset is only a noisy ground-truth. Models to further alleviate the bias in Internet search data and capture long-term COVID-19 trends could be an interesting future direction.

For national level COVID-19 death, the last week's COVID-19 death and all the Google search terms have significant effects on the 1 to 4 weeks ahead COVID-19 growth, shown in fig \ref{fig:Nat_Coef} in supplementary materials, which reflects a strong temporal auto-correlation and dependence on people search behaviors. Smoothing the penalized linear regression's coefficients with past three day's coefficients further leads to a smooth and continuous estimation curve and prevents undesired spikes, shown in Fig \ref{fig2}. ARGO also allows us to transparently understand how Google search information and historical COVID-19 information complement one another. For instance, past week COVID-19 death contribute positively to the current national COVID-19 death trend predictions, shown in fig \ref{fig:Nat_Coef}, which indicates that the current trend is likely to follow from the past week's growth/drop. Fig \ref{fig:Nat_Coef} also indicates the time-varying relationships between COVID-19 death trend and people's search behavior for COVID-19 related terms (general and symptom related searches). In the national COVID-19 1 to 4 weeks ahead predictions, time series models tend to have delay responses to sudden changes and are easily carried away by the changes, as shown in Fig \ref{fig2} from December 2020 to March 2021. Google search information, on the other hand, is better at reacting to sudden changes, but is also sensitive to public’s overreaction embedded in the search frequencies. Fortunately, the adaptive training can help ARGO achieves fast self-correction in the subsequent week.

For state level, besides producing ARGO state level predictions using the same framework as national level ARGO, we effectively combine state, regional, and national level publicly available data from Google searches and delayed COVID-19 cases and death to produce ARGOX-2Step state level estimations. ARGOX-NatConstraint improve upon ARGOX-2Step by restricting the sum of state level death predictions to be similar to national level death predictions, as ARGO national level predictions have already shown its strength. Both ARGOX-2Step and ARGOX-NatConstraint incorporate geographical and temporal correlation of COVID-19 death to provide accurate, reliable 1 to 4 weeks ahead estimations. To further improve accuracy and robustness, we combine all three methods and produce winner-takes-all ensemble forecast for 1 to 4 weeks ahead state level deaths. ARGO and ARGOX-2Step are unified frameworks adapted directly from influenza prediction with minimal changes, which demonstrates their robustness and general applicability, while reducing the possibility of over-fitting. Furthermore, the winner-takes-all ensemble approach efficiently combines all three frameworks and is able to outperform the constituent models for all states in all 1 to 4 weeks ahead predictions. Our national model and state-level performances are competitive to other state-of-arts models from CDC. Thus, we have shown that adapting ARGOX framework to COVID-19 can achieve accurate and robust results, and our model could serve as a valuable input for the CDC's current ensemble forecast.

\subsection*{Concluding Remarks}
In this paper, we demonstrate that methods for influenza prediction method using online search data \cite{ARGO, ARGO2_Regional, ARGOX} can be re-purposed for COVID-19 prediction. Specifically, by incorporating Google search information and autoregressive information, we could achieve strong performance on national level deaths predictions, while aggregating Google search information and cross-state-regional-national data could achieve competitive performance on state level death predictions, for 1 to 4 weeks ahead COVID-19 death forecasts, compared with other existing COVID-19 methods submitted to CDC. The combination of COVID-19 cases and deaths with optimally delayed Google search information, as well as the utilization of geographical structure, appear to be key factors in the enhanced accuracy of ARGO in national and state level predictions, demonstrating great additional insights which could assist and complement current CDC forecasts.

\clearpage


\newpage
\setcounter{page}{1}
\rfoot{\thepage}
\section*{Supporting Information for
Use Internet Search Data to Accurately Track State Level Influenza Epidemics}
\subsection*{Simin Ma, Shihao Yang}
\noindent Correspondence to:  shihao.yang@isye.gatech.edu
\noindent This PDF file includes:
\begin{itemize}
\item Supplementary Text
\item Supplementary Figs. \ref{fig:Nat_Constraint_1WeekProof} to \ref{fig:State_Ours_WY}
\item Supplementary Tables \ref{tab:23query} to \ref{tab:State_Ours_WY}
\end{itemize}

\setcounter{table}{0}
\renewcommand{\thetable}{S\arabic{table}}%
\setcounter{figure}{0}
\renewcommand{\thefigure}{S\arabic{figure}}%

This Supplementary Material is organized as following:

\subsection*{Implementation detail for ARGOX 2-Step}
In the first step, we use LASSO to aggregate the search volume information in the corresponding area. In the second step, we take a dichotomous approach for the 51 US states/districts, setting apart seven states: AK, HI, DE, KY, VT and ME. We first set apart AK and HI, since they are  geographically separated from the contiguous US. Then, we determine the rest by computing multiple correlation in COVID-19 incremental death count of each state to the COVID-19 incremental death counts of entire nation, the COVID-19 incremental death counts of the other regions (excluding the region that the state belongs) and the COVID-19 incremental death counts other states. DE, KY, VT and ME are the 4 states that have the lowest multiple correlations. A relatively low multiple correlation of a state implies that the state's COVID-19 death growth trend is not well aligned with other states', other regions' or the whole nation, indicating that information cross the other states or other regions might not help the stand-alone 7 states' death prediction. Therefore, we incorporate the dichotomous approach from ARGOX \cite{ARGOX} on the 45 ``joint'' states, and 6 ``alone'' states.

\subsubsection*{First Step}
For the first step, using the same notation as in section \ref{sec:ARGO_Nat}, we extract region/state level internet search information in region/state $m$ for day $T+l$ by estimating $\hat{y}_{T+l,m}$ using Google search terms with equation (\ref{eqn:argo_GTOnly_State}), for $l>0$. 

\begin{align}\label{eqn:argo_GTOnly_State}
    \hat{y}_{T+l,m} = \hat{\mu}_{y,m} + \sum^{K}_{k=1}\hat{\delta}_{k,m}X_{k,T+l-\hat{O}_k,m}+
    \sum^6_{r=1}\hat{\gamma}_{r,m}\mathbb{I}_{\{T+l, r\}}
\end{align}
where $X_{i,t,m}$ is the Google Trends data of search term $i$ day $t$ of region/state $m$ and $\mathbb{I}_{\{T+l,r\}}$ is a weekday $r$ indicator for the forecast date $T+l$. The coefficients $\{\mu_{y,m}, \bm{\delta}=(\delta_{1,m},\ldots,\delta_{K,m}), \bm{\gamma}=(\gamma_{1,m},\ldots, \gamma_{6,m}\}$ are obtained via
\begin{align}\label{eqn:argo_GTOnly_State_Obj}
\underset{\mu_{y,m},\bm{\delta}, \gamma_m,{\bm{\lambda}}}{\mathrm{argmin}} 
\sum_{t=T-M-l+1}^{T-l} &\left( y_{t+l,m}-\mu_{y,m}  - \sum^{27}_{k=1}{\delta}_{k,m}X_{k,t+l-\hat{O}_k,m}  +\sum^6_{r=1}\gamma_{r,m}\mathbb{I}_{\{T+l,r\}} \right)^2 +\lambda_{{\delta}} \|\bm{\delta}\|_1+\lambda_{{\gamma}}\|\bm{\gamma}\|_1
\end{align}
We set $M=56$ days for training, and $\bm{\lambda}=\{\lambda_{{\delta}}, \lambda_{{\gamma}}\}$ through cross-validation, where we let $\lambda_{{\delta}}= \lambda_{\gamma}$ for simplicity. Additionally, we use $K=23$ highly correlated Google search terms, and let $\hat{O}_k=\max\left(O_k,l\right)$ as the adjusted optimal lag of $k$th Google search term subject to $l^{\text{th}}$ day ahead prediction. Denote the regional estimates obtained as $(\hat{y}^{reg}_{T,1},\ldots,\hat{y}^{reg}_{T,10})$ and state estimates obtained as $(\hat{y}^{GT}_{T,1},\ldots,\hat{y}^{GT}_{T,51})$. 

Lastly, we obtain national COVID-19 death estimate $\hat{y}^{nat}_T$ using equation (\ref{eqn:argo_yhat}). Since all the estimates obtained above are daily COVID-19 incremental death, we aggregate them into 1 to 4 weeks total incremental death and work on 1 to 4 weeks ahead death forecast separately using the following steps. We denote  index $\tau$ for weekly indexing and $t$ for daily indexing.

\subsubsection*{Second Step}
For the 45 joint states, we gather the raw estimates for state/regional/national-level weekly COVID-19 incremental deaths from the first step to obtain the best linear predictor with ridge-regression inspired shrinkage for state-level COVID-19 increnental death estimate for week $\tau$ via ARGOX \cite{ARGOX}, using four predictors. Specifically, we use our raw estimates for the state-level weekly COVID-19 incremental deaths: $\hat{\bm{y}}^{GT}_{\tau} = (\hat{y}^{GT}_{\tau,1},\dots, \hat{y}^{GT}_{\tau,45})^\intercal$, expanded national and regional level COVID-19 death estimates:  $\hat{\bm{y}}_\tau^{nat}=(\hat{y}_\tau^{nat}, \dots,\hat{y}_\tau^{nat})^\intercal$ and $\hat{\bm{y}}_\tau^{reg}=(\hat{y}_{\tau, r_1}^{reg}, \dots,\hat{y}_{\tau, r_{45}}^{reg})^\intercal$, where $r_m$ is the region number for state $m$, and the previous week state-level groudtruth, with 30-weeks training window for parameter estimations.

For the 6 alone states, we take a stand-alone modeling approach \cite{ARGOX}, focusing on estimating the individual state's COVID-19 1 to 4 weeks ahead incremental death by integrating the within-state and national information in the second step. Specifically, we use 3 predictors, previous week state-level groundtruth, state level and national level COVID-19 estimates, $\hat{{y}}_{\tau, m}^{GT},  \hat{{y}}_{\tau}^{nat}$, for $m \in \{\text{AK, HI, DE, KY, VT and ME}\}$, where the regional terms are dropped. Similarly, we use the best linear predictor with ridge-regression inspired shrinkage to get the final estimates \cite{ARGOX}, with 30-weeks training period.

\clearpage
\subsection*{Detail Derivation for ARGOX-Nat-Constraint}
The detailed derivation for constrained optimization problem in equation (\ref{eqn:ARGOX_Nat_Constraint_Problem}) is shown as follows. For simplicity, we dropped week index $\tau$ and are solving the optimization problem in equation (\ref{eqn:ARGOX_Nat_Constraint_Problem}) to estimate the predictor for week $t$. We assume that $\bm{Z}_\tau,\bm{W}_\tau$ are demeaned in this case. Additionally, we rewrite the constraint in equation (\ref{eqn:ARGOX_Nat_Constraint_Problem}) as $\mathbf{1}^\intercal\bm{A}\bm{W}_\tau = \tilde{y}$ for simplicity, by denoting $\tilde{y} = \hat y^{*\text{ARGO}}_{\tau,nat}-\mathbf{1}^\intercal \bm{y}_{\tau-1}-\mathbf{1}^\intercal\mu_Z$. 

We first re-write the original optimization as a Lagrangian function after some simplification as :
\begin{equation} \label{eqn:ARGOX_Nat_Constraint_Lagrangian}
   f(A,\lambda)=\Tr(\Sigma_{ZZ}) + \Tr(\bm{A}\Sigma_{WW}\bm{A}^\intercal)- \Tr(2\Sigma_{ZW}A^\intercal)+ \lambda(\tilde{y}-\mathbf{1}^\intercal\bm{A}\bm{W})
\end{equation}
where $\Sigma_{ZZ}=\mathrm{Var}(\bm{Z})$, $\Sigma_{WW}=\mathrm{Var}(\bm{W})$ and $\Sigma_{ZW}=\mathrm{Cov}(\bm{Z}, \bm{W})$ and are constructed through equation ARGOX's setup \cite{ARGOX}. Then, we can re-write the original problem in equation (\ref{eqn:ARGOX_Nat_Constraint_Problem}) as:
\begin{equation} \label{eqn:ARGOX_Nat_Constraint_Soln}
\begin{split}
\min_{\bm{A},\lambda}\,&f(A) \\
\mathrm{s.t.}\;&\lambda(\tilde{y}-\mathbf{1}^\intercal\bm{A}\bm{W})=0
\end{split}
\end{equation}
By taking derivative with respect to $\bm{A}$ and $\lambda$, we have 
\begin{align*}
    \begin{cases}
    \nabla_{\bm{A}}f(A,\lambda)&=2\Sigma_{WW}A^\intercal-2\Sigma_{ZW}^\intercal-\lambda\bm{W}\mathbf{1}^\intercal\\
    \nabla_{\lambda}f(A,\lambda)&= \tilde{y}-\mathbf{1}^\intercal\bm{A}\bm{W}
    \end{cases}
\end{align*}
After setting them to $0$ for optimally condition and solve for $\bm{A}$ and $\lambda$, we have:
\begin{equation}\label{eqn:ARGOX_Nat_Constraint_Alambda}
 \begin{cases}
    \bm{A} &=\left(\Sigma_{ZW}+\frac{\lambda}{2}\mathbf{1}^\intercal\bm{W}^\intercal\right)\Sigma_{WW}^{-1}\\
    \lambda &=\frac{2}{n\bm{W}^T\Sigma_{WW}^{-1}\bm{W}}\left(\tilde{y}-\mathbf{1}^\intercal\Sigma{ZW}\Sigma_{WW}^{-1}\bm{W}\right)
     \end{cases}
\end{equation}
where $n=49$ is the length of vector $\mathbf{1}$. Let $\bm{\widetilde{W}}_\tau=\bm{W}_\tau-\mu_W$ be the demeaned predictor. Our estimate for the increment at week $t$ is
\begin{equation*}
   \bm{\hat{Z}}_{\tau}=\mu_Z + \hat{\bm{A}}\bm{\widetilde{W}}_\tau=\left(\Sigma_{ZW}+{\frac{1}{n\bm{\widetilde{W}}_{\tau}^T\Sigma_{WW}^{-1}\bm{\widetilde{W}}_{\tau}}\left(\tilde{y}-\mathbf{1}^\intercal\Sigma_{ZW}\Sigma_{WW}^{-1}\bm{\widetilde{W}}_{\tau}\right)}\mathbf{1}^\intercal\bm{\widetilde{W}}_{\tau}^\intercal\right)\Sigma_{WW}^{-1}\bm{\widetilde{W}}_\tau
\end{equation*}
Thus, our final prediction for state level COVID-19 week $t$ incremental death is 
\begin{equation}\label{eqn:ARGOX_Nat_Constraint_Blp}
    \bm{\hat{y}}_\tau = \bm{\hat{y}}_{\tau-1}+\mu_Z+\left(\Sigma_{ZW}+{\frac{1}{n\bm{\widetilde{W}}_{\tau}^T\Sigma_{WW}^{-1}\bm{\widetilde{W}}_{\tau}}\left(\tilde{y}-\mathbf{1}^\intercal\Sigma_{ZW}\Sigma_{WW}^{-1}\bm{\widetilde{W}}_{\tau}\right)}\mathbf{1}^\intercal\bm{\widetilde{W}}_{\tau}^\intercal\right)\Sigma_{WW}^{-1}\bm{\widetilde{W}}_\tau
\end{equation}

Moreover, we use the ridge-regression inspired shrinkage to modify the estimate, by replacing $\Sigma_{ZW}$ as $\frac{1}{2}\Sigma_{ZW}$ and $\Sigma_{WW}$ as $(\frac{1}{2}\Sigma_{WW}+\frac{1}{2}D_{WW})$ where $D_{WW}$ is the diagonal of the empirical covariance of $\bm{W}_\tau$:
\begin{align*}
   \bm{\hat{Z}}_{\tau}=\mu_Z+ \left(\frac{1}{2}\Sigma_{ZW}+{\frac{1}{n\bm{\widetilde{W}}_{\tau}^T\left(\frac{1}{2}\Sigma_{WW}+\frac{1}{2}D_{WW}\right)^{-1}\bm{\widetilde{W}}_{\tau}} \left(\tilde{y}-\frac{1}{2}\mathbf{1}^\intercal\Sigma_{ZW}\left(\frac{1}{2}\Sigma_{WW}+\frac{1}{2}D_{WW}\right)^{-1}\bm{\widetilde{W}}_{\tau}\right)}\mathbf{1}^\intercal\bm{\widetilde{W}}_{\tau}^\intercal\right)\\
   \left(\frac{1}{2}\Sigma_{WW}+\frac{1}{2}D_{WW}\right)^{-1}\bm{\widetilde{W}}_\tau
\end{align*}

Therefore, our final prediction for state-level COVID-19 week $t$ incremental death with ridge inspired shrinkage is:
\begin{align}\label{eqn:ARGOX_Nat_Constraint_Ridge_Blp}
    \bm{\hat{y}}_\tau = \bm{\hat{y}}_{\tau-1}+\mu_Z+ \left(\Sigma_{ZW}+{\frac{1}{n\bm{\widetilde{W}}_{\tau}^T\left(\Sigma_{WW}+D_{WW}\right)^{-1}\bm{\widetilde{W}}_{\tau}}\left(\tilde{y}-\mathbf{1}^\intercal\Sigma_{ZW}\left(\Sigma_{WW}+D_{WW}\right)^{-1}\bm{\widetilde{W}}_{\tau}\right)}\mathbf{1}^\intercal\bm{\widetilde{W}}_{\tau}^\intercal\right)\notag\\
    \left(\Sigma_{WW}+D_{WW}\right)^{-1}\bm{\widetilde{W}}_\tau
\end{align}

\clearpage
\subsection*{Selected important Google search queries}
Table \ref{tab:23query} lists the selected 23 important terms used in this study. They are selected through optimal lag selections.
\begin{table}[ht]
\footnotesize
\centering
\caption{selected 23 important terms}
\begin{tabular}{llll}
  \hline
coronavirus vaccine & cough  & covid 19 vaccine & coronavirus exposure \\
coronavirus cases & coronavirus test & covid 19 cases & covid 19 \\ 
exposed to coronavirus & fever &  headache & how long covid 19 \\
how long contagious & loss of smell & loss of taste & nausea \\ pneumonia & rapid covid 19 & rapid coronavirus & robitussin \\
sore throat & sinus &symptoms of the covid 19  &\\
\hline
\end{tabular}
\label{tab:23query}
\end{table}

\clearpage
\subsection*{Google search queries Optimal lagged Pearson correlations} 
Table \ref{tab:23query_PC} shows the optimal lag delayed Google search queries' Pearson correlation with COVID-19 daily incremental death. 

\begin{table}[ht]
\footnotesize
\centering
\caption{Optimal Lagged Google Query and COVID-19 Death Pearson Correlation from 2020-04-01 to 2020-06-30}\label{tab:23query_PC}
\begin{tabular}{llll}
  \hline
Google Query & Pearson Correlation & Google Query & Pearson Correlation \\ 
  \hline
loss of taste & 0.909 & covid 19 how long & 0.223 \\ 
  loss of smell & 0.877 & normal body & 0.223 \\ 
  how long contagious & 0.864 & body temperature & 0.222 \\ 
  covid 19 vaccine & 0.815 & cold vs coronavirus & 0.205 \\ 
  rapid covid 19 & 0.782 & coronavirus vs cold & 0.205 \\ 
  pneumonia & 0.761 & expectorant & 0.203 \\ 
  robitussin & 0.738 & acute bronchitis & 0.186 \\ 
  bronchitis & 0.724 & covid 19 hospital & 0.178 \\ 
  sinus & 0.711 & high fever & 0.157 \\ 
  cough & 0.699 & covid 19 relief & 0.153 \\ 
  covid 19 & 0.649 & human temperature & 0.153 \\ 
  fever & 0.649 & is coronavirus contagious & 0.147 \\ 
  symptoms of the covid 19 & 0.64 &normal body temperature & 0.14 \\ 
  how long covid 19 & 0.636 & signs of the coronavirus & 0.14 \\ 
  sore throat & 0.636 & contagious coronavirus & 0.129 \\ 
  coronavirus test & 0.628 & coronavirus contagious & 0.129 \\ 
  coronavirus cases & 0.622 & shortness of breath & 0.125 \\ 
   strep throat & 0.605 & coronavirus vitamin c & 0.089 \\ 
  coronavirus exposure & 0.573 & oseltamivir  & 0.082 \\ 
  coronavirus vaccine. & 0.549 & coronavirus test kit & 0.079 \\ 
  exposed to coronavirus & 0.523 & covid 19 treatment & 0.073 \\ 
  rapid coronavirus & 0.513 & cold and coronavirus & 0.061 \\ 
  upper respiratory & 0.51 & coronavirus and cold & 0.061 \\ 
  headache & 0.507 & how long does the coronavirus last & 0.054 \\ 
  the covid 19 & 0.506 & how long does coronavirus last & 0.052 \\ 
  covid 19 cases & 0.506 & symptoms of covid 19 & 0.049 \\ 
 nausea & 0.502 & coronavirus medication & 0.048 \\ 
  tessalon & 0.491  & coronavirus family & 0.039 \\ 
  symptoms of pneumonia & 0.482 & taking temperature & 0.037 \\ 
  oscillococcinum & 0.476 & do i have the coronavirus & 0.036 \\ 
  strep & 0.473 & respiratory coronavirus & 0.035 \\ 
  nasal congestion. & 0.468  & covid 19 what to do & 0.034 \\ 
  common cold & 0.446 & coronavirus hospital & 0.03 \\ 
  chest cold & 0.434 & i have the coronavirus & 0.018 \\ 
  walking pneumonia & 0.397 & coronavirus care & 0.013 \\ 
  coronavirus relief & 0.357  & covid 19 symptoms & 0.011 \\ 
  covid 19 test & 0.336 & ear thermometer & 0.01 \\ 
 cough fever & 0.331 & coronavirus how long & 0.01 \\ 
   fever cough & 0.331 & how long coronavirus & 0.01 \\ 
  covid 19 care & 0.24 & coronavirus recovery & 0.005 \\ 
   sinus infections & 0.23 & how to treat coronavirus & 0.004 \\ 
  reduce fever & 0.223 & coronavirus cough & 0.001 \\ 
   \hline
\end{tabular}
\end{table}

\clearpage
\subsection*{Google searches optimal lag}
Table \ref{tab:23query_optimallag} shows the selected 23 Google search queries' optimal lag (delay), selected through fitting regression of lagged terms against COVID-19 death trend and select the lag with minimal mean-squared error.
\begin{table}[ht]
\footnotesize
\centering
\caption{selected 23 important terms' optimal lag ranked by lags}\label{tab:23query_optimallag}
\begin{tabular}{llll}
Google Search Term & Optimal Lag \\
  \hline
coronavirus vaccine & 	5 \\
covid 19 vaccine&          			7  \\
coronavirus.exposure  &       		13 \\  
robitussin    &          15\\
sinus        &     			15\\
covid 19	&					21\\
how long covid 19 		&		21\\
symptoms of the covid 19   & 	21 \\
nausea				&		23\\
cough             	&			24\\
fever 				&		24\\
exposed to coronavirus       &		24\\
loss of taste 			&			24\\
loss of smell  			&		25\\
how long contagious       &    		25\\
rapid coronavirus       &	27\\	     
pneumonia 			&	28\\
rapid covid 19       	&		28\\
headache          		&		29 \\
sore throat        	&			30\\
coronavirus cases 		&		30	\\
coronavirus test           	&		30\\
covid 19 cases        	&			30\\
\hline
\end{tabular}
\end{table}

\clearpage
\subsection*{ARGOX-Nat-Constrained motivation}
\begin{figure}[htbp]
\centering
\subfloat[1 Week Ahead ARGO National Prediction and Summed ARGOX-2Step Predictions]
{\includegraphics[width=0.55\textwidth]{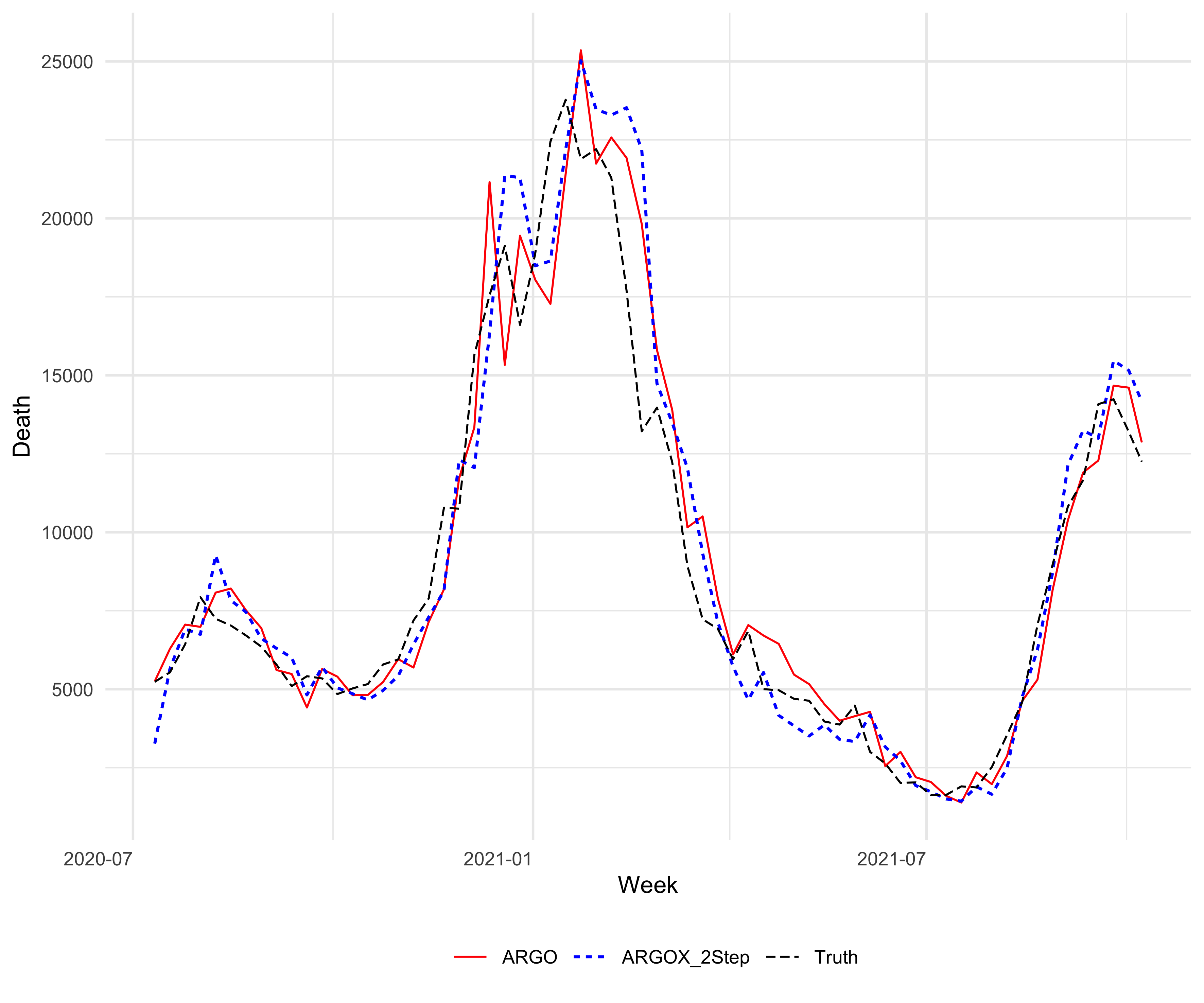}\label{fig:Nat_Constraint_1WeekProof}}
\hfill
\subfloat[2 Weeks Ahead ARGO National Prediction and Summed ARGOX-2Step Predictions]
{\includegraphics[width=0.55\textwidth]{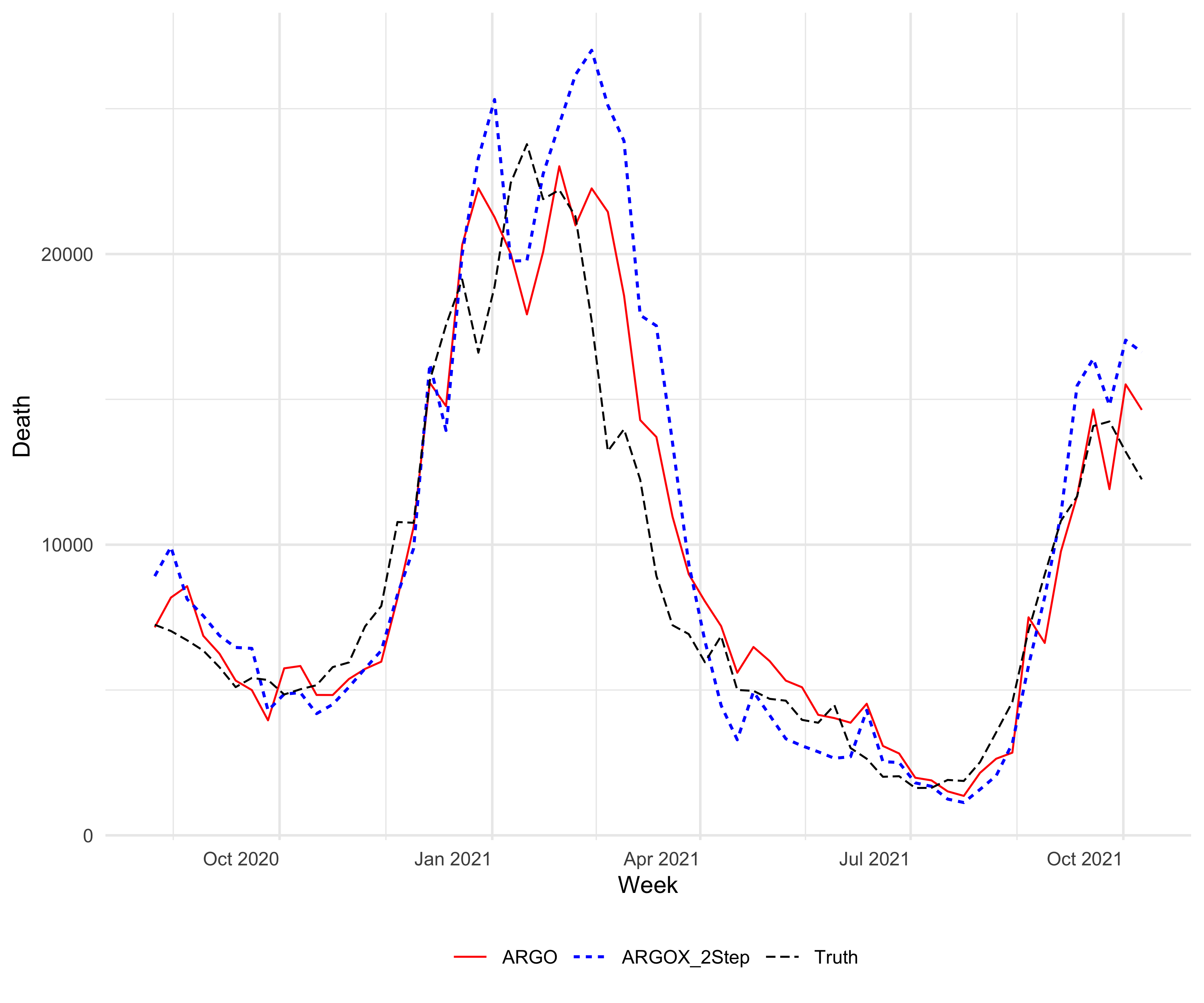}\label{fig:Nat_Constraint_2WeeksProof}}
\caption{The sum of ARGOX-2Step predictions is compared against ARGO national level prediction and JHU COVID-19 true deaths for 1 week ahead (up) and 2 weeks ahead (down) estimations. Illustrates the shortcoming of ARGOX-2Step prediction which doesn't sum up to reasonable national level predictions for both 1 and 2 weeks head prediction.}
\label{fig:Nat_ConstraintProof}
\end{figure}

\begin{figure}[htbp]
\centering
\subfloat[3 Weeks Ahead ARGO National Prediction and Summed ARGOX-2Step Predictions]
{\includegraphics[width=0.55\textwidth]{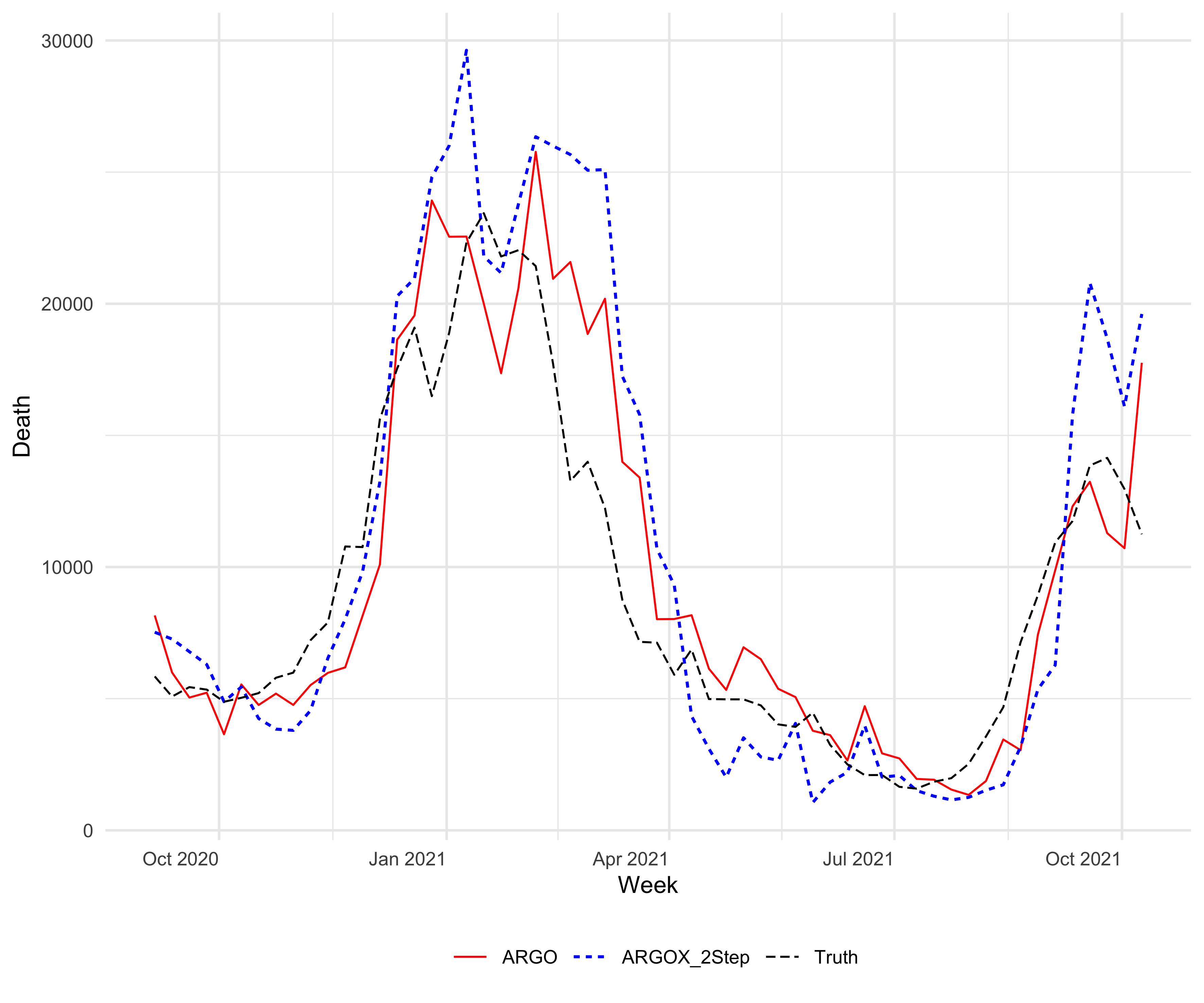}\label{fig:Nat_Constraint_3WeeksProof}}
\hfill
\subfloat[4 Weeks Ahead ARGO National Prediction and Summed ARGOX-2Step Predictions]
{\includegraphics[width=0.55\textwidth]{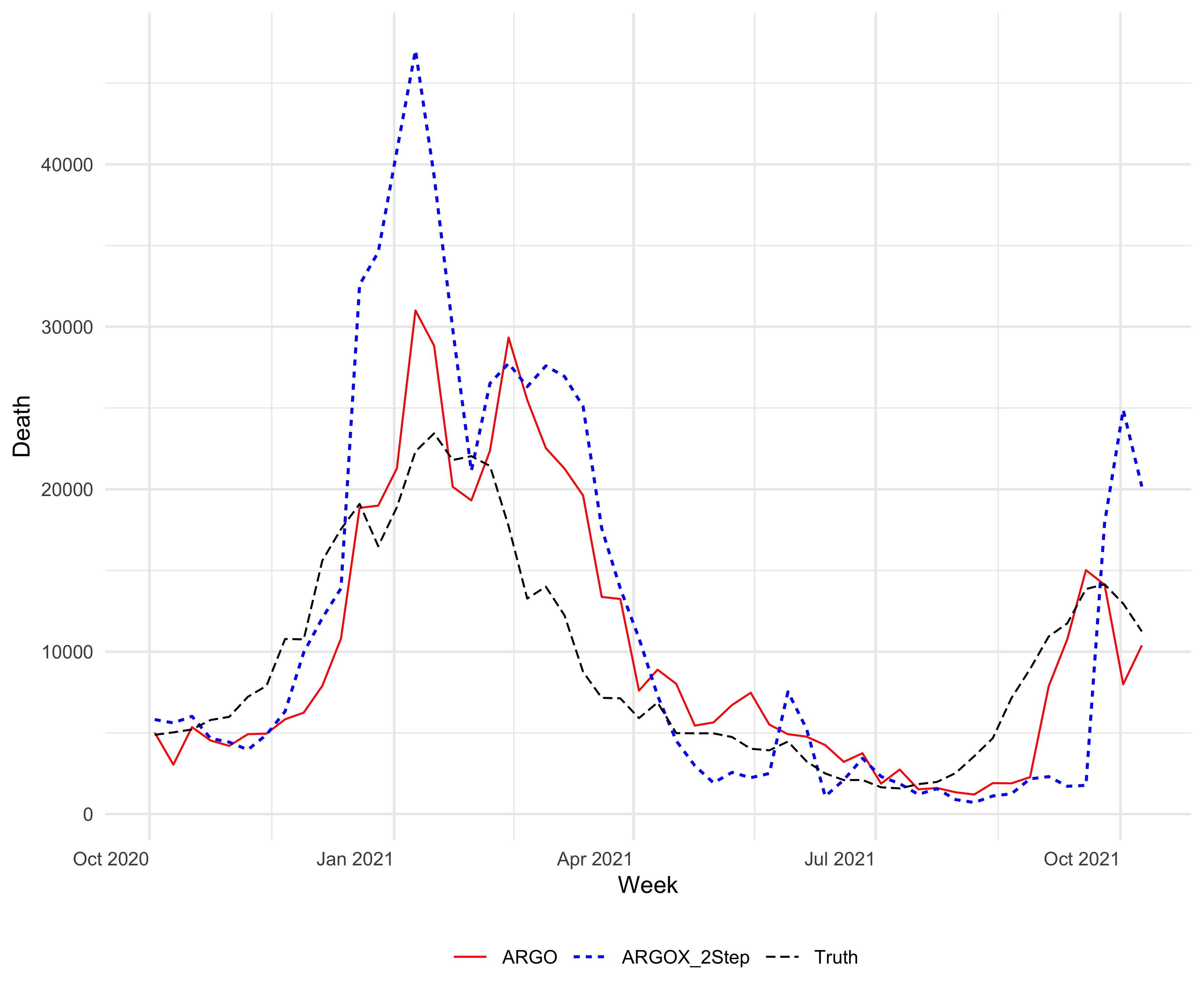}\label{fig:Nat_Constraint_4WeeksProof}}
\caption{The sum of ARGOX-2Step predictions is compared against ARGO national level prediction and JHU COVID-19 true deaths for 3 weeks ahead (up) and 4 weeks ahead (down) estimations. Illustrates the shortcoming of ARGOX-2Step prediction which doesn't sum up to reasonable national level predictions for both 3 and 4 weeks head prediction.}
\label{fig:Nat_ConstraintProof_3_4}
\end{figure}

Fig\ref{fig:Nat_Constraint_1WeekProof}, \ref{fig:Nat_Constraint_2WeeksProof}, \ref{fig:Nat_Constraint_3WeeksProof}, and \ref{fig:Nat_Constraint_4WeeksProof} show the inconsistency of aggregated state level ARGOX predictions comparing against ARGO national level predictions for future 1-4 weeks ahead predictions.

\clearpage

\clearpage
\subsection*{ARGO model parameter heatmaps}
\begin{figure}[htbp]
\centering
\subfloat[1 Week Ahead National Level ARGO Coefficients]
{\includegraphics[width=0.6\textwidth]{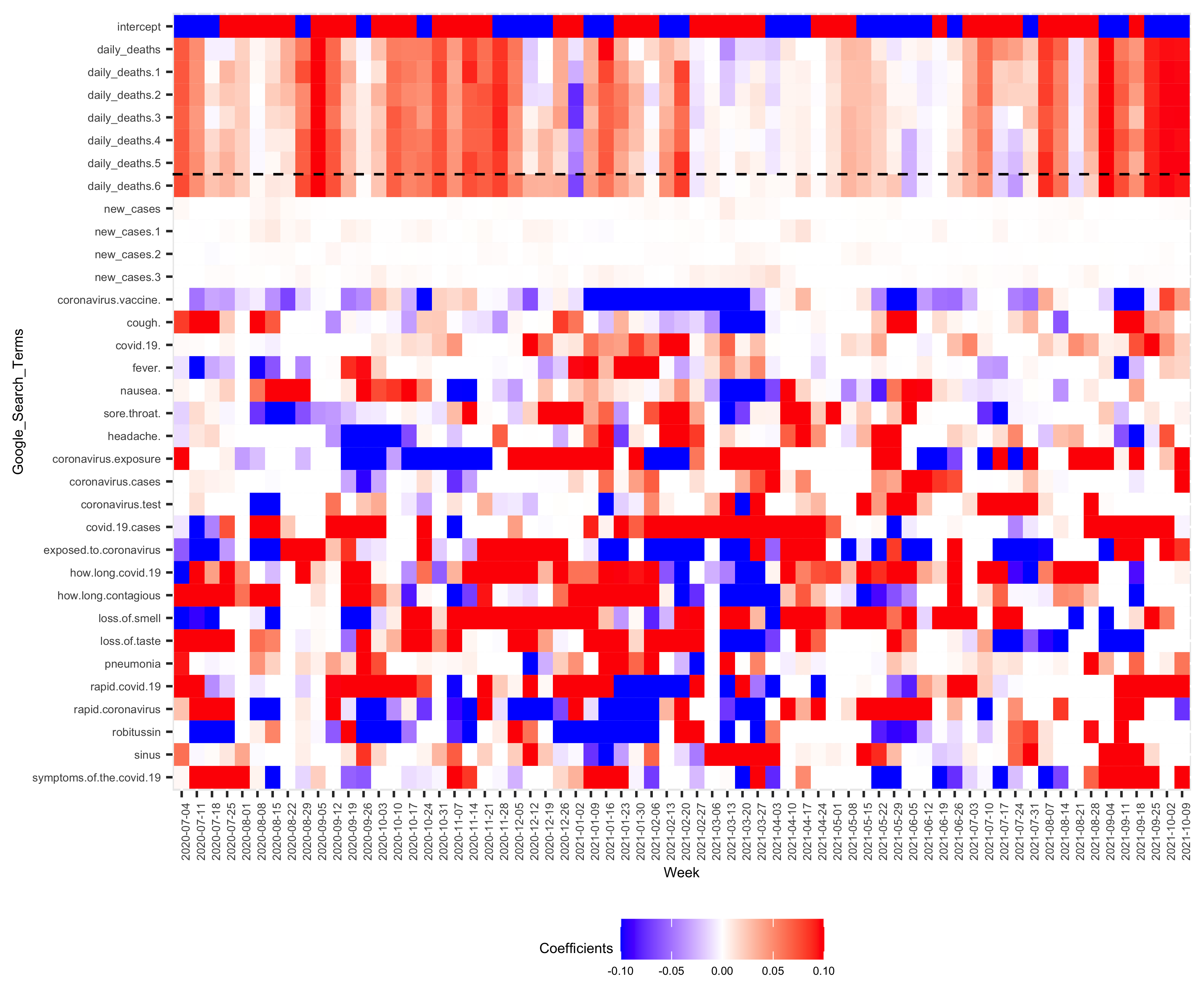}\label{fig:Nat_Coef_1Week}}
\hfill
\subfloat[2 Weeks Ahead National Level ARGO Coefficients]
{\includegraphics[width=0.6\textwidth]{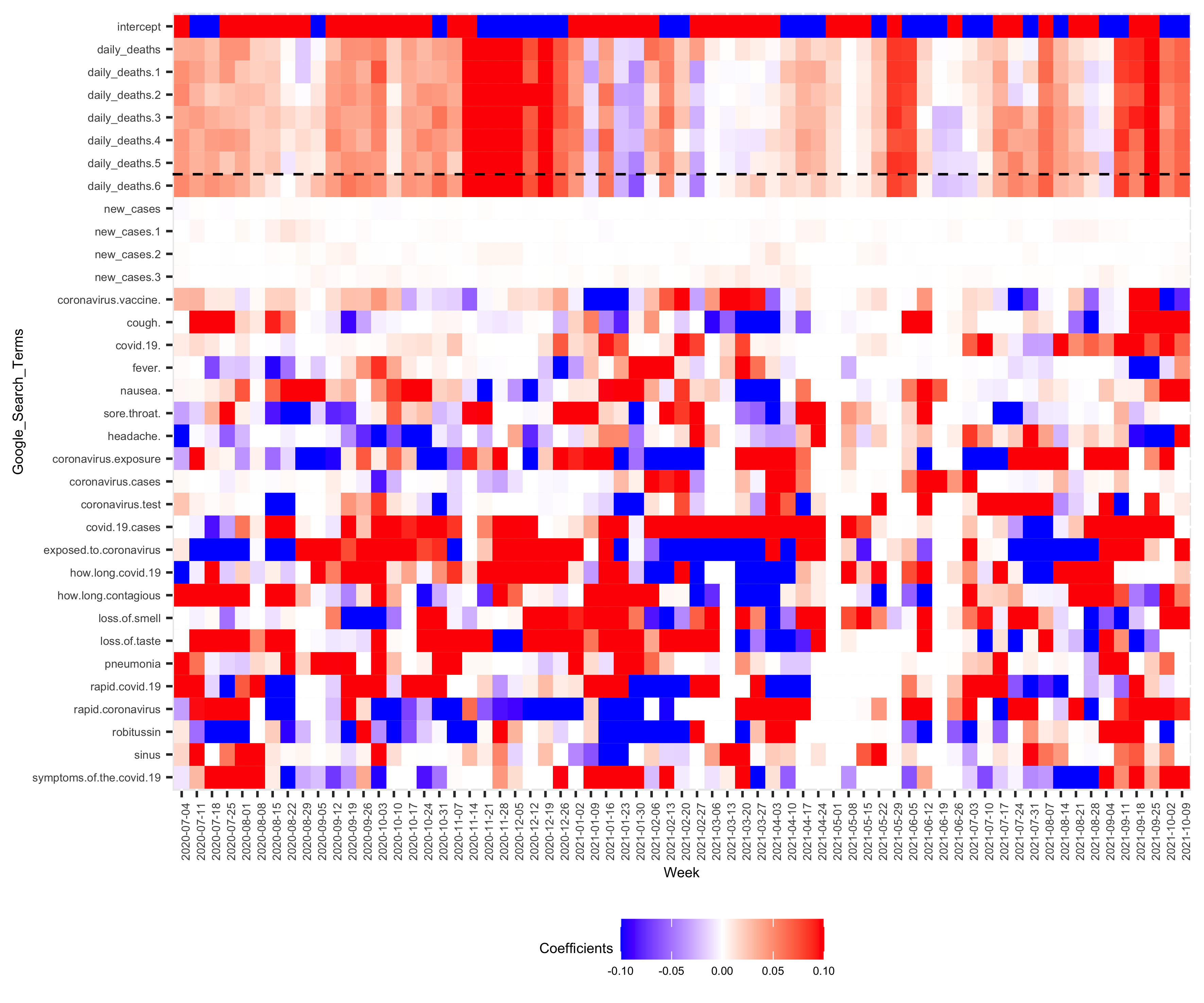}\label{fig:Nat_Coef_2Weeks}}
\caption{Smoothed coefficients for ARGO national level 1 week and 2 weeks ahead predictions. Coefficients larger than 0.1 are scaled to 0.1 and lower than -0.1 are scaled to -0.1, for simplicity. Red color represents positive coefficients, blue color represents negative coefficients and white color represents zero. Black horizontal dashed line separates Google query queries from autoregressive (cases and death) lags.}
\label{fig:Nat_Coef}
\end{figure}

\begin{figure}[htbp]
\centering
\subfloat[3 Weeks Ahead National Level ARGO Coefficients]
{\includegraphics[width=0.6\textwidth]{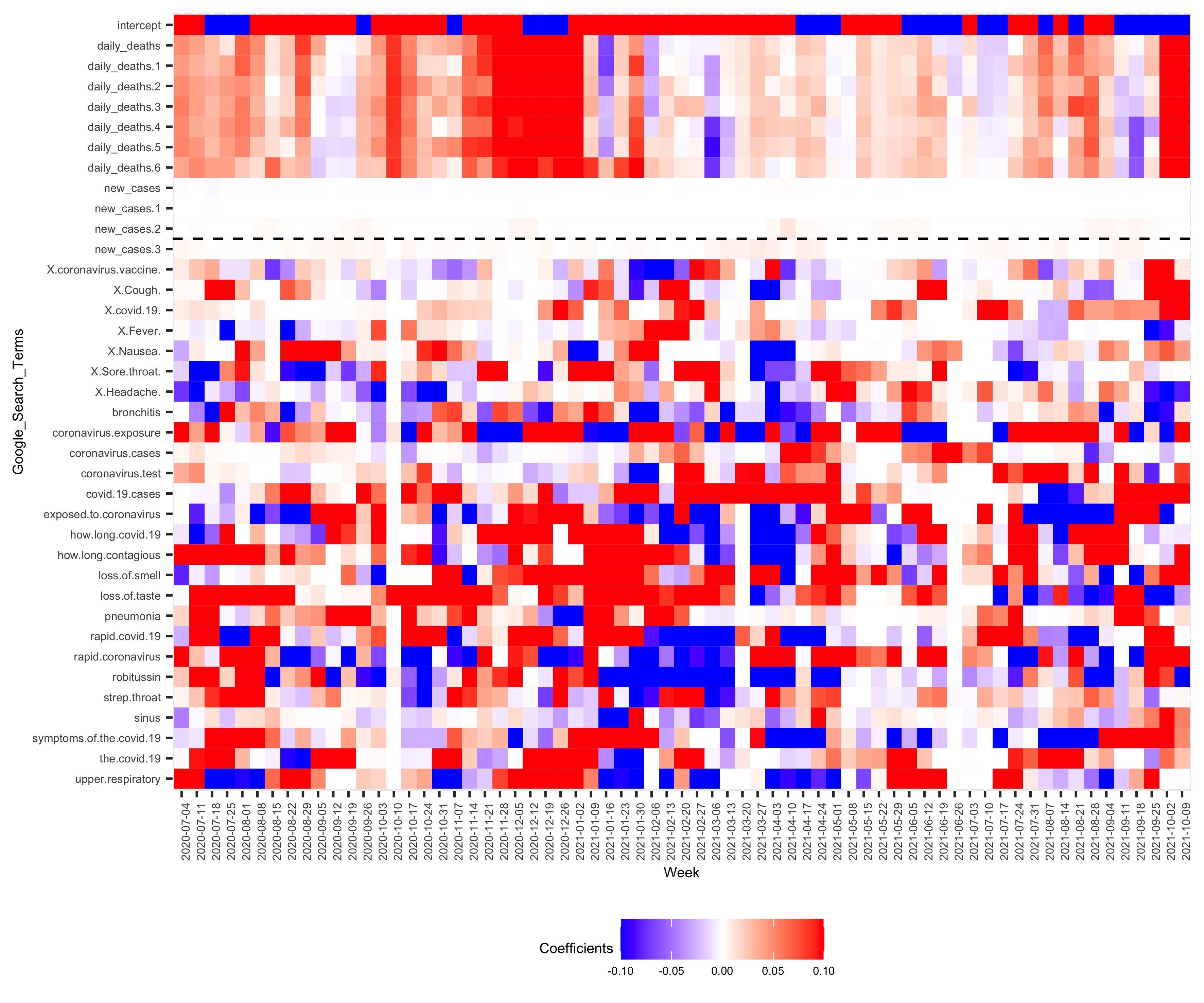}\label{fig:Nat_Coef_3Weeks}}
\hfill
\subfloat[4 Weeks Ahead National Level ARGO Coefficients]
{\includegraphics[width=0.6\textwidth]{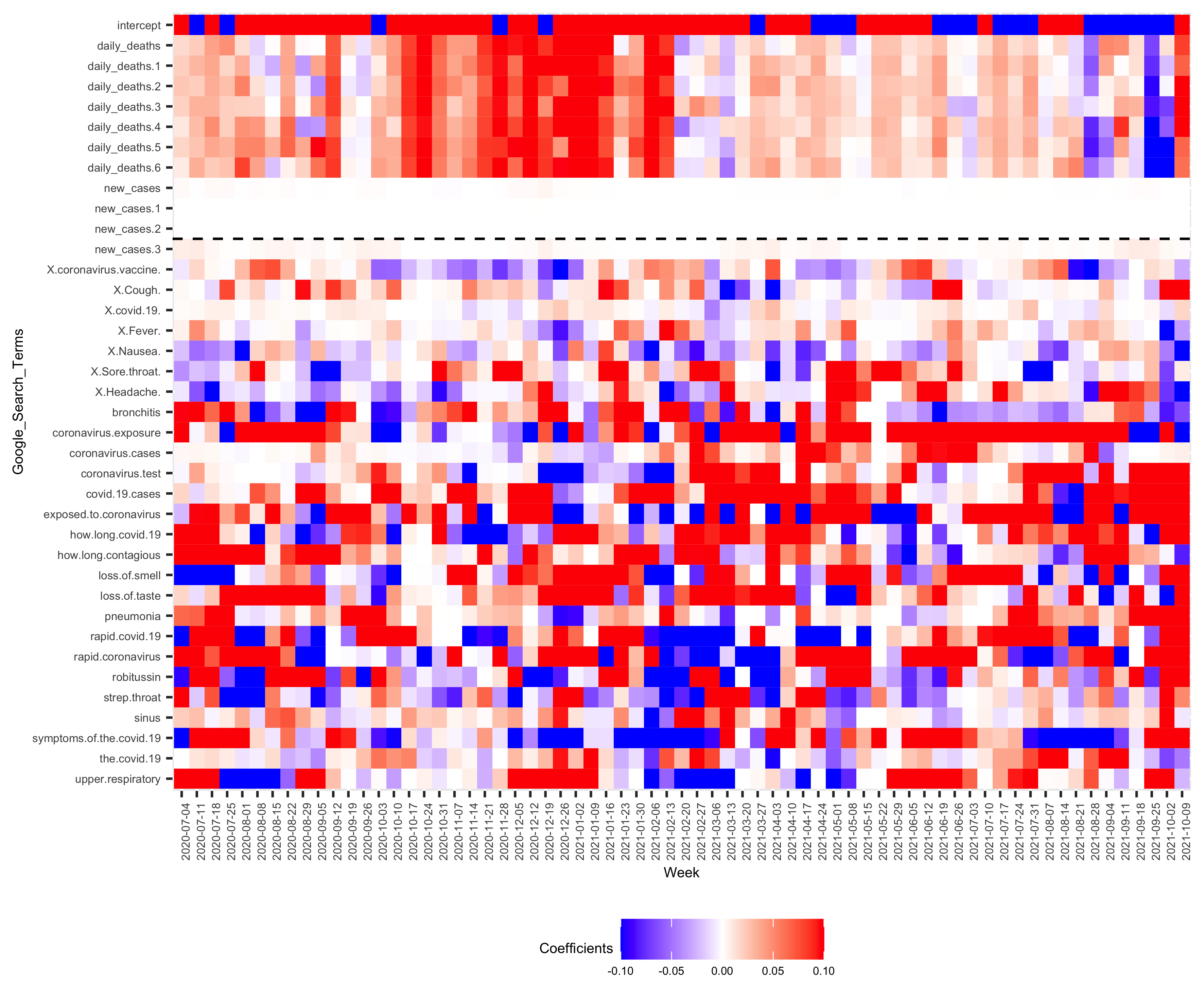}\label{fig:Nat_Coef_4Weeks}}
\caption{Smoothed coefficients for ARGO national level 3 weeks and 4 weeks ahead predictions. Coefficients larger than 0.1 are scaled to 0.1 and lower than -0.1 are scaled to -0.1, for simplicity. Red color represents positive coefficients, blue color represents negative coefficients and white color represents zero. Black horizontal dashed line separates Google query queries from autoregressive (cases and death) lags.}
\label{fig:Nat_Coef_3_4}
\end{figure}

Fig \ref{fig:Nat_Coef_1Week}, \ref{fig:Nat_Coef_2Weeks}, \ref{fig:Nat_Coef_3Weeks} and \ref{fig:Nat_Coef_4Weeks} show ARGO national level 1 to 4 weeks ahead forecasts' model parameter in heatmaps.

\clearpage
\subsection*{ARGO model parameter heatmaps (Continue)}

\clearpage
\subsection*{Winner-takes-all ensemble method selections}
\begin{figure}[htbp]
\centering
\subfloat[1 Week Ahead State Level Ensemble Selections]
{\includegraphics[width=0.7\textwidth]{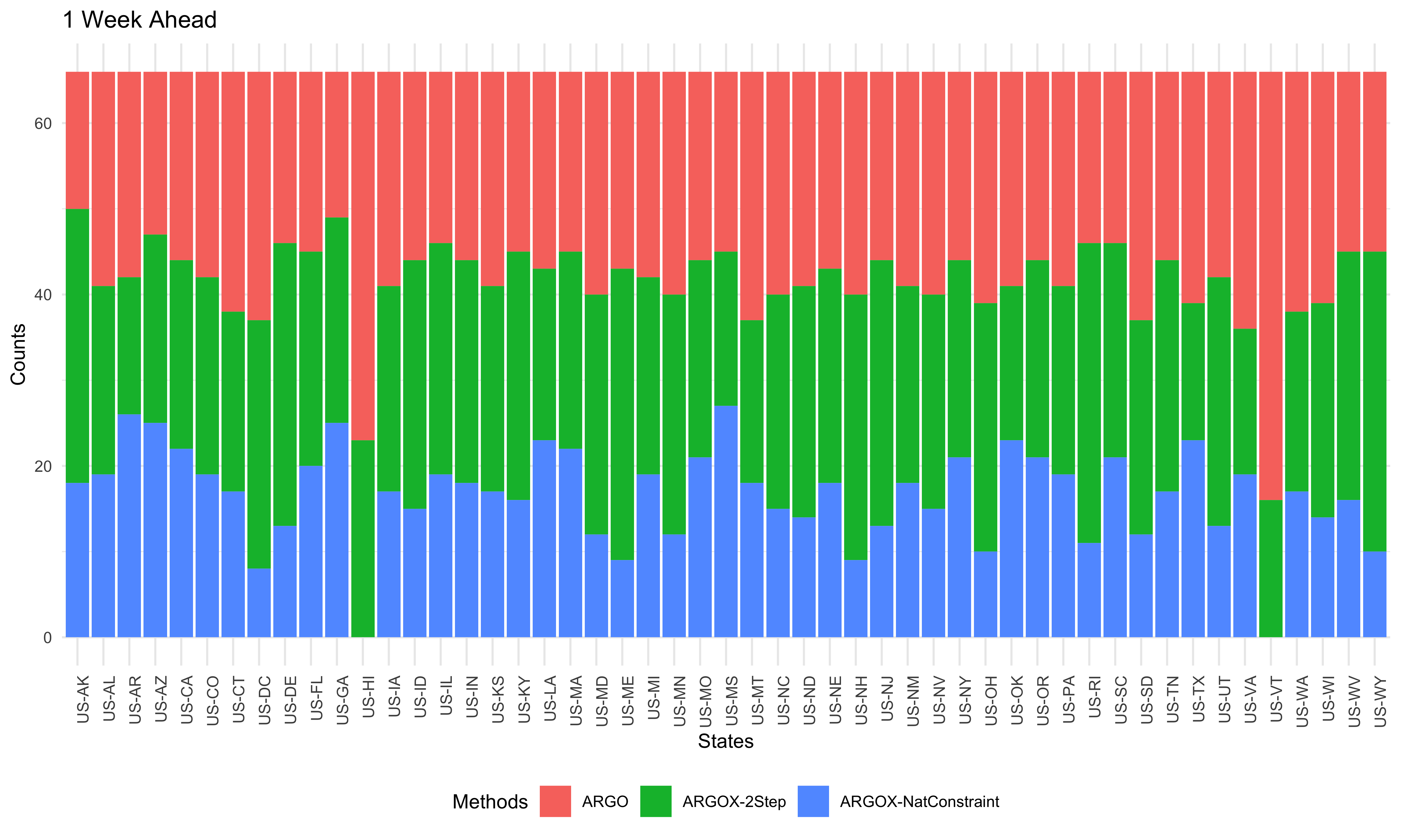}\label{fig:EnsembleSelect_1Week}}
\hfill
\subfloat[2 Weeks Ahead State Level Ensemble Selections]
{\includegraphics[width=0.7\textwidth]{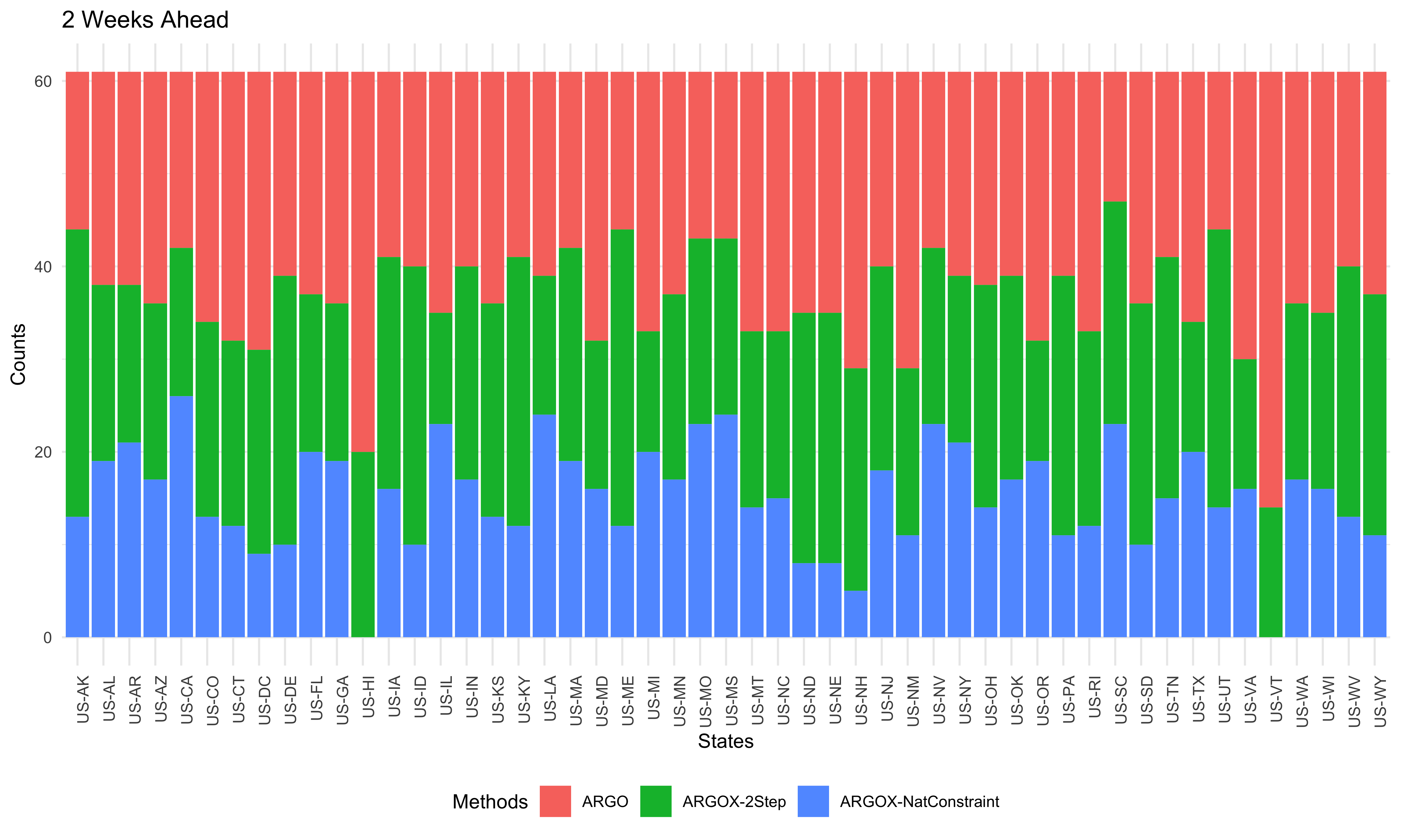}\label{fig:EnsembleSelect_2Weeks}}
\caption{Ensemble method's selection among ARGO, ARGOX-2Step and ARGOX-NatConstraint for all 51 U.S. States for 1 week and 2 weeks ahead predictions, from 2020-07-04 to 2021-10-09. }
\label{fig:EnsembleSelect_ALLStates}
\end{figure}

Fig \ref{fig:EnsembleSelect_1Week} and \ref{fig:EnsembleSelect_2Weeks} show the 1 week and 2 weeks ahead winner-takes-all state level forecasts selection among the three other methods: ARGO, ARGOX-2Step and ARGOX-NatCOnstraint.

\begin{figure}[htbp]
\centering
\subfloat[3 Weeks Ahead State Level Ensemble Selections]
{\includegraphics[width=0.7\textwidth]{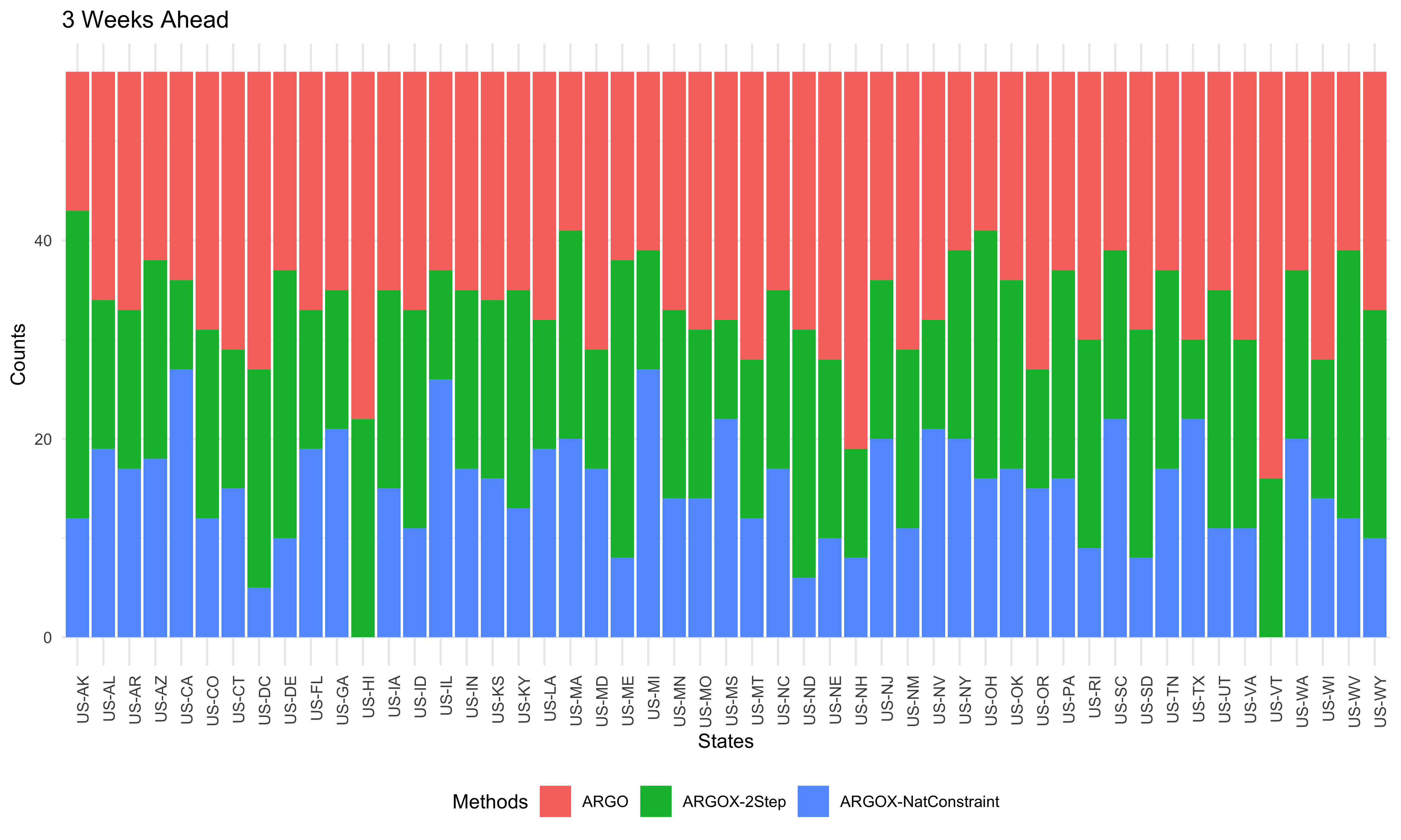}\label{fig:EnsembleSelect_3Weeks}}
\hfill
\subfloat[4 Weeks Ahead State Level Ensemble Selections]
{\includegraphics[width=0.7\textwidth]{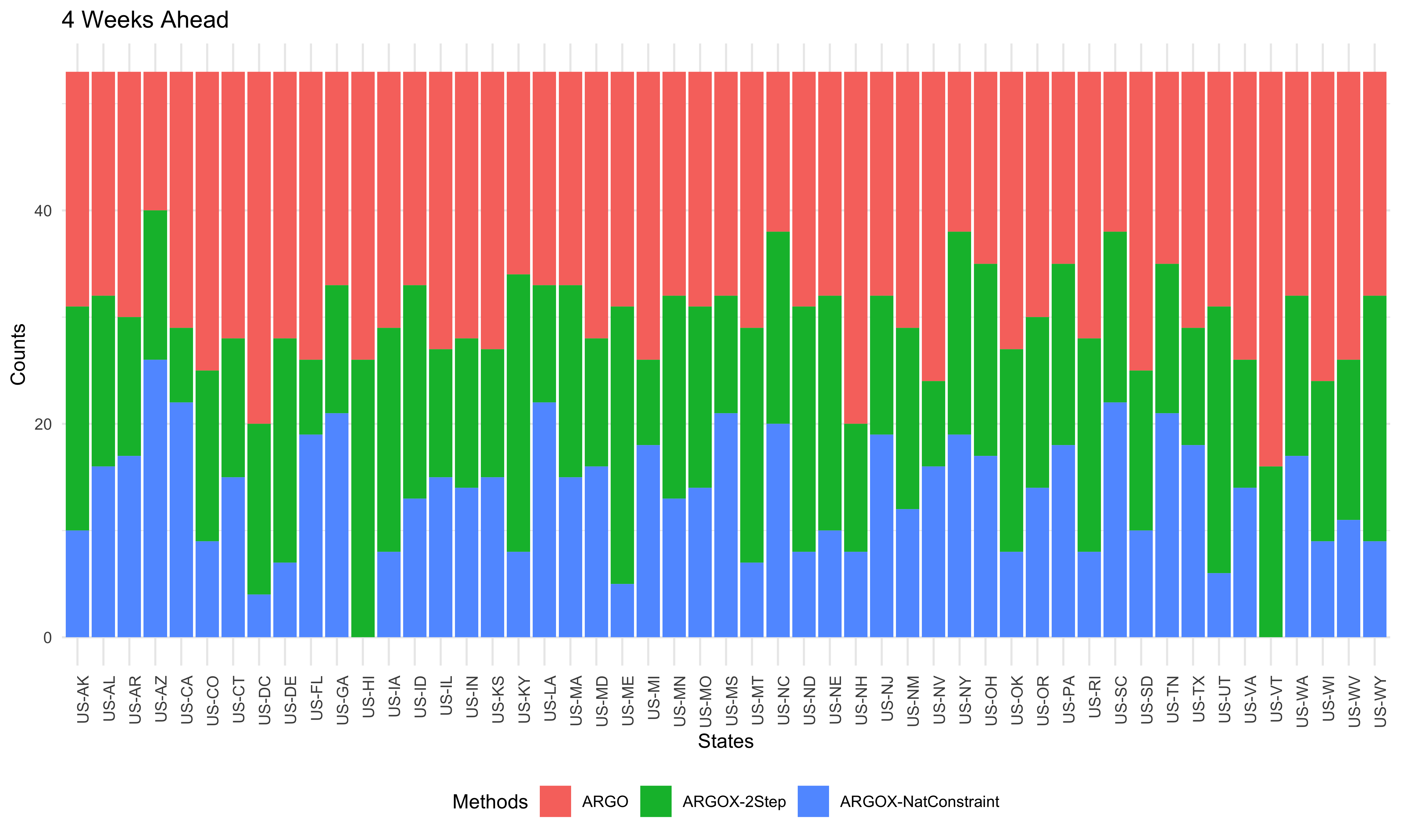}\label{fig:EnsembleSelect_4Weeks}}
\caption{Ensemble method's selection among ARGO, ARGOX-2Step and ARGOX-NatConstraint for all 51 U.S. States for 3 weeks and 4 weeks ahead predictions, from 2020-07-04 to 2021-10-09. }
\label{fig:EnsembleSelect_ALLStates_3_4}
\end{figure}

\clearpage
Fig \ref{fig:EnsembleSelect_3Weeks} and \ref{fig:EnsembleSelect_4Weeks} show the 3 weeks and 4 weeks ahead winner-takes-all state level forecasts selection among the three other methods: ARGO, ARGOX-2Step and ARGOX-NatCOnstraint.

\clearpage
\subsection*{State Level Forecasts Coverages}
Table \ref{tab:CI_Coverage} shows ARGOX-Ensemble 1 to 4 weeks ahead forecasts' actual coverage for all states. The coverage is for 95\% nominal confidence interval. The averages coverage over all states are 0.88\%, 0.86\% , 0.83\%, and 0.78\% for 1-4 weeks ahead predictions.
\begin{table}[ht]
\centering
\scalebox{0.95}{
\begin{tabular}{lrrrr}
  \hline
 States & 1 week ahead & 2 weeks ahead& 4 weeks ahead & 4 weeks ahead\\ 
  \hline
  US-AK & 0.82 & 0.77 & 0.76 & 0.81 \\ 
  US-AL & 0.85 & 0.79 & 0.78 & 0.79 \\ 
  US-AR & 0.89 & 0.82 & 0.82 & 0.76 \\ 
  US-AZ & 0.87 & 0.83 & 0.82 & 0.75 \\ 
  US-CA & 0.88 & 0.88 & 0.82 & 0.75 \\ 
  US-CO & 0.88 & 0.87 & 0.86 & 0.80 \\ 
  US-CT & 0.84 & 0.81 & 0.80 & 0.74 \\ 
  US-DC & 0.90 & 0.90 & 0.87 & 0.89 \\ 
  US-DE & 0.91 & 0.91 & 0.90 & 0.85 \\ 
  US-FL & 0.82 & 0.85 & 0.78 & 0.76 \\ 
  US-GA & 0.94 & 0.92 & 0.90 & 0.81 \\ 
  US-HI & 0.89 & 0.88 & 0.86 & 0.86 \\ 
  US-IA & 0.86 & 0.84 & 0.80 & 0.74 \\ 
  US-ID & 0.88 & 0.87 & 0.85 & 0.79 \\ 
  US-IL & 0.89 & 0.88 & 0.83 & 0.75 \\ 
  US-IN & 0.89 & 0.84 & 0.76 & 0.72 \\ 
  US-KS & 0.86 & 0.81 & 0.81 & 0.72 \\ 
  US-KY & 0.88 & 0.87 & 0.89 & 0.91 \\ 
  US-LA & 0.88 & 0.85 & 0.79 & 0.81 \\ 
  US-MA & 0.88 & 0.90 & 0.83 & 0.80 \\ 
  US-MD & 0.91 & 0.89 & 0.84 & 0.84 \\ 
  US-ME & 0.88 & 0.85 & 0.81 & 0.78 \\ 
  US-MI & 0.93 & 0.92 & 0.82 & 0.85 \\ 
  US-MN & 0.89 & 0.87 & 0.82 & 0.78 \\ 
  US-MO & 0.91 & 0.89 & 0.86 & 0.80 \\ 
  US-MS & 0.85 & 0.84 & 0.80 & 0.72 \\ 
  US-MT & 0.83 & 0.87 & 0.81 & 0.82 \\ 
  US-NC & 0.91 & 0.88 & 0.87 & 0.70 \\ 
  US-ND & 0.84 & 0.83 & 0.81 & 0.75 \\ 
  US-NE & 0.87 & 0.88 & 0.86 & 0.84 \\ 
  US-NH & 0.89 & 0.88 & 0.83 & 0.79 \\ 
  US-NJ & 0.92 & 0.91 & 0.88 & 0.82 \\ 
  US-NM & 0.90 & 0.92 & 0.88 & 0.86 \\ 
  US-NV & 0.86 & 0.88 & 0.85 & 0.75 \\ 
  US-NY & 0.89 & 0.85 & 0.84 & 0.72 \\ 
  US-OH & 0.91 & 0.87 & 0.82 & 0.77 \\ 
  US-OK & 0.89 & 0.81 & 0.85 & 0.81 \\ 
  US-OR & 0.90 & 0.86 & 0.85 & 0.79 \\ 
  US-PA & 0.90 & 0.87 & 0.78 & 0.72 \\ 
  US-RI & 0.91 & 0.91 & 0.85 & 0.81 \\ 
  US-SC & 0.87 & 0.80 & 0.79 & 0.76 \\ 
  US-SD & 0.89 & 0.86 & 0.86 & 0.78 \\ 
  US-TN & 0.82 & 0.86 & 0.80 & 0.71 \\ 
  US-TX & 0.90 & 0.86 & 0.82 & 0.72 \\ 
  US-UT & 0.87 & 0.90 & 0.88 & 0.86 \\ 
  US-VA & 0.89 & 0.85 & 0.86 & 0.79 \\ 
  US-VT & 0.91 & 0.88 & 0.85 & 0.81 \\ 
  US-WA & 0.90 & 0.88 & 0.83 & 0.81 \\ 
  US-WI & 0.90 & 0.88 & 0.86 & 0.79 \\ 
  US-WV & 0.83 & 0.82 & 0.76 & 0.74 \\ 
  US-WY & 0.79 & 0.76 & 0.76 & 0.73 \\ 
   \hline
\end{tabular}
}
\caption{All States' 1-4 weeks ahead forecast coverage.}
\label{tab:CI_Coverage}
\end{table}

\clearpage
\subsection*{More state-level model comparisons}
Table \ref{tab:State_Other} shows 1 to 4 weeks ahead state level forecast result comparison the 8 different methods through 3 error metrics. Fig \ref{fig:State_Compare_Other_RMSE_Heatmap_12} to fig \ref{fig:State_Compare_Other_PC_Heatmap_34} shows the teams comparison in three error metrics as heatmaps throughout 1 to 4 weeks ahead forecasts.

\begin{figure}[htbp]
\centering
\includegraphics[width=1\textwidth]{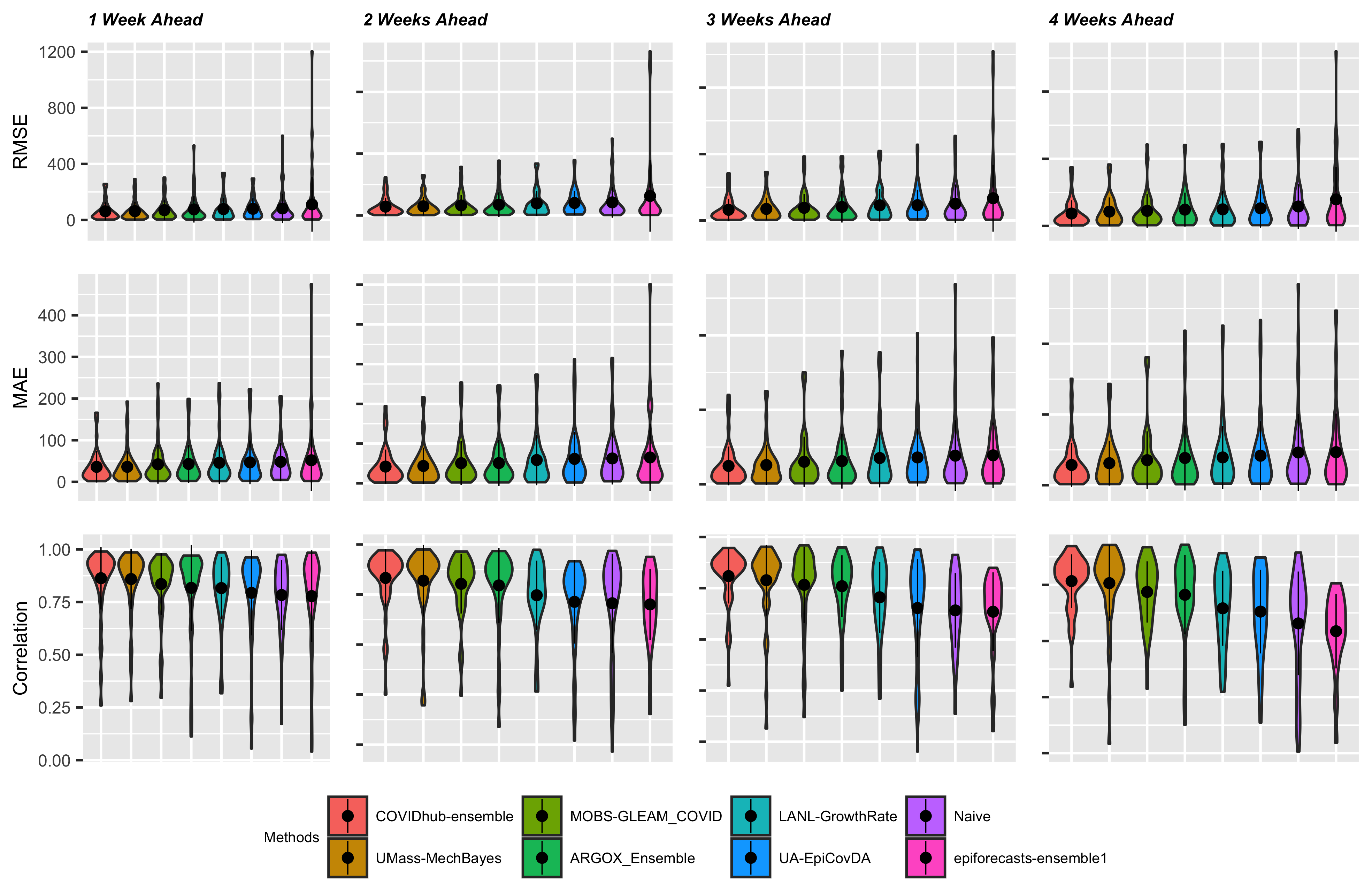}
\caption{Comparison among different models' 1 to 4 weeks (from left to right) ahead U.S. states level weekly incremental death predictions (from 2020-07-04 to 2021-10-10). The RMSE, MAE and Pearson correlation for each method across all states are reported in the violin plot. The methods (x-axis) are sorted based on their RMSE.}
\label{fig4}
\end{figure}

\begin{figure}[htbp]
\centering
\includegraphics[width=0.8\textwidth]{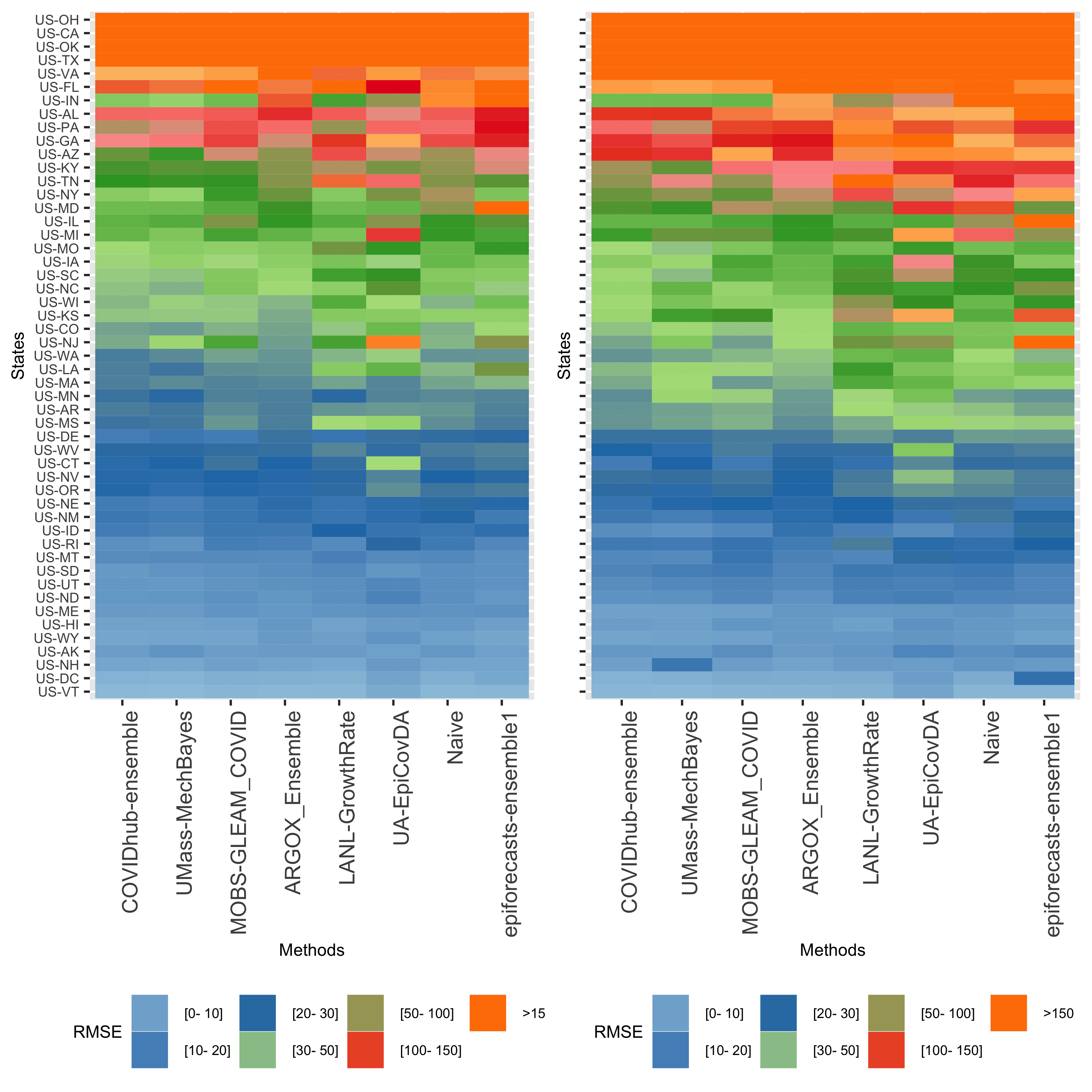}
\caption{State Level 1 week (left) and 2 weeks (right) ahead all teams RMSE comparison heatmap. States (y-axis) are sorted based on the best performing method's RMSE, in this case ARGOX-Ensemble. RMSE greater than 200 are scaled to 200.}
\label{fig:State_Compare_Other_RMSE_Heatmap_12}
\end{figure}

\begin{figure}[htbp]
\centering
\includegraphics[width=0.8\textwidth]{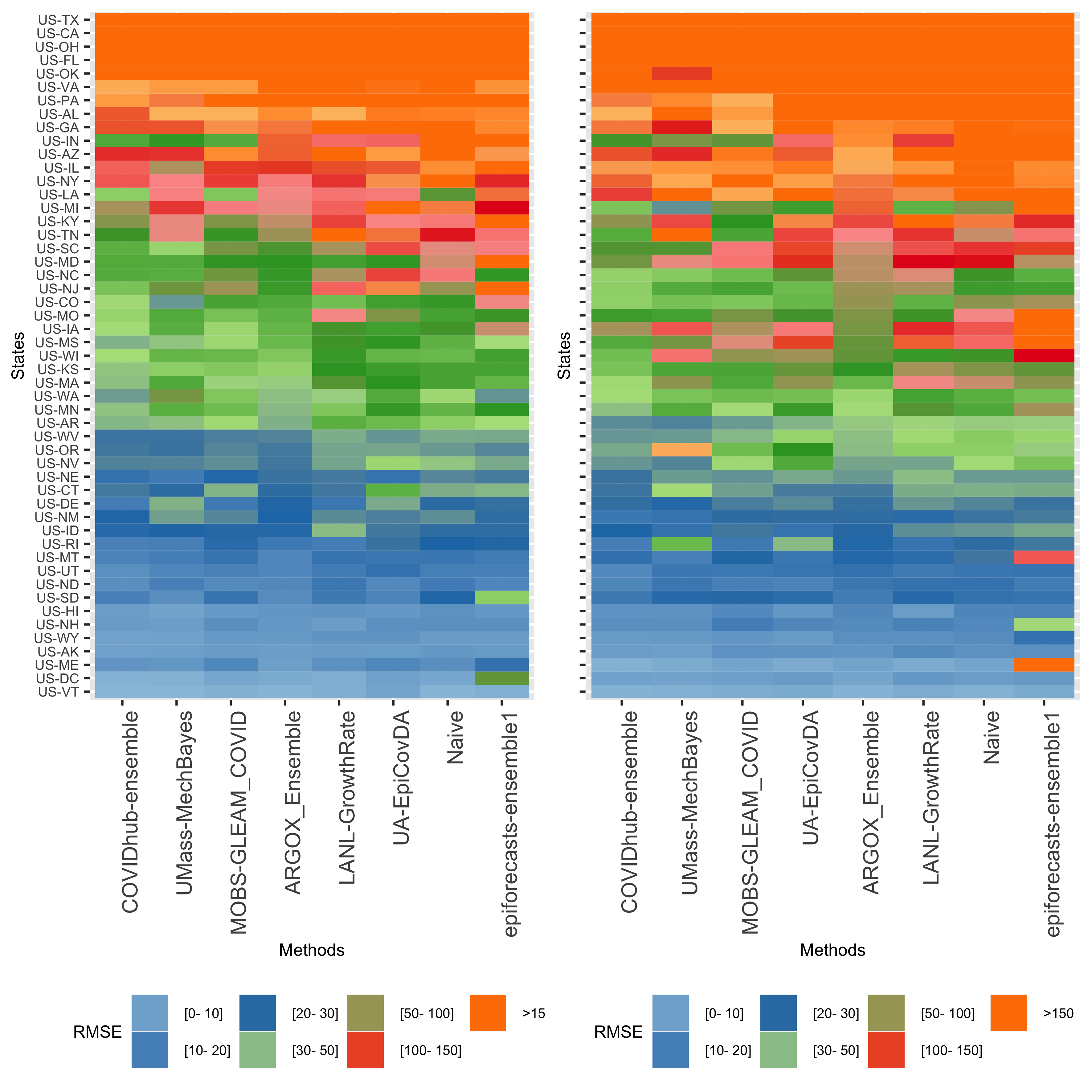}
\caption{State Level 3 weeks (left) and 4 weeks (right) ahead all teams RMSE comparison heatmap. States (y-axis) are sorted based on the best performing method's RMSE, in this case ARGOX-Ensemble. RMSE greater than 200 are scaled to 200.}
\label{fig:State_Compare_Other_RMSE_Heatmap_34}
\end{figure}

\begin{figure}[htbp]
\centering
\includegraphics[width=0.8\textwidth]{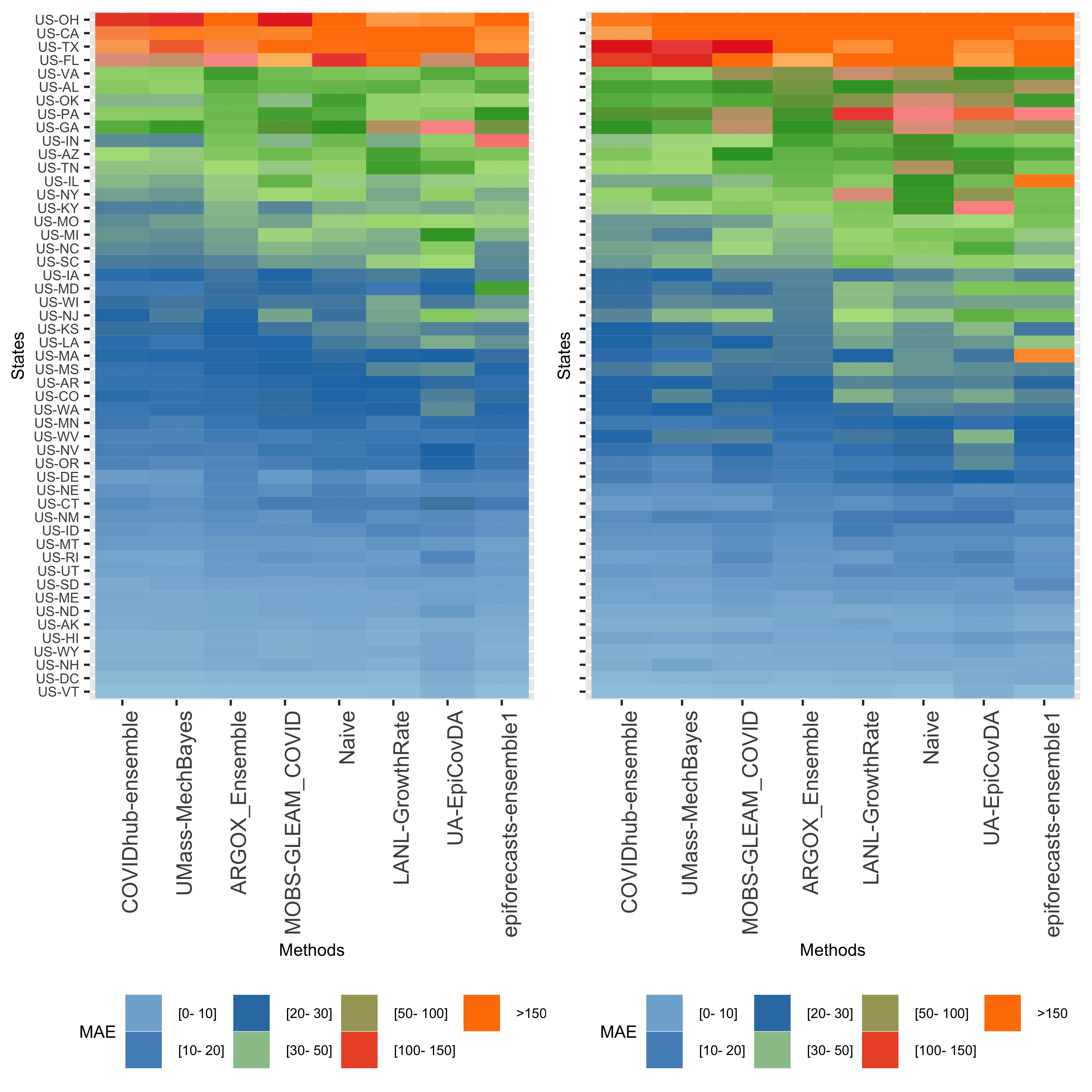}
\caption{State Level 1 week (left) and 2 weeks (right) ahead all teams MAE comparison heatmap. States (y-axis) are sorted based on the best performing method's MAE, in this case ARGOX-Ensemble. MAE greater than 200 are scaled to 200.}
\label{fig:State_Compare_Other_MAE_Heatmap_12}
\end{figure}

\begin{figure}[htbp]
\centering
\includegraphics[width=0.8\textwidth]{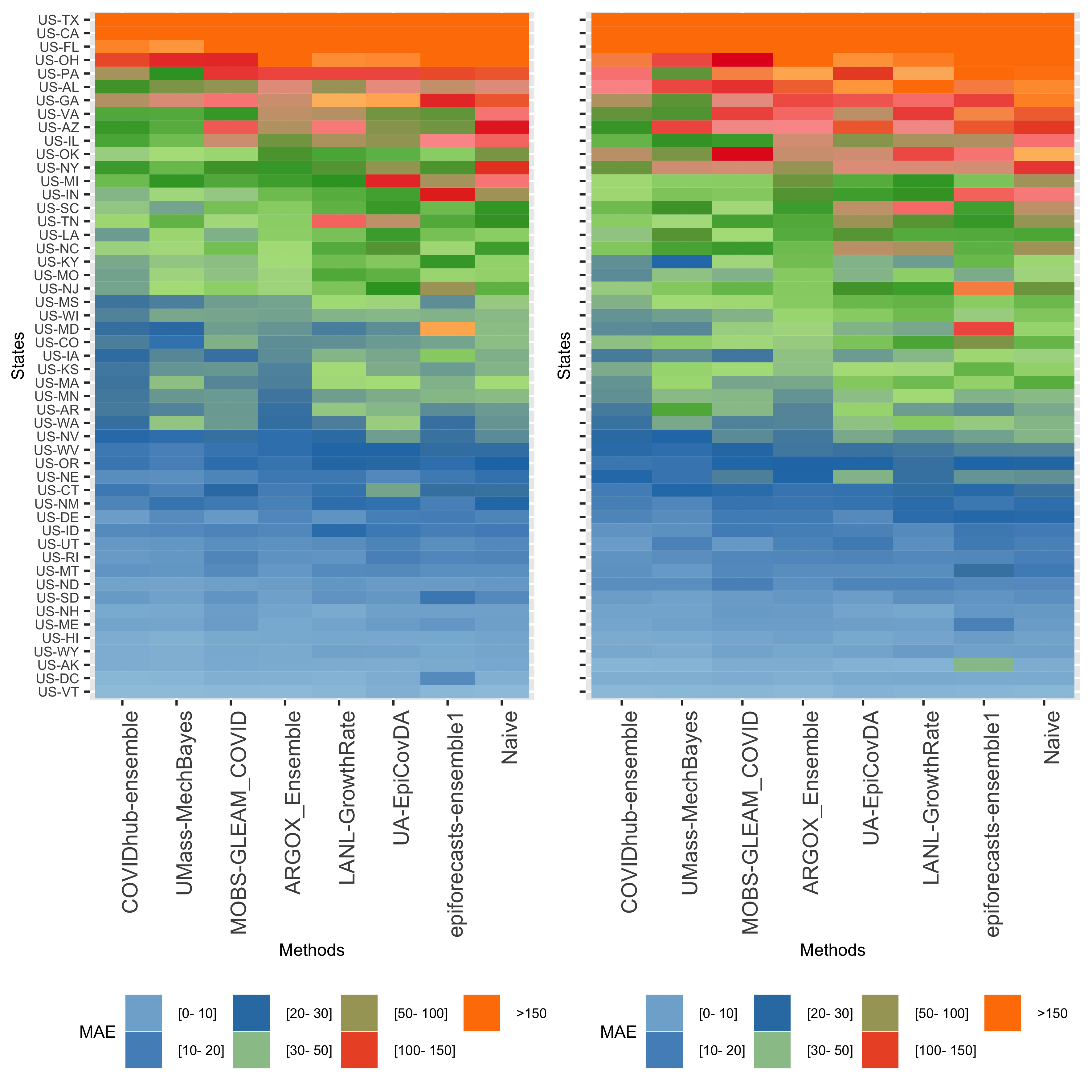}
\caption{State Level 3 weeks (left) and 4 weeks (right) ahead all teams MAE comparison heatmap. States (y-axis) are sorted based on the best performing method's MAE, in this case ARGOX-Ensemble. MAE greater than 200 are scaled to 200.}
\label{fig:State_Compare_Other_MAE_Heatmap_34}
\end{figure}

\begin{figure}[htbp]
\centering
\includegraphics[width=0.8\textwidth]{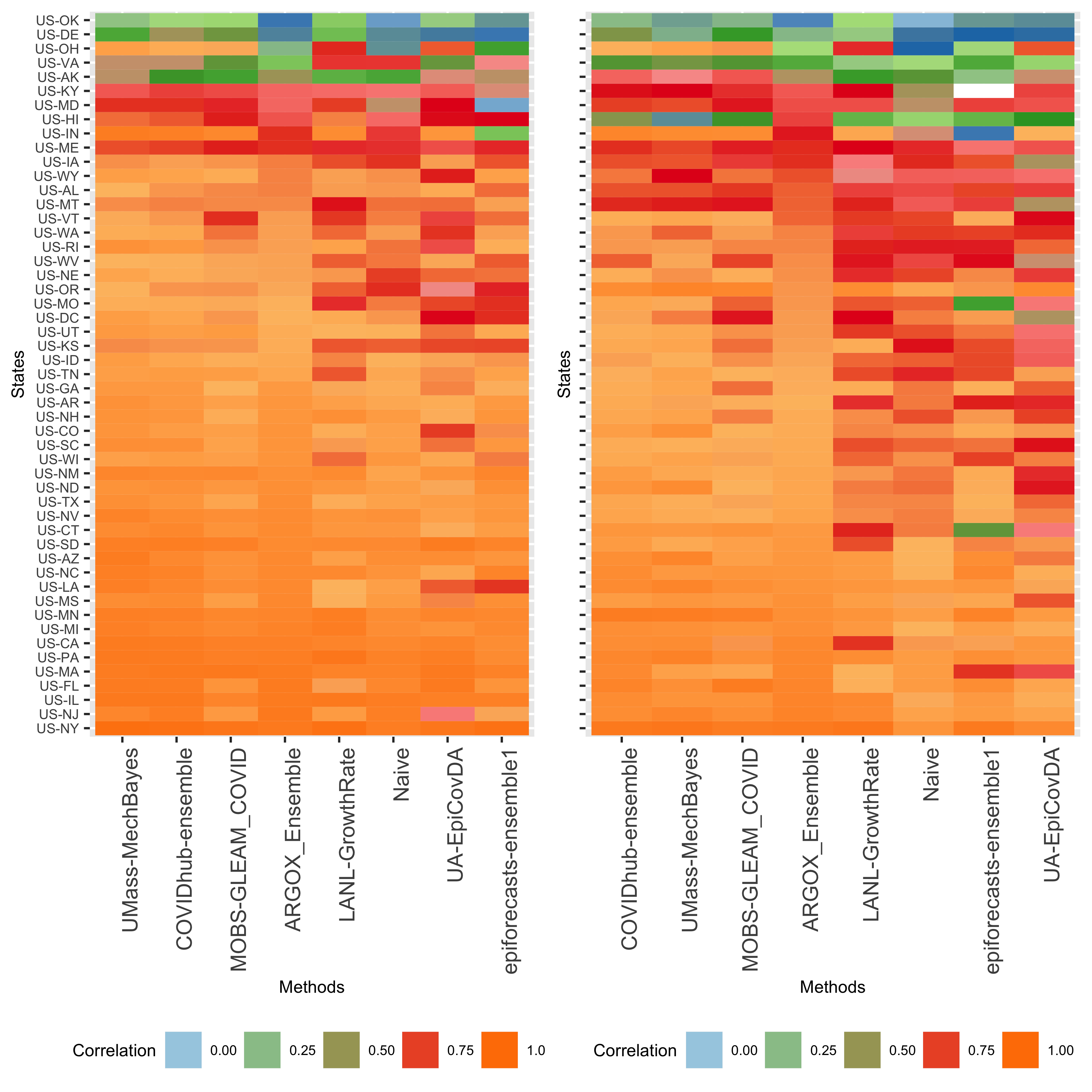}
\caption{State Level 1 week (left) and 2 weeks (right) ahead all teams Pearson correlation against JHU groundtruth comparison heatmap. States (y-axis) are sorted based on the best performing method's correlation, in this case ARGOX-Ensemble}
\label{fig:State_Compare_Other_PC_Heatmap_12}
\end{figure}

\begin{figure}[htbp]
\centering
\includegraphics[width=0.8\textwidth]{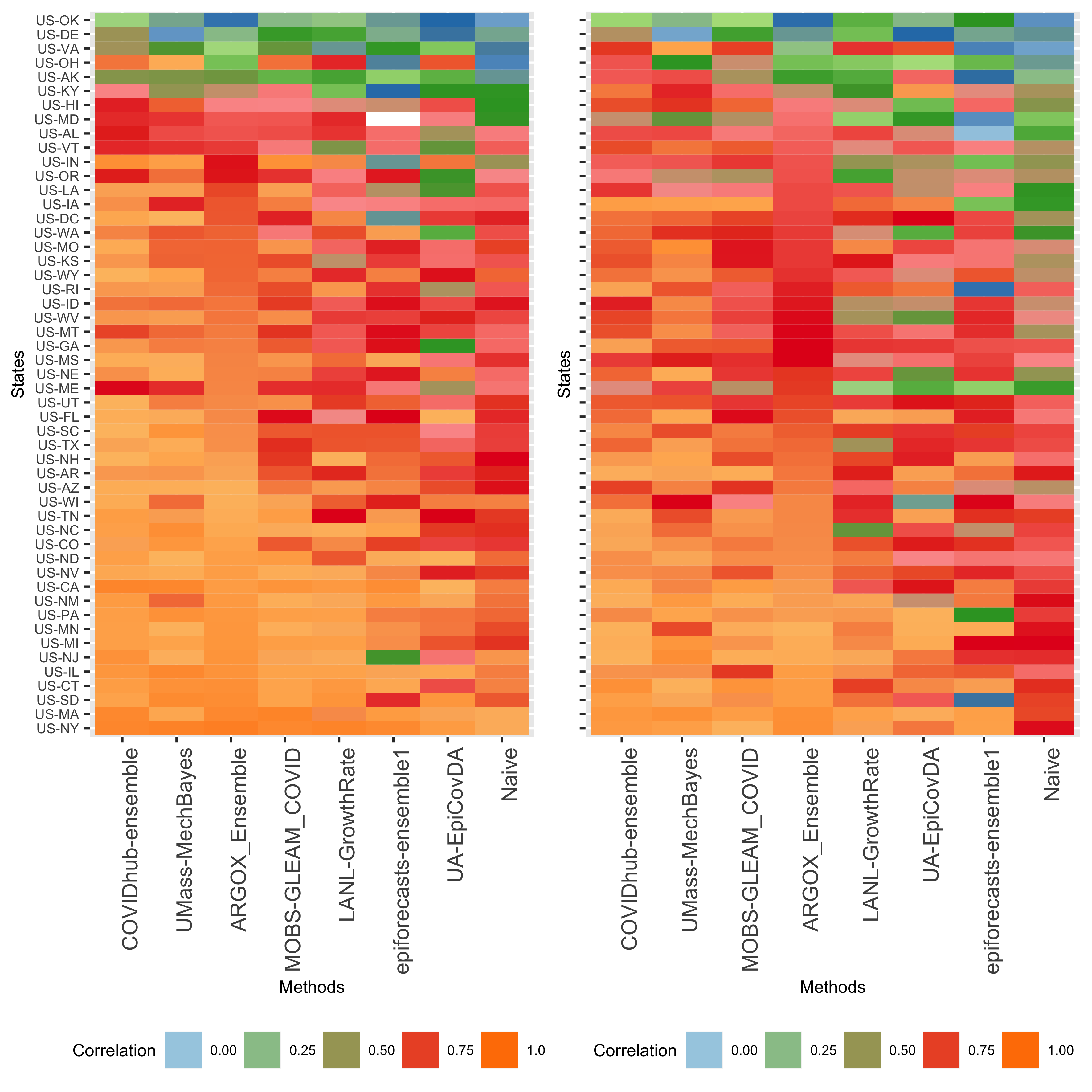}
\caption{State Level 3 weeks (left) and 4 weeks (right) ahead all teams Pearson correlation against JHU groundtruth comparison heatmap. States (y-axis) are sorted based on the best performing method's correlation, in this case ARGOX-Ensemble}
\label{fig:State_Compare_Other_PC_Heatmap_34}
\end{figure}

\clearpage
\newgeometry{left=1.5cm,bottom=2cm}
\subsection*{Detailed estimation results for each state}
\begin{table}[ht]
\renewrobustcmd{\bfseries}{\fontseries{b}\selectfont}
\renewrobustcmd{\boldmath}{}
\newrobustcmd{\B}{\bfseries}
\centering
\begin{tabular}{lrrrr}
  \hline
& 1 Week Ahead & 2 Weeks Ahead & 3 Weeks Ahead & 4 Weeks Ahead \\ 
    \hline \multicolumn{1}{l}{RMSE} \\
 \hspace{1em} ARGO & 13.64 & 11.66 & 12.85 & 10.55 \\ 
   \hspace{1em} ARGOX 2Step & 11.99 & 10.51 & 12.37 & 12.35 \\ 
   \hspace{1em} ARGOX NatConstraint & 21.46 & 24.47 & 41.14 & 72.79 \\ 
   \hspace{1em} Ensemble & \B11.04 & \B9.08 & \B11.14 & \B8.70 \\ 
   \hspace{1em} Naive& 12.25 & 13.04 & 12.99 & 11.89 \\ 
   \multicolumn{1}{l}{MAE} \\
  \hspace{1em} ARGO &8.63 & 7.95 & 9.07 & 7.77 \\ 
   \hspace{1em} ARGOX 2Step & \B6.81 & 6.60 & 8.20 & 8.07 \\ 
   \hspace{1em} ARGOX NatConstraint & 13.41 & 17.49 & 26.03 & 40.23 \\ 
   \hspace{1em} Ensemble & 7.08 & \B6.49 & \B6.96 & \B6.13 \\ 
   \hspace{1em} Naive& 8.02 & 7.79 & 8.67 & 8.09 \\ 
   \multicolumn{1}{l}{Correlation} \\
 \hspace{1em} ARGO &0.21 & 0.36 & 0.21 & 0.52 \\ 
   \hspace{1em} ARGOX 2Step & 0.50 & 0.60 & 0.35 & 0.55 \\ 
   \hspace{1em} ARGOX NatConstraint & 0.20 & 0.09 & 0.19 & 0.08 \\ 
   \hspace{1em} Ensemble & \B0.50 & \B0.64 & \B0.44 & \B0.73 \\ 
   \hspace{1em} Naive& 0.38 & 0.27 & 0.15 & 0.36 \\ 
   \hline
\end{tabular}
\caption{Comparison of different methods for state-level COVID-19 1 to 4 weeks ahead incremental death in Alaska (AK). The RMSE, MAE, and correlation are reported and best performed method is highlighted in boldface.} 
\label{tab:State_Ours_AK}
\end{table}

\begin{figure}[!h] 
  \centering 
\includegraphics[width=0.6\linewidth, page=1]{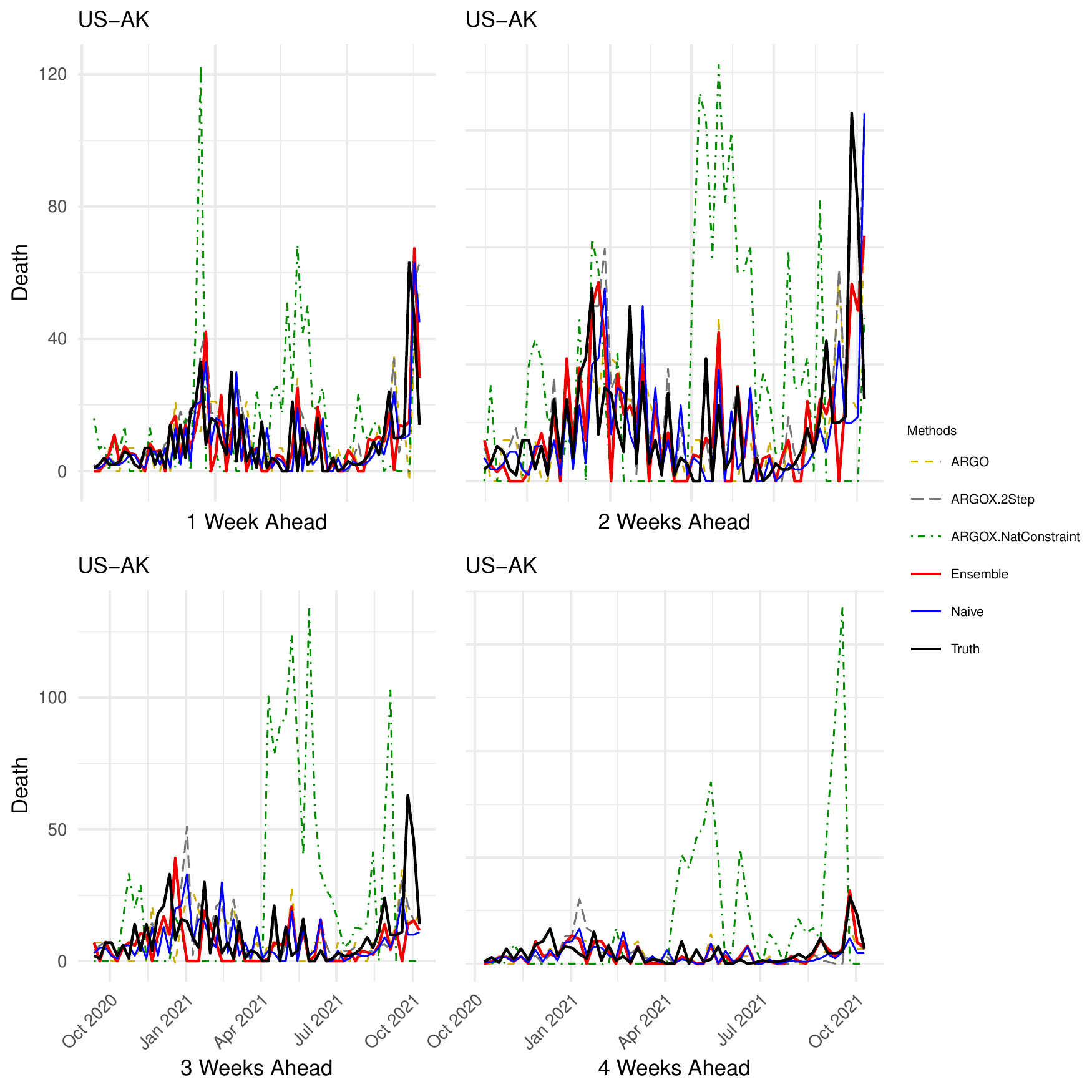} 
\caption{Plots of the COVID-19 1 week (top left), 2 weeks (top right), 3 weeks (bottom left), and 4 weeks (bottom right) ahead estimates for Alaska (AK).}
\label{fig:State_Ours_AK}
\end{figure}
\newpage

\begin{table}[ht]
\renewrobustcmd{\bfseries}{\fontseries{b}\selectfont}
\renewrobustcmd{\boldmath}{}
\newrobustcmd{\B}{\bfseries}
\centering
\begin{tabular}{lrrrr}
  \hline
& 1 Week Ahead & 2 Weeks Ahead & 3 Weeks Ahead & 4 Weeks Ahead \\ 
    \hline \multicolumn{1}{l}{RMSE} \\
 \hspace{1em} ARGO & 160.87 & 185.67 & 210.86 & 215.09 \\ 
   \hspace{1em} ARGOX 2Step & 149.04 & 214.96 & 249.36 & 314.00 \\ 
   \hspace{1em} ARGOX NatConstraint & 145.77 & 192.07 & 223.76 & 273.01 \\ 
   \hspace{1em} Ensemble & 138.56 & \B174.31 & \B201.40 & \B187.28 \\ 
   \hspace{1em} Naive& \B126.95 & 178.26 & 207.14 & 248.44 \\ 
   \multicolumn{1}{l}{MAE} \\
  \hspace{1em} ARGO &95.75 & 104.82 & 124.89 & 133.58 \\ 
   \hspace{1em} ARGOX 2Step & 83.25 & 123.00 & 144.25 & 199.11 \\ 
   \hspace{1em} ARGOX NatConstraint & 88.10 & 113.82 & 146.04 & 184.81 \\ 
   \hspace{1em} Ensemble & 76.45 & 101.02 & \B121.38 & \B110.89 \\ 
   \hspace{1em} Naive & \B73.54 & \B100.87 & 123.47 & 157.06 \\ 
   \multicolumn{1}{l}{Correlation} \\
 \hspace{1em} ARGO &0.74 & 0.66 & 0.61 & 0.50 \\ 
   \hspace{1em} ARGOX 2Step & 0.81 & 0.69 & 0.59 & 0.41 \\ 
   \hspace{1em} ARGOX NatConstraint & 0.78 & 0.64 & 0.50 & 0.25 \\ 
   \hspace{1em} Ensemble & 0.81 & \B0.73 & \B0.63 & \B0.64 \\ 
   \hspace{1em} Naive& \B0.83 & 0.68 & 0.58 & 0.42 \\ 
   \hline
\end{tabular}
\caption{Comparison of different methods for state-level COVID-19 1 to 4 weeks ahead incremental death in Alabama (AL). The MSE, MAE, and correlation are reported and best performed method is highlighted in boldface.} 
\end{table}

\begin{figure}[!h] 
  \centering 
\includegraphics[width=0.6\linewidth, page=2]{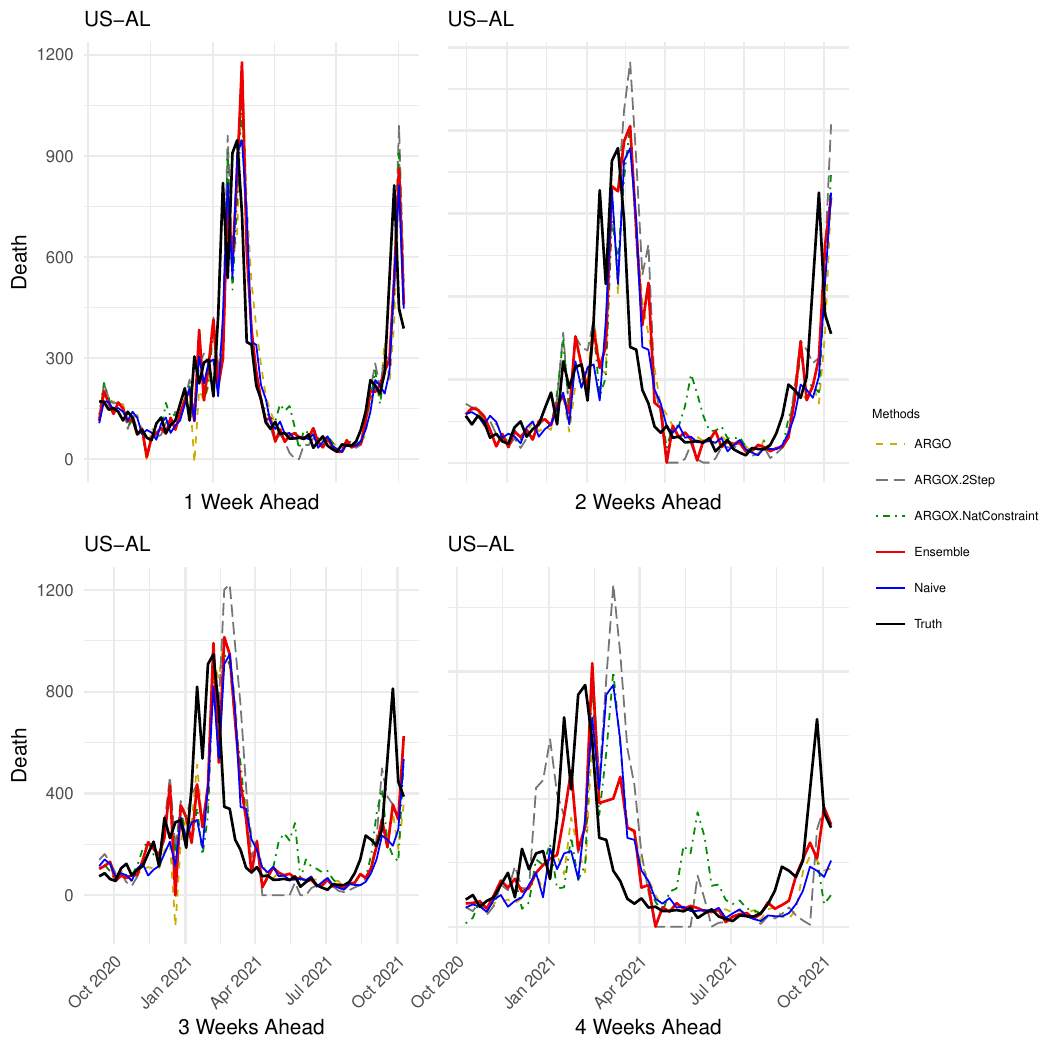} 
\caption{Plots of the COVID-19 1 week (top left), 2 weeks (top right), 3 weeks (bottom left), and 4 weeks (bottom right) ahead estimates for Alabama (AL).}
\end{figure}
\newpage

\begin{table}[ht]
\renewrobustcmd{\bfseries}{\fontseries{b}\selectfont}
\renewrobustcmd{\boldmath}{}
\newrobustcmd{\B}{\bfseries}
\centering
\begin{tabular}{lrrrr}
  \hline
& 1 Week Ahead & 2 Weeks Ahead & 3 Weeks Ahead & 4 Weeks Ahead \\ 
    \hline \multicolumn{1}{l}{RMSE} \\
 \hspace{1em} ARGO & 46.95 & 58.91 & 70.05 & 84.36 \\ 
   \hspace{1em} ARGOX 2Step & 47.28 & 65.06 & 83.08 & 120.70 \\ 
   \hspace{1em} ARGOX NatConstraint & 45.91 & 57.09 & 82.67 & 114.78 \\ 
   \hspace{1em} Ensemble & \B36.49 & \B43.85 & \B50.68 & \B63.11 \\ 
   \hspace{1em} Naive& 42.52 & 53.64 & 67.23 & 80.52 \\ 
   \multicolumn{1}{l}{MAE} \\
  \hspace{1em} ARGO &32.74 & 40.89 & 47.44 & 58.77 \\ 
   \hspace{1em} ARGOX 2Step & 34.17 & 46.85 & 60.76 & 90.22 \\ 
   \hspace{1em} ARGOX NatConstraint & 34.11 & 45.74 & 64.54 & 87.17 \\ 
   \hspace{1em} Ensemble & \B25.31 & \B30.35 & \B33.95 & \B44.16 \\ 
   \hspace{1em} Naive& 28.86 & 37.77 & 46.63 & 55.38 \\ 
   \multicolumn{1}{l}{Correlation} \\
 \hspace{1em} ARGO &0.84 & 0.76 & 0.68 & 0.54 \\ 
   \hspace{1em} ARGOX 2Step & 0.88 & 0.81 & 0.72 & 0.60 \\ 
   \hspace{1em} ARGOX NatConstraint & 0.84 & 0.76 & 0.49 & 0.25 \\ 
   \hspace{1em} Ensemble & \B0.91 & \B0.88 & \B0.83 & \B0.75 \\ 
   \hspace{1em} Naive& 0.87 & 0.80 & 0.71 & 0.60 \\ 
   \hline
\end{tabular}
\caption{Comparison of different methods for state-level COVID-19 1 to 4 weeks ahead incremental death in Arkansas (AR). The MSE, MAE, and correlation are reported and best performed method is highlighted in boldface.} 
\end{table}

\begin{figure}[!h] 
  \centering 
\includegraphics[width=0.6\linewidth, page=3]{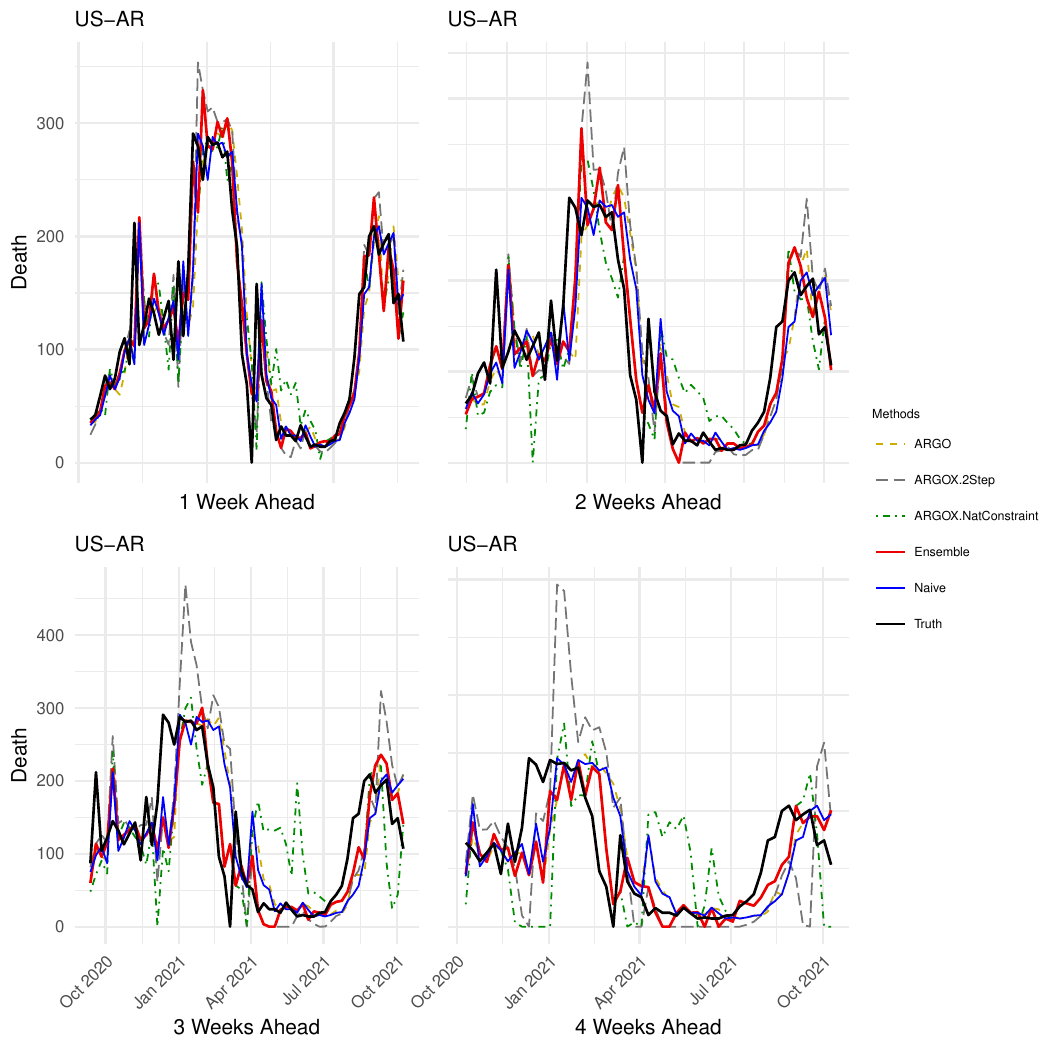} 
\caption{Plots of the COVID-19 1 week (top left), 2 weeks (top right), 3 weeks (bottom left), and 4 weeks (bottom right) ahead estimates for Arkansas (AR).}
\end{figure}
\newpage

\begin{table}[ht]
\renewrobustcmd{\bfseries}{\fontseries{b}\selectfont}
\renewrobustcmd{\boldmath}{}
\newrobustcmd{\B}{\bfseries}
\centering
\begin{tabular}{lrrrr}
  \hline
& 1 Week Ahead & 2 Weeks Ahead & 3 Weeks Ahead & 4 Weeks Ahead \\ 
    \hline \multicolumn{1}{l}{RMSE} \\
 \hspace{1em} ARGO & 120.67 & 162.85 & 182.19 & 263.49 \\ 
   \hspace{1em} ARGOX 2Step & 115.87 & 203.04 & 219.93 & 283.83 \\ 
   \hspace{1em} ARGOX NatConstraint & 110.58 & 165.66 & 168.03 & 173.78 \\ 
   \hspace{1em} Ensemble & \B100.29 & \B144.54 & \B146.15 & \B155.19 \\ 
   \hspace{1em} Naive& 102.39 & 155.81 & 201.28 & 245.29 \\ 
   \multicolumn{1}{l}{MAE} \\
  \hspace{1em} ARGO &77.43 & 106.25 & 116.32 & 165.34 \\ 
   \hspace{1em} ARGOX 2Step & 76.09 & 133.03 & 134.95 & 180.84 \\ 
   \hspace{1em} ARGOX NatConstraint & 73.22 & 116.64 & 121.96 & 124.81 \\ 
   \hspace{1em} Ensemble & 66.13 & \B91.65 & \B91.67 & \B94.12 \\ 
   \hspace{1em} Naive& \B65.22 & 95.67 & 131.16 & 165.77 \\ 
   \multicolumn{1}{l}{Correlation} \\
 \hspace{1em} ARGO &0.92 & 0.86 & 0.83 & 0.77 \\ 
   \hspace{1em} ARGOX 2Step & 0.93 & 0.85 & 0.84 & 0.89 \\ 
   \hspace{1em} ARGOX NatConstraint & 0.92 & 0.83 & 0.82 & 0.89 \\ 
   \hspace{1em} Ensemble & \B0.94 & \B0.89 & \B0.89 & \B0.93 \\ 
   \hspace{1em} Naive& 0.93 & 0.84 & 0.75 & 0.65 \\ 
   \hline
\end{tabular}
\caption{Comparison of different methods for state-level COVID-19 1 to 4 weeks ahead incremental death in Arizona (AZ). The MSE, MAE, and correlation are reported and best performed method is highlighted in boldface.} 
\end{table}

\begin{figure}[!h] 
  \centering 
\includegraphics[width=0.6\linewidth, page=4]{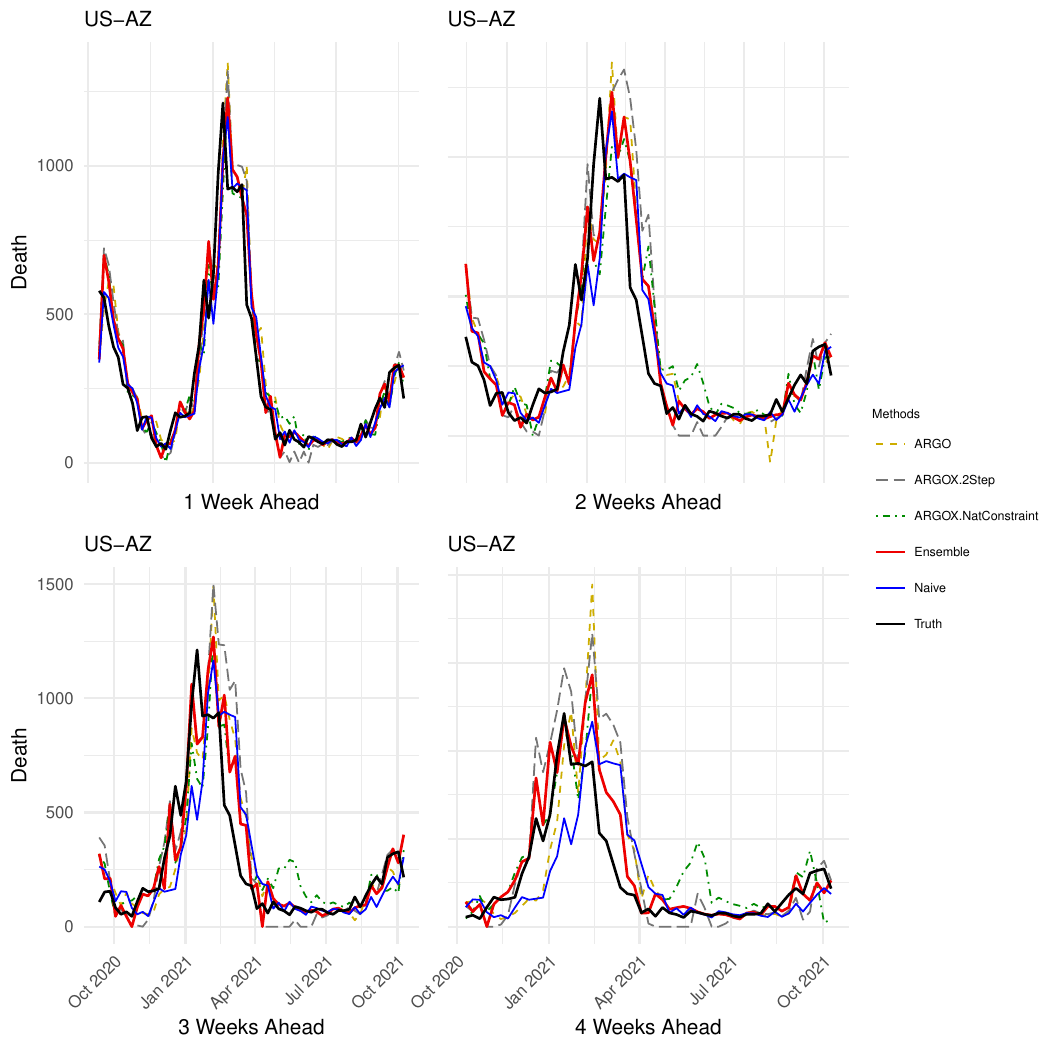} 
\caption{Plots of the COVID-19 1 week (top left), 2 weeks (top right), 3 weeks (bottom left), and 4 weeks (bottom right) ahead estimates for Arizona (AZ).}
\end{figure}
\newpage

\begin{table}[ht]
\renewrobustcmd{\bfseries}{\fontseries{b}\selectfont}
\renewrobustcmd{\boldmath}{}
\newrobustcmd{\B}{\bfseries}
\centering
\begin{tabular}{lrrrr}
  \hline
& 1 Week Ahead & 2 Weeks Ahead & 3 Weeks Ahead & 4 Weeks Ahead \\ 
    \hline \multicolumn{1}{l}{RMSE} \\
 \hspace{1em} ARGO & 
377.26 & 490.77 & 604.88 & 709.10 \\ 
   \hspace{1em} ARGOX 2Step & 347.91 & 588.98 & 758.07 & 863.25 \\ 
   \hspace{1em} ARGOX NatConstraint & 329.73 & 487.32 & 542.22 & 510.15 \\ 
   \hspace{1em} Ensemble & \B300.45 & \B413.08 & \B494.40 & \B475.20 \\ 
   \hspace{1em} Naive& 329.46 & 446.78 & 629.90 & 800.76 \\ 
   \multicolumn{1}{l}{MAE} \\
  \hspace{1em} ARGO &244.46 & 319.77 & 411.04 & 475.13 \\ 
   \hspace{1em} ARGOX 2Step & 234.92 & 361.36 & 495.85 & 622.93 \\ 
   \hspace{1em} ARGOX NatConstraint & 217.31 & 292.63 & 358.01 & 403.04 \\ 
   \hspace{1em} Ensemble & \B193.23 & \B250.30 & \B299.53 & \B349.12 \\ 
   \hspace{1em} Naive& 212.88 & 280.21 & 422.28 & 566.66 \\ 
   \multicolumn{1}{l}{Correlation} \\
 \hspace{1em} ARGO &0.93 & 0.89 & 0.89 & 0.88 \\ 
   \hspace{1em} ARGOX 2Step & 0.95 & 0.90 & 0.86 & 0.89 \\ 
   \hspace{1em} ARGOX NatConstraint & 0.94 & 0.89 & 0.87 & 0.92 \\ 
   \hspace{1em} Ensemble & \B0.95 & \B0.93 & \B0.90 & \B0.93 \\ 
   \hspace{1em} Naive& 0.94 & 0.90 & 0.81 & 0.71 \\ 
   \hline
\end{tabular}
\caption{Comparison of different methods for state-level COVID-19 1 to 4 weeks ahead incremental death in California (CA). The MSE, MAE, and correlation are reported and best performed method is highlighted in boldface.} 
\end{table}

\begin{figure}[!h] 
  \centering 
\includegraphics[width=0.6\linewidth, page=5]{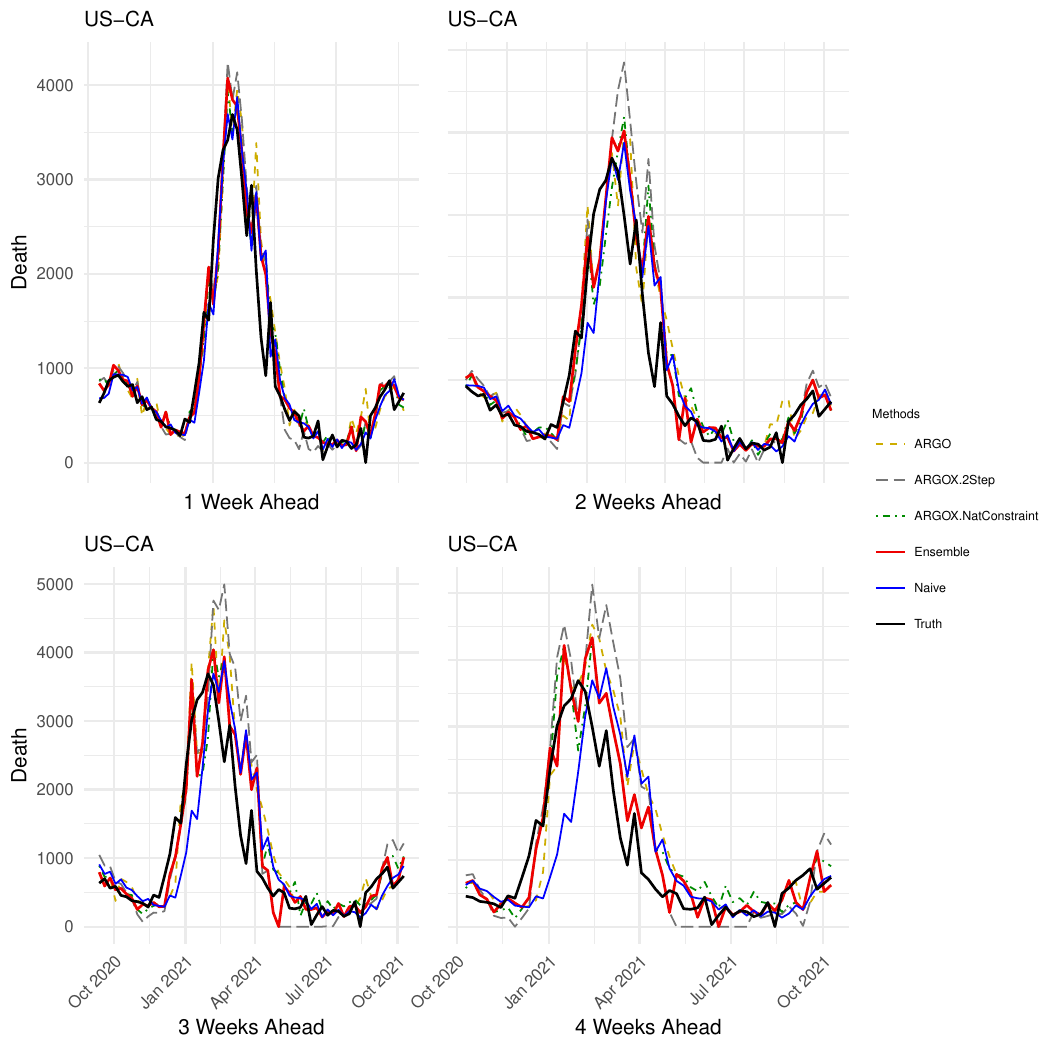} 
\caption{Plots of the COVID-19 1 week (top left), 2 weeks (top right), 3 weeks (bottom left), and 4 weeks (bottom right) ahead estimates for California (CA).}
\end{figure}
\newpage

\begin{table}[ht]
\renewrobustcmd{\bfseries}{\fontseries{b}\selectfont}
\renewrobustcmd{\boldmath}{}
\newrobustcmd{\B}{\bfseries}
\centering
\begin{tabular}{lrrrr}
  \hline
&1 Week Ahead & 2 Weeks Ahead & 3 Weeks Ahead & 4 Weeks Ahead \\ 
    \hline \multicolumn{1}{l}{RMSE} \\
 \hspace{1em} ARGO & 
51.21 & 79.20 & 91.59 & 177.94 \\ 
   \hspace{1em} ARGOX 2Step & 60.44 & 97.89 & 145.27 & 315.33 \\ 
   \hspace{1em} ARGOX NatConstraint & 51.69 & 80.45 & 117.17 & 217.44 \\ 
   \hspace{1em} Ensemble & \B45.85 & 77.74 & \B82.17 & 174.87 \\ 
   \hspace{1em} Naive& 48.58 & \B71.48 & 89.94 & \B109.26 \\ 
   \multicolumn{1}{l}{MAE} \\
  \hspace{1em} ARGO &28.32 & 40.34 & 53.07 & 85.70 \\ 
   \hspace{1em} ARGOX 2Step & 31.54 & 50.00 & 66.25 & 140.10 \\ 
   \hspace{1em} ARGOX NatConstraint & 32.31 & 50.45 & 71.14 & 122.59 \\ 
   \hspace{1em} Ensemble & \B24.92 & \B39.42 & \B42.83 & 81.01 \\ 
   \hspace{1em} Naive& 28.80 & 40.54 & 56.68 & \B69.51 \\ 
   \multicolumn{1}{l}{Correlation} \\
 \hspace{1em} ARGO &0.90 & 0.82 & 0.79 & 0.62 \\ 
   \hspace{1em} ARGOX 2Step & 0.89 & 0.83 & 0.81 & 0.59 \\ 
   \hspace{1em} ARGOX NatConstraint & 0.89 & 0.82 & 0.77 & 0.49 \\ 
   \hspace{1em} Ensemble & \B0.92 & \B0.84 & \B0.89 & \B0.66 \\ 
   \hspace{1em} Naive& 0.89 & 0.77 & 0.66 & 0.51 \\ 
   \hline
\end{tabular}
\caption{Comparison of different methods for state-level COVID-19 1 to 4 weeks ahead incremental death in Colorado (CO). The MSE, MAE, and correlation are reported and best performed method is highlighted in boldface.} 
\end{table}

\begin{figure}[!h] 
  \centering 
\includegraphics[width=0.6\linewidth, page=6]{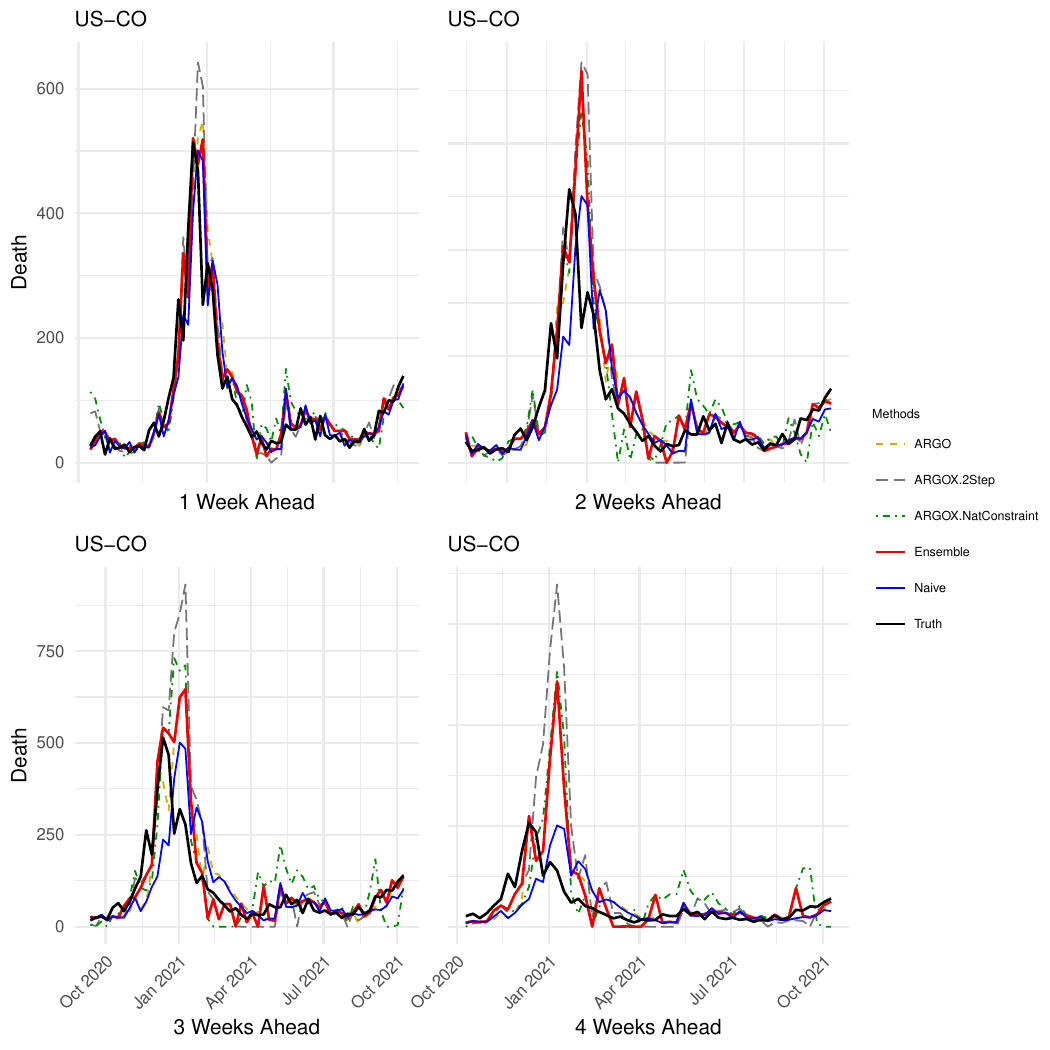} 
\caption{Plots of the COVID-19 1 week (top left), 2 weeks (top right), 3 weeks (bottom left), and 4 weeks (bottom right) ahead estimates for Colorado (CO).}
\end{figure}
\newpage

\begin{table}[ht]
\renewrobustcmd{\bfseries}{\fontseries{b}\selectfont}
\renewrobustcmd{\boldmath}{}
\newrobustcmd{\B}{\bfseries}
\centering
\begin{tabular}{lrrrr}
  \hline
& 1 Week Ahead & 2 Weeks Ahead & 3 Weeks Ahead & 4 Weeks Ahead \\ 
    \hline \multicolumn{1}{l}{RMSE} \\
 \hspace{1em} ARGO & 
26.55 & 28.46 & 42.39 & 67.06 \\ 
   \hspace{1em} ARGOX 2Step & 33.03 & 37.83 & 54.87 & 119.30 \\ 
   \hspace{1em} ARGOX NatConstraint & 39.46 & 42.78 & 61.55 & 88.40 \\ 
   \hspace{1em} Ensemble & 27.53 & 31.38 & 29.23 & 42.46 \\ 
   \hspace{1em} Naive& 32.60 & 34.24 & 48.35 & 60.28 \\ 
   \multicolumn{1}{l}{MAE} \\
  \hspace{1em} ARGO &16.15 & 19.33 & 27.96 & 44.30 \\ 
   \hspace{1em} ARGOX 2Step & 19.27 & 24.96 & 37.12 & 71.17 \\ 
   \hspace{1em} ARGOX NatConstraint & 27.45 & 32.28 & 45.94 & 65.55 \\ 
   \hspace{1em} Ensemble & 15.79 & 19.30 & 18.63 & 26.97 \\ 
   \hspace{1em} Naive& 18.71 & 22.33 & 33.56 & 43.66 \\ 
   \multicolumn{1}{l}{Correlation} \\
 \hspace{1em} ARGO &0.94 & 0.94 & 0.90 & 0.85 \\ 
   \hspace{1em} ARGOX 2Step & 0.93 & 0.94 & 0.92 & 0.90 \\ 
   \hspace{1em} ARGOX NatConstraint & 0.87 & 0.84 & 0.71 & 0.50 \\ 
   \hspace{1em} Ensemble & 0.94 & 0.92 & 0.94 & 0.87 \\ 
   \hspace{1em} Naive& 0.91 & 0.91 & 0.82 & 0.73 \\ 
   \hline
\end{tabular}
\caption{Comparison of different methods for state-level COVID-19 1 to 4 weeks ahead incremental death in Connecticut (CT). The MSE, MAE, and correlation are reported and best performed method is highlighted in boldface.} 
\end{table}

\begin{figure}[!h] 
  \centering 
\includegraphics[width=0.6\linewidth, page=7]{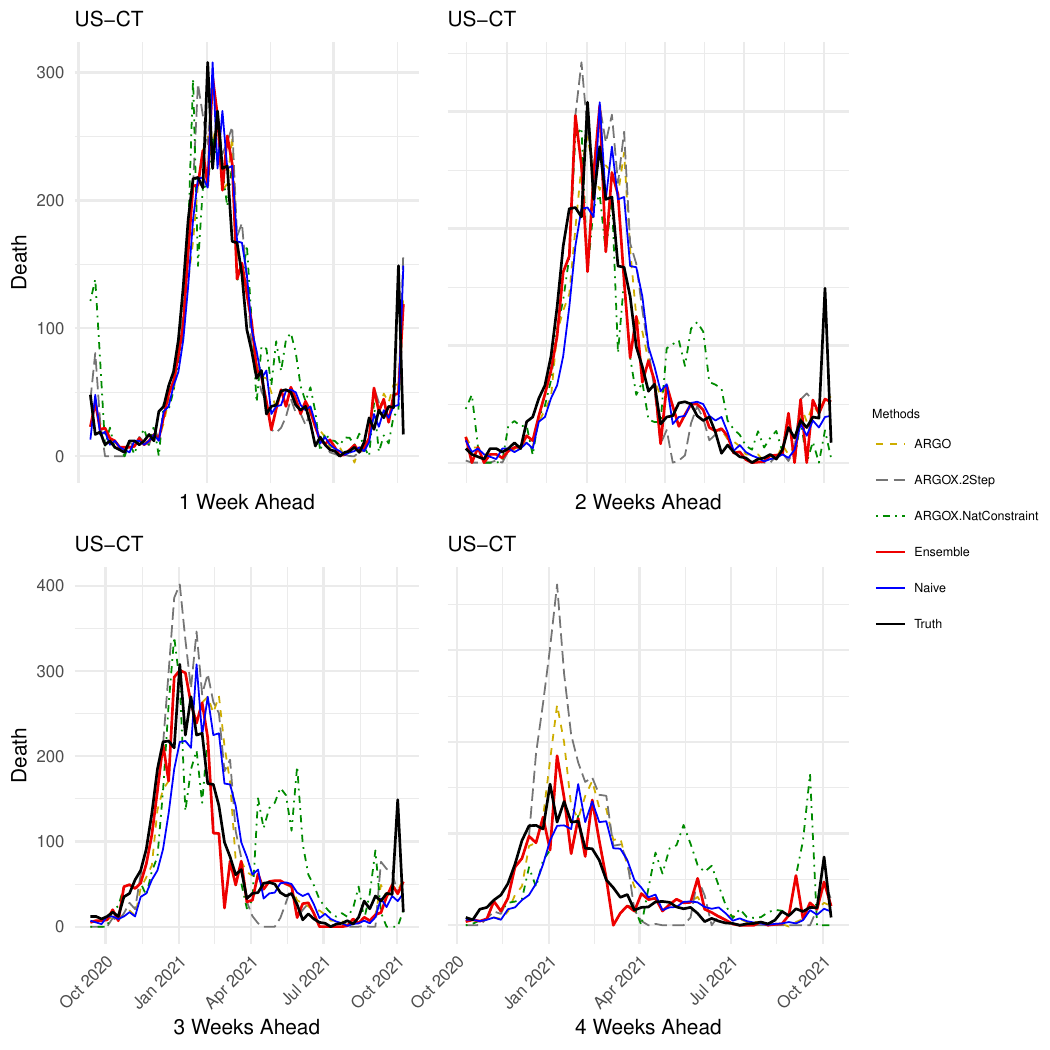} 
\caption{Plots of the COVID-19 1 week (top left), 2 weeks (top right), 3 weeks (bottom left), and 4 weeks (bottom right) ahead estimates for Connecticut (CT).}
\end{figure}
\newpage

\begin{table}[ht]
\renewrobustcmd{\bfseries}{\fontseries{b}\selectfont}
\renewrobustcmd{\boldmath}{}
\newrobustcmd{\B}{\bfseries}
\centering
\begin{tabular}{lrrrr}
  \hline
& 1 Week Ahead & 2 Weeks Ahead & 3 Weeks Ahead & 4 Weeks Ahead \\ 
    \hline \multicolumn{1}{l}{RMSE} \\
 \hspace{1em} ARGO &
5.50 & 7.15 & 8.13 & 8.54 \\ 
   \hspace{1em} ARGOX 2Step & 5.74 & 7.12 & 8.46 & 13.68 \\ 
   \hspace{1em} ARGOX NatConstraint & 21.56 & 23.38 & 42.11 & 78.14 \\ 
   \hspace{1em} Ensemble & 5.20 & 5.73 & 6.74 & 8.46 \\ 
   \hspace{1em} Naive& 5.48 & 6.27 & 7.46 & 8.57 \\ 
   \multicolumn{1}{l}{MAE} \\
  \hspace{1em} ARGO &4.06 & 4.92 & 5.65 & 6.34 \\ 
   \hspace{1em} ARGOX 2Step & 4.21 & 4.93 & 5.78 & 8.97 \\ 
   \hspace{1em} ARGOX NatConstraint & 13.91 & 17.77 & 28.34 & 43.88 \\ 
   \hspace{1em} Ensemble & 3.62 & 3.84 & 4.72 & 5.75 \\ 
   \hspace{1em} Naive& 3.80 & 4.34 & 5.32 & 6.06 \\ 
   \multicolumn{1}{l}{Correlation} \\
 \hspace{1em} ARGO &0.83 & 0.77 & 0.69 & 0.64 \\ 
   \hspace{1em} ARGOX 2Step & 0.85 & 0.81 & 0.78 & 0.76 \\ 
   \hspace{1em} ARGOX NatConstraint & 0.17 & 0.09 & 0.23 & 0.27 \\ 
   \hspace{1em} Ensemble & 0.86 & 0.85 & 0.81 & 0.74 \\ 
   \hspace{1em} Naive& 0.84 & 0.80 & 0.72 & 0.65 \\ 
   \hline
\end{tabular}
\caption{Comparison of different methods for state-level COVID-19 1 to 4 weeks ahead incremental death in District of Columbia (DC). The MSE, MAE, and correlation are reported and best performed method is highlighted in boldface.} 
\end{table}

\begin{figure}[!h] 
  \centering 
\includegraphics[width=0.6\linewidth, page=8]{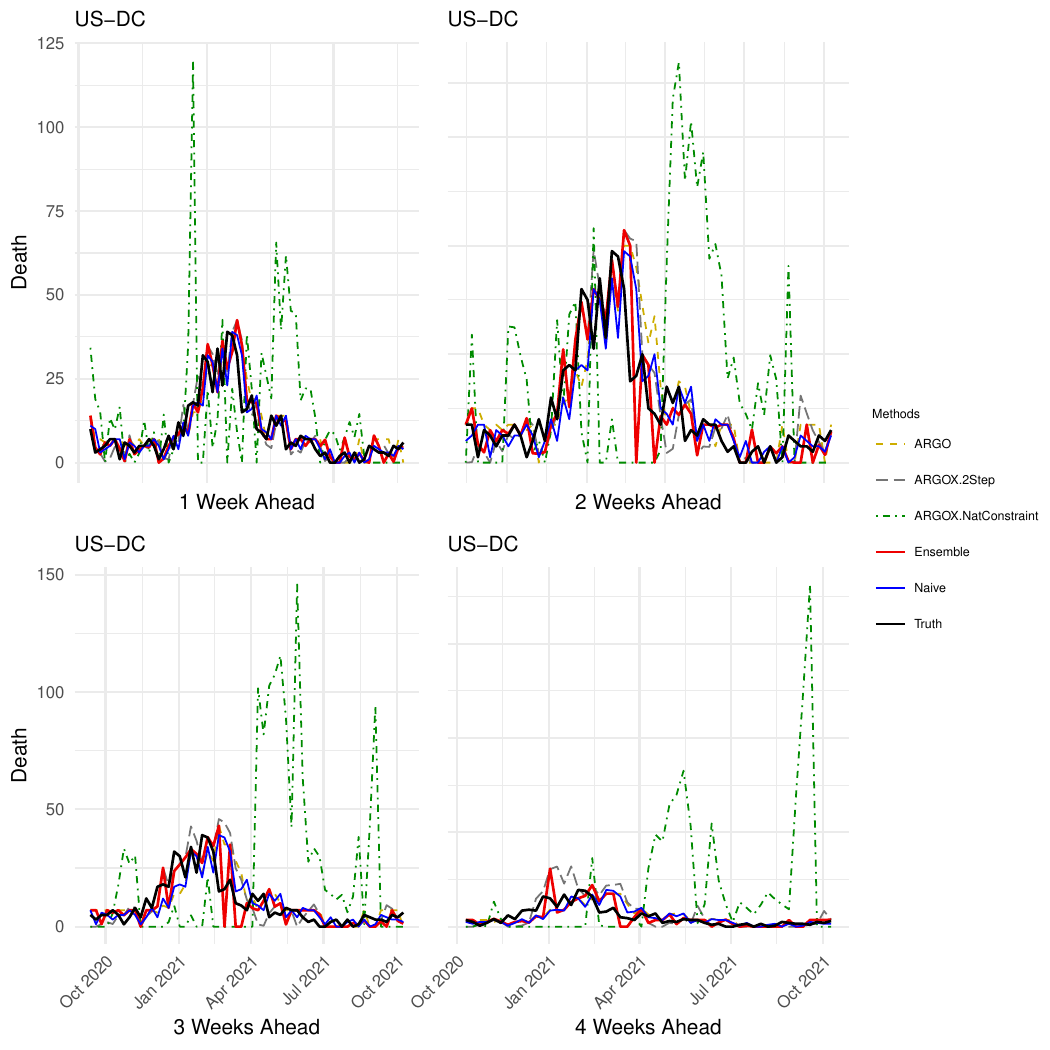} 
\caption{Plots of the COVID-19 1 week (top left), 2 weeks (top right), 3 weeks (bottom left), and 4 weeks (bottom right) ahead estimates for District of Columbia (DC).}
\end{figure}
\newpage

\begin{table}[ht]
\renewrobustcmd{\bfseries}{\fontseries{b}\selectfont}
\renewrobustcmd{\boldmath}{}
\newrobustcmd{\B}{\bfseries}
\centering
\begin{tabular}{lrrrr}
  \hline
&1 Week Ahead & 2 Weeks Ahead & 3 Weeks Ahead & 4 Weeks Ahead \\ 
    \hline \multicolumn{1}{l}{RMSE} \\
 \hspace{1em} ARGO &  
34.19 & 34.97 & 36.15 & 37.13 \\ 
   \hspace{1em} ARGOX 2Step & 28.66 & 30.86 & 29.68 & 31.31 \\ 
   \hspace{1em} ARGOX NatConstraint & 36.70 & 40.77 & 52.88 & 65.22 \\ 
   \hspace{1em} Ensemble & 32.27 & 29.91 & 29.56 & 28.24 \\ 
   \hspace{1em} Naive& 30.57 & 31.46 & 30.99 & 32.73 \\ 
   \multicolumn{1}{l}{MAE} \\
  \hspace{1em} ARGO &17.08 & 17.54 & 20.58 & 21.87 \\ 
   \hspace{1em} ARGOX 2Step & 13.65 & 14.74 & 16.85 & 20.47 \\ 
   \hspace{1em} ARGOX NatConstraint & 24.46 & 28.05 & 38.60 & 49.33 \\ 
   \hspace{1em} Ensemble & 16.54 & 15.30 & 16.94 & 19.09 \\ 
   \hspace{1em} Naive& 15.78 & 16.41 & 17.86 & 19.98 \\ 
   \multicolumn{1}{l}{Correlation} \\
 \hspace{1em} ARGO &0.16 & 0.14 & 0.15 & 0.10 \\ 
   \hspace{1em} ARGOX 2Step & 0.31 & 0.32 & 0.37 & 0.27 \\ 
   \hspace{1em} ARGOX NatConstraint & 0.06 & 0.15 & 0.31 & 0.28 \\ 
   \hspace{1em} Ensemble & 0.20 & 0.27 & 0.28 & 0.24 \\ 
   \hspace{1em} Naive& 0.21 & 0.19 & 0.24 & 0.19 \\ 
   \hline
\end{tabular}
\caption{Comparison of different methods for state-level COVID-19 1 to 4 weeks ahead incremental death in Delaware (DE). The MSE, MAE, and correlation are reported and best performed method is highlighted in boldface.} 
\end{table}

\begin{figure}[!h] 
  \centering 
\includegraphics[width=0.6\linewidth, page=9]{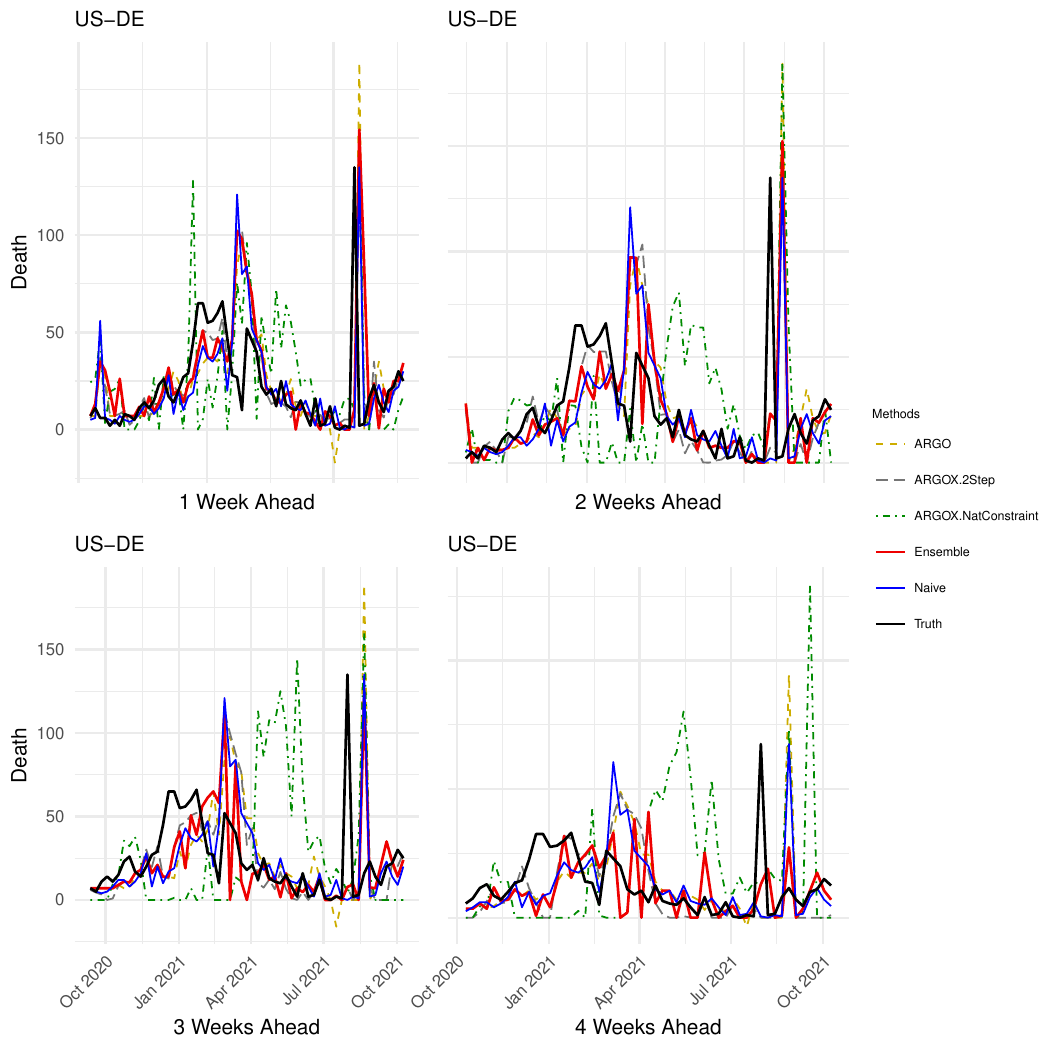} 
\caption{Plots of the COVID-19 1 week (top left), 2 weeks (top right), 3 weeks (bottom left), and 4 weeks (bottom right) ahead estimates for Delaware (DE).}
\end{figure}
\newpage

\begin{table}[ht]
\renewrobustcmd{\bfseries}{\fontseries{b}\selectfont}
\renewrobustcmd{\boldmath}{}
\newrobustcmd{\B}{\bfseries}
\centering
\begin{tabular}{lrrrr}
  \hline
&1 Week Ahead & 2 Weeks Ahead & 3 Weeks Ahead & 4 Weeks Ahead \\ 
    \hline \multicolumn{1}{l}{RMSE} \\
 \hspace{1em} ARGO &  
336.79 & 409.16 & 471.29 & 563.07 \\ 
   \hspace{1em} ARGOX 2Step & 204.84 & 377.77 & 622.75 & 1003.53 \\ 
   \hspace{1em} ARGOX NatConstraint & 169.06 & 270.12 & 487.53 & 856.66 \\ 
   \hspace{1em} Ensemble & 162.17 & 235.75 & 377.36 & 553.32 \\ 
   \hspace{1em} Naive& 189.58 & 313.18 & 421.49 & 529.94 \\ 
   \multicolumn{1}{l}{MAE} \\
  \hspace{1em} ARGO &185.22 & 248.95 & 299.82 & 375.79 \\ 
   \hspace{1em} ARGOX 2Step & 149.02 & 266.21 & 423.51 & 630.38 \\ 
   \hspace{1em} ARGOX NatConstraint & 128.47 & 206.57 & 343.59 & 523.00 \\ 
   \hspace{1em} Ensemble & 115.56 & 167.16 & 237.20 & 353.87 \\ 
   \hspace{1em} Naive& 135.29 & 206.89 & 276.21 & 364.36 \\ 
   \multicolumn{1}{l}{Correlation} \\
 \hspace{1em} ARGO &0.85 & 0.81 & 0.69 & 0.58 \\ 
   \hspace{1em} ARGOX 2Step & 0.96 & 0.92 & 0.81 & 0.37 \\ 
   \hspace{1em} ARGOX NatConstraint & 0.96 & 0.92 & 0.80 & 0.34 \\ 
   \hspace{1em} Ensemble & 0.97 & 0.95 & 0.85 & 0.59 \\ 
   \hspace{1em} Naive& 0.94 & 0.84 & 0.72 & 0.56 \\ 
   \hline
\end{tabular}
\caption{Comparison of different methods for state-level COVID-19 1 to 4 weeks ahead incremental death in Florida (FL). The MSE, MAE, and correlation are reported and best performed method is highlighted in boldface.} 
\end{table}

\begin{figure}[!h] 
  \centering 
\includegraphics[width=0.6\linewidth, page=10]{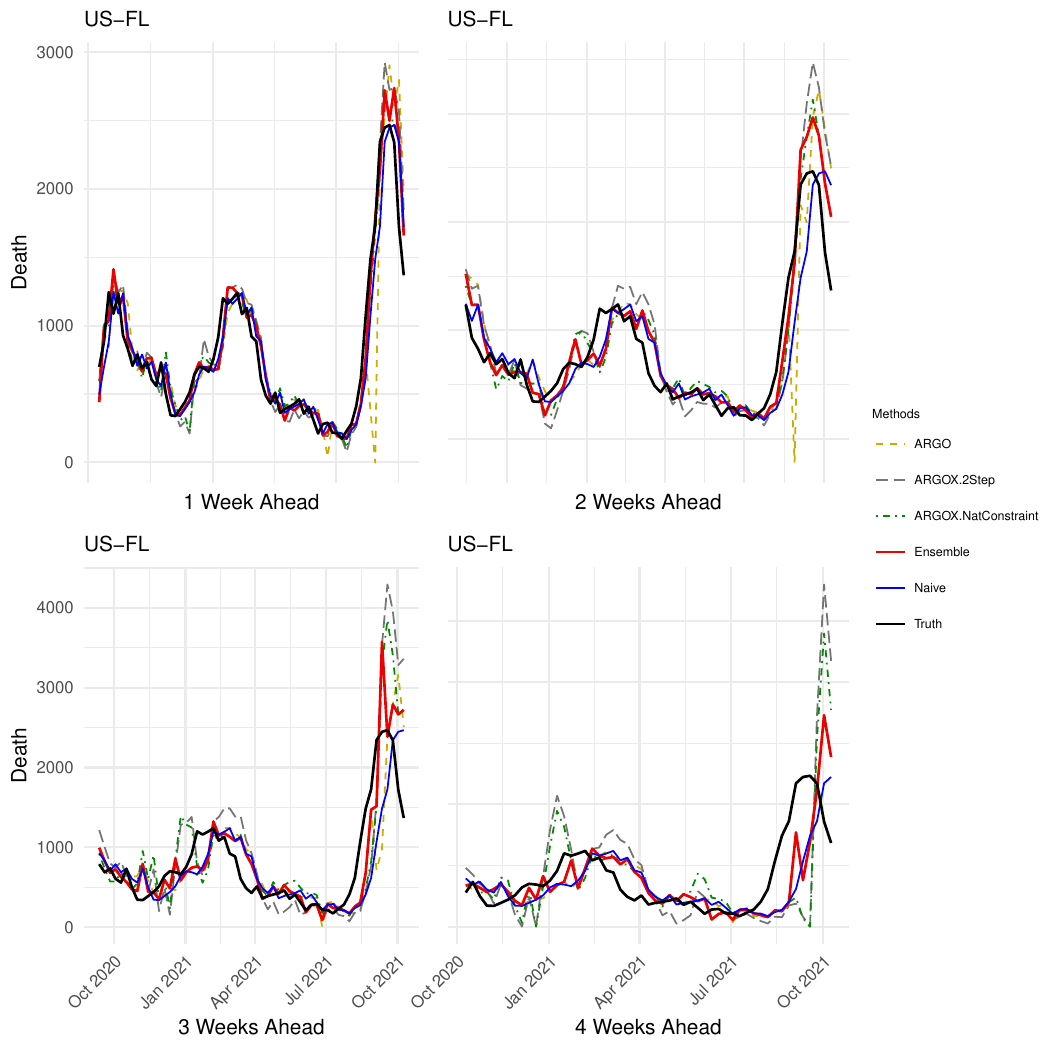} 
\caption{Plots of the COVID-19 1 week (top left), 2 weeks (top right), 3 weeks (bottom left), and 4 weeks (bottom right) ahead estimates for Florida (FL).}
\end{figure}
\newpage

\begin{table}[ht]
\renewrobustcmd{\bfseries}{\fontseries{b}\selectfont}
\renewrobustcmd{\boldmath}{}
\newrobustcmd{\B}{\bfseries}
\centering
\begin{tabular}{lrrrr}
  \hline
&1 Week Ahead & 2 Weeks Ahead & 3 Weeks Ahead & 4 Weeks Ahead \\ 
    \hline \multicolumn{1}{l}{RMSE} \\
 \hspace{1em} ARGO &  
117.07 & 162.23 & 229.27 & 253.81 \\ 
   \hspace{1em} ARGOX 2Step & 119.39 & 203.95 & 268.96 & 334.83 \\ 
   \hspace{1em} ARGOX NatConstraint & 120.73 & 191.81 & 238.50 & 284.16 \\ 
   \hspace{1em} Ensemble & 109.71 & 144.68 & 162.73 & 227.49 \\ 
   \hspace{1em} Naive& 130.96 & 175.65 & 226.76 & 275.53 \\ 
   \multicolumn{1}{l}{MAE} \\
  \hspace{1em} ARGO &80.09 & 105.87 & 152.91 & 182.00 \\ 
   \hspace{1em} ARGOX 2Step & 80.51 & 134.36 & 198.02 & 247.92 \\ 
   \hspace{1em} ARGOX NatConstraint & 80.04 & 130.19 & 167.76 & 214.41 \\ 
   \hspace{1em} Ensemble & 69.63 & 85.01 & 104.26 & 159.24 \\ 
   \hspace{1em} Naive& 87.20 & 117.39 & 156.63 & 196.66 \\ 
   \multicolumn{1}{l}{Correlation} \\
 \hspace{1em} ARGO &0.89 & 0.82 & 0.70 & 0.63 \\ 
   \hspace{1em} ARGOX 2Step & 0.90 & 0.77 & 0.70 & 0.55 \\ 
   \hspace{1em} ARGOX NatConstraint & 0.88 & 0.72 & 0.63 & 0.49 \\ 
   \hspace{1em} Ensemble & 0.90 & 0.84 & 0.82 & 0.68 \\ 
   \hspace{1em} Naive& 0.87 & 0.76 & 0.61 & 0.46 \\ 
   \hline
\end{tabular}
\caption{Comparison of different methods for state-level COVID-19 1 to 4 weeks ahead incremental death in Georgia (GA). The MSE, MAE, and correlation are reported and best performed method is highlighted in boldface.} 
\end{table}

\begin{figure}[!h] 
  \centering 
\includegraphics[width=0.6\linewidth, page=11]{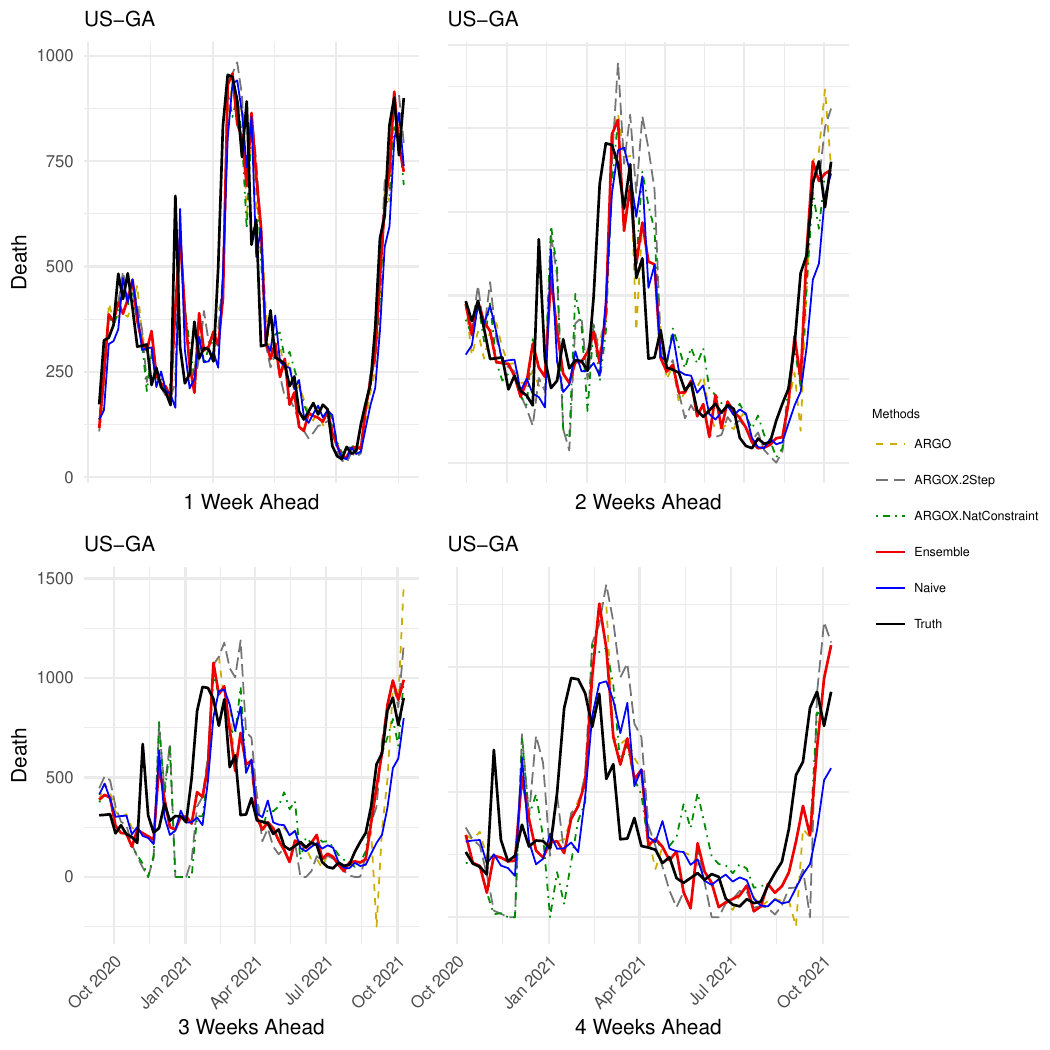} 
\caption{Plots of the COVID-19 1 week (top left), 2 weeks (top right), 3 weeks (bottom left), and 4 weeks (bottom right) ahead estimates for Georgia (GA).}
\end{figure}
\newpage

\begin{table}[ht]
\renewrobustcmd{\bfseries}{\fontseries{b}\selectfont}
\renewrobustcmd{\boldmath}{}
\newrobustcmd{\B}{\bfseries}
\centering
\begin{tabular}{lrrrr}
  \hline
&1 Week Ahead & 2 Weeks Ahead & 3 Weeks Ahead & 4 Weeks Ahead \\ 
    \hline \multicolumn{1}{l}{RMSE} \\
 \hspace{1em} ARGO &  
12.37 & 15.56 & 17.08 & 16.84 \\ 
   \hspace{1em} ARGOX 2Step & 11.36 & 12.19 & 12.84 & 14.48 \\ 
   \hspace{1em} ARGOX NatConstraint & 12.37 & 15.56 & 17.08 & 16.84 \\ 
   \hspace{1em} Ensemble & 12.12 & 12.61 & 13.14 & 15.72 \\ 
   \hspace{1em} Naive& 12.23 & 13.46 & 14.85 & 16.91 \\ 
   \multicolumn{1}{l}{MAE} \\
  \hspace{1em} ARGO &6.98 & 9.15 & 10.46 & 10.53 \\ 
   \hspace{1em} ARGOX 2Step & 5.41 & 6.34 & 7.17 & 8.54 \\ 
   \hspace{1em} ARGOX NatConstraint & 6.98 & 9.15 & 10.46 & 10.53 \\ 
   \hspace{1em} Ensemble & 6.85 & 7.01 & 7.68 & 9.79 \\ 
   \hspace{1em} Naive& 6.66 & 7.39 & 8.89 & 9.75 \\ 
   \multicolumn{1}{l}{Correlation} \\
 \hspace{1em} ARGO &0.59 & 0.34 & 0.15 & 0.17 \\ 
   \hspace{1em} ARGOX 2Step & 0.66 & 0.63 & 0.57 & 0.39 \\ 
   \hspace{1em} ARGOX NatConstraint & 0.59 & 0.34 & 0.15 & 0.17 \\ 
   \hspace{1em} Ensemble & 0.63 & 0.62 & 0.56 & 0.38 \\ 
   \hspace{1em} Naive& 0.60 & 0.50 & 0.39 & 0.21 \\ 
   \hline
\end{tabular}
\caption{Comparison of different methods for state-level COVID-19 1 to 4 weeks ahead incremental death in Hawaii (HI). The MSE, MAE, and correlation are reported and best performed method is highlighted in boldface.} 
\end{table}

\begin{figure}[!h] 
  \centering 
\includegraphics[width=0.6\linewidth, page=12]{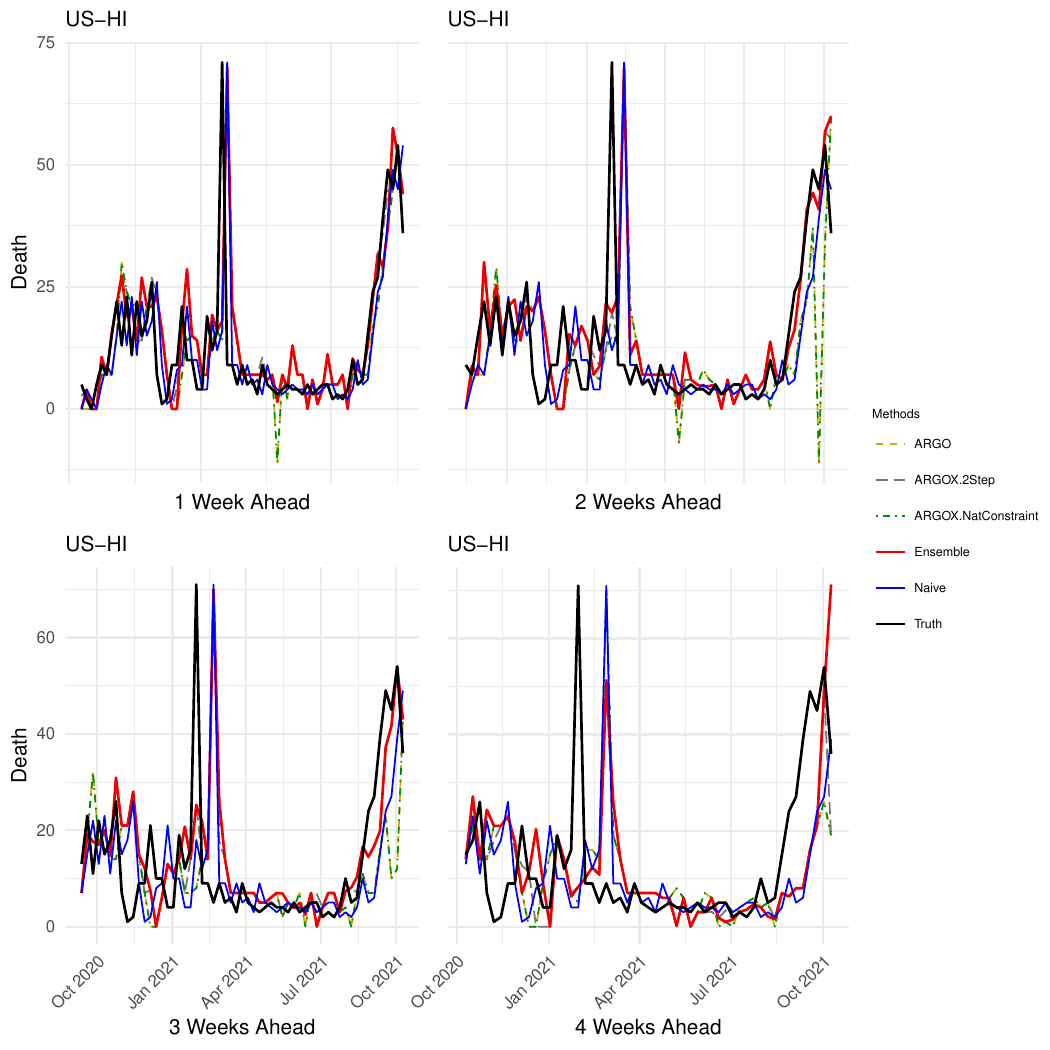} 
\caption{Plots of the COVID-19 1 week (top left), 2 weeks (top right), 3 weeks (bottom left), and 4 weeks (bottom right) ahead estimates for Hawaii (HI).}
\end{figure}
\newpage

\begin{table}[ht]
\renewrobustcmd{\bfseries}{\fontseries{b}\selectfont}
\renewrobustcmd{\boldmath}{}
\newrobustcmd{\B}{\bfseries}
\centering
\begin{tabular}{lrrrr}
  \hline
&1 Week Ahead & 2 Weeks Ahead & 3 Weeks Ahead & 4 Weeks Ahead \\ 
    \hline \multicolumn{1}{l}{RMSE} \\
 \hspace{1em} ARGO &  
71.96 & 78.46 & 90.22 & 117.39 \\ 
   \hspace{1em} ARGOX 2Step & 86.29 & 105.25 & 137.77 & 240.74 \\ 
   \hspace{1em} ARGOX NatConstraint & 73.07 & 98.38 & 118.05 & 174.90 \\ 
   \hspace{1em} Ensemble & 63.60 & 70.71 & 77.19 & 111.52 \\ 
   \hspace{1em} Naive& 72.52 & 90.20 & 95.31 & 109.89 \\ 
   \multicolumn{1}{l}{MAE} \\
  \hspace{1em} ARGO &36.14 & 44.61 & 55.12 & 66.34 \\ 
   \hspace{1em} ARGOX 2Step & 43.45 & 49.59 & 67.69 & 114.26 \\ 
   \hspace{1em} ARGOX NatConstraint & 42.32 & 54.95 & 75.55 & 110.54 \\ 
   \hspace{1em} Ensemble & 34.30 & 37.02 & 43.60 & 60.25 \\ 
   \hspace{1em} Naive& 34.78 & 43.05 & 53.12 & 64.06 \\ 
   \multicolumn{1}{l}{Correlation} \\
 \hspace{1em} ARGO &0.72 & 0.70 & 0.64 & 0.48 \\ 
   \hspace{1em} ARGOX 2Step & 0.71 & 0.66 & 0.61 & 0.49 \\ 
   \hspace{1em} ARGOX NatConstraint & 0.75 & 0.59 & 0.51 & 0.25 \\ 
   \hspace{1em} Ensemble & 0.81 & 0.76 & 0.76 & 0.64 \\ 
   \hspace{1em} Naive& 0.74 & 0.61 & 0.59 & 0.49 \\ 
   \hline
\end{tabular}
\caption{Comparison of different methods for state-level COVID-19 1 to 4 weeks ahead incremental death in Iowa (IA). The MSE, MAE, and correlation are reported and best performed method is highlighted in boldface.} 
\end{table}

\begin{figure}[!h] 
  \centering 
\includegraphics[width=0.6\linewidth, page=13]{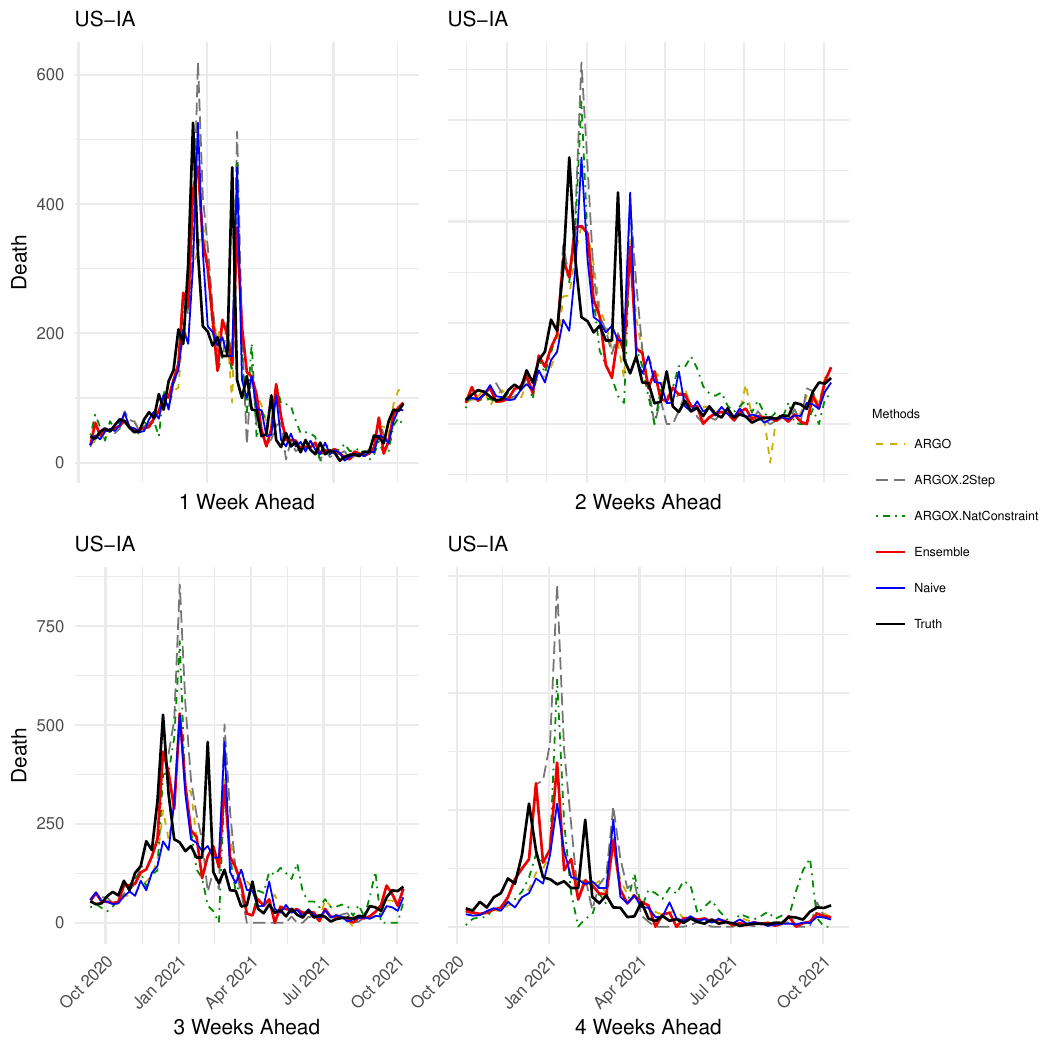} 
\caption{Plots of the COVID-19 1 week (top left), 2 weeks (top right), 3 weeks (bottom left), and 4 weeks (bottom right) ahead estimates for Iowa (IA).}
\end{figure}
\newpage

\begin{table}[ht]
\renewrobustcmd{\bfseries}{\fontseries{b}\selectfont}
\renewrobustcmd{\boldmath}{}
\newrobustcmd{\B}{\bfseries}
\centering
\begin{tabular}{lrrrr}
  \hline
&1 Week Ahead & 2 Weeks Ahead & 3 Weeks Ahead & 4 Weeks Ahead \\ 
    \hline \multicolumn{1}{l}{RMSE} \\
 \hspace{1em} ARGO &  
24.41 & 31.66 & 35.46 & 40.57 \\ 
   \hspace{1em} ARGOX 2Step & 22.39 & 25.41 & 31.87 & 70.04 \\ 
   \hspace{1em} ARGOX NatConstraint & 31.45 & 37.98 & 60.45 & 78.91 \\ 
   \hspace{1em} Ensemble & 21.58 & 21.16 & 27.35 & 33.50 \\ 
   \hspace{1em} Naive& 22.76 & 25.02 & 32.75 & 40.89 \\ 
   \multicolumn{1}{l}{MAE} \\
  \hspace{1em} ARGO &15.31 & 20.84 & 22.56 & 28.11 \\ 
   \hspace{1em} ARGOX 2Step & 14.51 & 16.32 & 22.75 & 41.86 \\ 
   \hspace{1em} ARGOX NatConstraint & 22.20 & 28.55 & 42.08 & 58.78 \\ 
   \hspace{1em} Ensemble & 13.93 & 13.72 & 16.61 & 23.13 \\ 
   \hspace{1em} Naive& 14.45 & 16.98 & 21.88 & 27.26 \\ 
   \multicolumn{1}{l}{Correlation} \\
 \hspace{1em} ARGO &0.83 & 0.71 & 0.67 & 0.56 \\ 
   \hspace{1em} ARGOX 2Step & 0.86 & 0.85 & 0.82 & 0.55 \\ 
   \hspace{1em} ARGOX NatConstraint & 0.70 & 0.57 & 0.16 & 0.14 \\ 
   \hspace{1em} Ensemble & 0.87 & 0.88 & 0.81 & 0.72 \\ 
   \hspace{1em} Naive& 0.85 & 0.82 & 0.70 & 0.53 \\ 
   \hline
\end{tabular}
\caption{Comparison of different methods for state-level COVID-19 1 to 4 weeks ahead incremental death in Idaho (ID). The MSE, MAE, and correlation are reported and best performed method is highlighted in boldface.} 
\end{table}

\begin{figure}[!h] 
  \centering 
\includegraphics[width=0.6\linewidth, page=14]{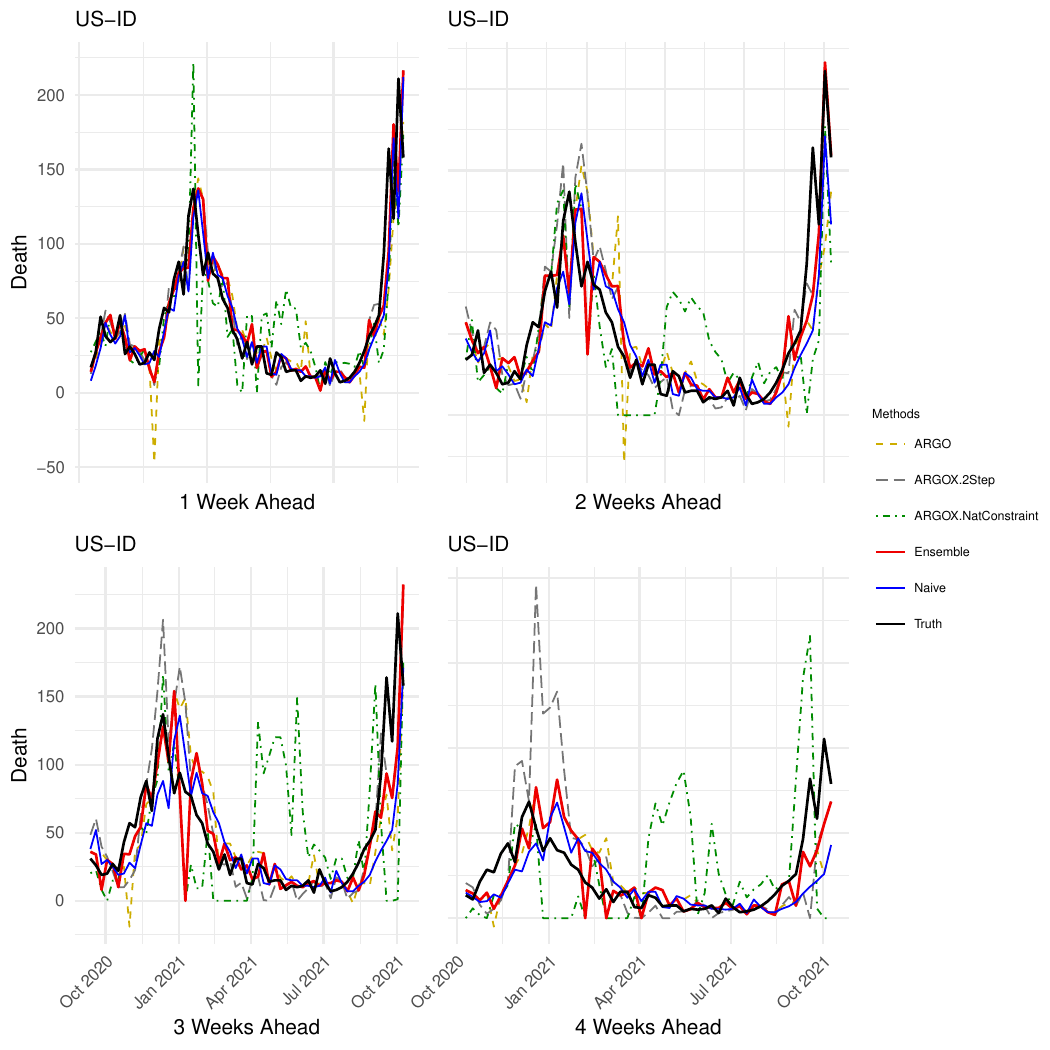} 
\caption{Plots of the COVID-19 1 week (top left), 2 weeks (top right), 3 weeks (bottom left), and 4 weeks (bottom right) ahead estimates for Idaho (ID).}
\end{figure}
\newpage

\begin{table}[ht]
\renewrobustcmd{\bfseries}{\fontseries{b}\selectfont}
\renewrobustcmd{\boldmath}{}
\newrobustcmd{\B}{\bfseries}
\centering
\begin{tabular}{lrrrr}
  \hline
&1 Week Ahead & 2 Weeks Ahead & 3 Weeks Ahead & 4 Weeks Ahead \\ 
    \hline \multicolumn{1}{l}{RMSE} \\
 \hspace{1em} ARGO &  
93.20 & 126.32 & 196.58 & 249.61 \\ 
   \hspace{1em} ARGOX 2Step & 106.04 & 178.81 & 331.37 & 726.77 \\ 
   \hspace{1em} ARGOX NatConstraint & 81.56 & 128.50 & 259.79 & 559.56 \\ 
   \hspace{1em} Ensemble & 80.80 & 114.90 & 144.90 & 177.10 \\ 
   \hspace{1em} Naive& 80.40 & 135.82 & 191.37 & 245.79 \\ 
   \multicolumn{1}{l}{MAE} \\
  \hspace{1em} ARGO &60.40 & 82.23 & 136.35 & 183.09 \\ 
   \hspace{1em} ARGOX 2Step & 65.71 & 115.34 & 200.42 & 376.82 \\ 
   \hspace{1em} ARGOX NatConstraint & 55.74 & 82.28 & 139.31 & 279.90 \\ 
   \hspace{1em} Ensemble & 50.10 & 72.20 & 93.19 & 127.58 \\ 
   \hspace{1em} Naive& 51.00 & 90.74 & 131.07 & 173.11 \\ 
   \multicolumn{1}{l}{Correlation} \\
 \hspace{1em} ARGO &0.96 & 0.93 & 0.88 & 0.76 \\ 
   \hspace{1em} ARGOX 2Step & 0.97 & 0.94 & 0.89 & 0.76 \\ 
   \hspace{1em} ARGOX NatConstraint & 0.97 & 0.94 & 0.86 & 0.73 \\ 
   \hspace{1em} Ensemble & 0.97 & 0.95 & 0.93 & 0.88 \\ 
   \hspace{1em} Naive& 0.96 & 0.90 & 0.82 & 0.71 \\ 
   \hline
\end{tabular}
\caption{Comparison of different methods for state-level COVID-19 1 to 4 weeks ahead incremental death in Illinois (IL). The MSE, MAE, and correlation are reported and best performed method is highlighted in boldface.} 
\end{table}

\begin{figure}[!h] 
  \centering 
\includegraphics[width=0.6\linewidth, page=15]{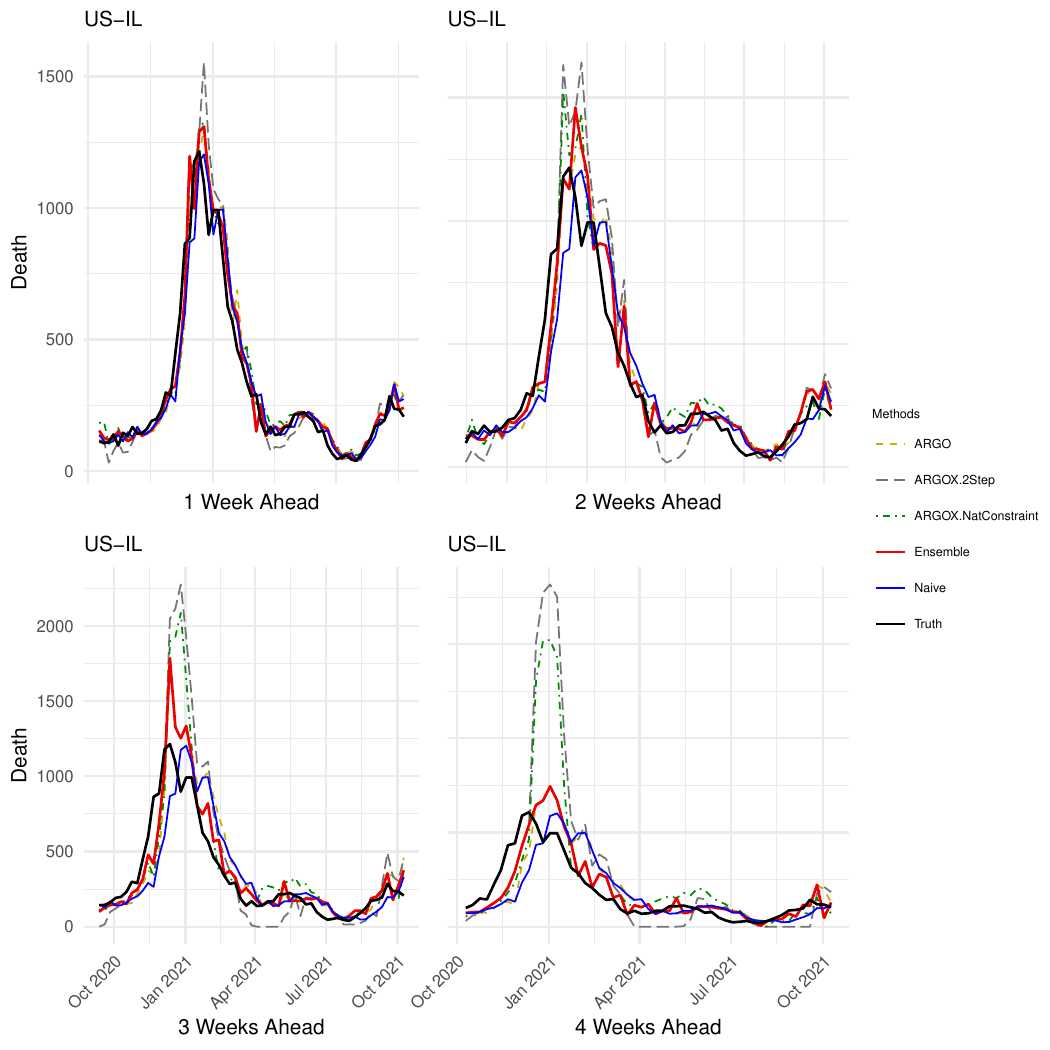} 
\caption{Plots of the COVID-19 1 week (top left), 2 weeks (top right), 3 weeks (bottom left), and 4 weeks (bottom right) ahead estimates for Illinois (IL).}
\end{figure}
\newpage

\begin{table}[ht]
\renewrobustcmd{\bfseries}{\fontseries{b}\selectfont}
\renewrobustcmd{\boldmath}{}
\newrobustcmd{\B}{\bfseries}
\centering
\begin{tabular}{lrrrr}
  \hline
&1 Week Ahead & 2 Weeks Ahead & 3 Weeks Ahead & 4 Weeks Ahead \\ 
    \hline \multicolumn{1}{l}{RMSE} \\
 \hspace{1em} ARGO &  
155.29 & 186.05 & 212.60 & 268.18 \\ 
   \hspace{1em} ARGOX 2Step & 190.70 & 216.88 & 221.87 & 324.23 \\ 
   \hspace{1em} ARGOX NatConstraint & 180.71 & 195.64 & 196.47 & 245.83 \\ 
   \hspace{1em} Ensemble & 156.74 & 176.05 & 168.57 & 208.61 \\ 
   \hspace{1em} Naive& 189.95 & 226.54 & 243.79 & 271.18 \\ 
   \multicolumn{1}{l}{MAE} \\
  \hspace{1em} ARGO &68.58 & 91.67 & 111.91 & 157.49 \\ 
   \hspace{1em} ARGOX 2Step & 80.46 & 96.98 & 109.27 & 205.65 \\ 
   \hspace{1em} ARGOX NatConstraint & 83.63 & 98.87 & 108.81 & 153.51 \\ 
   \hspace{1em} Ensemble & 68.60 & 75.94 & 77.81 & 105.62 \\ 
   \hspace{1em} Naive& 72.20 & 92.05 & 115.32 & 141.72 \\ 
   \multicolumn{1}{l}{Correlation} \\
 \hspace{1em} ARGO &0.74 & 0.65 & 0.60 & 0.48 \\ 
   \hspace{1em} ARGOX 2Step & 0.70 & 0.66 & 0.72 & 0.68 \\ 
   \hspace{1em} ARGOX NatConstraint & 0.67 & 0.60 & 0.64 & 0.54 \\ 
   \hspace{1em} Ensemble & 0.74 & 0.69 & 0.71 & 0.62 \\ 
   \hspace{1em} Naive& 0.67 & 0.53 & 0.48 & 0.38 \\ 
   \hline
\end{tabular}
\caption{Comparison of different methods for state-level COVID-19 1 to 4 weeks ahead incremental death in Indiana (IN). The MSE, MAE, and correlation are reported and best performed method is highlighted in boldface.} 
\end{table}

\begin{figure}[!h] 
  \centering 
\includegraphics[width=0.6\linewidth, page=16]{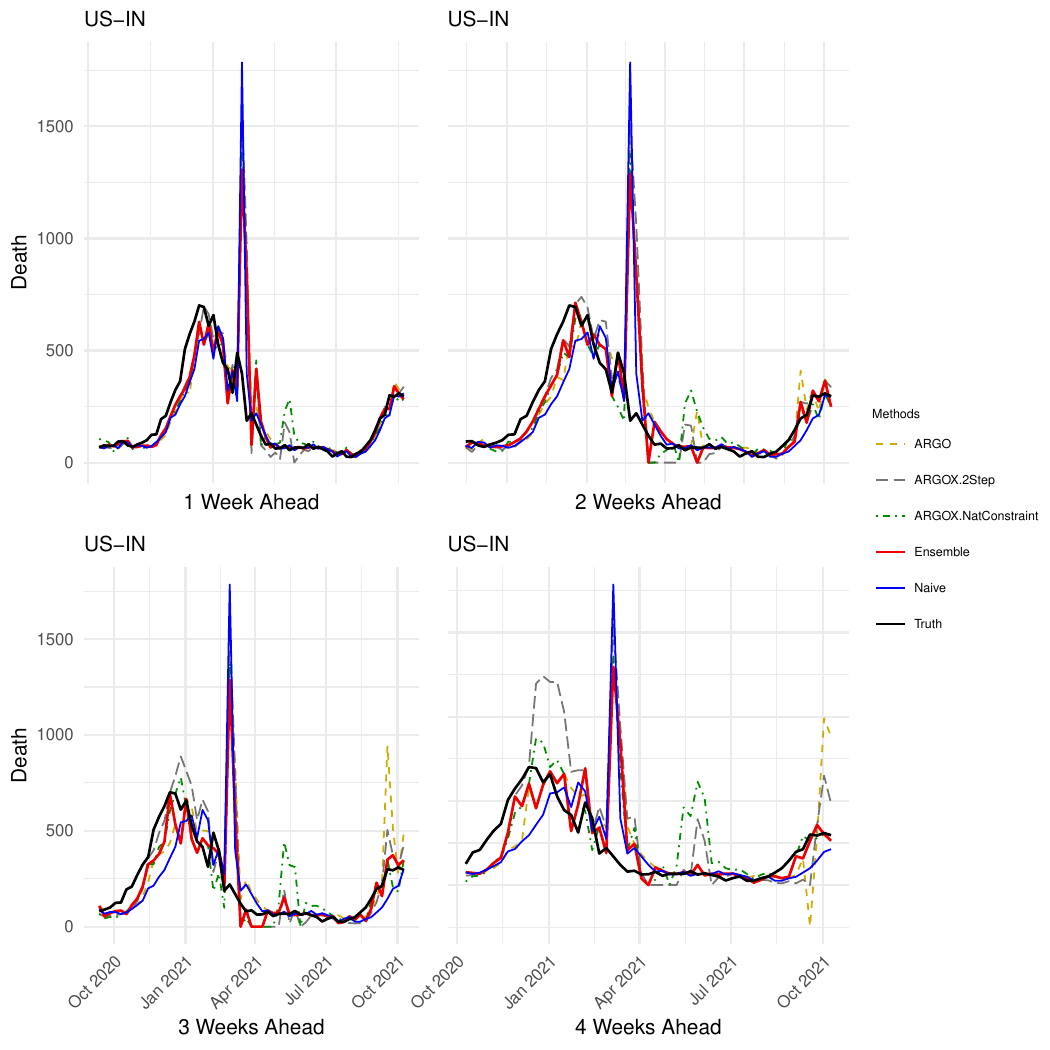} 
\caption{Plots of the COVID-19 1 week (top left), 2 weeks (top right), 3 weeks (bottom left), and 4 weeks (bottom right) ahead estimates for Indiana (IN).}
\end{figure}
\newpage

\begin{table}[ht]
\renewrobustcmd{\bfseries}{\fontseries{b}\selectfont}
\renewrobustcmd{\boldmath}{}
\newrobustcmd{\B}{\bfseries}
\centering
\begin{tabular}{lrrrr}
  \hline
&1 Week Ahead & 2 Weeks Ahead & 3 Weeks Ahead & 4 Weeks Ahead \\ 
    \hline \multicolumn{1}{l}{RMSE} \\
 \hspace{1em} ARGO &  
56.83 & 67.34 & 72.82 & 79.02 \\ 
   \hspace{1em} ARGOX 2Step & 59.31 & 76.82 & 79.39 & 120.84 \\ 
   \hspace{1em} ARGOX NatConstraint & 60.77 & 74.88 & 86.21 & 108.72 \\ 
   \hspace{1em} Ensemble & 47.65 & 60.67 & 65.80 & 63.25 \\ 
   \hspace{1em} Naive& 62.58 & 69.24 & 85.85 & 86.46 \\ 
   \multicolumn{1}{l}{MAE} \\
  \hspace{1em} ARGO &35.52 & 42.20 & 46.21 & 51.64 \\ 
   \hspace{1em} ARGOX 2Step & 35.70 & 45.40 & 53.22 & 80.53 \\ 
   \hspace{1em} ARGOX NatConstraint & 39.29 & 55.04 & 68.65 & 83.53 \\ 
   \hspace{1em} Ensemble & 29.53 & 37.19 & 42.13 & 43.93 \\ 
   \hspace{1em} Naive& 39.57 & 45.11 & 56.28 & 58.49 \\ 
   \multicolumn{1}{l}{Correlation} \\
 \hspace{1em} ARGO &0.82 & 0.76 & 0.76 & 0.79 \\ 
   \hspace{1em} ARGOX 2Step & 0.83 & 0.78 & 0.77 & 0.76 \\ 
   \hspace{1em} ARGOX NatConstraint & 0.78 & 0.67 & 0.52 & 0.39 \\ 
   \hspace{1em} Ensemble & 0.87 & 0.80 & 0.76 & 0.80 \\ 
   \hspace{1em} Naive& 0.77 & 0.73 & 0.60 & 0.61 \\ 
   \hline
\end{tabular}
\caption{Comparison of different methods for state-level COVID-19 1 to 4 weeks ahead incremental death in Kansas (KS). The MSE, MAE, and correlation are reported and best performed method is highlighted in boldface.} 
\end{table}

\begin{figure}[!h] 
  \centering 
\includegraphics[width=0.6\linewidth, page=17]{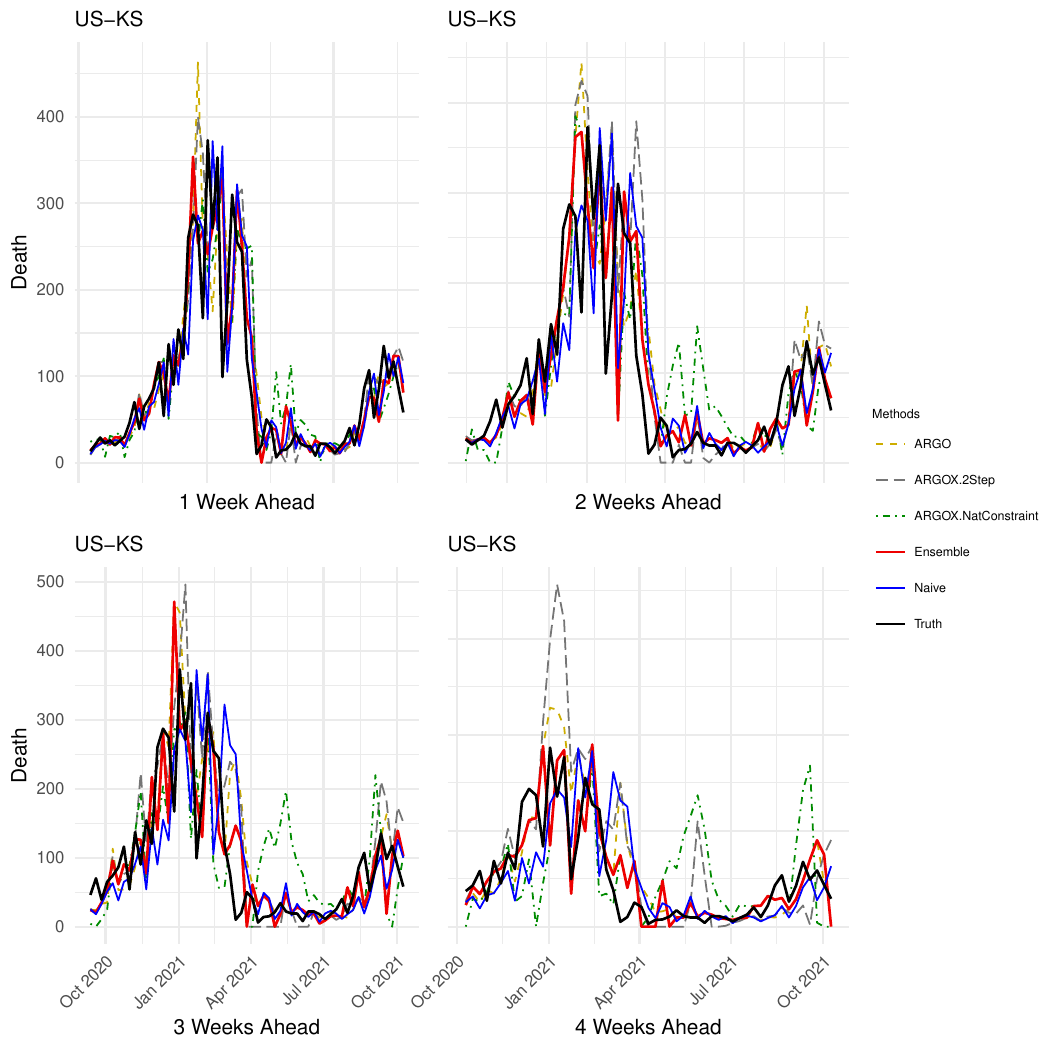} 
\caption{Plots of the COVID-19 1 week (top left), 2 weeks (top right), 3 weeks (bottom left), and 4 weeks (bottom right) ahead estimates for Kansas (KS).}
\end{figure}
\newpage

\begin{table}[ht]
\renewrobustcmd{\bfseries}{\fontseries{b}\selectfont}
\renewrobustcmd{\boldmath}{}
\newrobustcmd{\B}{\bfseries}
\centering
\begin{tabular}{lrrrr}
  \hline
&1 Week Ahead & 2 Weeks Ahead & 3 Weeks Ahead & 4 Weeks Ahead \\ 
    \hline \multicolumn{1}{l}{RMSE} \\
 \hspace{1em} ARGO &  
109.01 & 159.93 & 121.45 & 145.58 \\ 
   \hspace{1em} ARGOX 2Step & 102.91 & 121.52 & 126.54 & 132.04 \\ 
   \hspace{1em} ARGOX NatConstraint & 106.88 & 138.09 & 175.75 & 182.99 \\ 
   \hspace{1em} Ensemble & 99.48 & 110.72 & 114.87 & 106.74 \\ 
   \hspace{1em} Naive& 101.12 & 120.15 & 127.37 & 142.19 \\ 
   \multicolumn{1}{l}{MAE} \\
  \hspace{1em} ARGO &54.55 & 74.72 & 70.95 & 90.70 \\ 
   \hspace{1em} ARGOX 2Step & 48.64 & 63.38 & 76.14 & 84.65 \\ 
   \hspace{1em} ARGOX NatConstraint & 58.72 & 82.50 & 109.62 & 123.48 \\ 
   \hspace{1em} Ensemble & 50.26 & 57.10 & 64.13 & 62.55 \\ 
   \hspace{1em} Naive& 47.80 & 59.66 & 69.68 & 86.13 \\ 
   \multicolumn{1}{l}{Correlation} \\
 \hspace{1em} ARGO &0.52 & 0.23 & 0.40 & 0.31 \\ 
   \hspace{1em} ARGOX 2Step & 0.58 & 0.45 & 0.46 & 0.45 \\ 
   \hspace{1em} ARGOX NatConstraint & 0.52 & 0.23 & 0.09 & 0.09 \\ 
   \hspace{1em} Ensemble & 0.60 & 0.48 & 0.48 & 0.53 \\ 
   \hspace{1em} Naive& 0.59 & 0.42 & 0.36 & 0.23 \\ 
   \hline
\end{tabular}
\caption{Comparison of different methods for state-level COVID-19 1 to 4 weeks ahead incremental death in Kentucky (KY). The MSE, MAE, and correlation are reported and best performed method is highlighted in boldface.} 
\end{table}

\begin{figure}[!h] 
  \centering 
\includegraphics[width=0.6\linewidth, page=18]{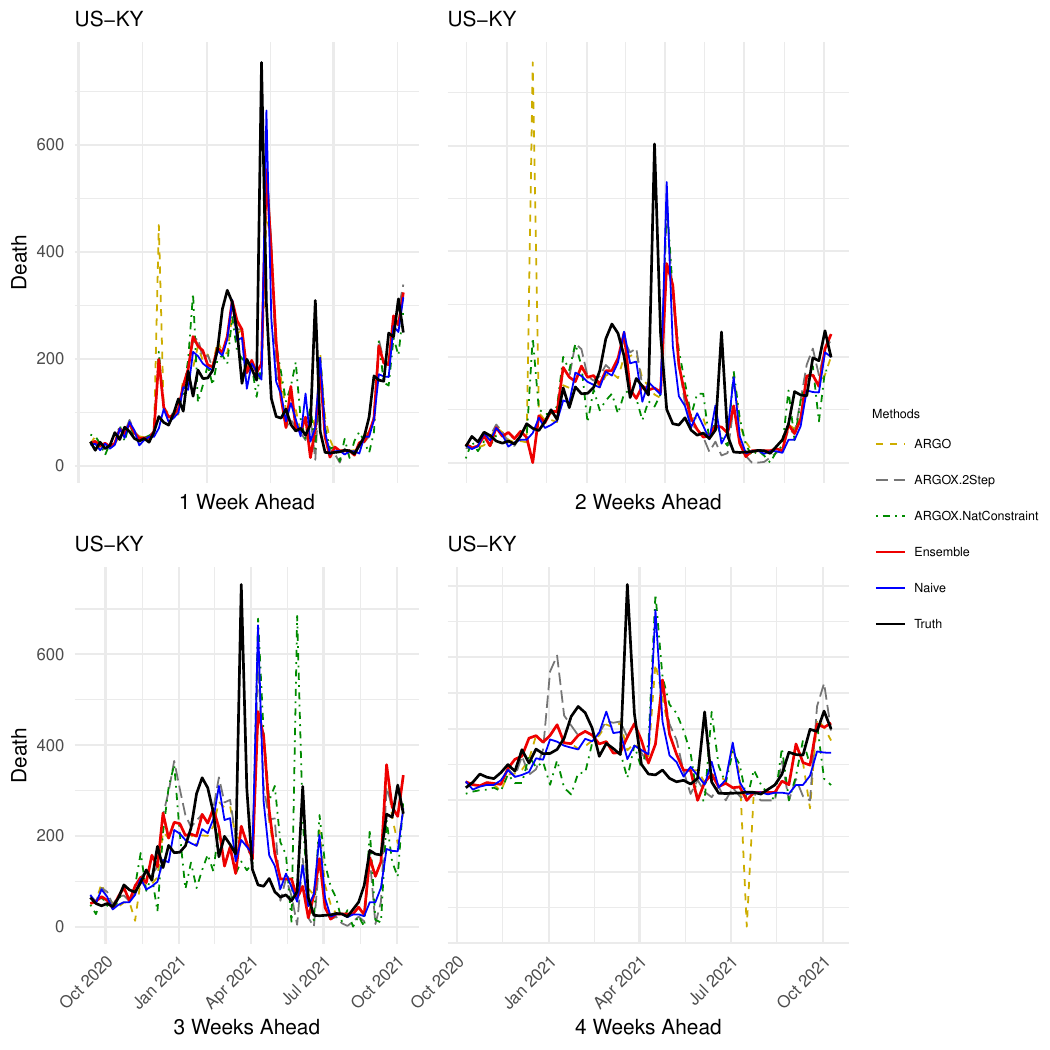} 
\caption{Plots of the COVID-19 1 week (top left), 2 weeks (top right), 3 weeks (bottom left), and 4 weeks (bottom right) ahead estimates for Kentucky (KY).}
\end{figure}
\newpage

\begin{table}[ht]
\renewrobustcmd{\bfseries}{\fontseries{b}\selectfont}
\renewrobustcmd{\boldmath}{}
\newrobustcmd{\B}{\bfseries}
\centering
\begin{tabular}{lrrrr}
  \hline
&1 Week Ahead & 2 Weeks Ahead & 3 Weeks Ahead & 4 Weeks Ahead \\ 
    \hline \multicolumn{1}{l}{RMSE} \\
 \hspace{1em} ARGO &  
49.97 & 86.52 & 171.76 & 267.26 \\ 
   \hspace{1em} ARGOX 2Step & 49.88 & 93.87 & 158.02 & 189.82 \\ 
   \hspace{1em} ARGOX NatConstraint & 49.73 & 71.77 & 128.16 & 128.25 \\ 
   \hspace{1em} Ensemble & 40.83 & 55.96 & 118.68 & 118.89 \\ 
   \hspace{1em} Naive& 50.15 & 67.27 & 93.41 & 110.63 \\ 
   \multicolumn{1}{l}{MAE} \\
  \hspace{1em} ARGO &33.89 & 54.03 & 94.37 & 140.58 \\ 
   \hspace{1em} ARGOX 2Step & 34.31 & 58.91 & 96.11 & 134.13 \\ 
   \hspace{1em} ARGOX NatConstraint & 38.11 & 51.11 & 82.20 & 95.80 \\ 
   \hspace{1em} Ensemble & 27.27 & 34.87 & 63.37 & 74.11 \\ 
   \hspace{1em} Naive& 35.32 & 45.80 & 63.37 & 77.26 \\ 
   \multicolumn{1}{l}{Correlation} \\
 \hspace{1em} ARGO &0.92 & 0.87 & 0.69 & 0.54 \\ 
   \hspace{1em} ARGOX 2Step & 0.93 & 0.85 & 0.72 & 0.53 \\ 
   \hspace{1em} ARGOX NatConstraint & 0.89 & 0.82 & 0.61 & 0.46 \\ 
   \hspace{1em} Ensemble & 0.94 & 0.92 & 0.78 & 0.63 \\ 
   \hspace{1em} Naive& 0.89 & 0.82 & 0.66 & 0.54 \\ 
   \hline
\end{tabular}
\caption{Comparison of different methods for state-level COVID-19 1 to 4 weeks ahead incremental death in Louisiana (LA). The MSE, MAE, and correlation are reported and best performed method is highlighted in boldface.} 
\end{table}

\begin{figure}[!h] 
  \centering 
\includegraphics[width=0.6\linewidth, page=19]{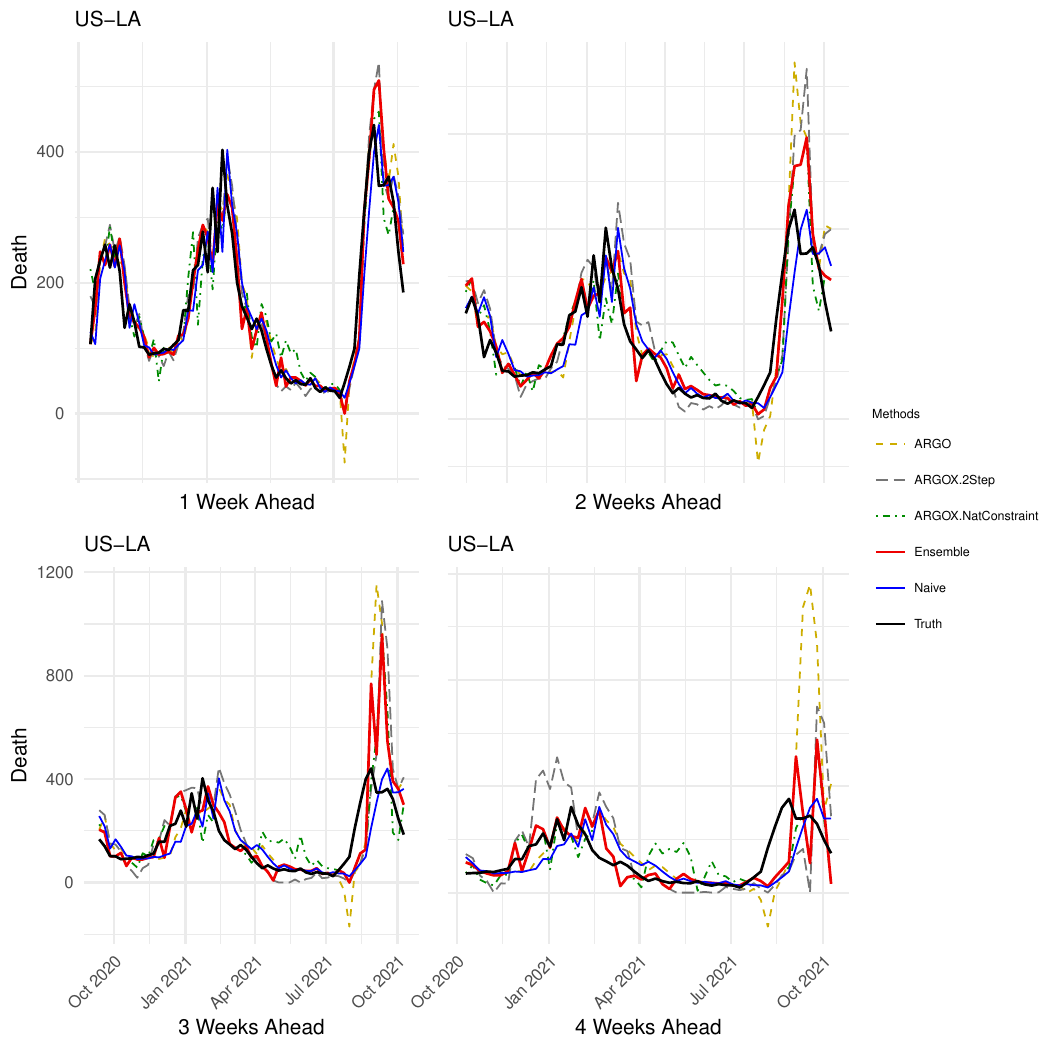} 
\caption{Plots of the COVID-19 1 week (top left), 2 weeks (top right), 3 weeks (bottom left), and 4 weeks (bottom right) ahead estimates for Louisiana (LA).}
\end{figure}
\newpage

\begin{table}[ht]
\renewrobustcmd{\bfseries}{\fontseries{b}\selectfont}
\renewrobustcmd{\boldmath}{}
\newrobustcmd{\B}{\bfseries}
\centering
\begin{tabular}{lrrrr}
  \hline
&1 Week Ahead & 2 Weeks Ahead & 3 Weeks Ahead & 4 Weeks Ahead \\ 
    \hline \multicolumn{1}{l}{RMSE} \\
 \hspace{1em} ARGO &  
50.82 & 72.55 & 81.25 & 98.91 \\ 
   \hspace{1em} ARGOX 2Step & 47.67 & 68.93 & 74.30 & 114.95 \\ 
   \hspace{1em} ARGOX NatConstraint & 56.79 & 66.43 & 93.37 & 113.33 \\ 
   \hspace{1em} Ensemble & 36.42 & 43.63 & 43.71 & 54.23 \\ 
   \hspace{1em} Naive& 44.56 & 63.25 & 79.82 & 102.37 \\ 
   \multicolumn{1}{l}{MAE} \\
  \hspace{1em} ARGO &33.05 & 47.30 & 58.51 & 73.08 \\ 
   \hspace{1em} ARGOX 2Step & 35.05 & 45.67 & 53.75 & 79.19 \\ 
   \hspace{1em} ARGOX NatConstraint & 37.69 & 45.55 & 69.38 & 85.26 \\ 
   \hspace{1em} Ensemble & 25.44 & 25.95 & 30.75 & 39.00 \\ 
   \hspace{1em} Naive& 30.45 & 39.64 & 56.89 & 75.57 \\ 
   \multicolumn{1}{l}{Correlation} \\
 \hspace{1em} ARGO &0.94 & 0.88 & 0.88 & 0.81 \\ 
   \hspace{1em} ARGOX 2Step & 0.96 & 0.93 & 0.94 & 0.95 \\ 
   \hspace{1em} ARGOX NatConstraint & 0.92 & 0.90 & 0.80 & 0.72 \\ 
   \hspace{1em} Ensemble & 0.97 & 0.96 & 0.96 & 0.95 \\ 
   \hspace{1em} Naive& 0.95 & 0.91 & 0.86 & 0.79 \\ 
   \hline
\end{tabular}
\caption{Comparison of different methods for state-level COVID-19 1 to 4 weeks ahead incremental death in Massachusetts (MA). The MSE, MAE, and correlation are reported and best performed method is highlighted in boldface.} 
\end{table}

\begin{figure}[!h] 
  \centering 
\includegraphics[width=0.6\linewidth, page=20]{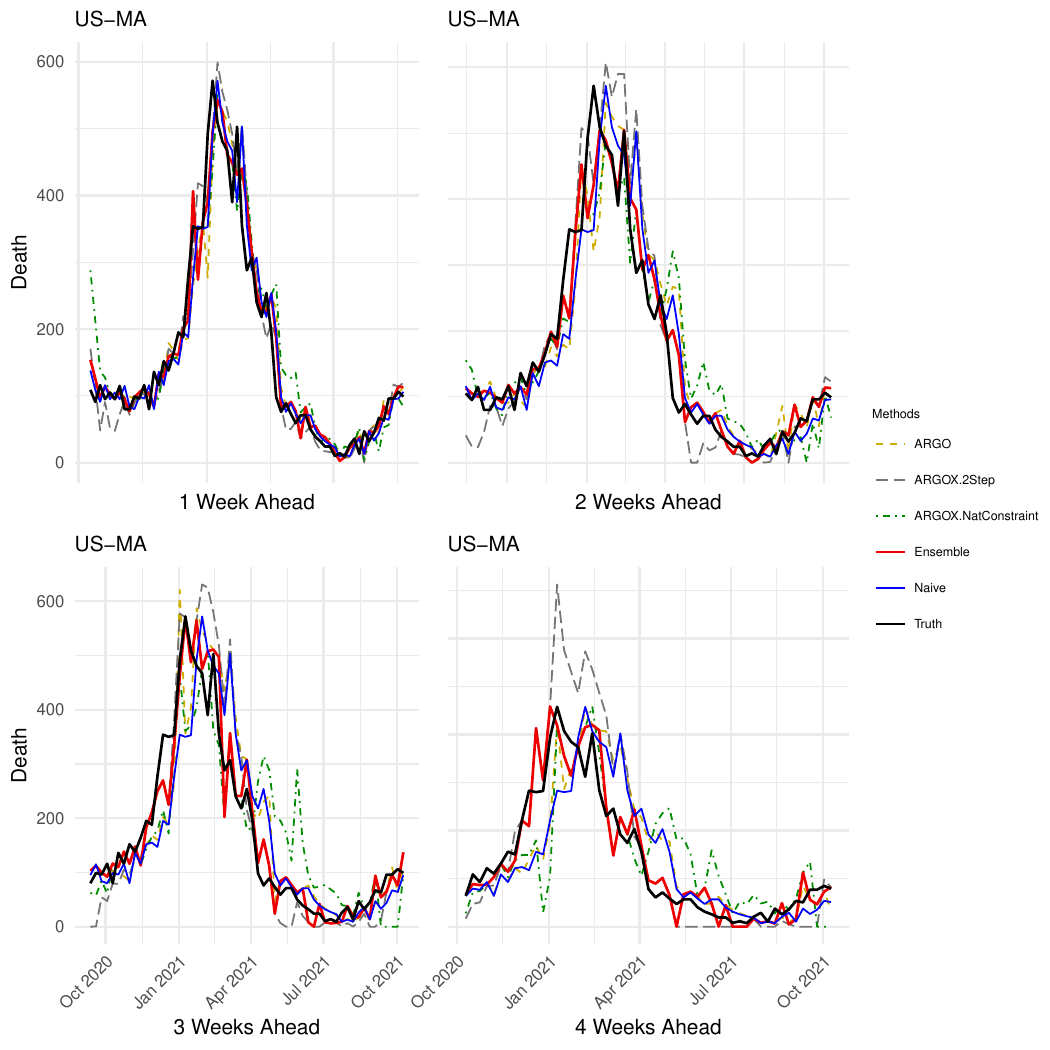} 
\caption{Plots of the COVID-19 1 week (top left), 2 weeks (top right), 3 weeks (bottom left), and 4 weeks (bottom right) ahead estimates for Massachusetts (MA).}
\end{figure}
\newpage

\begin{table}[ht]
\renewrobustcmd{\bfseries}{\fontseries{b}\selectfont}
\renewrobustcmd{\boldmath}{}
\newrobustcmd{\B}{\bfseries}
\centering
\begin{tabular}{lrrrr}
  \hline
&1 Week Ahead & 2 Weeks Ahead & 3 Weeks Ahead & 4 Weeks Ahead \\ 
    \hline \multicolumn{1}{l}{RMSE} \\
 \hspace{1em} ARGO &  
89.96 & 94.19 & 104.69 & 123.73 \\ 
   \hspace{1em} ARGOX 2Step & 92.83 & 120.88 & 149.63 & 201.18 \\ 
   \hspace{1em} ARGOX NatConstraint & 92.24 & 120.60 & 148.08 & 158.75 \\ 
   \hspace{1em} Ensemble & 88.72 & 88.22 & 90.56 & 104.40 \\ 
   \hspace{1em} Naive& 100.24 & 105.47 & 116.68 & 124.54 \\ 
   \multicolumn{1}{l}{MAE} \\
  \hspace{1em} ARGO &33.94 & 42.33 & 55.07 & 75.91 \\ 
   \hspace{1em} ARGOX 2Step & 36.52 & 62.55 & 91.21 & 126.49 \\ 
   \hspace{1em} ARGOX NatConstraint & 44.86 & 66.61 & 95.26 & 106.52 \\ 
   \hspace{1em} Ensemble & 33.05 & 36.58 & 43.02 & 60.55 \\ 
   \hspace{1em} Naive& 32.68 & 43.23 & 54.40 & 67.09 \\ 
   \multicolumn{1}{l}{Correlation} \\
 \hspace{1em} ARGO &0.60 & 0.59 & 0.55 & 0.45 \\ 
   \hspace{1em} ARGOX 2Step & 0.60 & 0.49 & 0.44 & 0.57 \\ 
   \hspace{1em} ARGOX NatConstraint & 0.55 & 0.36 & 0.31 & 0.36 \\ 
   \hspace{1em} Ensemble & 0.61 & 0.63 & 0.63 & 0.61 \\ 
   \hspace{1em} Naive& 0.53 & 0.51 & 0.43 & 0.37 \\ 
   \hline
\end{tabular}
\caption{Comparison of different methods for state-level COVID-19 1 to 4 weeks ahead incremental death in Maryland (MD). The MSE, MAE, and correlation are reported and best performed method is highlighted in boldface.} 
\end{table}

\begin{figure}[!h] 
  \centering 
\includegraphics[width=0.6\linewidth, page=21]{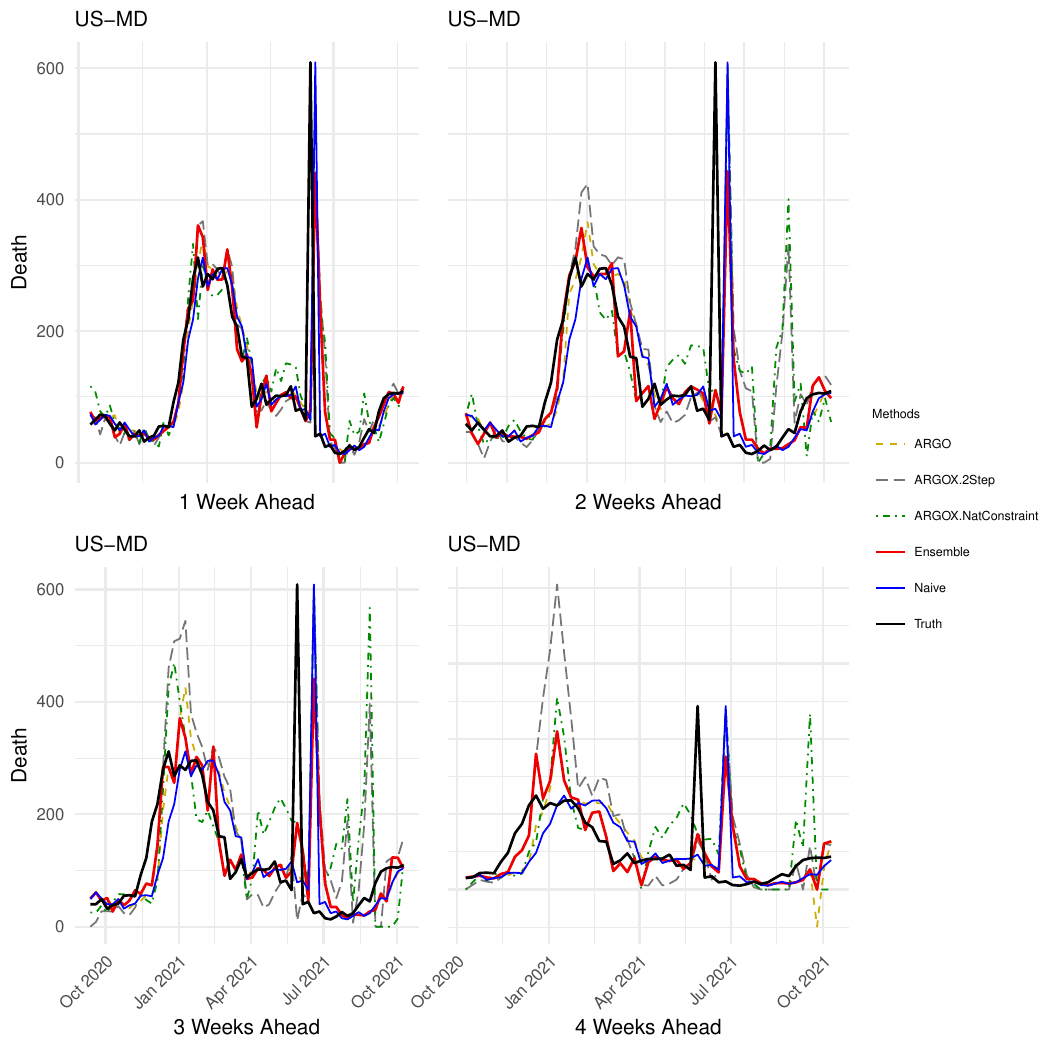} 
\caption{Plots of the COVID-19 1 week (top left), 2 weeks (top right), 3 weeks (bottom left), and 4 weeks (bottom right) ahead estimates for Maryland (MD).}
\end{figure}
\newpage

\begin{table}[ht]
\renewrobustcmd{\bfseries}{\fontseries{b}\selectfont}
\renewrobustcmd{\boldmath}{}
\newrobustcmd{\B}{\bfseries}
\centering
\begin{tabular}{lrrrr}
  \hline
&1 Week Ahead & 2 Weeks Ahead & 3 Weeks Ahead & 4 Weeks Ahead \\ 
    \hline \multicolumn{1}{l}{RMSE} \\
 \hspace{1em} ARGO &  
12.41 & 14.06 & 15.06 & 28.44 \\ 
   \hspace{1em} ARGOX 2Step & 12.10 & 15.14 & 16.50 & 22.86 \\ 
   \hspace{1em} ARGOX NatConstraint & 26.37 & 27.20 & 41.86 & 80.66 \\ 
   \hspace{1em} Ensemble & 12.11 & 11.22 & 10.77 & 17.73 \\ 
   \hspace{1em} Naive& 13.33 & 14.87 & 16.03 & 18.51 \\ 
   \multicolumn{1}{l}{MAE} \\
  \hspace{1em} ARGO &8.26 & 9.90 & 10.82 & 17.21 \\ 
   \hspace{1em} ARGOX 2Step & 7.24 & 8.98 & 10.92 & 14.30 \\ 
   \hspace{1em} ARGOX NatConstraint & 18.32 & 19.91 & 31.21 & 50.91 \\ 
   \hspace{1em} Ensemble & 7.82 & 7.64 & 8.15 & 11.97 \\ 
   \hspace{1em} Naive& 8.62 & 9.90 & 11.39 & 12.81 \\ 
   \multicolumn{1}{l}{Correlation} \\
 \hspace{1em} ARGO &0.71 & 0.63 & 0.61 & 0.19 \\ 
   \hspace{1em} ARGOX 2Step & 0.75 & 0.72 & 0.67 & 0.64 \\ 
   \hspace{1em} ARGOX NatConstraint & 0.22 & 0.04 & 0.21 & 0.16 \\ 
   \hspace{1em} Ensemble & 0.73 & 0.77 & 0.80 & 0.59 \\ 
   \hspace{1em} Naive& 0.68 & 0.61 & 0.56 & 0.43 \\ 
   \hline
\end{tabular}
\caption{Comparison of different methods for state-level COVID-19 1 to 4 weeks ahead incremental death in Maine (ME). The MSE, MAE, and correlation are reported and best performed method is highlighted in boldface.} 
\end{table}

\begin{figure}[!h] 
  \centering 
\includegraphics[width=0.6\linewidth, page=22]{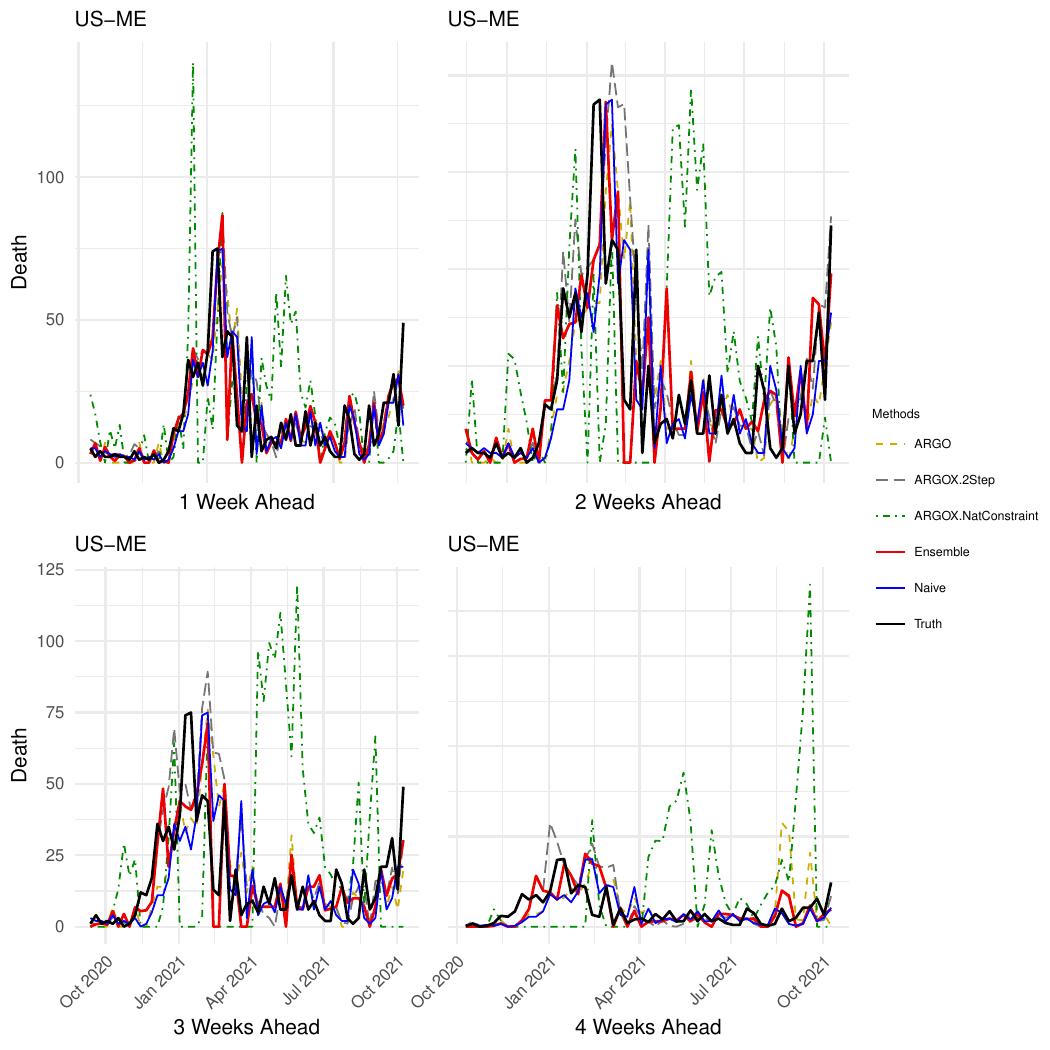} 
\caption{Plots of the COVID-19 1 week (top left), 2 weeks (top right), 3 weeks (bottom left), and 4 weeks (bottom right) ahead estimates for Maine (ME).}
\end{figure}
\newpage

\begin{table}[ht]
\renewrobustcmd{\bfseries}{\fontseries{b}\selectfont}
\renewrobustcmd{\boldmath}{}
\newrobustcmd{\B}{\bfseries}
\centering
\begin{tabular}{lrrrr}
  \hline
&1 Week Ahead & 2 Weeks Ahead & 3 Weeks Ahead & 4 Weeks Ahead \\ 
    \hline \multicolumn{1}{l}{RMSE} \\
 \hspace{1em} ARGO &  
90.19 & 109.94 & 164.40 & 209.86 \\ 
   \hspace{1em} ARGOX 2Step & 91.07 & 159.06 & 258.31 & 526.52 \\ 
   \hspace{1em} ARGOX NatConstraint & 80.32 & 111.77 & 193.49 & 370.44 \\ 
   \hspace{1em} Ensemble & 73.89 & 87.16 & 118.81 & 171.83 \\ 
   \hspace{1em} Naive& 85.11 & 128.85 & 174.53 & 222.53 \\ 
   \multicolumn{1}{l}{MAE} \\
  \hspace{1em} ARGO &59.82 & 82.07 & 127.56 & 165.25 \\ 
   \hspace{1em} ARGOX 2Step & 60.36 & 105.94 & 174.32 & 328.23 \\ 
   \hspace{1em} ARGOX NatConstraint & 53.32 & 78.59 & 125.07 & 230.39 \\ 
   \hspace{1em} Ensemble & 45.28 & 63.77 & 89.20 & 130.82 \\ 
   \hspace{1em} Naive& 51.98 & 93.39 & 135.02 & 176.26 \\ 
   \multicolumn{1}{l}{Correlation} \\
 \hspace{1em} ARGO &0.92 & 0.90 & 0.81 & 0.68 \\ 
   \hspace{1em} ARGOX 2Step & 0.93 & 0.90 & 0.83 & 0.69 \\ 
   \hspace{1em} ARGOX NatConstraint & 0.94 & 0.92 & 0.84 & 0.73 \\ 
   \hspace{1em} Ensemble & 0.95 & 0.95 & 0.91 & 0.81 \\ 
   \hspace{1em} Naive& 0.93 & 0.84 & 0.71 & 0.54 \\ 
   \hline
\end{tabular}
\caption{Comparison of different methods for state-level COVID-19 1 to 4 weeks ahead incremental death in Michigan (MI). The MSE, MAE, and correlation are reported and best performed method is highlighted in boldface.} 
\end{table}

\begin{figure}[!h] 
  \centering 
\includegraphics[width=0.6\linewidth, page=23]{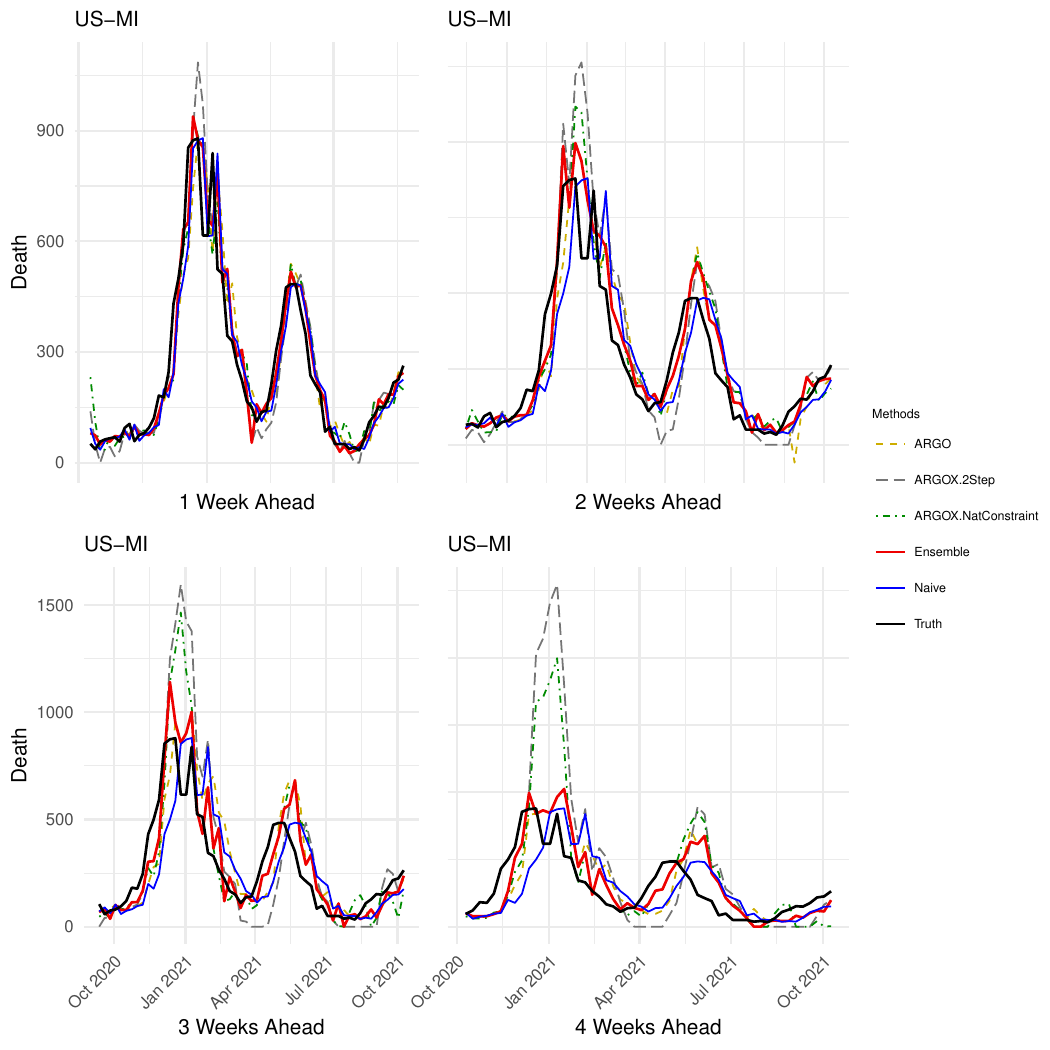} 
\caption{Plots of the COVID-19 1 week (top left), 2 weeks (top right), 3 weeks (bottom left), and 4 weeks (bottom right) ahead estimates for Michigan (MI).}
\end{figure}
\newpage

\begin{table}[ht]
\renewrobustcmd{\bfseries}{\fontseries{b}\selectfont}
\renewrobustcmd{\boldmath}{}
\newrobustcmd{\B}{\bfseries}
\centering
\begin{tabular}{lrrrr}
  \hline
&1 Week Ahead & 2 Weeks Ahead & 3 Weeks Ahead & 4 Weeks Ahead \\ 
    \hline \multicolumn{1}{l}{RMSE} \\
 \hspace{1em} ARGO &  
39.16 & 54.65 & 75.47 & 121.31 \\ 
   \hspace{1em} ARGOX 2Step & 44.02 & 73.59 & 117.78 & 251.69 \\ 
   \hspace{1em} ARGOX NatConstraint & 44.43 & 60.41 & 96.04 & 153.01 \\ 
   \hspace{1em} Ensemble & 36.49 & 41.12 & 50.74 & 100.91 \\ 
   \hspace{1em} Naive& 39.47 & 57.86 & 76.28 & 96.07 \\ 
   \multicolumn{1}{l}{MAE} \\
  \hspace{1em} ARGO &24.77 & 36.10 & 51.86 & 82.15 \\ 
   \hspace{1em} ARGOX 2Step & 27.15 & 44.89 & 73.05 & 127.43 \\ 
   \hspace{1em} ARGOX NatConstraint & 30.59 & 44.29 & 71.09 & 105.99 \\ 
   \hspace{1em} Ensemble & 22.74 & 26.41 & 33.48 & 55.16 \\ 
   \hspace{1em} Naive& 25.83 & 39.79 & 54.18 & 69.51 \\ 
   \multicolumn{1}{l}{Correlation} \\
 \hspace{1em} ARGO &0.93 & 0.87 & 0.83 & 0.77 \\ 
   \hspace{1em} ARGOX 2Step & 0.93 & 0.89 & 0.88 & 0.75 \\ 
   \hspace{1em} ARGOX NatConstraint & 0.91 & 0.88 & 0.82 & 0.69 \\ 
   \hspace{1em} Ensemble & 0.95 & 0.93 & 0.92 & 0.83 \\ 
   \hspace{1em} Naive& 0.93 & 0.85 & 0.76 & 0.64 \\ 
   \hline
\end{tabular}
\caption{Comparison of different methods for state-level COVID-19 1 to 4 weeks ahead incremental death in Minnesota (MN). The MSE, MAE, and correlation are reported and best performed method is highlighted in boldface.} 
\end{table}

\begin{figure}[!h] 
  \centering 
\includegraphics[width=0.6\linewidth, page=24]{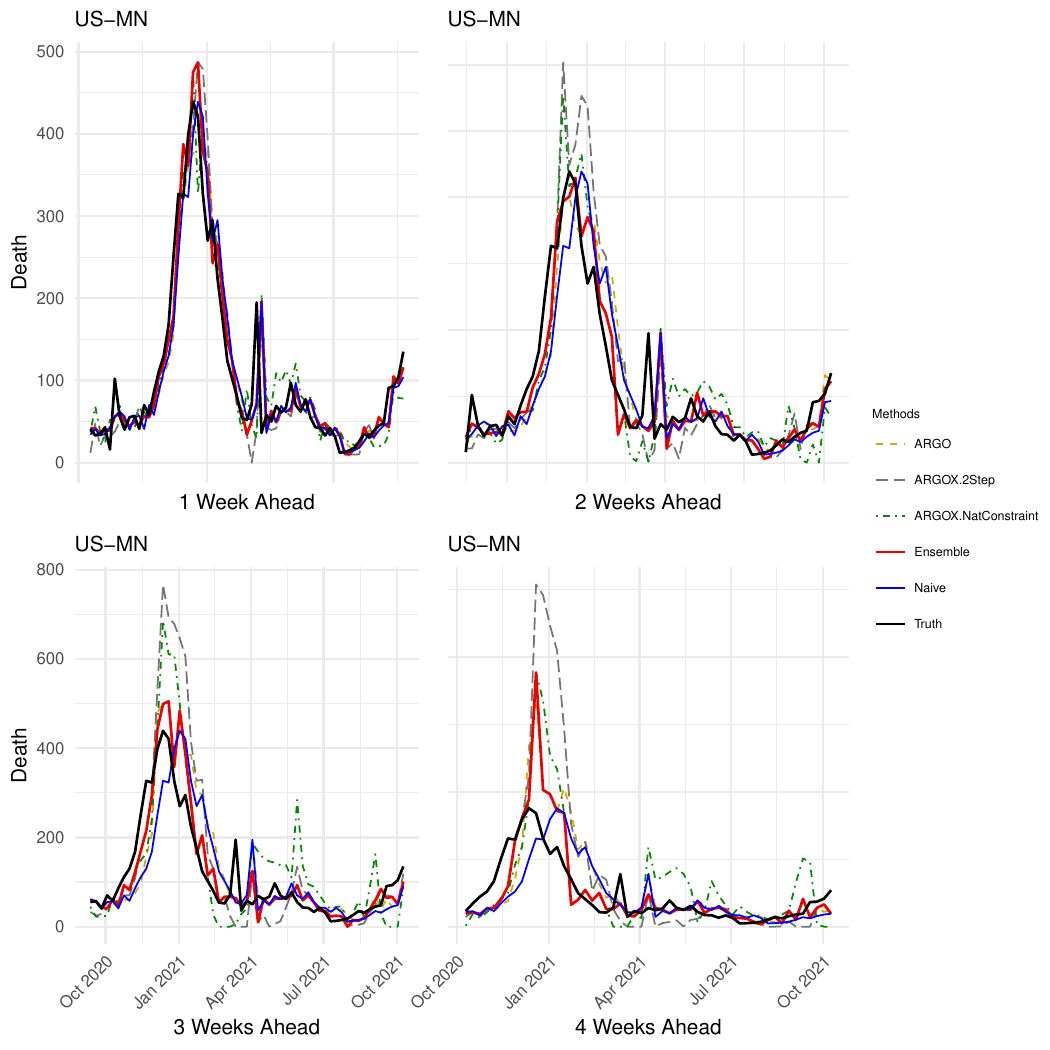} 
\caption{Plots of the COVID-19 1 week (top left), 2 weeks (top right), 3 weeks (bottom left), and 4 weeks (bottom right) ahead estimates for Minnesota (MN).}
\end{figure}
\newpage

\begin{table}[ht]
\renewrobustcmd{\bfseries}{\fontseries{b}\selectfont}
\renewrobustcmd{\boldmath}{}
\newrobustcmd{\B}{\bfseries}
\centering
\begin{tabular}{lrrrr}
  \hline
&1 Week Ahead & 2 Weeks Ahead & 3 Weeks Ahead & 4 Weeks Ahead \\ 
    \hline \multicolumn{1}{l}{RMSE} \\
 \hspace{1em} ARGO &  
63.44 & 82.54 & 85.75 & 107.92 \\ 
   \hspace{1em} ARGOX 2Step & 68.67 & 87.34 & 118.50 & 177.62 \\ 
   \hspace{1em} ARGOX NatConstraint & 72.77 & 95.61 & 117.44 & 153.57 \\ 
   \hspace{1em} Ensemble & 65.08 & 65.04 & 78.67 & 92.20 \\ 
   \hspace{1em} Naive& 71.69 & 73.98 & 85.03 & 106.37 \\ 
   \multicolumn{1}{l}{MAE} \\
  \hspace{1em} ARGO &49.00 & 60.07 & 66.32 & 85.66 \\ 
   \hspace{1em} ARGOX 2Step & 51.52 & 63.48 & 96.73 & 128.71 \\ 
   \hspace{1em} ARGOX NatConstraint & 53.38 & 69.65 & 92.26 & 118.06 \\ 
   \hspace{1em} Ensemble & 49.15 & 47.75 & 60.16 & 72.60 \\ 
   \hspace{1em} Naive& 55.92 & 56.57 & 67.35 & 81.58 \\ 
   \multicolumn{1}{l}{Correlation} \\
 \hspace{1em} ARGO &0.84 & 0.77 & 0.70 & 0.60 \\ 
   \hspace{1em} ARGOX 2Step & 0.84 & 0.78 & 0.66 & 0.51 \\ 
   \hspace{1em} ARGOX NatConstraint & 0.78 & 0.64 & 0.43 & 0.14 \\ 
   \hspace{1em} Ensemble & 0.85 & 0.85 & 0.75 & 0.67 \\ 
   \hspace{1em} Naive& 0.80 & 0.78 & 0.72 & 0.57 \\ 
   \hline
\end{tabular}
\caption{Comparison of different methods for state-level COVID-19 1 to 4 weeks ahead incremental death in Missouri (MO). The MSE, MAE, and correlation are reported and best performed method is highlighted in boldface.} 
\end{table}

\begin{figure}[!h] 
  \centering 
\includegraphics[width=0.6\linewidth, page=25]{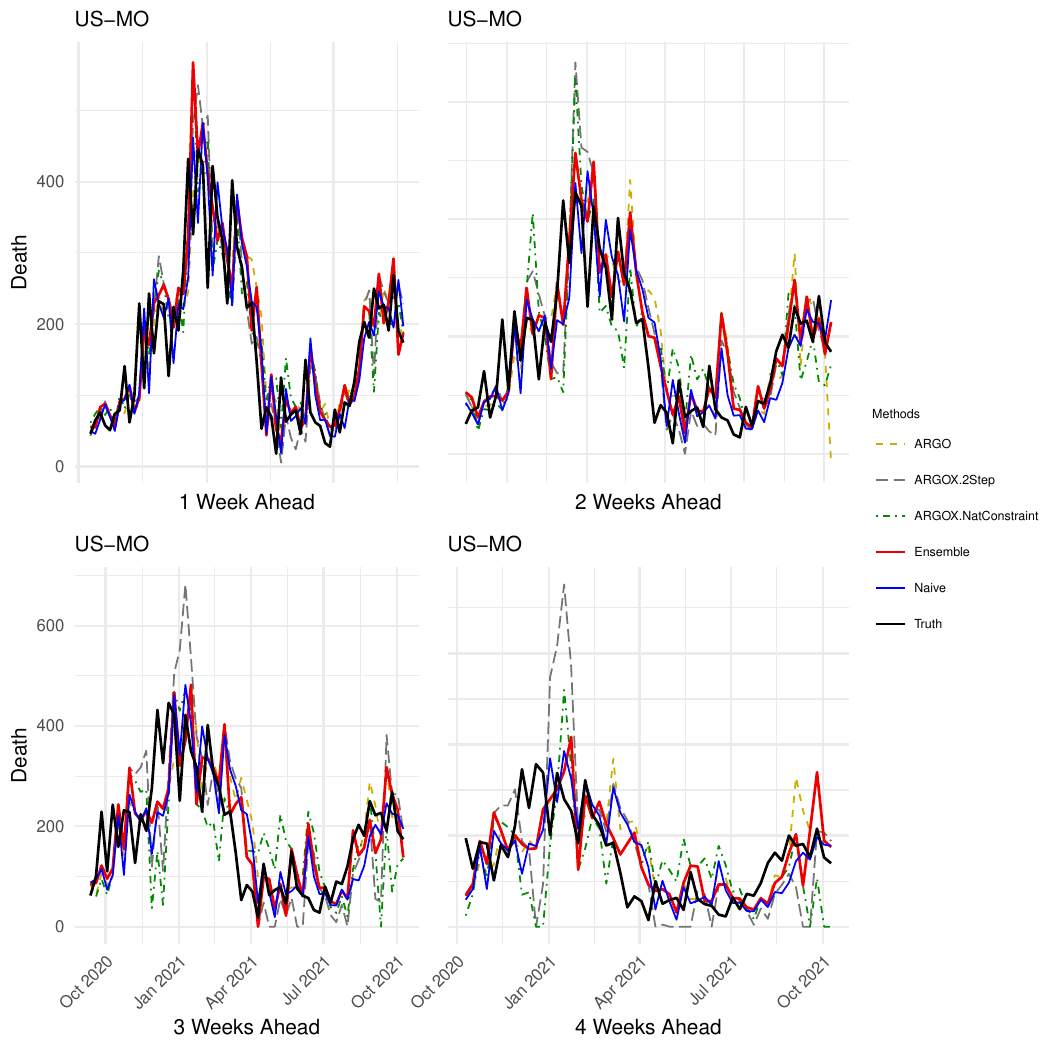} 
\caption{Plots of the COVID-19 1 week (top left), 2 weeks (top right), 3 weeks (bottom left), and 4 weeks (bottom right) ahead estimates for Missouri (MO).}
\end{figure}
\newpage

\begin{table}[ht]
\centering
\renewrobustcmd{\bfseries}{\fontseries{b}\selectfont}
\renewrobustcmd{\boldmath}{}
\newrobustcmd{\B}{\bfseries}
\begin{tabular}{lrrrr}
  \hline
&1 Week Ahead & 2 Weeks Ahead & 3 Weeks Ahead & 4 Weeks Ahead \\ 
    \hline \multicolumn{1}{l}{RMSE} \\
 \hspace{1em} ARGO &  
121.48 & 163.95 & 153.48 & 202.54 \\ 
   \hspace{1em} ARGOX 2Step & 46.93 & 75.91 & 120.38 & 177.29 \\ 
   \hspace{1em} ARGOX NatConstraint & 41.83 & 54.35 & 82.93 & 118.96 \\ 
   \hspace{1em} Ensemble & 34.80 & 52.64 & 72.49 & 113.19 \\ 
   \hspace{1em} Naive& 40.21 & 58.52 & 75.87 & 92.56 \\ 
   \multicolumn{1}{l}{MAE} \\
  \hspace{1em} ARGO &48.58 & 64.38 & 74.60 & 96.17 \\ 
   \hspace{1em} ARGOX 2Step & 34.02 & 50.77 & 74.70 & 113.52 \\ 
   \hspace{1em} ARGOX NatConstraint & 32.06 & 42.06 & 59.82 & 82.44 \\ 
   \hspace{1em} Ensemble & 25.65 & 35.16 & 43.58 & 60.04 \\ 
   \hspace{1em} Naive& 29.20 & 40.72 & 53.51 & 67.55 \\ 
   \multicolumn{1}{l}{Correlation} \\
 \hspace{1em} ARGO &0.71 & 0.61 & 0.67 & 0.51 \\ 
   \hspace{1em} ARGOX 2Step & 0.91 & 0.87 & 0.80 & 0.61 \\ 
   \hspace{1em} ARGOX NatConstraint & 0.90 & 0.84 & 0.73 & 0.52 \\ 
   \hspace{1em} Ensemble & 0.94 & 0.89 & 0.83 & 0.70 \\ 
   \hspace{1em} Naive& 0.91 & 0.81 & 0.69 & 0.56 \\ 
   \hline
\end{tabular}
\caption{Comparison of different methods for state-level COVID-19 1 to 4 weeks ahead incremental death in Mississippi (MS). The MSE, MAE, and correlation are reported and best performed method is highlighted in boldface.} 
\end{table}

\begin{figure}[!h] 
  \centering 
\includegraphics[width=0.6\linewidth, page=26]{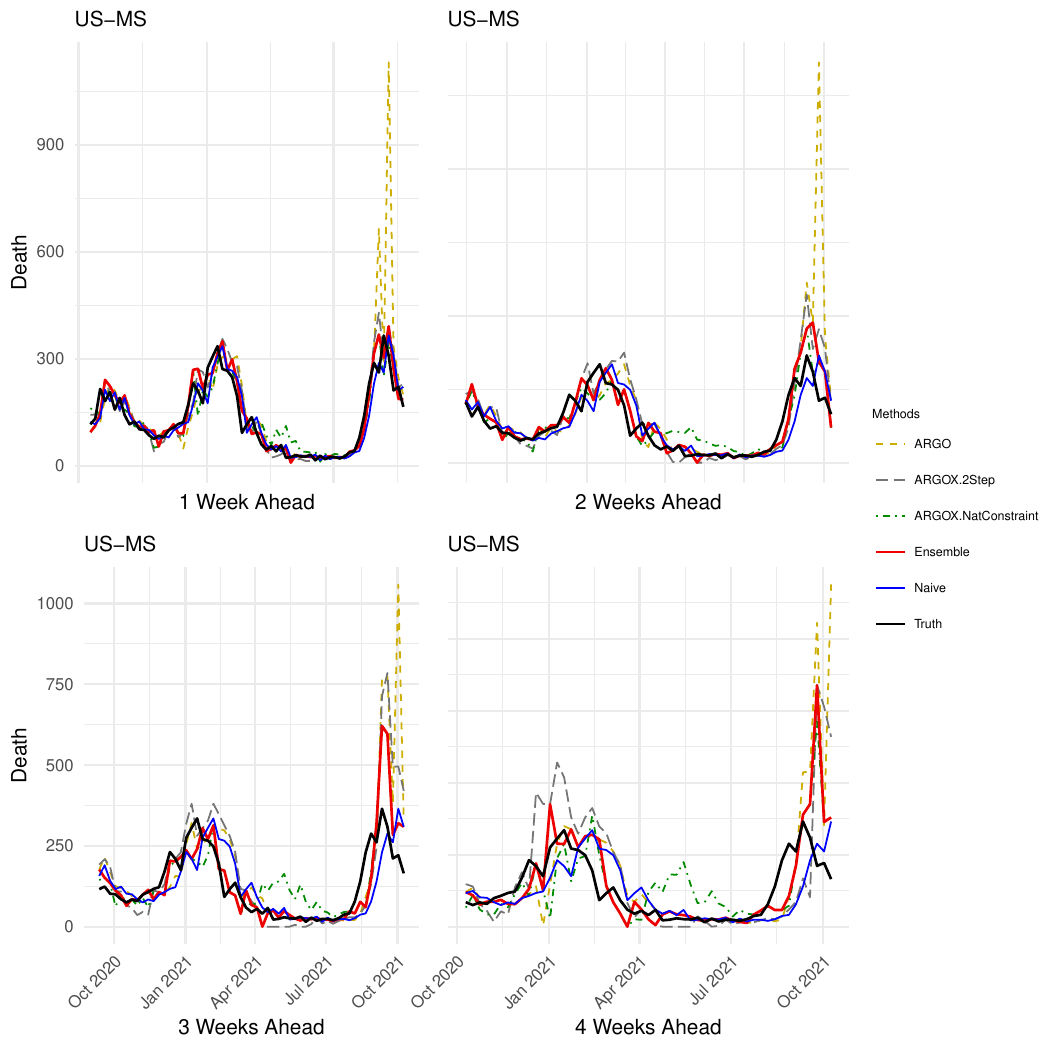} 
\caption{Plots of the COVID-19 1 week (top left), 2 weeks (top right), 3 weeks (bottom left), and 4 weeks (bottom right) ahead estimates for Mississippi (MS).}
\end{figure}
\newpage

\begin{table}[ht]
\centering
\renewrobustcmd{\bfseries}{\fontseries{b}\selectfont}
\renewrobustcmd{\boldmath}{}
\newrobustcmd{\B}{\bfseries}
\begin{tabular}{lrrrr}
  \hline
&1 Week Ahead & 2 Weeks Ahead & 3 Weeks Ahead & 4 Weeks Ahead \\ 
    \hline \multicolumn{1}{l}{RMSE} \\
 \hspace{1em} ARGO &  
17.33 & 24.23 & 26.52 & 24.63 \\ 
   \hspace{1em} ARGOX 2Step & 17.43 & 28.02 & 34.04 & 50.59 \\ 
   \hspace{1em} ARGOX NatConstraint & 27.17 & 35.61 & 48.82 & 85.82 \\ 
   \hspace{1em} Ensemble & 15.71 & 18.18 & 18.52 & 22.47 \\ 
   \hspace{1em} Naive& 16.67 & 22.44 & 24.44 & 26.46 \\ 
   \multicolumn{1}{l}{MAE} \\
  \hspace{1em} ARGO &12.45 & 16.36 & 18.04 & 17.83 \\ 
   \hspace{1em} ARGOX 2Step & 12.51 & 20.05 & 24.22 & 29.88 \\ 
   \hspace{1em} ARGOX NatConstraint & 19.85 & 27.84 & 38.00 & 58.84 \\ 
   \hspace{1em} Ensemble & 11.68 & 13.29 & 13.55 & 16.35 \\ 
   \hspace{1em} Naive& 11.11 & 15.54 & 16.42 & 18.89 \\ 
   \multicolumn{1}{l}{Correlation} \\
 \hspace{1em} ARGO &0.77 & 0.61 & 0.58 & 0.58 \\ 
   \hspace{1em} ARGOX 2Step & 0.77 & 0.55 & 0.54 & 0.40 \\ 
   \hspace{1em} ARGOX NatConstraint & 0.53 & 0.26 & 0.05 & 0.18 \\ 
   \hspace{1em} Ensemble & 0.81 & 0.76 & 0.78 & 0.67 \\ 
   \hspace{1em} Naive& 0.79 & 0.62 & 0.55 & 0.50 \\ 
   \hline
\end{tabular}
\caption{Comparison of different methods for state-level COVID-19 1 to 4 weeks ahead incremental death in Montana (MT). The MSE, MAE, and correlation are reported and best performed method is highlighted in boldface.} 
\end{table}

\begin{figure}[!h] 
  \centering 
\includegraphics[width=0.6\linewidth, page=27]{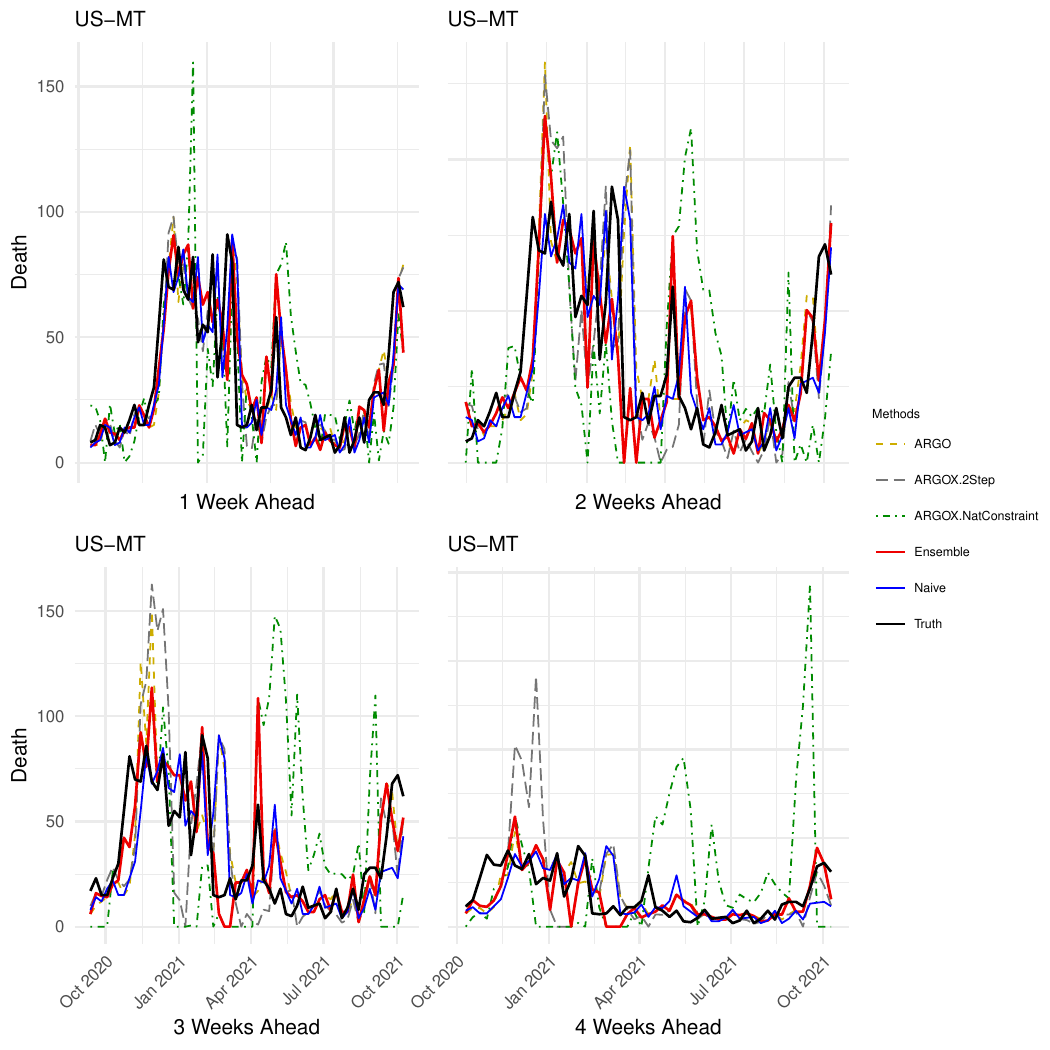} 
\caption{Plots of the COVID-19 1 week (top left), 2 weeks (top right), 3 weeks (bottom left), and 4 weeks (bottom right) ahead estimates for Montana (MT).}
\end{figure}
\newpage

\begin{table}[ht]
\centering
\renewrobustcmd{\bfseries}{\fontseries{b}\selectfont}
\renewrobustcmd{\boldmath}{}
\newrobustcmd{\B}{\bfseries}
\begin{tabular}{lrrrr}
  \hline
&1 Week Ahead & 2 Weeks Ahead & 3 Weeks Ahead & 4 Weeks Ahead \\ 
    \hline \multicolumn{1}{l}{RMSE} \\
 \hspace{1em} ARGO &  
68.35 & 85.59 & 122.18 & 230.37 \\ 
   \hspace{1em} ARGOX 2Step & 69.13 & 111.84 & 149.51 & 204.46 \\ 
   \hspace{1em} ARGOX NatConstraint & 74.04 & 93.69 & 125.76 & 166.04 \\ 
   \hspace{1em} Ensemble & 59.07 & 72.21 & 87.87 & 115.43 \\ 
   \hspace{1em} Naive& 67.12 & 90.71 & 126.15 & 159.80 \\ 
   \multicolumn{1}{l}{MAE} \\
  \hspace{1em} ARGO &48.66 & 58.26 & 83.32 & 134.98 \\ 
   \hspace{1em} ARGOX 2Step & 50.91 & 75.74 & 104.88 & 138.47 \\ 
   \hspace{1em} ARGOX NatConstraint & 56.35 & 71.22 & 95.96 & 127.45 \\ 
   \hspace{1em} Ensemble & 44.87 & 48.03 & 60.27 & 77.53 \\ 
   \hspace{1em} Naive& 47.94 & 65.16 & 88.93 & 117.70 \\ 
   \multicolumn{1}{l}{Correlation} \\
 \hspace{1em} ARGO &0.91 & 0.87 & 0.78 & 0.43 \\ 
   \hspace{1em} ARGOX 2Step & 0.92 & 0.85 & 0.79 & 0.65 \\ 
   \hspace{1em} ARGOX NatConstraint & 0.89 & 0.83 & 0.71 & 0.52 \\ 
   \hspace{1em} Ensemble & 0.94 & 0.91 & 0.88 & 0.79 \\ 
   \hspace{1em} Naive& 0.92 & 0.85 & 0.73 & 0.58 \\ 
   \hline
\end{tabular}
\caption{Comparison of different methods for state-level COVID-19 1 to 4 weeks ahead incremental death in North Carolina (NC). The MSE, MAE, and correlation are reported and best performed method is highlighted in boldface.} 
\end{table}

\begin{figure}[!h] 
  \centering 
\includegraphics[width=0.6\linewidth, page=28]{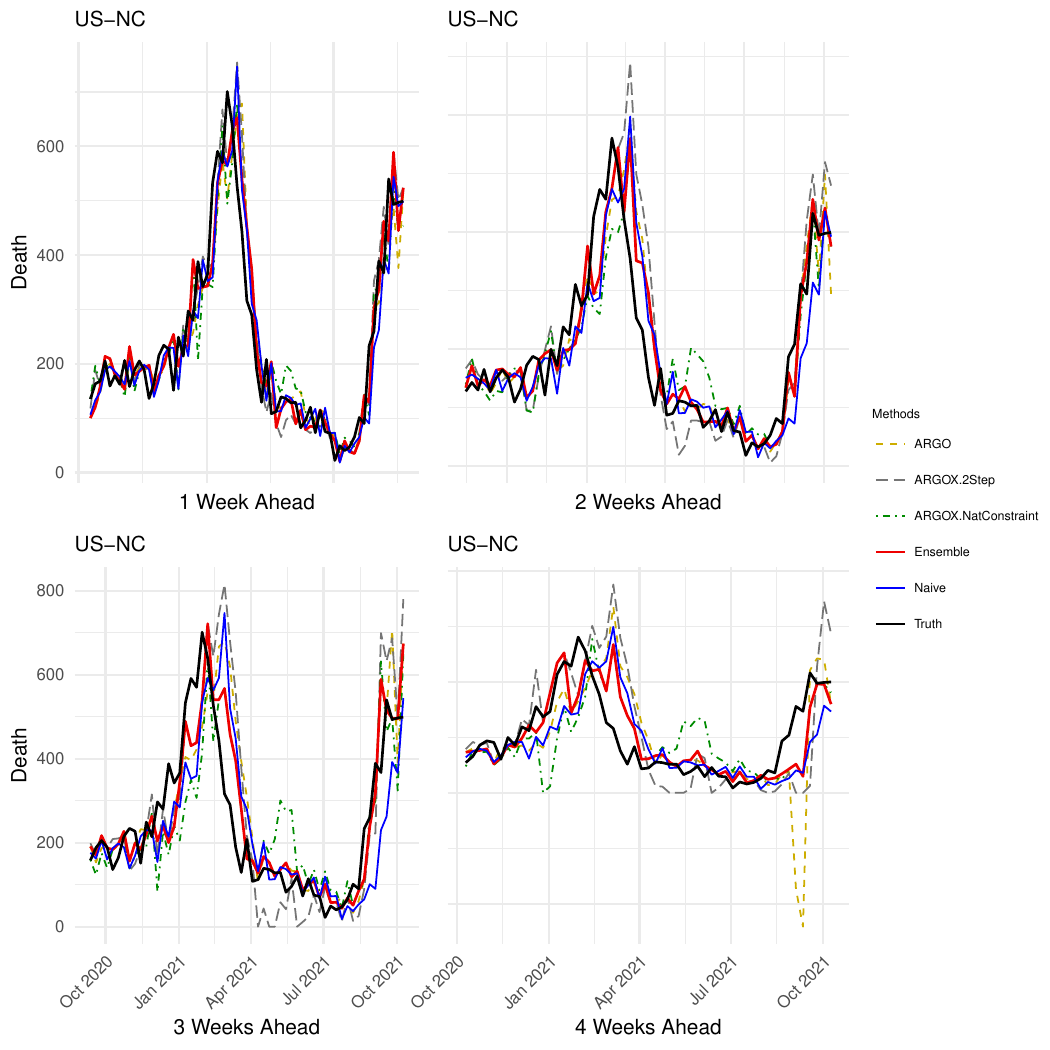} 
\caption{Plots of the COVID-19 1 week (top left), 2 weeks (top right), 3 weeks (bottom left), and 4 weeks (bottom right) ahead estimates for North Carolina (NC).}
\end{figure}
\newpage

\begin{table}[ht]
\centering
\renewrobustcmd{\bfseries}{\fontseries{b}\selectfont}
\renewrobustcmd{\boldmath}{}
\newrobustcmd{\B}{\bfseries}
\begin{tabular}{lrrrr}
  \hline
&1 Week Ahead & 2 Weeks Ahead & 3 Weeks Ahead & 4 Weeks Ahead \\ 
    \hline \multicolumn{1}{l}{RMSE} \\
 \hspace{1em} ARGO &  
28.83 & 36.30 & 23.77 & 35.50 \\ 
   \hspace{1em} ARGOX 2Step & 18.06 & 22.74 & 19.65 & 26.04 \\ 
   \hspace{1em} ARGOX NatConstraint & 20.96 & 28.42 & 45.88 & 82.47 \\ 
   \hspace{1em} Ensemble & 12.84 & 16.84 & 18.29 & 23.80 \\ 
   \hspace{1em} Naive& 14.68 & 20.73 & 22.43 & 26.03 \\ 
   \multicolumn{1}{l}{MAE} \\
  \hspace{1em} ARGO &12.26 & 14.98 & 14.96 & 19.58 \\ 
   \hspace{1em} ARGOX 2Step & 9.61 & 12.07 & 12.92 & 14.55 \\ 
   \hspace{1em} ARGOX NatConstraint & 15.97 & 21.56 & 33.42 & 51.74 \\ 
   \hspace{1em} Ensemble & 7.42 & 8.78 & 11.84 & 13.00 \\ 
   \hspace{1em} Naive& 8.42 & 10.80 & 13.58 & 15.91 \\ 
   \multicolumn{1}{l}{Correlation} \\
 \hspace{1em} ARGO &0.78 & 0.76 & 0.82 & 0.71 \\ 
   \hspace{1em} ARGOX 2Step & 0.86 & 0.81 & 0.90 & 0.83 \\ 
   \hspace{1em} ARGOX NatConstraint & 0.81 & 0.65 & 0.34 & 0.06 \\ 
   \hspace{1em} Ensemble & 0.93 & 0.89 & 0.89 & 0.81 \\ 
   \hspace{1em} Naive& 0.90 & 0.80 & 0.78 & 0.72 \\ 
   \hline
\end{tabular}
\caption{Comparison of different methods for state-level COVID-19 1 to 4 weeks ahead incremental death in North Dakota (ND). The MSE, MAE, and correlation are reported and best performed method is highlighted in boldface.} 
\end{table}

\begin{figure}[!h] 
  \centering 
\includegraphics[width=0.6\linewidth, page=29]{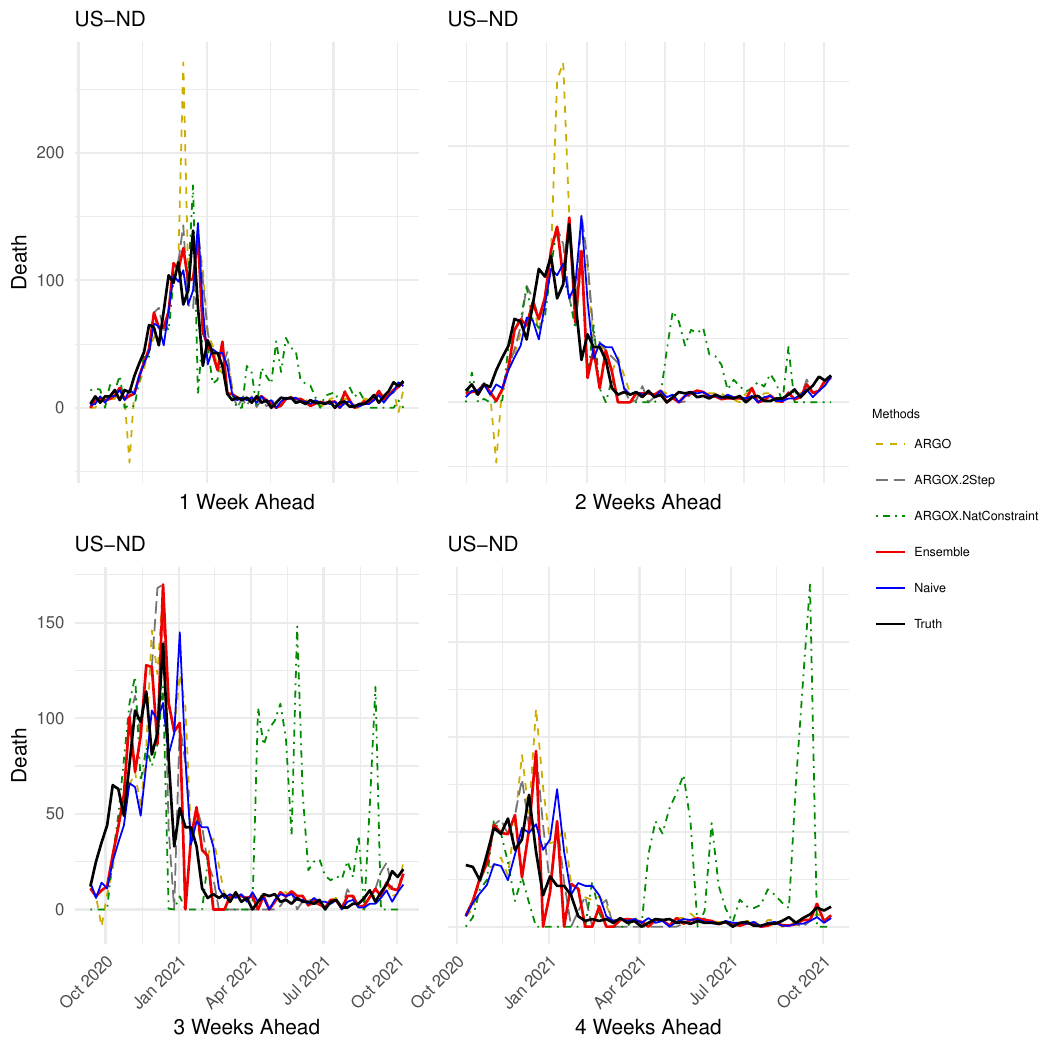} 
\caption{Plots of the COVID-19 1 week (top left), 2 weeks (top right), 3 weeks (bottom left), and 4 weeks (bottom right) ahead estimates for North Dakota (ND).}
\end{figure}
\newpage

\begin{table}[ht]
\centering
\renewrobustcmd{\bfseries}{\fontseries{b}\selectfont}
\renewrobustcmd{\boldmath}{}
\newrobustcmd{\B}{\bfseries}
\begin{tabular}{lrrrr}
  \hline
&1 Week Ahead & 2 Weeks Ahead & 3 Weeks Ahead & 4 Weeks Ahead \\ 
    \hline \multicolumn{1}{l}{RMSE} \\
 \hspace{1em} ARGO &  
26.49 & 30.57 & 35.04 & 45.47 \\ 
   \hspace{1em} ARGOX 2Step & 30.64 & 40.63 & 53.74 & 114.57 \\ 
   \hspace{1em} ARGOX NatConstraint & 30.50 & 44.26 & 64.27 & 82.46 \\ 
   \hspace{1em} Ensemble & 25.75 & 26.53 & 35.15 & 32.96 \\ 
   \hspace{1em} Naive& 31.06 & 35.32 & 41.02 & 46.38 \\ 
   \multicolumn{1}{l}{MAE} \\
  \hspace{1em} ARGO &18.52 & 21.20 & 25.19 & 31.64 \\ 
   \hspace{1em} ARGOX 2Step & 19.86 & 23.80 & 33.84 & 53.80 \\ 
   \hspace{1em} ARGOX NatConstraint & 22.01 & 34.32 & 49.42 & 56.98 \\ 
   \hspace{1em} Ensemble & 16.73 & 18.39 & 24.20 & 23.75 \\ 
   \hspace{1em} Naive& 19.12 & 21.23 & 27.16 & 31.08 \\ 
   \multicolumn{1}{l}{Correlation} \\
 \hspace{1em} ARGO &0.83 & 0.79 & 0.73 & 0.61 \\ 
   \hspace{1em} ARGOX 2Step & 0.81 & 0.78 & 0.73 & 0.60 \\ 
   \hspace{1em} ARGOX NatConstraint & 0.79 & 0.63 & 0.37 & 0.10 \\ 
   \hspace{1em} Ensemble & 0.84 & 0.85 & 0.81 & 0.80 \\ 
   \hspace{1em} Naive& 0.75 & 0.69 & 0.59 & 0.51 \\ 
   \hline
\end{tabular}
\caption{Comparison of different methods for state-level COVID-19 1 to 4 weeks ahead incremental death in Nebraska (NE). The MSE, MAE, and correlation are reported and best performed method is highlighted in boldface.} 
\end{table}

\begin{figure}[!h] 
  \centering 
\includegraphics[width=0.6\linewidth, page=30]{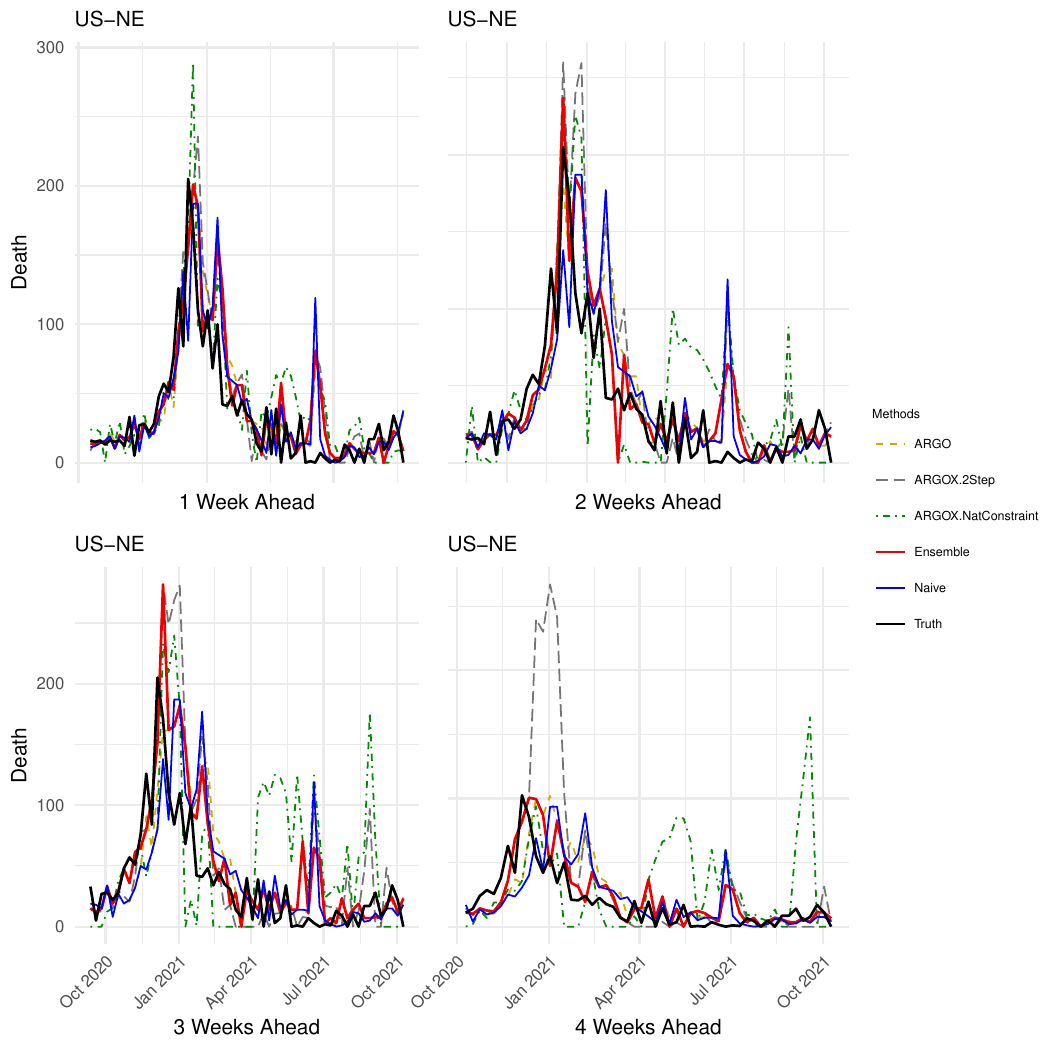} 
\caption{Plots of the COVID-19 1 week (top left), 2 weeks (top right), 3 weeks (bottom left), and 4 weeks (bottom right) ahead estimates for Nebraska (NE).}
\end{figure}
\newpage

\begin{table}[ht]
\centering
\renewrobustcmd{\bfseries}{\fontseries{b}\selectfont}
\renewrobustcmd{\boldmath}{}
\newrobustcmd{\B}{\bfseries}
\begin{tabular}{lrrrr}
  \hline
&1 Week Ahead & 2 Weeks Ahead & 3 Weeks Ahead & 4 Weeks Ahead \\ 
    \hline \multicolumn{1}{l}{RMSE} \\
 \hspace{1em} ARGO &  
10.41 & 13.22 & 15.46 & 20.62 \\ 
   \hspace{1em} ARGOX 2Step & 10.07 & 14.09 & 16.84 & 26.31 \\ 
   \hspace{1em} ARGOX NatConstraint & 24.17 & 27.10 & 46.26 & 77.22 \\ 
   \hspace{1em} Ensemble & 8.41 & 11.51 & 11.35 & 17.33 \\ 
   \hspace{1em} Naive& 8.30 & 12.17 & 15.00 & 18.52 \\ 
   \multicolumn{1}{l}{MAE} \\
  \hspace{1em} ARGO &6.82 & 8.59 & 9.72 & 13.66 \\ 
   \hspace{1em} ARGOX 2Step & 6.53 & 9.59 & 12.91 & 19.03 \\ 
   \hspace{1em} ARGOX NatConstraint & 16.42 & 20.90 & 32.74 & 49.33 \\ 
   \hspace{1em} Ensemble & 5.81 & 7.53 & 7.81 & 10.70 \\ 
   \hspace{1em} Naive& 5.83 & 7.75 & 10.02 & 12.26 \\ 
   \multicolumn{1}{l}{Correlation} \\
 \hspace{1em} ARGO &0.86 & 0.82 & 0.75 & 0.71 \\ 
   \hspace{1em} ARGOX 2Step & 0.89 & 0.85 & 0.82 & 0.75 \\ 
   \hspace{1em} ARGOX NatConstraint & 0.47 & 0.24 & 0.13 & 0.16 \\ 
   \hspace{1em} Ensemble & 0.92 & 0.86 & 0.87 & 0.79 \\ 
   \hspace{1em} Naive& 0.91 & 0.81 & 0.72 & 0.59 \\ 
   \hline
\end{tabular}
\caption{Comparison of different methods for state-level COVID-19 1 to 4 weeks ahead incremental death in New Hampshire (NH). The MSE, MAE, and correlation are reported and best performed method is highlighted in boldface.} 
\end{table}

\begin{figure}[!h] 
  \centering 
\includegraphics[width=0.6\linewidth, page=31]{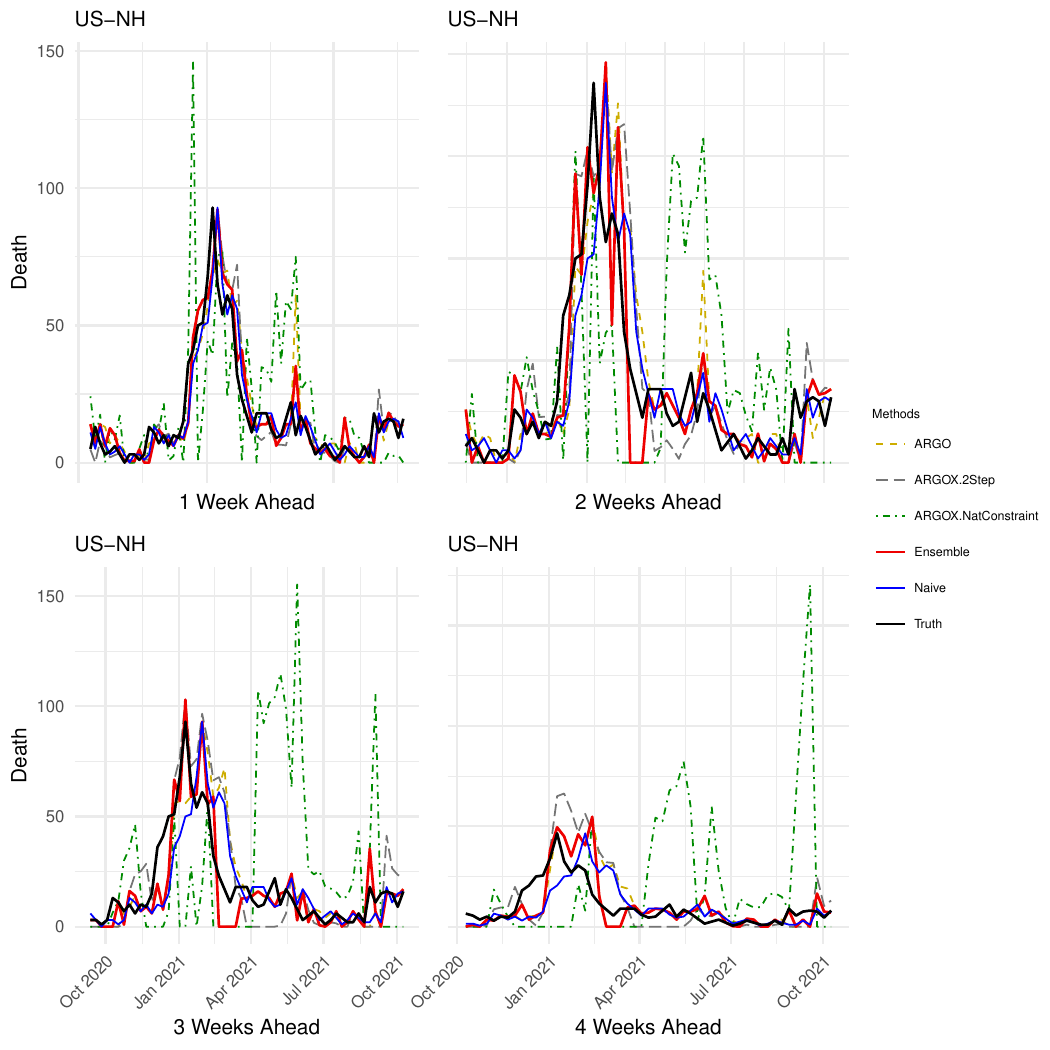} 
\caption{Plots of the COVID-19 1 week (top left), 2 weeks (top right), 3 weeks (bottom left), and 4 weeks (bottom right) ahead estimates for New Hampshire (NH).}
\end{figure}
\newpage

\begin{table}[ht]
\centering
\renewrobustcmd{\bfseries}{\fontseries{b}\selectfont}
\renewrobustcmd{\boldmath}{}
\newrobustcmd{\B}{\bfseries}
\begin{tabular}{lrrrr}
  \hline
&1 Week Ahead & 2 Weeks Ahead & 3 Weeks Ahead & 4 Weeks Ahead \\ 
    \hline \multicolumn{1}{l}{RMSE} \\
 \hspace{1em} ARGO &  
69.48 & 135.53 & 128.70 & 130.52 \\ 
   \hspace{1em} ARGOX 2Step & 63.82 & 71.47 & 103.62 & 218.04 \\ 
   \hspace{1em} ARGOX NatConstraint & 86.14 & 60.37 & 90.07 & 123.40 \\ 
   \hspace{1em} Ensemble & 42.30 & 39.54 & 61.98 & 81.39 \\ 
   \hspace{1em} Naive& 48.43 & 67.47 & 91.05 & 117.82 \\ 
   \multicolumn{1}{l}{MAE} \\
  \hspace{1em} ARGO &42.69 & 63.93 & 78.42 & 89.96 \\ 
   \hspace{1em} ARGOX 2Step & 38.61 & 49.76 & 76.32 & 141.41 \\ 
   \hspace{1em} ARGOX NatConstraint & 45.78 & 49.21 & 67.73 & 90.89 \\ 
   \hspace{1em} Ensemble & 28.29 & 28.18 & 42.75 & 54.64 \\ 
   \hspace{1em} Naive& 32.12 & 46.77 & 66.93 & 91.45 \\ 
   \multicolumn{1}{l}{Correlation} \\
 \hspace{1em} ARGO &0.94 & 0.84 & 0.88 & 0.90 \\ 
   \hspace{1em} ARGOX 2Step & 0.95 & 0.96 & 0.96 & 0.93 \\ 
   \hspace{1em} ARGOX NatConstraint & 0.89 & 0.94 & 0.88 & 0.86 \\ 
   \hspace{1em} Ensemble & 0.97 & 0.97 & 0.96 & 0.94 \\ 
   \hspace{1em} Naive& 0.96 & 0.93 & 0.87 & 0.78 \\ 
   \hline
\end{tabular}
\caption{Comparison of different methods for state-level COVID-19 1 to 4 weeks ahead incremental death in New Jersey (NJ). The MSE, MAE, and correlation are reported and best performed method is highlighted in boldface.} 
\end{table}

\begin{figure}[!h] 
  \centering 
\includegraphics[width=0.6\linewidth, page=32]{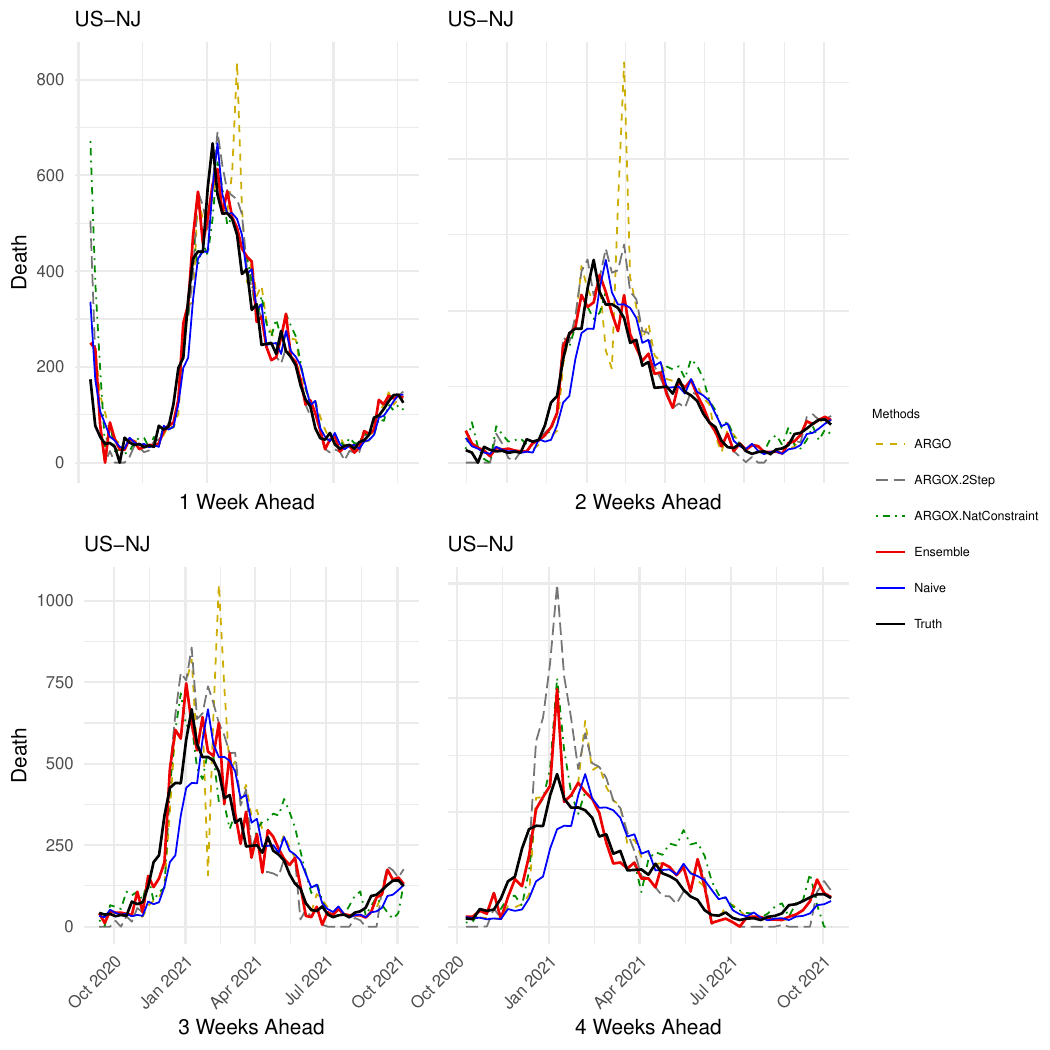} 
\caption{Plots of the COVID-19 1 week (top left), 2 weeks (top right), 3 weeks (bottom left), and 4 weeks (bottom right) ahead estimates for New Jersey (NJ).}
\end{figure}
\newpage

\begin{table}[ht]
\centering
\renewrobustcmd{\bfseries}{\fontseries{b}\selectfont}
\renewrobustcmd{\boldmath}{}
\newrobustcmd{\B}{\bfseries}
\begin{tabular}{lrrrr}
  \hline
&1 Week Ahead & 2 Weeks Ahead & 3 Weeks Ahead & 4 Weeks Ahead \\ 
    \hline \multicolumn{1}{l}{RMSE} \\
 \hspace{1em} ARGO &  
26.49 & 29.74 & 47.00 & 67.01 \\ 
   \hspace{1em} ARGOX 2Step & 29.05 & 36.52 & 64.02 & 137.66 \\ 
   \hspace{1em} ARGOX NatConstraint & 38.80 & 44.00 & 72.21 & 104.43 \\ 
   \hspace{1em} Ensemble & 24.75 & 26.70 & 30.48 & 38.62 \\ 
   \hspace{1em} Naive& 30.12 & 34.23 & 42.78 & 53.50 \\ 
   \multicolumn{1}{l}{MAE} \\
  \hspace{1em} ARGO &16.62 & 19.87 & 31.61 & 42.72 \\ 
   \hspace{1em} ARGOX 2Step & 17.91 & 24.78 & 42.39 & 73.09 \\ 
   \hspace{1em} ARGOX NatConstraint & 24.90 & 35.14 & 58.43 & 77.50 \\ 
   \hspace{1em} Ensemble & 15.23 & 16.42 & 20.97 & 26.87 \\ 
   \hspace{1em} Naive& 18.29 & 22.75 & 30.93 & 39.02 \\ 
   \multicolumn{1}{l}{Correlation} \\
 \hspace{1em} ARGO &0.91 & 0.89 & 0.80 & 0.64 \\ 
   \hspace{1em} ARGOX 2Step & 0.91 & 0.91 & 0.85 & 0.77 \\ 
   \hspace{1em} ARGOX NatConstraint & 0.80 & 0.76 & 0.54 & 0.32 \\ 
   \hspace{1em} Ensemble & 0.92 & 0.91 & 0.90 & 0.85 \\ 
   \hspace{1em} Naive& 0.88 & 0.86 & 0.79 & 0.68 \\ 
   \hline
\end{tabular}
\caption{Comparison of different methods for state-level COVID-19 1 to 4 weeks ahead incremental death in New Mexico (NM). The MSE, MAE, and correlation are reported and best performed method is highlighted in boldface.} 
\end{table}

\begin{figure}[!h] 
  \centering 
\includegraphics[width=0.6\linewidth, page=33]{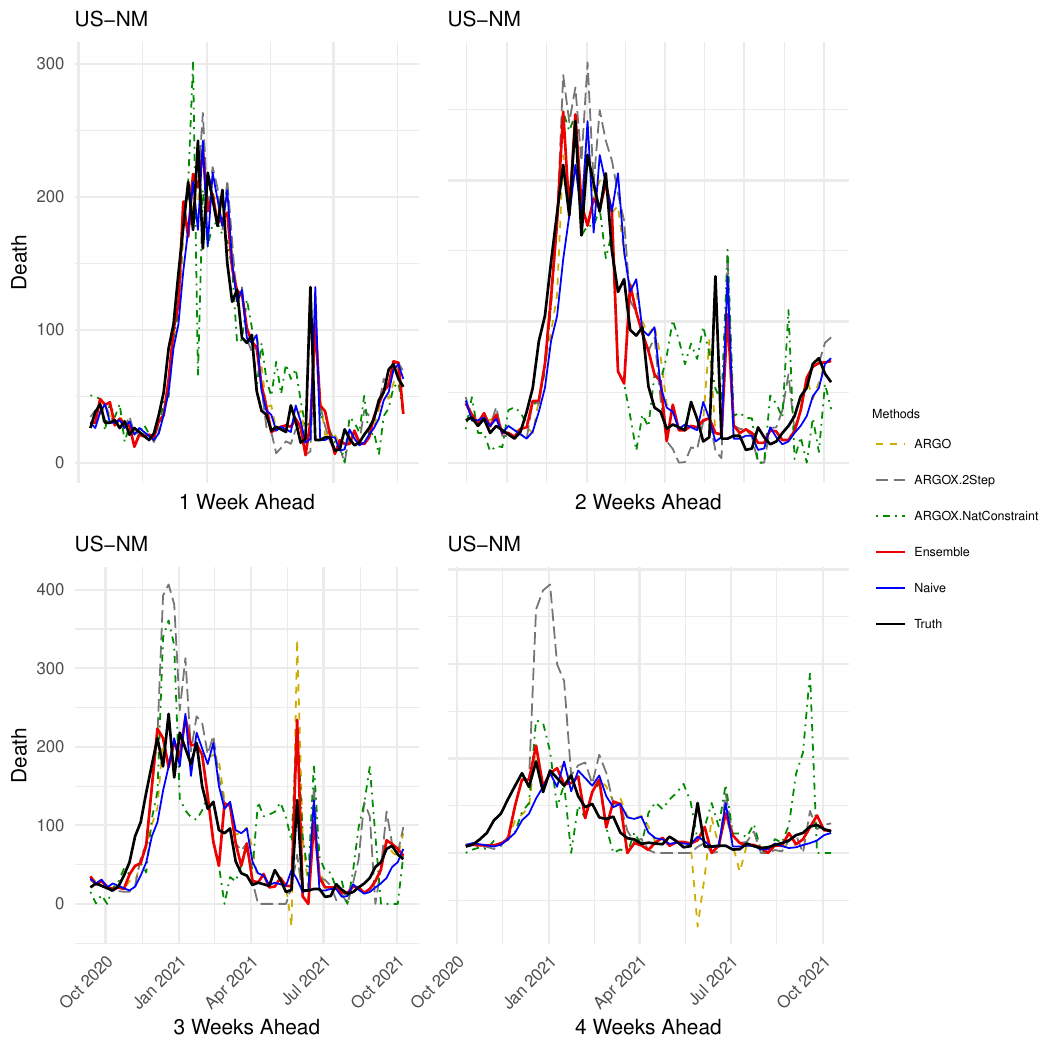} 
\caption{Plots of the COVID-19 1 week (top left), 2 weeks (top right), 3 weeks (bottom left), and 4 weeks (bottom right) ahead estimates for New Mexico (NM).}
\end{figure}
\newpage

\begin{table}[ht]
\centering
\renewrobustcmd{\bfseries}{\fontseries{b}\selectfont}
\renewrobustcmd{\boldmath}{}
\newrobustcmd{\B}{\bfseries}
\begin{tabular}{lrrrr}
  \hline
&1 Week Ahead & 2 Weeks Ahead & 3 Weeks Ahead & 4 Weeks Ahead \\ 
    \hline \multicolumn{1}{l}{RMSE} \\
 \hspace{1em} ARGO &  
29.58 & 37.11 & 48.11 & 64.45 \\ 
   \hspace{1em} ARGOX 2Step & 34.42 & 54.39 & 69.42 & 150.09 \\ 
   \hspace{1em} ARGOX NatConstraint & 39.73 & 47.01 & 73.93 & 86.24 \\ 
   \hspace{1em} Ensemble & 26.62 & 29.58 & 32.48 & 52.39 \\ 
   \hspace{1em} Naive& 28.99 & 42.02 & 54.64 & 66.68 \\ 
   \multicolumn{1}{l}{MAE} \\
  \hspace{1em} ARGO &21.37 & 28.39 & 34.79 & 43.64 \\ 
   \hspace{1em} ARGOX 2Step & 25.25 & 39.29 & 50.60 & 100.14 \\ 
   \hspace{1em} ARGOX NatConstraint & 30.68 & 34.79 & 55.91 & 69.26 \\ 
   \hspace{1em} Ensemble & 18.25 & 20.96 & 23.73 & 33.11 \\ 
   \hspace{1em} Naive& 21.69 & 30.70 & 39.84 & 49.89 \\ 
   \multicolumn{1}{l}{Correlation} \\
 \hspace{1em} ARGO &0.92 & 0.88 & 0.81 & 0.71 \\ 
   \hspace{1em} ARGOX 2Step & 0.91 & 0.86 & 0.85 & 0.75 \\ 
   \hspace{1em} ARGOX NatConstraint & 0.84 & 0.79 & 0.58 & 0.58 \\ 
   \hspace{1em} Ensemble & 0.93 & 0.92 & 0.92 & 0.84 \\ 
   \hspace{1em} Naive& 0.92 & 0.84 & 0.75 & 0.65 \\ 
   \hline
\end{tabular}
\caption{Comparison of different methods for state-level COVID-19 1 to 4 weeks ahead incremental death in Nevada (NV). The MSE, MAE, and correlation are reported and best performed method is highlighted in boldface.} 
\end{table}

\begin{figure}[!h] 
  \centering 
\includegraphics[width=0.6\linewidth, page=34]{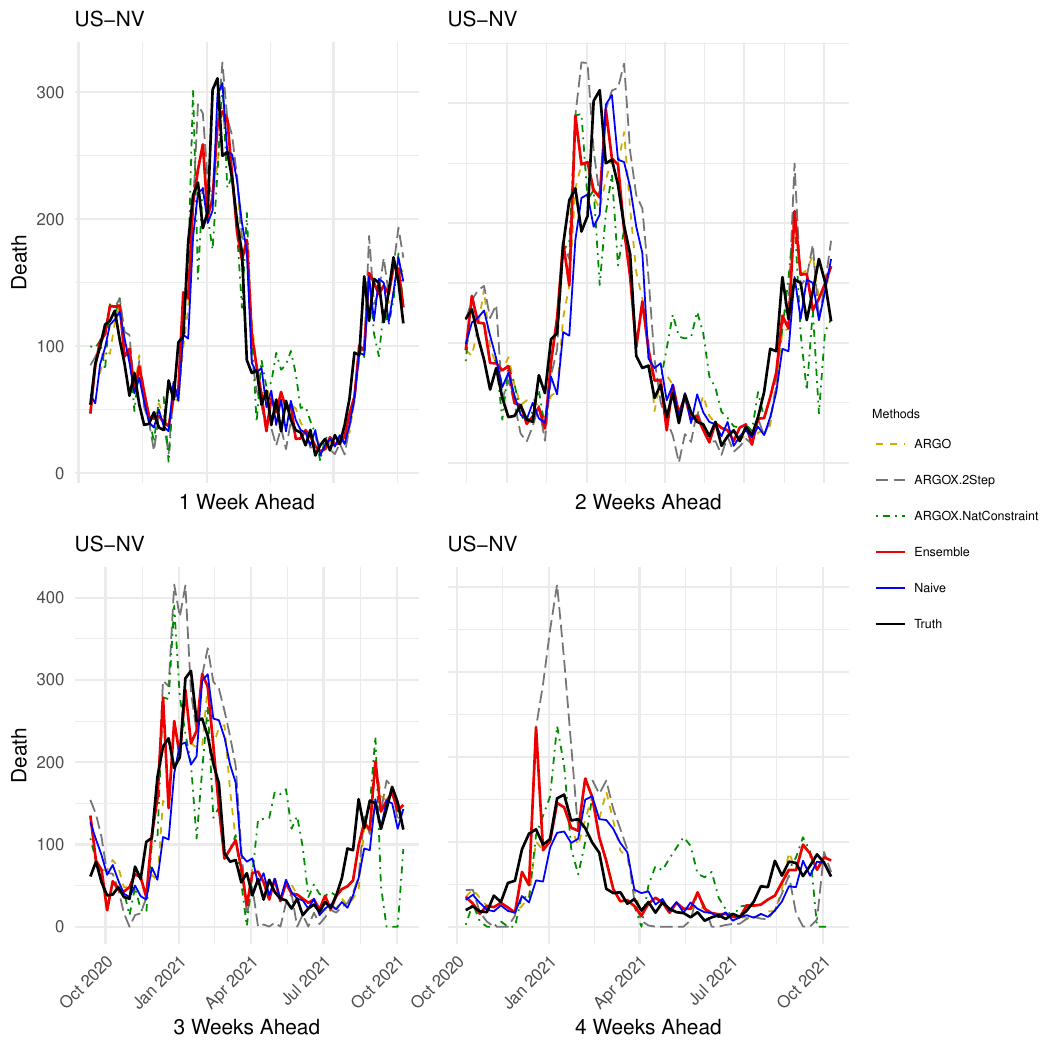} 
\caption{Plots of the COVID-19 1 week (top left), 2 weeks (top right), 3 weeks (bottom left), and 4 weeks (bottom right) ahead estimates for Nevada (NV).}
\end{figure}
\newpage

\begin{table}[ht]
\centering
\renewrobustcmd{\bfseries}{\fontseries{b}\selectfont}
\renewrobustcmd{\boldmath}{}
\newrobustcmd{\B}{\bfseries}
\begin{tabular}{lrrrr}
  \hline
&1 Week Ahead & 2 Weeks Ahead & 3 Weeks Ahead & 4 Weeks Ahead \\ 
    \hline \multicolumn{1}{l}{RMSE} \\
 \hspace{1em} ARGO &  
79.52 & 110.01 & 164.41 & 260.63 \\ 
   \hspace{1em} ARGOX 2Step & 80.66 & 168.89 & 223.40 & 335.95 \\ 
   \hspace{1em} ARGOX NatConstraint & 82.05 & 134.51 & 150.09 & 221.51 \\ 
   \hspace{1em} Ensemble & 59.80 & 84.78 & 99.24 & 163.50 \\ 
   \hspace{1em} Naive& 74.46 & 135.35 & 199.56 & 266.89 \\ 
   \multicolumn{1}{l}{MAE} \\
  \hspace{1em} ARGO &59.35 & 81.72 & 120.51 & 198.40 \\ 
   \hspace{1em} ARGOX 2Step & 59.26 & 107.96 & 140.45 & 221.12 \\ 
   \hspace{1em} ARGOX NatConstraint & 59.52 & 99.04 & 115.50 & 171.33 \\ 
   \hspace{1em} Ensemble & 43.37 & 63.45 & 73.21 & 117.39 \\ 
   \hspace{1em} Naive& 52.45 & 97.13 & 146.91 & 201.23 \\ 
   \multicolumn{1}{l}{Correlation} \\
 \hspace{1em} ARGO &0.98 & 0.97 & 0.94 & 0.86 \\ 
   \hspace{1em} ARGOX 2Step & 0.99 & 0.95 & 0.94 & 0.94 \\ 
   \hspace{1em} ARGOX NatConstraint & 0.98 & 0.94 & 0.93 & 0.89 \\ 
   \hspace{1em} Ensemble & 0.99 & 0.98 & 0.97 & 0.96 \\ 
   \hspace{1em} Naive& 0.98 & 0.94 & 0.87 & 0.77 \\ 
   \hline
\end{tabular}
\caption{Comparison of different methods for state-level COVID-19 1 to 4 weeks ahead incremental death in New York (NY). The MSE, MAE, and correlation are reported and best performed method is highlighted in boldface.} 
\end{table}

\begin{figure}[!h] 
  \centering 
\includegraphics[width=0.6\linewidth, page=35]{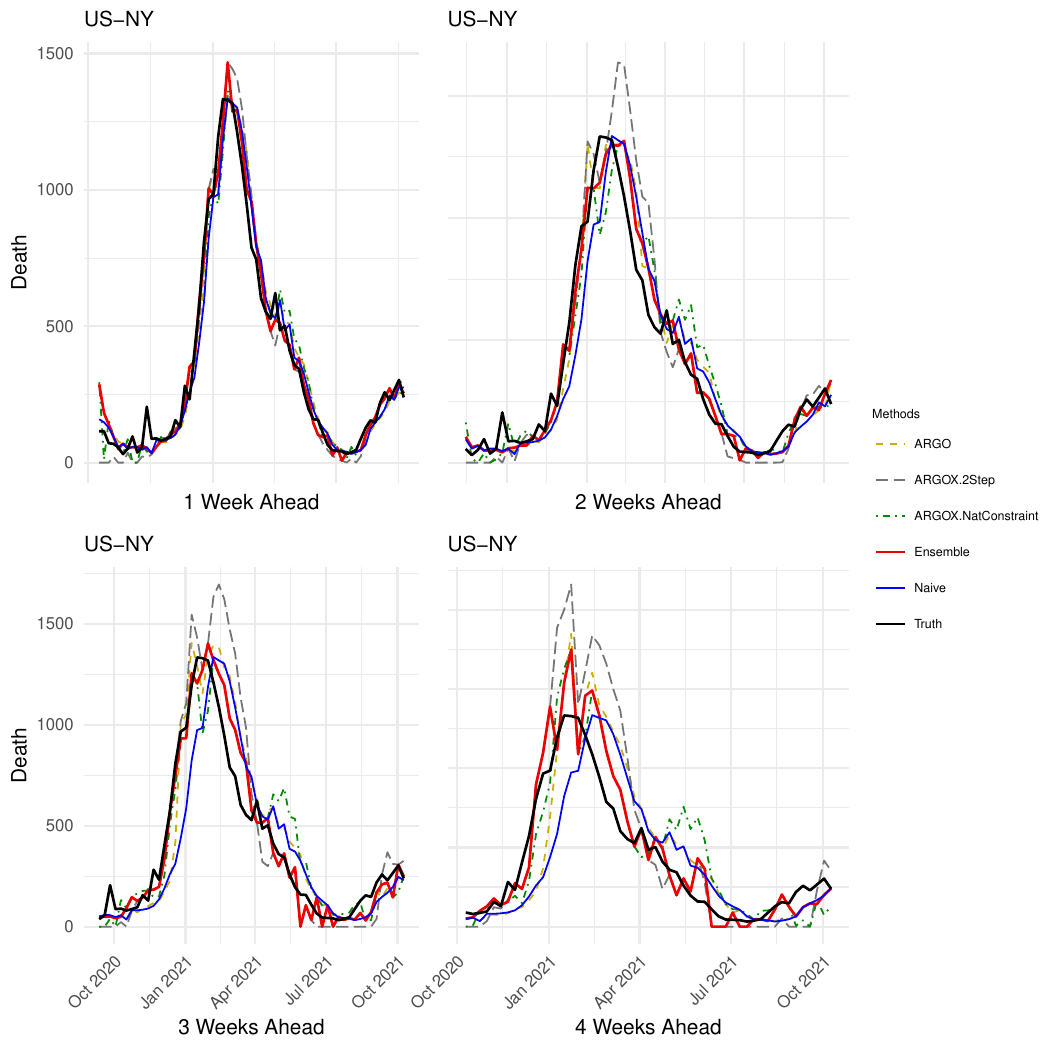} 
\caption{Plots of the COVID-19 1 week (top left), 2 weeks (top right), 3 weeks (bottom left), and 4 weeks (bottom right) ahead estimates for New York (NY).}
\end{figure}
\newpage

\begin{table}[ht]
\centering
\renewrobustcmd{\bfseries}{\fontseries{b}\selectfont}
\renewrobustcmd{\boldmath}{}
\newrobustcmd{\B}{\bfseries}
\begin{tabular}{lrrrr}
  \hline
&1 Week Ahead & 2 Weeks Ahead & 3 Weeks Ahead & 4 Weeks Ahead \\ 
    \hline \multicolumn{1}{l}{RMSE} \\
 \hspace{1em} ARGO &  
454.01 & 469.57 & 501.70 & 753.12 \\ 
   \hspace{1em} ARGOX 2Step & 514.29 & 544.14 & 558.18 & 713.08 \\ 
   \hspace{1em} ARGOX NatConstraint & 501.97 & 515.81 & 535.90 & 672.49 \\ 
   \hspace{1em} Ensemble & 538.82 & 463.90 & 436.55 & 511.72 \\ 
   \hspace{1em} Naive& 610.52 & 650.75 & 689.46 & 730.02 \\ 
   \multicolumn{1}{l}{MAE} \\
  \hspace{1em} ARGO &244.48 & 254.18 & 288.07 & 350.94 \\ 
   \hspace{1em} ARGOX 2Step & 227.61 & 275.04 & 259.89 & 392.09 \\ 
   \hspace{1em} ARGOX NatConstraint & 241.01 & 287.84 & 301.64 & 369.31 \\ 
   \hspace{1em} Ensemble & 242.43 & 246.86 & 244.53 & 282.77 \\ 
   \hspace{1em} Naive& 244.28 & 272.02 & 309.28 & 347.91 \\ 
   \multicolumn{1}{l}{Correlation} \\
 \hspace{1em} ARGO &0.27 & 0.25 & 0.22 & 0.08 \\ 
   \hspace{1em} ARGOX 2Step & 0.29 & 0.24 & 0.29 & 0.31 \\ 
   \hspace{1em} ARGOX NatConstraint & 0.25 & 0.18 & 0.19 & 0.18 \\ 
   \hspace{1em} Ensemble & 0.24 & 0.27 & 0.32 & 0.22 \\ 
   \hspace{1em} Naive& 0.20 & 0.13 & 0.07 & 0.02 \\ 
   \hline
\end{tabular}
\caption{Comparison of different methods for state-level COVID-19 1 to 4 weeks ahead incremental death in Ohio (OH). The MSE, MAE, and correlation are reported and best performed method is highlighted in boldface.} 
\end{table}

\begin{figure}[!h] 
  \centering 
\includegraphics[width=0.6\linewidth, page=36]{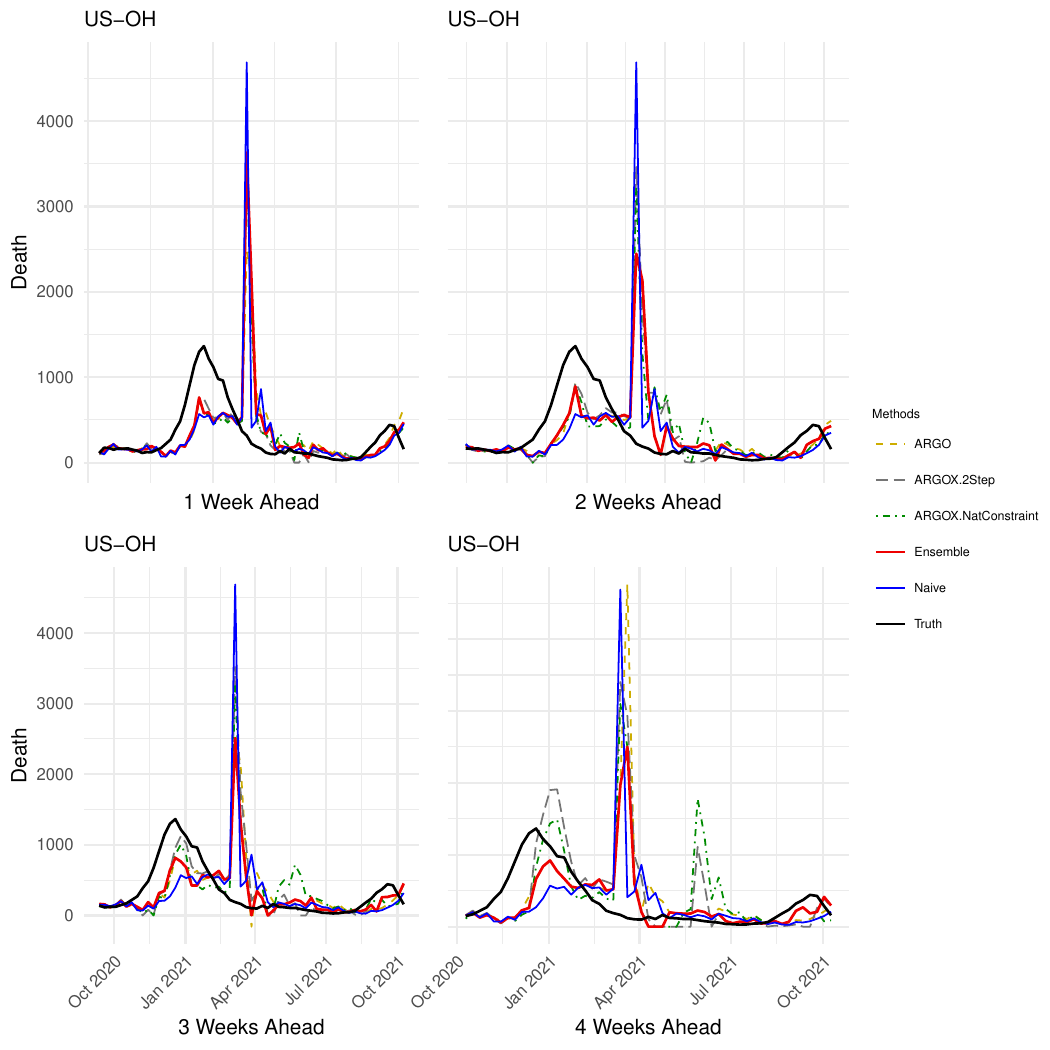} 
\caption{Plots of the COVID-19 1 week (top left), 2 weeks (top right), 3 weeks (bottom left), and 4 weeks (bottom right) ahead estimates for Ohio (OH).}
\end{figure}
\newpage

\begin{table}[ht]
\centering
\renewrobustcmd{\bfseries}{\fontseries{b}\selectfont}
\renewrobustcmd{\boldmath}{}
\newrobustcmd{\B}{\bfseries}
\begin{tabular}{lrrrr}
  \hline
&1 Week Ahead & 2 Weeks Ahead & 3 Weeks Ahead & 4 Weeks Ahead \\ 
    \hline \multicolumn{1}{l}{RMSE} \\
 \hspace{1em} ARGO &  
278.83 & 294.47 & 295.34 & 301.56 \\ 
   \hspace{1em} ARGOX 2Step & 280.92 & 317.45 & 312.31 & 325.56 \\ 
   \hspace{1em} ARGOX NatConstraint & 272.73 & 312.19 & 317.71 & 350.49 \\ 
   \hspace{1em} Ensemble & 270.49 & 281.81 & 291.13 & 301.83 \\ 
   \hspace{1em} Naive& 297.15 & 312.39 & 317.74 & 326.24 \\ 
   \multicolumn{1}{l}{MAE} \\
  \hspace{1em} ARGO &84.78 & 97.31 & 106.74 & 118.60 \\ 
   \hspace{1em} ARGOX 2Step & 88.28 & 111.17 & 127.64 & 149.85 \\ 
   \hspace{1em} ARGOX NatConstraint & 93.17 & 117.88 & 149.62 & 189.11 \\ 
   \hspace{1em} Ensemble & 73.09 & 86.60 & 102.57 & 114.49 \\ 
   \hspace{1em} Naive& 82.54 & 95.30 & 110.00 & 123.92 \\ 
   \multicolumn{1}{l}{Correlation} \\
 \hspace{1em} ARGO &0.07 & 0.01 & 0.05 & 0.06 \\ 
   \hspace{1em} ARGOX 2Step & 0.04 & 0.00 & 0.00 & 0.05 \\ 
   \hspace{1em} ARGOX NatConstraint & 0.06 & 0.01 & 0.04 & 0.00 \\ 
   \hspace{1em} Ensemble & 0.10 & 0.06 & 0.09 & 0.08 \\ 
   \hspace{1em} Naive& 0.05 & 0.00 & 0.02 & 0.03 \\ 
   \hline
\end{tabular}
\caption{Comparison of different methods for state-level COVID-19 1 to 4 weeks ahead incremental death in Oklahoma (OK). The MSE, MAE, and correlation are reported and best performed method is highlighted in boldface.} 
\end{table}

\begin{figure}[!h] 
  \centering 
\includegraphics[width=0.6\linewidth, page=37]{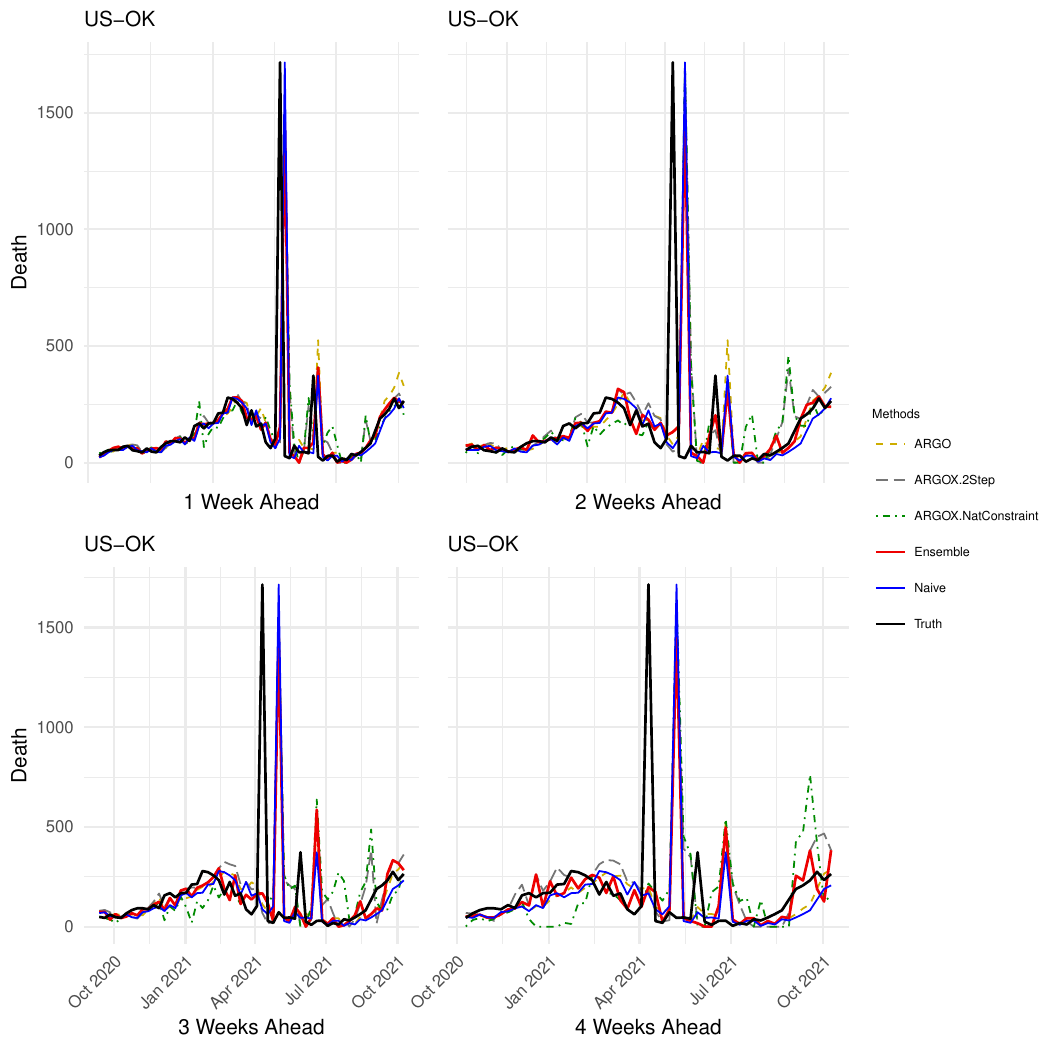} 
\caption{Plots of the COVID-19 1 week (top left), 2 weeks (top right), 3 weeks (bottom left), and 4 weeks (bottom right) ahead estimates for Oklahoma (OK).}
\end{figure}
\newpage

\begin{table}[ht]
\centering
\renewrobustcmd{\bfseries}{\fontseries{b}\selectfont}
\renewrobustcmd{\boldmath}{}
\newrobustcmd{\B}{\bfseries}
\begin{tabular}{lrrrr}
  \hline
&1 Week Ahead & 2 Weeks Ahead & 3 Weeks Ahead & 4 Weeks Ahead \\ 
    \hline \multicolumn{1}{l}{RMSE} \\
 \hspace{1em} ARGO &  
29.29 & 35.04 & 43.92 & 58.05 \\ 
   \hspace{1em} ARGOX 2Step & 33.17 & 43.03 & 61.75 & 97.12 \\ 
   \hspace{1em} ARGOX NatConstraint & 40.60 & 44.46 & 59.51 & 80.28 \\ 
   \hspace{1em} Ensemble & 26.08 & 28.81 & 37.00 & 47.49 \\ 
   \hspace{1em} Naive& 33.95 & 39.91 & 45.82 & 47.48 \\ 
   \multicolumn{1}{l}{MAE} \\
  \hspace{1em} ARGO &20.26 & 25.23 & 30.95 & 40.74 \\ 
   \hspace{1em} ARGOX 2Step & 22.05 & 29.81 & 41.89 & 62.90 \\ 
   \hspace{1em} ARGOX NatConstraint & 26.94 & 31.89 & 44.98 & 63.85 \\ 
   \hspace{1em} Ensemble & 17.79 & 20.66 & 25.50 & 32.94 \\ 
   \hspace{1em} Naive& 22.85 & 27.89 & 32.07 & 34.74 \\ 
   \multicolumn{1}{l}{Correlation} \\
 \hspace{1em} ARGO &0.81 & 0.75 & 0.70 & 0.60 \\ 
   \hspace{1em} ARGOX 2Step & 0.78 & 0.73 & 0.66 & 0.54 \\ 
   \hspace{1em} ARGOX NatConstraint & 0.61 & 0.54 & 0.37 & 0.27 \\ 
   \hspace{1em} Ensemble & 0.84 & 0.81 & 0.71 & 0.68 \\ 
   \hspace{1em} Naive& 0.73 & 0.64 & 0.54 & 0.52 \\ 
   \hline
\end{tabular}
\caption{Comparison of different methods for state-level COVID-19 1 to 4 weeks ahead incremental death in Oregon (OR). The MSE, MAE, and correlation are reported and best performed method is highlighted in boldface.} 
\end{table}

\begin{figure}[!h] 
  \centering 
\includegraphics[width=0.6\linewidth, page=38]{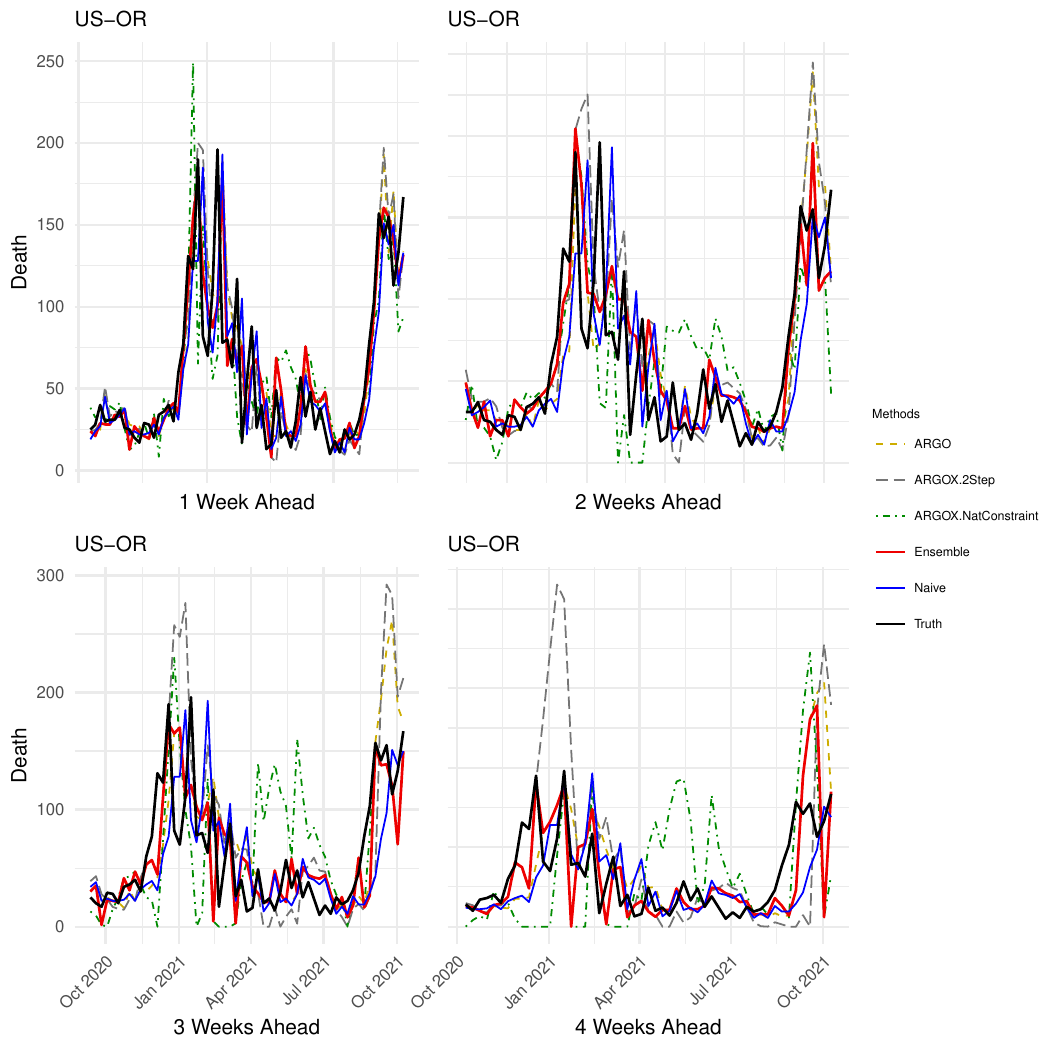} 
\caption{Plots of the COVID-19 1 week (top left), 2 weeks (top right), 3 weeks (bottom left), and 4 weeks (bottom right) ahead estimates for Oregon (OR).}
\end{figure}
\newpage

\begin{table}[ht]
\centering
\renewrobustcmd{\bfseries}{\fontseries{b}\selectfont}
\renewrobustcmd{\boldmath}{}
\newrobustcmd{\B}{\bfseries}
\begin{tabular}{lrrrr}
  \hline
&1 Week Ahead & 2 Weeks Ahead & 3 Weeks Ahead & 4 Weeks Ahead \\ 
    \hline \multicolumn{1}{l}{RMSE} \\
 \hspace{1em} ARGO &  
143.30 & 172.61 & 251.55 & 427.57 \\ 
   \hspace{1em} ARGOX 2Step & 142.55 & 247.80 & 343.32 & 714.24 \\ 
   \hspace{1em} ARGOX NatConstraint & 123.13 & 189.82 & 247.18 & 537.98 \\ 
   \hspace{1em} Ensemble & 122.67 & 140.66 & 233.51 & 369.62 \\ 
   \hspace{1em} Naive& 122.56 & 192.65 & 276.34 & 348.00 \\ 
   \multicolumn{1}{l}{MAE} \\
  \hspace{1em} ARGO &83.08 & 119.39 & 172.02 & 273.79 \\ 
   \hspace{1em} ARGOX 2Step & 85.02 & 136.64 & 211.45 & 405.54 \\ 
   \hspace{1em} ARGOX NatConstraint & 77.20 & 126.19 & 166.88 & 321.04 \\ 
   \hspace{1em} Ensemble & 71.08 & 87.44 & 142.54 & 212.83 \\ 
   \hspace{1em} Naive& 77.75 & 117.59 & 172.47 & 234.11 \\ 
   \multicolumn{1}{l}{Correlation} \\
 \hspace{1em} ARGO &0.94 & 0.91 & 0.84 & 0.79 \\ 
   \hspace{1em} ARGOX 2Step & 0.95 & 0.91 & 0.89 & 0.85 \\ 
   \hspace{1em} ARGOX NatConstraint & 0.95 & 0.91 & 0.88 & 0.82 \\ 
   \hspace{1em} Ensemble & 0.96 & 0.95 & 0.91 & 0.86 \\ 
   \hspace{1em} Naive& 0.95 & 0.89 & 0.78 & 0.66 \\ 
   \hline
\end{tabular}
\caption{Comparison of different methods for state-level COVID-19 1 to 4 weeks ahead incremental death in Pennsylvania (PA). The MSE, MAE, and correlation are reported and best performed method is highlighted in boldface.} 
\end{table}

\begin{figure}[!h] 
  \centering 
\includegraphics[width=0.6\linewidth, page=39]{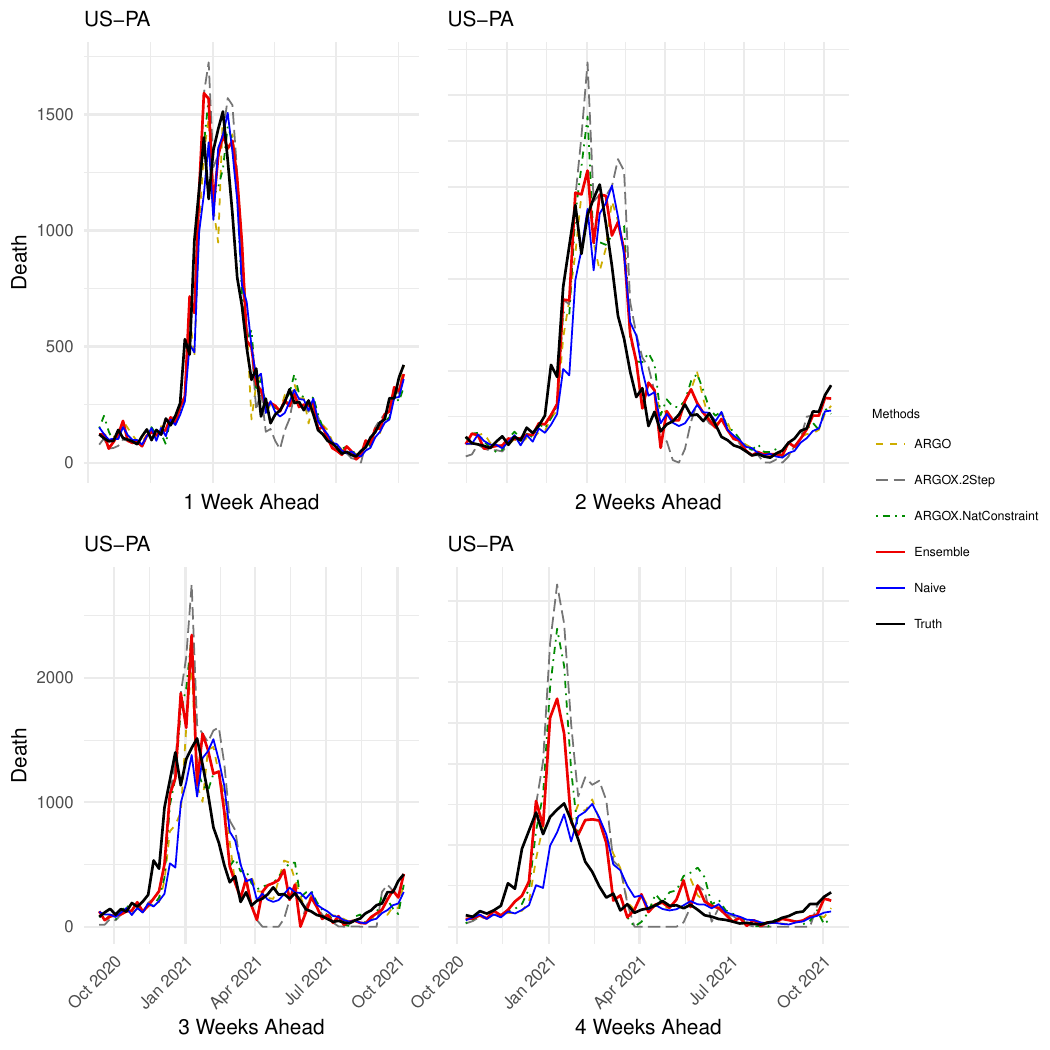} 
\caption{Plots of the COVID-19 1 week (top left), 2 weeks (top right), 3 weeks (bottom left), and 4 weeks (bottom right) ahead estimates for Pennsylvania (PA).}
\end{figure}
\newpage

\begin{table}[ht]
\centering
\renewrobustcmd{\bfseries}{\fontseries{b}\selectfont}
\renewrobustcmd{\boldmath}{}
\newrobustcmd{\B}{\bfseries}
\begin{tabular}{lrrrr}
  \hline
&1 Week Ahead & 2 Weeks Ahead & 3 Weeks Ahead & 4 Weeks Ahead \\ 
    \hline \multicolumn{1}{l}{RMSE} \\
 \hspace{1em} ARGO &  
19.51 & 29.62 & 33.64 & 39.45 \\ 
   \hspace{1em} ARGOX 2Step & 19.32 & 26.02 & 29.53 & 49.71 \\ 
   \hspace{1em} ARGOX NatConstraint & 29.65 & 29.91 & 50.79 & 86.17 \\ 
   \hspace{1em} Ensemble & 19.21 & 18.01 & 21.17 & 32.32 \\ 
   \hspace{1em} Naive& 21.24 & 25.32 & 29.53 & 34.99 \\ 
   \multicolumn{1}{l}{MAE} \\
  \hspace{1em} ARGO &11.17 & 15.20 & 19.14 & 24.45 \\ 
   \hspace{1em} ARGOX 2Step & 10.99 & 15.59 & 18.21 & 29.49 \\ 
   \hspace{1em} ARGOX NatConstraint & 19.30 & 23.17 & 38.33 & 61.30 \\ 
   \hspace{1em} Ensemble & 10.93 & 11.22 & 12.83 & 20.50 \\ 
   \hspace{1em} Naive& 11.71 & 14.84 & 17.63 & 21.96 \\ 
   \multicolumn{1}{l}{Correlation} \\
 \hspace{1em} ARGO &0.85 & 0.73 & 0.70 & 0.58 \\ 
   \hspace{1em} ARGOX 2Step & 0.85 & 0.81 & 0.82 & 0.74 \\ 
   \hspace{1em} ARGOX NatConstraint & 0.63 & 0.59 & 0.10 & 0.28 \\ 
   \hspace{1em} Ensemble & 0.84 & 0.86 & 0.82 & 0.69 \\ 
   \hspace{1em} Naive& 0.80 & 0.73 & 0.65 & 0.53 \\ 
   \hline
\end{tabular}
\caption{Comparison of different methods for state-level COVID-19 1 to 4 weeks ahead incremental death in Rhode Island (RI). The MSE, MAE, and correlation are reported and best performed method is highlighted in boldface.} 
\end{table}

\begin{figure}[!h] 
  \centering 
\includegraphics[width=0.6\linewidth, page=40]{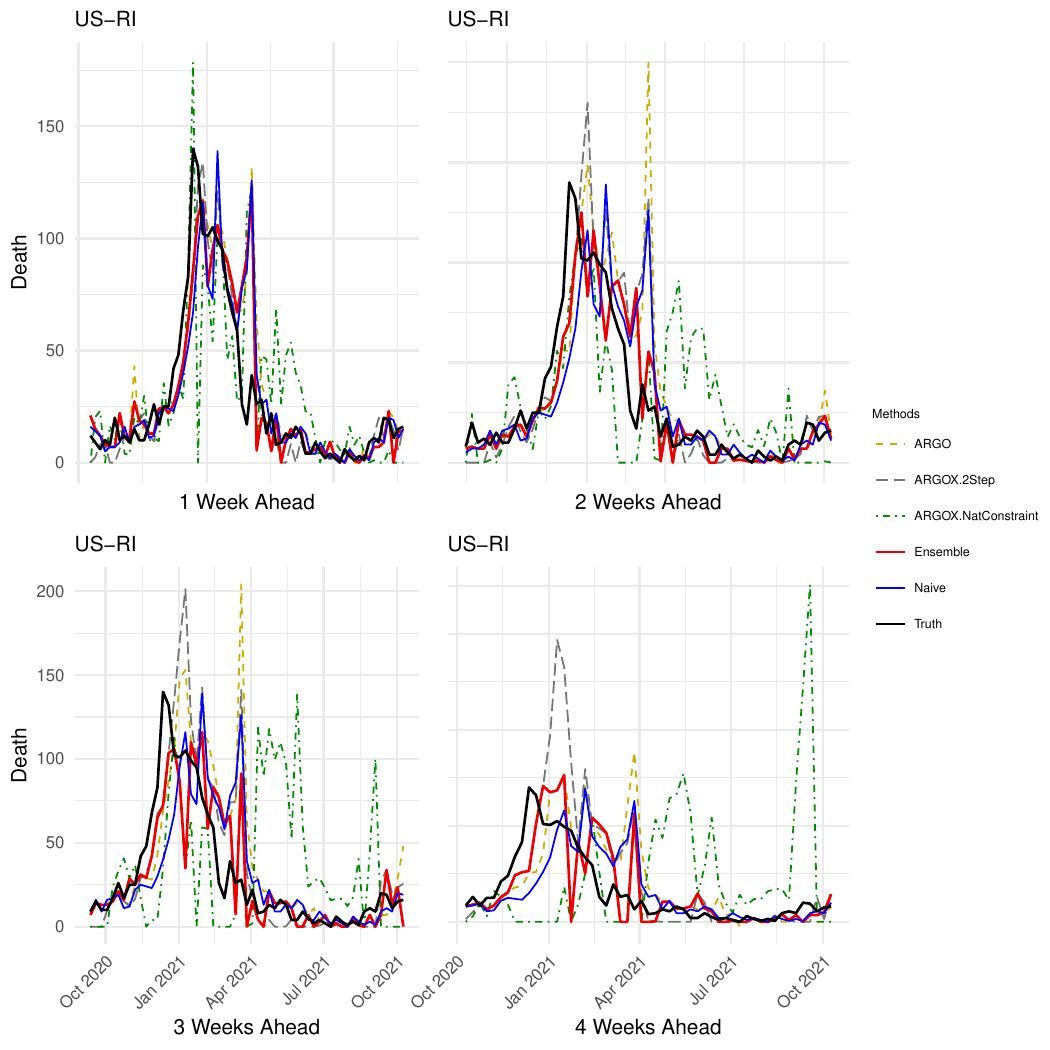} 
\caption{Plots of the COVID-19 1 week (top left), 2 weeks (top right), 3 weeks (bottom left), and 4 weeks (bottom right) ahead estimates for Rhode Island (RI).}
\end{figure}
\newpage

\begin{table}[ht]
\centering
\renewrobustcmd{\bfseries}{\fontseries{b}\selectfont}
\renewrobustcmd{\boldmath}{}
\newrobustcmd{\B}{\bfseries}
\begin{tabular}{lrrrr}
  \hline
&1 Week Ahead & 2 Weeks Ahead & 3 Weeks Ahead & 4 Weeks Ahead \\ 
    \hline \multicolumn{1}{l}{RMSE} \\
 \hspace{1em} ARGO &  
64.02 & 93.01 & 127.33 & 170.70 \\ 
   \hspace{1em} ARGOX 2Step & 75.18 & 99.29 & 138.46 & 193.03 \\ 
   \hspace{1em} ARGOX NatConstraint & 72.23 & 82.94 & 103.95 & 145.52 \\ 
   \hspace{1em} Ensemble & 59.94 & 65.72 & 76.91 & 102.81 \\ 
   \hspace{1em} Naive& 65.09 & 86.89 & 111.93 & 138.95 \\ 
   \multicolumn{1}{l}{MAE} \\
  \hspace{1em} ARGO &42.38 & 65.79 & 88.61 & 117.25 \\ 
   \hspace{1em} ARGOX 2Step & 43.31 & 69.37 & 98.11 & 134.52 \\ 
   \hspace{1em} ARGOX NatConstraint & 45.57 & 58.75 & 78.46 & 113.01 \\ 
   \hspace{1em} Ensemble & 37.33 & 45.68 & 53.45 & 67.91 \\ 
   \hspace{1em} Naive& 43.05 & 62.98 & 82.09 & 99.17 \\ 
   \multicolumn{1}{l}{Correlation} \\
 \hspace{1em} ARGO &0.90 & 0.83 & 0.76 & 0.66 \\ 
   \hspace{1em} ARGOX 2Step & 0.89 & 0.85 & 0.78 & 0.61 \\ 
   \hspace{1em} ARGOX NatConstraint & 0.87 & 0.82 & 0.74 & 0.53 \\ 
   \hspace{1em} Ensemble & 0.92 & 0.90 & 0.88 & 0.80 \\ 
   \hspace{1em} Naive& 0.89 & 0.81 & 0.69 & 0.54 \\ 
   \hline
\end{tabular}
\caption{Comparison of different methods for state-level COVID-19 1 to 4 weeks ahead incremental death in South Carolina (SC). The MSE, MAE, and correlation are reported and best performed method is highlighted in boldface.} 
\end{table}

\begin{figure}[!h] 
  \centering 
\includegraphics[width=0.6\linewidth, page=41]{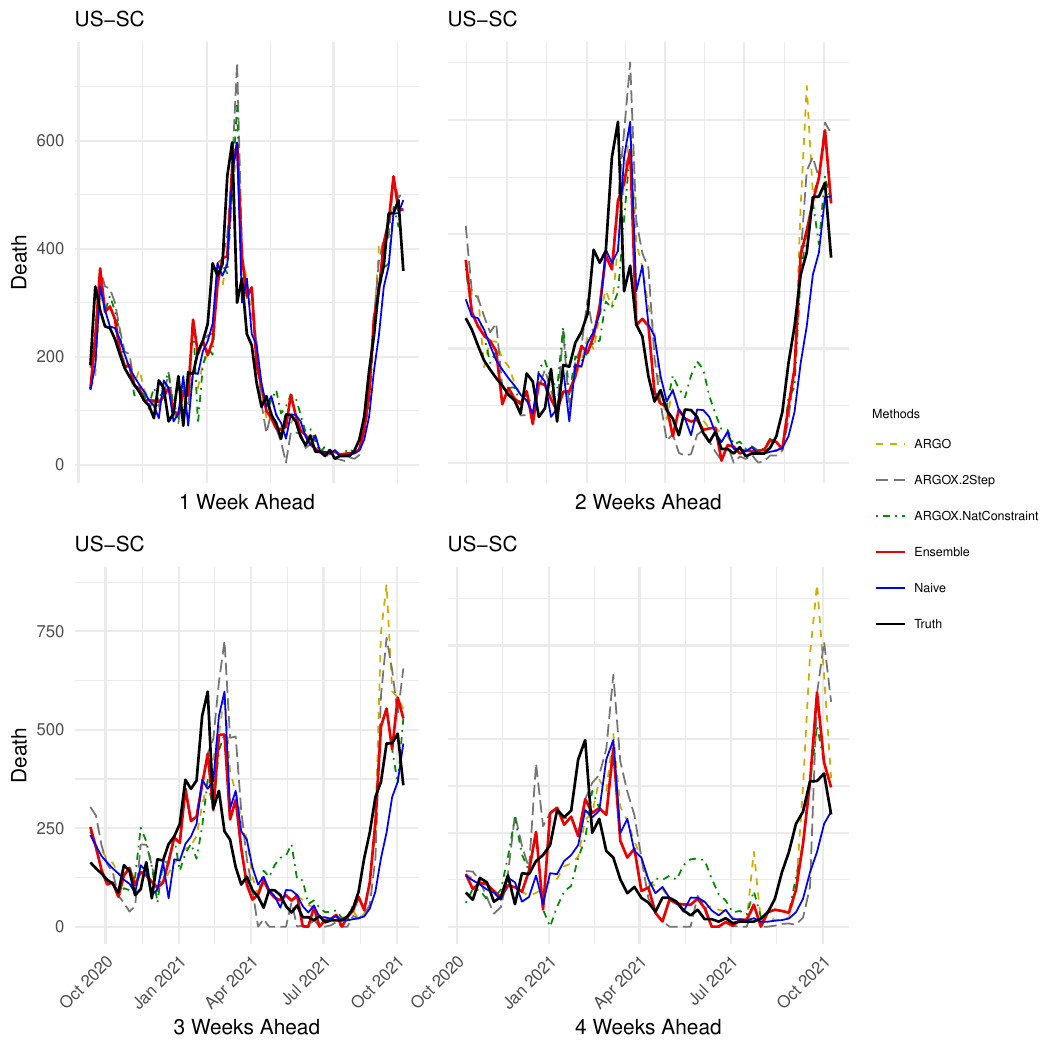} 
\caption{Plots of the COVID-19 1 week (top left), 2 weeks (top right), 3 weeks (bottom left), and 4 weeks (bottom right) ahead estimates for South Carolina (SC).}
\end{figure}
\newpage 

\begin{table}[ht]
\centering
\renewrobustcmd{\bfseries}{\fontseries{b}\selectfont}
\renewrobustcmd{\boldmath}{}
\newrobustcmd{\B}{\bfseries}
\begin{tabular}{lrrrr}
  \hline
&1 Week Ahead & 2 Weeks Ahead & 3 Weeks Ahead & 4 Weeks Ahead \\ 
    \hline \multicolumn{1}{l}{RMSE} \\
 \hspace{1em} ARGO &  
16.91 & 37.23 & 21.86 & 42.98 \\ 
   \hspace{1em} ARGOX 2Step & 17.53 & 20.40 & 46.68 & 109.20 \\ 
   \hspace{1em} ARGOX NatConstraint & 26.76 & 28.82 & 56.74 & 83.92 \\ 
   \hspace{1em} Ensemble & 15.38 & 24.76 & 16.64 & 30.26 \\ 
   \hspace{1em} Naive& 14.69 & 20.94 & 29.93 & 37.60 \\ 
   \multicolumn{1}{l}{MAE} \\
  \hspace{1em} ARGO &10.15 & 15.70 & 14.26 & 24.57 \\ 
   \hspace{1em} ARGOX 2Step & 10.15 & 11.48 & 20.22 & 43.47 \\ 
   \hspace{1em} ARGOX NatConstraint & 19.82 & 23.20 & 41.35 & 58.57 \\ 
   \hspace{1em} Ensemble & 8.75 & 11.81 & 10.75 & 18.06 \\ 
   \hspace{1em} Naive& 8.37 & 12.85 & 18.56 & 25.06 \\ 
   \multicolumn{1}{l}{Correlation} \\
 \hspace{1em} ARGO &0.91 & 0.60 & 0.87 & 0.71 \\ 
   \hspace{1em} ARGOX 2Step & 0.93 & 0.93 & 0.82 & 0.56 \\ 
   \hspace{1em} ARGOX NatConstraint & 0.82 & 0.79 & 0.49 & 0.12 \\ 
   \hspace{1em} Ensemble & 0.94 & 0.83 & 0.93 & 0.83 \\ 
   \hspace{1em} Naive& 0.94 & 0.88 & 0.76 & 0.64 \\ 
   \hline
\end{tabular}
\caption{Comparison of different methods for state-level COVID-19 1 to 4 weeks ahead incremental death in South Dakota (SD). The MSE, MAE, and correlation are reported and best performed method is highlighted in boldface.} 
\end{table}

\begin{figure}[!h] 
  \centering 
\includegraphics[width=0.6\linewidth, page=42]{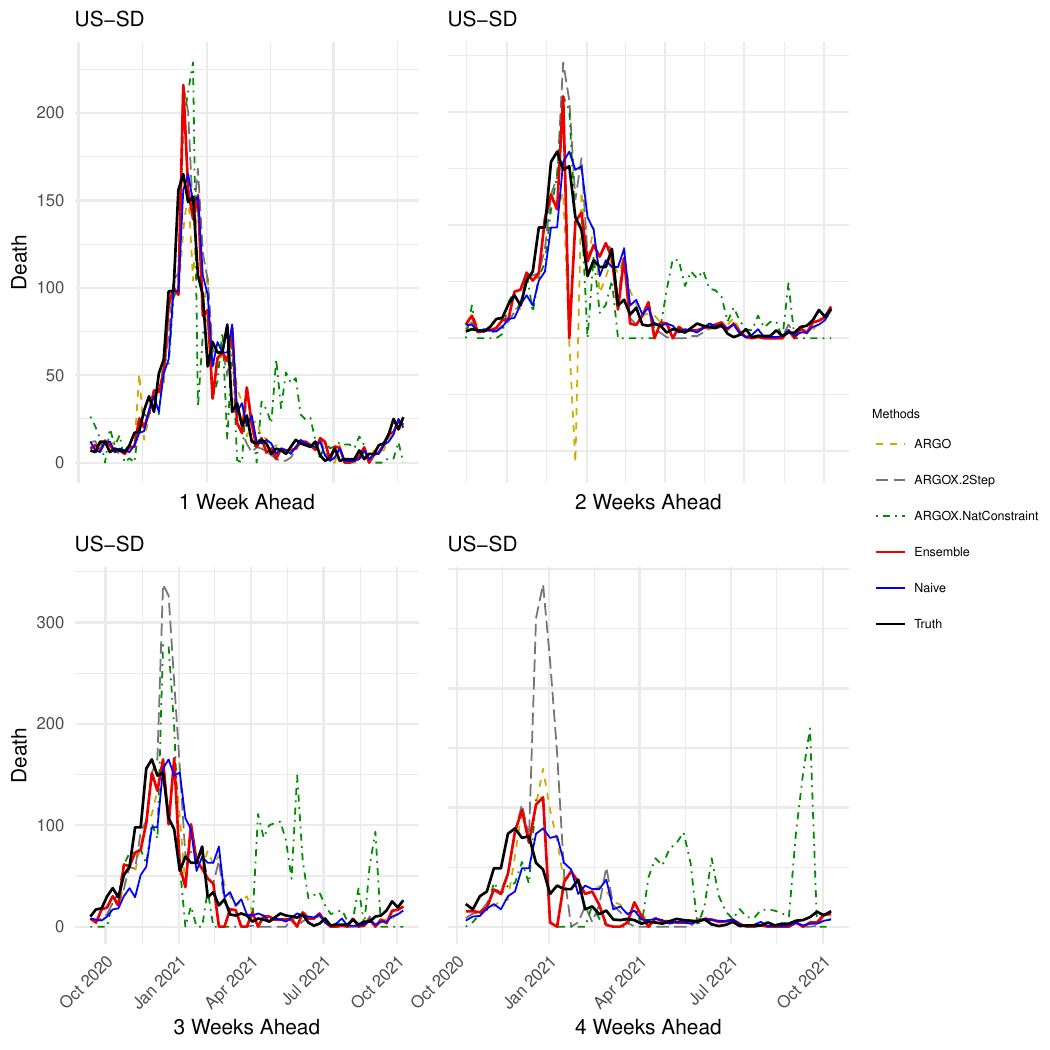} 
\caption{Plots of the COVID-19 1 week (top left), 2 weeks (top right), 3 weeks (bottom left), and 4 weeks (bottom right) ahead estimates for South Dakota (SD).}
\end{figure}
\newpage

\begin{table}[ht]
\centering
\renewrobustcmd{\bfseries}{\fontseries{b}\selectfont}
\renewrobustcmd{\boldmath}{}
\newrobustcmd{\B}{\bfseries}
\begin{tabular}{lrrrr}
  \hline
&1 Week Ahead & 2 Weeks Ahead & 3 Weeks Ahead & 4 Weeks Ahead \\ 
    \hline \multicolumn{1}{l}{RMSE} \\
 \hspace{1em} ARGO &  
111.04 & 131.32 & 143.62 & 173.33 \\ 
   \hspace{1em} ARGOX 2Step & 115.04 & 166.14 & 166.39 & 224.79 \\ 
   \hspace{1em} ARGOX NatConstraint & 110.68 & 147.38 & 135.09 & 181.58 \\ 
   \hspace{1em} Ensemble & 99.06 & 119.89 & 107.77 & 144.30 \\ 
   \hspace{1em} Naive& 98.73 & 144.47 & 151.02 & 178.85 \\ 
   \multicolumn{1}{l}{MAE} \\
  \hspace{1em} ARGO &71.69 & 82.20 & 90.91 & 121.96 \\ 
   \hspace{1em} ARGOX 2Step & 66.75 & 95.37 & 98.64 & 155.41 \\ 
   \hspace{1em} ARGOX NatConstraint & 67.64 & 94.03 & 97.47 & 134.18 \\ 
   \hspace{1em} Ensemble & 58.68 & 68.75 & 67.51 & 91.52 \\ 
   \hspace{1em} Naive& 62.51 & 89.61 & 95.53 & 121.94 \\ 
   \multicolumn{1}{l}{Correlation} \\
 \hspace{1em} ARGO &0.83 & 0.77 & 0.75 & 0.69 \\ 
   \hspace{1em} ARGOX 2Step & 0.85 & 0.77 & 0.80 & 0.73 \\ 
   \hspace{1em} ARGOX NatConstraint & 0.84 & 0.72 & 0.77 & 0.61 \\ 
   \hspace{1em} Ensemble & 0.88 & 0.82 & 0.86 & 0.83 \\ 
   \hspace{1em} Naive& 0.88 & 0.74 & 0.73 & 0.65 \\ 
   \hline
\end{tabular}
\caption{Comparison of different methods for state-level COVID-19 1 to 4 weeks ahead incremental death in Tennessee (TN). The MSE, MAE, and correlation are reported and best performed method is highlighted in boldface.} 
\end{table}

\begin{figure}[!h] 
  \centering 
\includegraphics[width=0.6\linewidth, page=43]{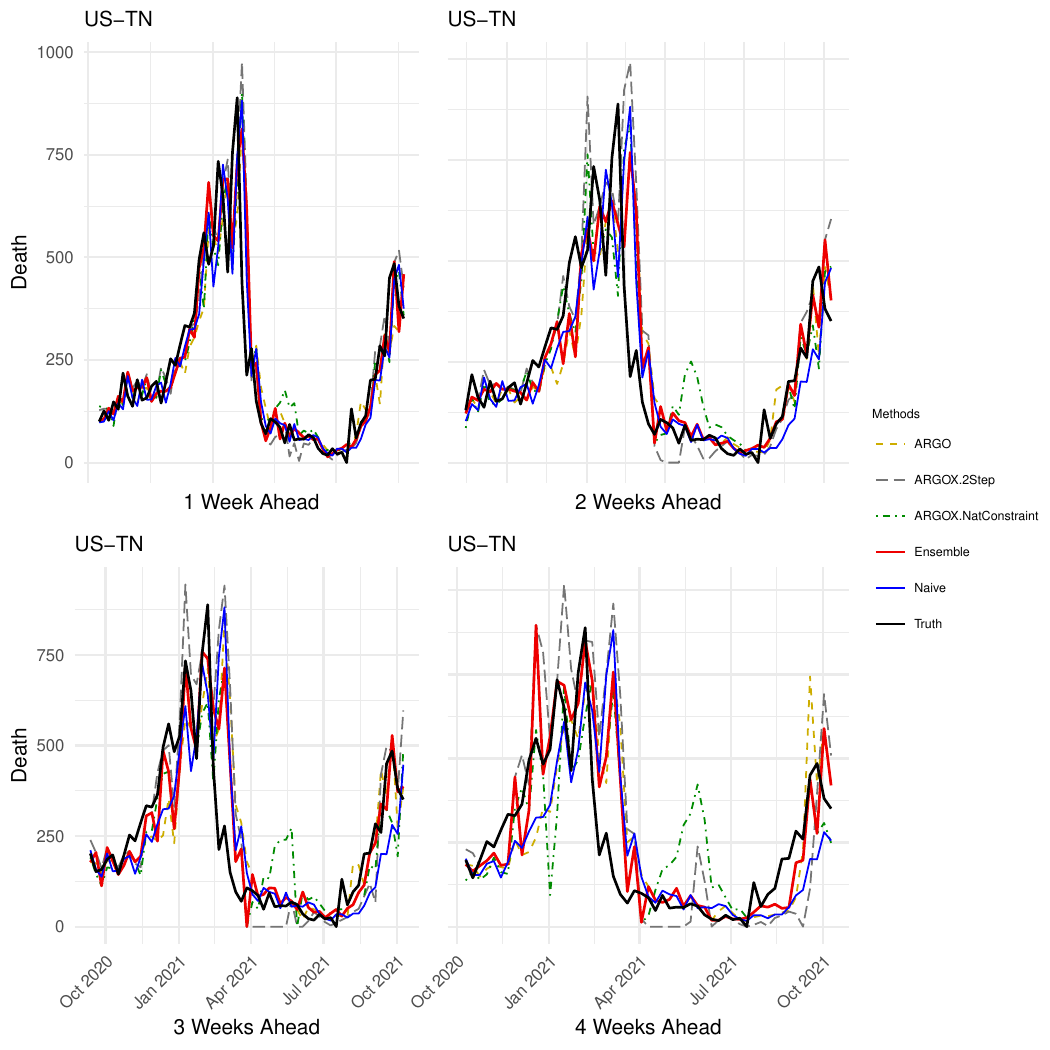} 
\caption{Plots of the COVID-19 1 week (top left), 2 weeks (top right), 3 weeks (bottom left), and 4 weeks (bottom right) ahead estimates for Tennessee (TN).}
\end{figure}
\newpage

\begin{table}[ht]
\centering
\renewrobustcmd{\bfseries}{\fontseries{b}\selectfont}
\renewrobustcmd{\boldmath}{}
\newrobustcmd{\B}{\bfseries}
\begin{tabular}{lrrrr}
  \hline
&1 Week Ahead & 2 Weeks Ahead & 3 Weeks Ahead & 4 Weeks Ahead \\ 
    \hline \multicolumn{1}{l}{RMSE} \\
 \hspace{1em} ARGO &  
273.16 & 385.13 & 500.99 & 719.81 \\ 
   \hspace{1em} ARGOX 2Step & 325.74 & 519.08 & 645.13 & 1017.57 \\ 
   \hspace{1em} ARGOX NatConstraint & 289.95 & 446.70 & 513.92 & 836.86 \\ 
   \hspace{1em} Ensemble & 242.15 & 340.47 & 444.25 & 554.74 \\ 
   \hspace{1em} Naive& 299.93 & 406.20 & 461.48 & 562.10 \\ 
   \multicolumn{1}{l}{MAE} \\
  \hspace{1em} ARGO &181.02 & 269.28 & 347.95 & 479.51 \\ 
   \hspace{1em} ARGOX 2Step & 221.73 & 368.83 & 494.77 & 716.01 \\ 
   \hspace{1em} ARGOX NatConstraint & 193.36 & 324.47 & 390.53 & 599.32 \\ 
   \hspace{1em} Ensemble & 159.38 & 233.79 & 311.64 & 370.09 \\ 
   \hspace{1em} Naive& 194.68 & 282.46 & 342.60 & 428.34 \\ 
   \multicolumn{1}{l}{Correlation} \\
 \hspace{1em} ARGO &0.92 & 0.86 & 0.80 & 0.69 \\ 
   \hspace{1em} ARGOX 2Step & 0.91 & 0.85 & 0.82 & 0.68 \\ 
   \hspace{1em} ARGOX NatConstraint & 0.90 & 0.81 & 0.79 & 0.63 \\ 
   \hspace{1em} Ensemble & 0.93 & 0.89 & 0.86 & 0.79 \\ 
   \hspace{1em} Naive& 0.89 & 0.80 & 0.74 & 0.63 \\ 
   \hline
\end{tabular}
\caption{Comparison of different methods for state-level COVID-19 1 to 4 weeks ahead incremental death in Texas (TX). The MSE, MAE, and correlation are reported and best performed method is highlighted in boldface.} 
\end{table}

\begin{figure}[!h] 
  \centering 
\includegraphics[width=0.6\linewidth, page=44]{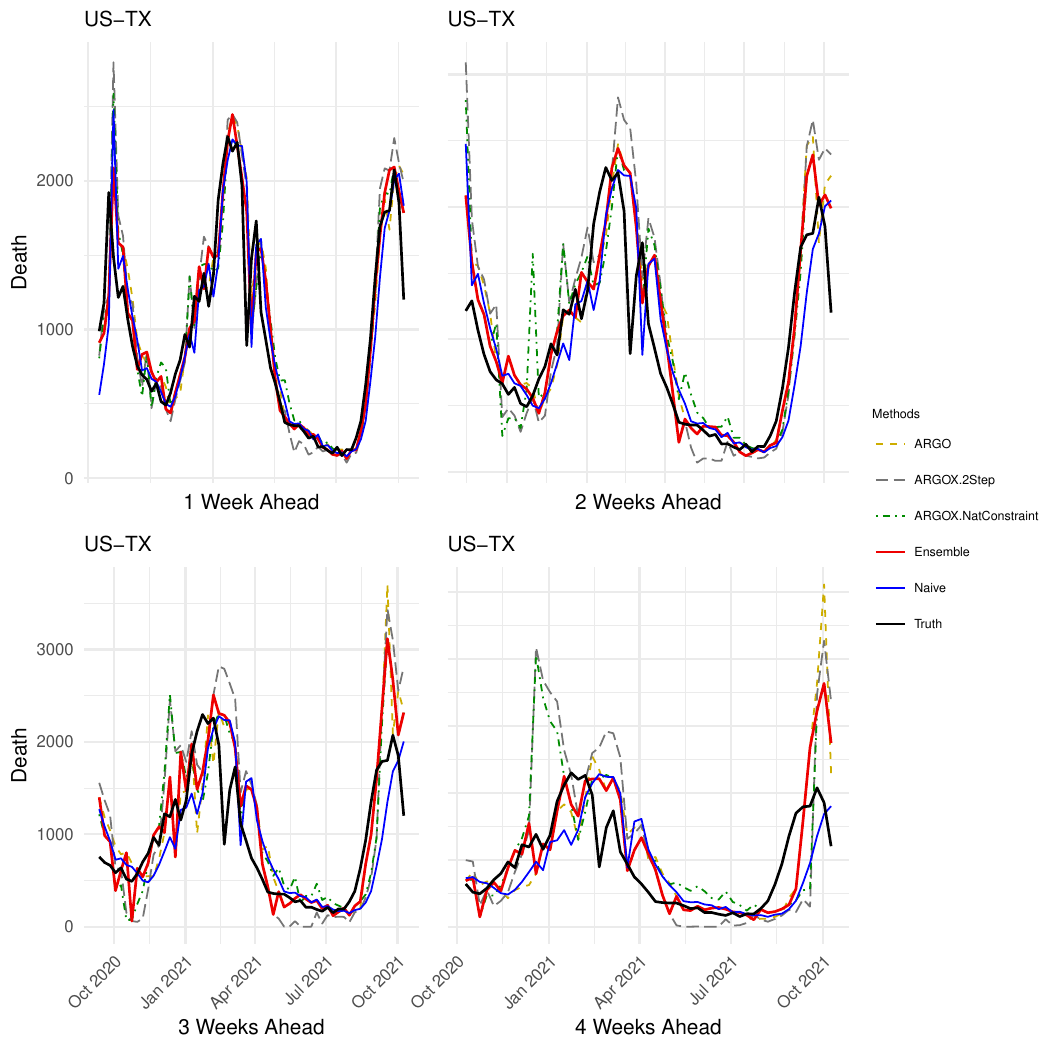} 
\caption{Plots of the COVID-19 1 week (top left), 2 weeks (top right), 3 weeks (bottom left), and 4 weeks (bottom right) ahead estimates for Texas (TX).}
\end{figure}
\newpage

\begin{table}[ht]
\centering
\renewrobustcmd{\bfseries}{\fontseries{b}\selectfont}
\renewrobustcmd{\boldmath}{}
\newrobustcmd{\B}{\bfseries}
\begin{tabular}{lrrrr}
  \hline
&1 Week Ahead & 2 Weeks Ahead & 3 Weeks Ahead & 4 Weeks Ahead \\ 
    \hline \multicolumn{1}{l}{RMSE} \\
 \hspace{1em} ARGO &  
16.90 & 20.70 & 26.79 & 27.06 \\ 
   \hspace{1em} ARGOX 2Step & 16.85 & 21.93 & 26.94 & 44.16 \\ 
   \hspace{1em} ARGOX NatConstraint & 29.64 & 37.37 & 55.88 & 80.03 \\ 
   \hspace{1em} Ensemble & 13.90 & 13.42 & 17.90 & 25.91 \\ 
   \hspace{1em} Naive& 14.45 & 18.31 & 20.63 & 24.40 \\ 
   \multicolumn{1}{l}{MAE} \\
  \hspace{1em} ARGO &12.71 & 16.26 & 20.26 & 21.74 \\ 
   \hspace{1em} ARGOX 2Step & 12.67 & 16.34 & 20.21 & 29.89 \\ 
   \hspace{1em} ARGOX NatConstraint & 22.53 & 29.02 & 44.06 & 61.34 \\ 
   \hspace{1em} Ensemble & 10.85 & 11.14 & 14.55 & 18.56 \\ 
   \hspace{1em} Naive& 10.54 & 15.13 & 17.53 & 21.00 \\ 
   \multicolumn{1}{l}{Correlation} \\
 \hspace{1em} ARGO &0.80 & 0.72 & 0.63 & 0.62 \\ 
   \hspace{1em} ARGOX 2Step & 0.83 & 0.81 & 0.83 & 0.74 \\ 
   \hspace{1em} ARGOX NatConstraint & 0.48 & 0.21 & 0.13 & 0.25 \\ 
   \hspace{1em} Ensemble & 0.87 & 0.88 & 0.82 & 0.77 \\ 
   \hspace{1em} Naive& 0.86 & 0.78 & 0.73 & 0.63 \\ 
   \hline
\end{tabular}
\caption{Comparison of different methods for state-level COVID-19 1 to 4 weeks ahead incremental death in Utah (UT). The MSE, MAE, and correlation are reported and best performed method is highlighted in boldface.} 
\end{table}

\begin{figure}[!h] 
  \centering 
\includegraphics[width=0.6\linewidth, page=45]{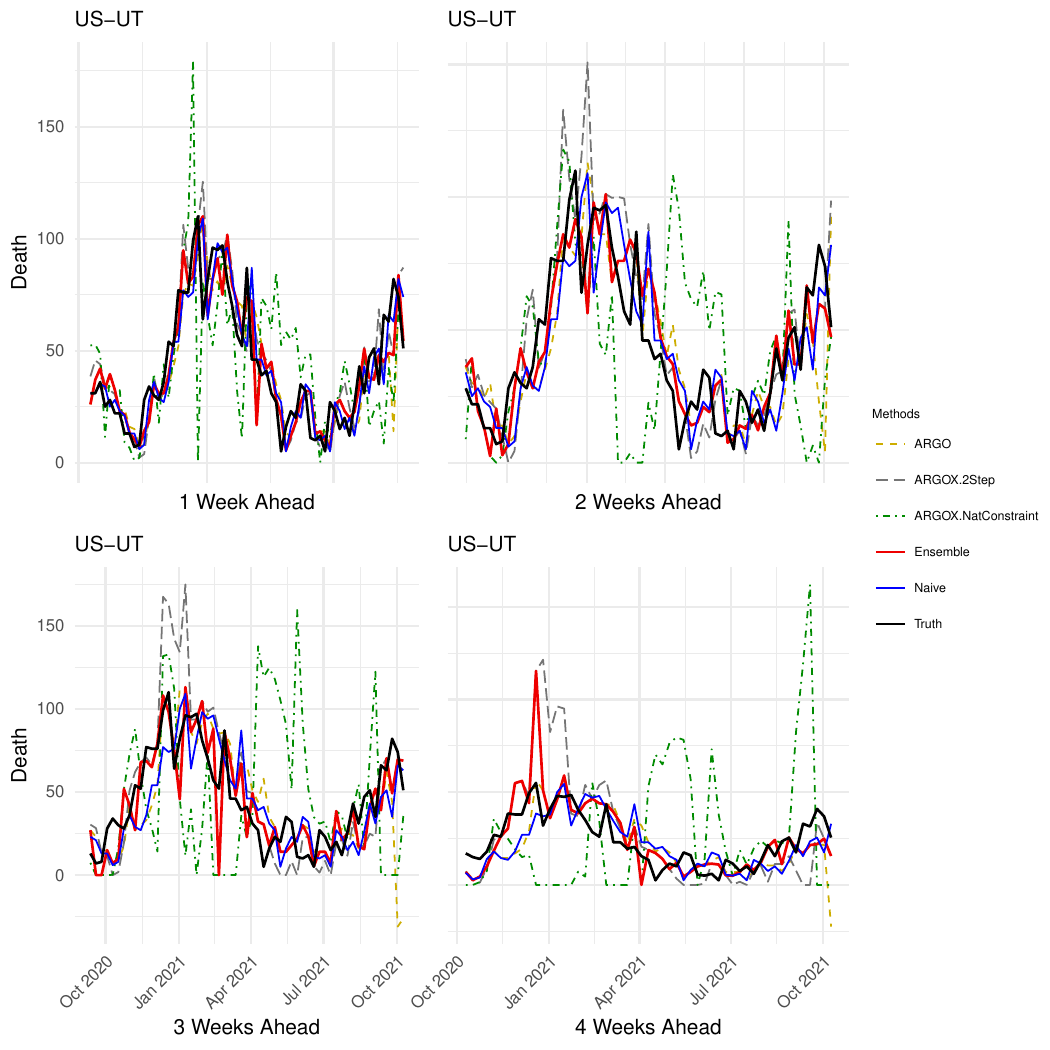} 
\caption{Plots of the COVID-19 1 week (top left), 2 weeks (top right), 3 weeks (bottom left), and 4 weeks (bottom right) ahead estimates for Utah (UT).}
\end{figure}
\newpage

\begin{table}[ht]
\centering
\renewrobustcmd{\bfseries}{\fontseries{b}\selectfont}
\renewrobustcmd{\boldmath}{}
\newrobustcmd{\B}{\bfseries}
\begin{tabular}{lrrrr}
  \hline
&1 Week Ahead & 2 Weeks Ahead & 3 Weeks Ahead & 4 Weeks Ahead \\ 
    \hline \multicolumn{1}{l}{RMSE} \\
 \hspace{1em} ARGO &  
221.80 & 213.93 & 276.08 & 263.32 \\ 
   \hspace{1em} ARGOX 2Step & 197.21 & 249.07 & 320.41 & 322.40 \\ 
   \hspace{1em} ARGOX NatConstraint & 189.60 & 240.74 & 310.94 & 315.27 \\ 
   \hspace{1em} Ensemble & 207.71 & 211.39 & 258.33 & 263.34 \\ 
   \hspace{1em} Naive& 163.07 & 248.77 & 275.91 & 278.63 \\ 
   \multicolumn{1}{l}{MAE} \\
  \hspace{1em} ARGO &89.77 & 101.31 & 131.39 & 131.34 \\ 
   \hspace{1em} ARGOX 2Step & 90.04 & 128.60 & 157.04 & 174.71 \\ 
   \hspace{1em} ARGOX NatConstraint & 91.51 & 122.53 & 162.62 & 183.44 \\ 
   \hspace{1em} Ensemble & 83.60 & 96.66 & 118.49 & 127.35 \\ 
   \hspace{1em} Naive& 68.60 & 115.08 & 132.18 & 142.55 \\ 
   \multicolumn{1}{l}{Correlation} \\
 \hspace{1em} ARGO &0.25 & 0.31 & 0.19 & 0.32 \\ 
   \hspace{1em} ARGOX 2Step & 0.54 & 0.32 & 0.22 & 0.27 \\ 
   \hspace{1em} ARGOX NatConstraint & 0.52 & 0.20 & 0.05 & 0.04 \\ 
   \hspace{1em} Ensemble & 0.33 & 0.38 & 0.26 & 0.32 \\ 
   \hspace{1em} Naive& 0.67 & 0.28 & 0.16 & 0.19 \\ 
   \hline
\end{tabular}
\caption{Comparison of different methods for state-level COVID-19 1 to 4 weeks ahead incremental death in Virginia (VA). The MSE, MAE, and correlation are reported and best performed method is highlighted in boldface.} 
\end{table}

\begin{figure}[!h] 
  \centering 
\includegraphics[width=0.6\linewidth, page=46]{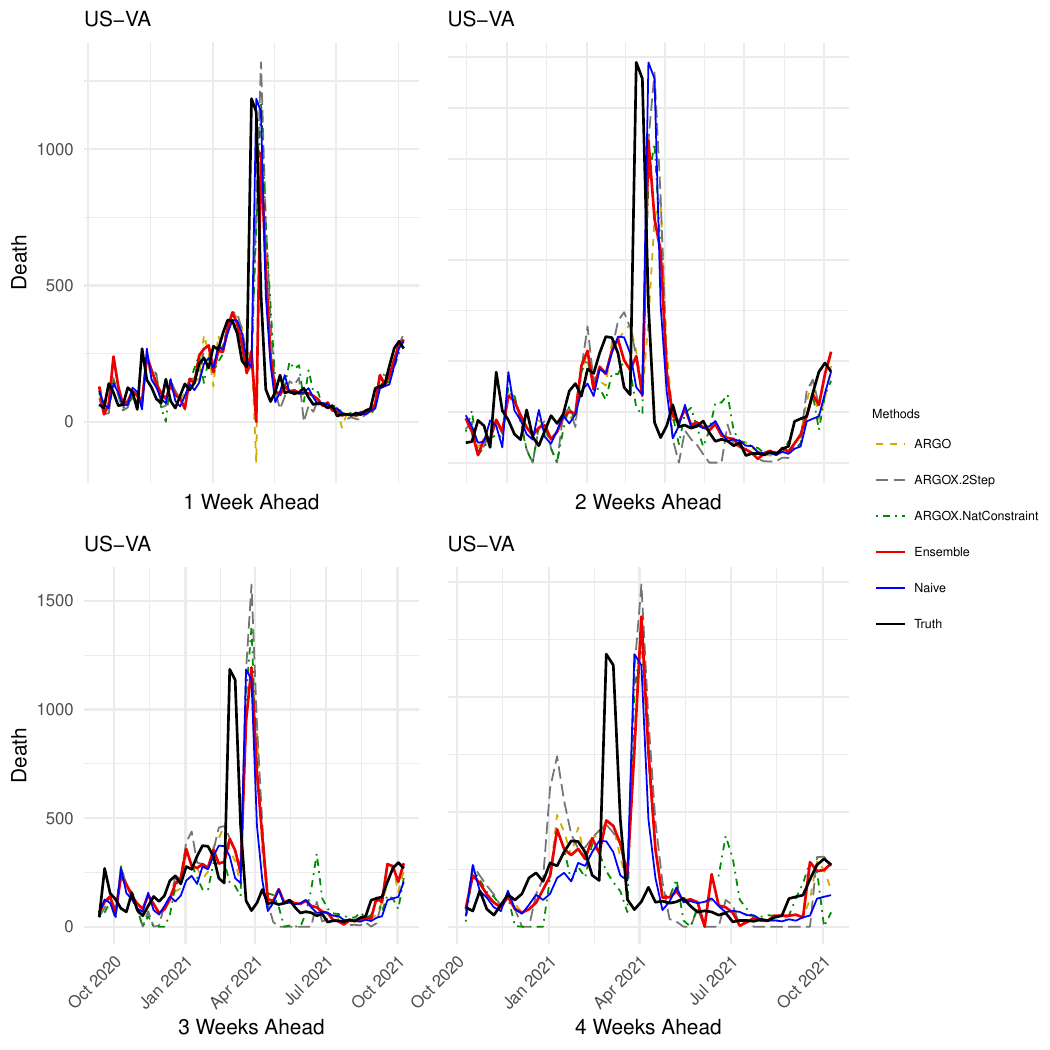} 
\caption{Plots of the COVID-19 1 week (top left), 2 weeks (top right), 3 weeks (bottom left), and 4 weeks (bottom right) ahead estimates for Virginia (VA).}
\end{figure}
\newpage

\begin{table}[ht]
\centering
\renewrobustcmd{\bfseries}{\fontseries{b}\selectfont}
\renewrobustcmd{\boldmath}{}
\newrobustcmd{\B}{\bfseries}
\begin{tabular}{lrrrr}
  \hline
&1 Week Ahead & 2 Weeks Ahead & 3 Weeks Ahead & 4 Weeks Ahead \\ 
    \hline \multicolumn{1}{l}{RMSE} \\
 \hspace{1em} ARGO &  
8.64 & 7.36 & 5.26 & 5.48 \\ 
   \hspace{1em} ARGOX 2Step & 2.76 & 3.47 & 4.22 & 4.83 \\ 
   \hspace{1em} ARGOX NatConstraint & 8.64 & 7.36 & 5.26 & 5.48 \\ 
   \hspace{1em} Ensemble & 3.25 & 4.17 & 4.59 & 5.05 \\ 
   \hspace{1em} Naive& 3.07 & 3.68 & 4.52 & 5.27 \\ 
   \multicolumn{1}{l}{MAE} \\
  \hspace{1em} ARGO &3.20 & 3.21 & 3.35 & 3.94 \\ 
   \hspace{1em} ARGOX 2Step & 1.57 & 2.08 & 2.48 & 3.21 \\ 
   \hspace{1em} ARGOX NatConstraint & 3.20 & 3.21 & 3.35 & 3.94 \\ 
   \hspace{1em} Ensemble & 1.90 & 2.67 & 3.00 & 3.54 \\ 
   \hspace{1em} Naive& 1.82 & 2.34 & 2.89 & 3.70 \\ 
   \multicolumn{1}{l}{Correlation} \\
 \hspace{1em} ARGO &0.41 & 0.45 & 0.57 & 0.48 \\ 
   \hspace{1em} ARGOX 2Step & 0.88 & 0.80 & 0.72 & 0.62 \\ 
   \hspace{1em} ARGOX NatConstraint & 0.41 & 0.45 & 0.57 & 0.48 \\ 
   \hspace{1em} Ensemble & 0.83 & 0.72 & 0.63 & 0.58 \\ 
   \hspace{1em} Naive& 0.80 & 0.72 & 0.58 & 0.43 \\ 
   \hline
\end{tabular}
\caption{Comparison of different methods for state-level COVID-19 1 to 4 weeks ahead incremental death in Vermont (VT). The MSE, MAE, and correlation are reported and best performed method is highlighted in boldface.} 
\end{table}

\begin{figure}[!h] 
  \centering 
\includegraphics[width=0.6\linewidth, page=47]{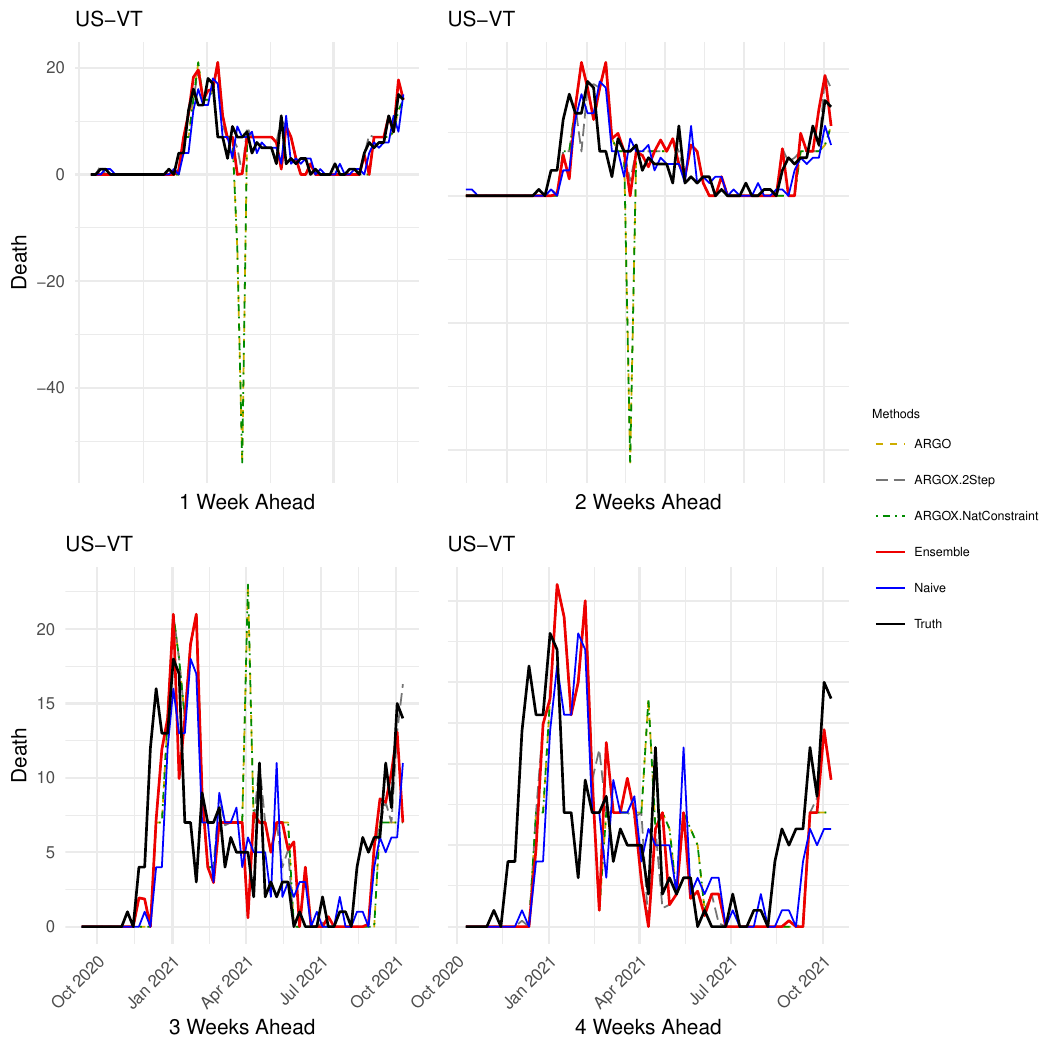} 
\caption{Plots of the COVID-19 1 week (top left), 2 weeks (top right), 3 weeks (bottom left), and 4 weeks (bottom right) ahead estimates for Vermont (VT).}
\end{figure}
\newpage

\begin{table}[ht]
\centering
\renewrobustcmd{\bfseries}{\fontseries{b}\selectfont}
\renewrobustcmd{\boldmath}{}
\newrobustcmd{\B}{\bfseries}
\begin{tabular}{lrrrr}
  \hline
&1 Week Ahead & 2 Weeks Ahead & 3 Weeks Ahead & 4 Weeks Ahead \\ 
    \hline \multicolumn{1}{l}{RMSE} \\
 \hspace{1em} ARGO &  
46.00 & 53.79 & 65.19 & 68.98 \\ 
   \hspace{1em} ARGOX 2Step & 45.06 & 59.67 & 69.13 & 97.48 \\ 
   \hspace{1em} ARGOX NatConstraint & 58.13 & 66.44 & 84.37 & 85.54 \\ 
   \hspace{1em} Ensemble & 42.26 & 44.87 & 54.70 & 54.92 \\ 
   \hspace{1em} Naive& 41.50 & 45.88 & 62.63 & 68.29 \\ 
   \multicolumn{1}{l}{MAE} \\
  \hspace{1em} ARGO &27.31 & 35.67 & 42.77 & 51.70 \\ 
   \hspace{1em} ARGOX 2Step & 28.83 & 37.59 & 49.50 & 70.45 \\ 
   \hspace{1em} ARGOX NatConstraint & 36.50 & 44.31 & 61.72 & 64.17 \\ 
   \hspace{1em} Ensemble & 24.45 & 24.90 & 34.47 & 40.93 \\ 
   \hspace{1em} Naive& 27.03 & 33.85 & 45.42 & 52.17 \\ 
   \multicolumn{1}{l}{Correlation} \\
 \hspace{1em} ARGO &0.80 & 0.78 & 0.68 & 0.64 \\ 
   \hspace{1em} ARGOX 2Step & 0.83 & 0.77 & 0.76 & 0.68 \\ 
   \hspace{1em} ARGOX NatConstraint & 0.68 & 0.57 & 0.39 & 0.42 \\ 
   \hspace{1em} Ensemble & 0.84 & 0.83 & 0.78 & 0.82 \\ 
   \hspace{1em} Naive& 0.84 & 0.80 & 0.64 & 0.57 \\ 
   \hline
\end{tabular}
\caption{Comparison of different methods for state-level COVID-19 1 to 4 weeks ahead incremental death in Washington (WA). The MSE, MAE, and correlation are reported and best performed method is highlighted in boldface.} 
\end{table}

\begin{figure}[!h] 
  \centering 
\includegraphics[width=0.6\linewidth, page=48]{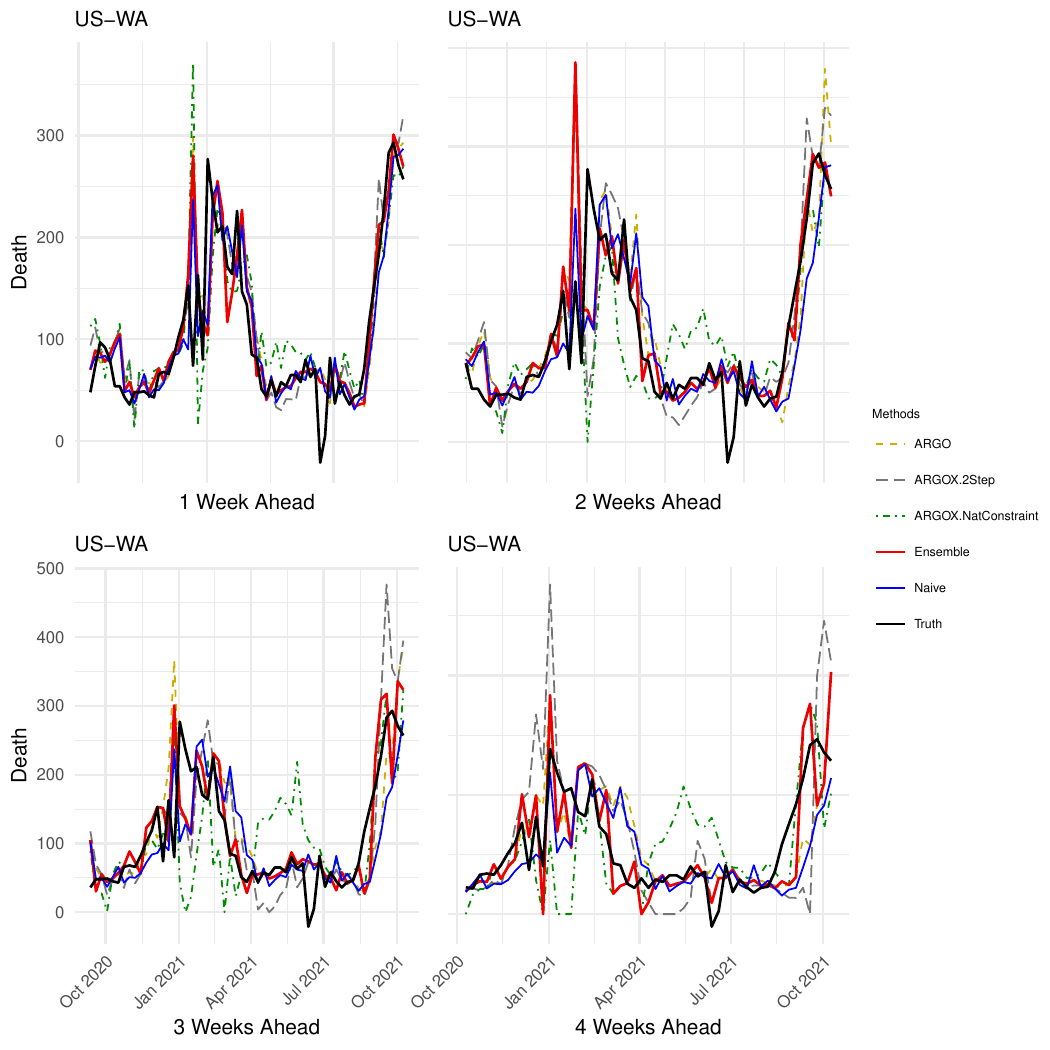} 
\caption{Plots of the COVID-19 1 week (top left), 2 weeks (top right), 3 weeks (bottom left), and 4 weeks (bottom right) ahead estimates for Washington (WA).}
\end{figure}
\newpage

\begin{table}[ht]
\centering
\renewrobustcmd{\bfseries}{\fontseries{b}\selectfont}
\renewrobustcmd{\boldmath}{}
\newrobustcmd{\B}{\bfseries}
\begin{tabular}{lrrrr}
  \hline
&1 Week Ahead & 2 Weeks Ahead & 3 Weeks Ahead & 4 Weeks Ahead \\ 
    \hline \multicolumn{1}{l}{RMSE} \\
 \hspace{1em} ARGO &  
48.10 & 57.70 & 76.39 & 111.45 \\ 
   \hspace{1em} ARGOX 2Step & 55.62 & 68.56 & 108.31 & 229.51 \\ 
   \hspace{1em} ARGOX NatConstraint & 58.61 & 62.93 & 99.19 & 186.40 \\ 
   \hspace{1em} Ensemble & 49.94 & 52.61 & 70.04 & 112.62 \\ 
   \hspace{1em} Naive& 49.54 & 64.24 & 76.17 & 91.21 \\ 
   \multicolumn{1}{l}{MAE} \\
  \hspace{1em} ARGO &33.38 & 41.07 & 54.63 & 79.13 \\ 
   \hspace{1em} ARGOX 2Step & 36.94 & 48.53 & 75.03 & 125.25 \\ 
   \hspace{1em} ARGOX NatConstraint & 42.94 & 48.30 & 72.87 & 132.51 \\ 
   \hspace{1em} Ensemble & 32.73 & 38.69 & 49.11 & 74.21 \\ 
   \hspace{1em} Naive& 34.80 & 45.93 & 55.46 & 64.45 \\ 
   \multicolumn{1}{l}{Correlation} \\
 \hspace{1em} ARGO &0.91 & 0.88 & 0.80 & 0.81 \\ 
   \hspace{1em} ARGOX 2Step & 0.90 & 0.90 & 0.85 & 0.74 \\ 
   \hspace{1em} ARGOX NatConstraint & 0.87 & 0.86 & 0.74 & 0.56 \\ 
   \hspace{1em} Ensemble & 0.92 & 0.90 & 0.86 & 0.81 \\ 
   \hspace{1em} Naive& 0.91 & 0.85 & 0.80 & 0.72 \\ 
   \hline
\end{tabular}
\caption{Comparison of different methods for state-level COVID-19 1 to 4 weeks ahead incremental death in Wisconsin (WI). The MSE, MAE, and correlation are reported and best performed method is highlighted in boldface.} 
\end{table}

\begin{figure}[!h] 
  \centering 
\includegraphics[width=0.6\linewidth, page=49]{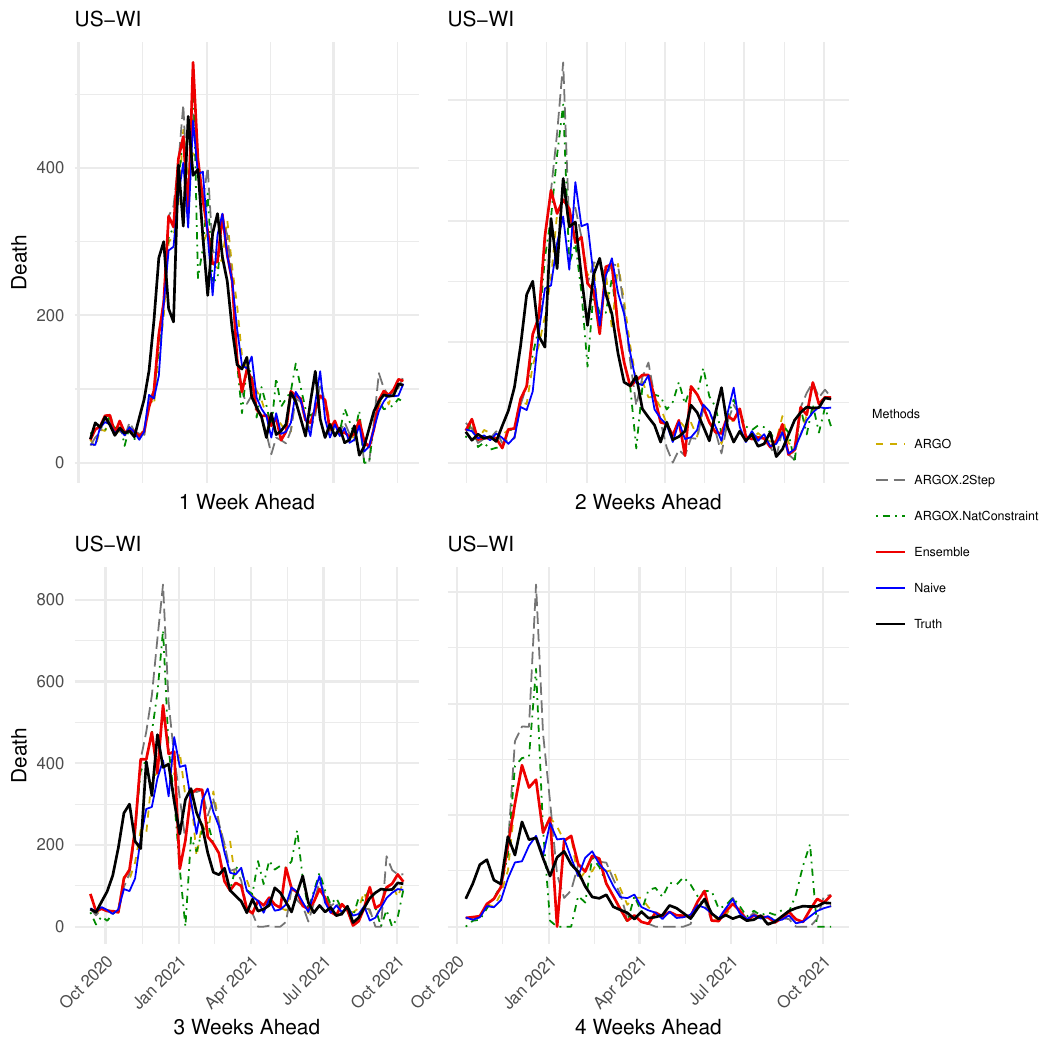} 
\caption{Plots of the COVID-19 1 week (top left), 2 weeks (top right), 3 weeks (bottom left), and 4 weeks (bottom right) ahead estimates for Wisconsin (WI).}
\end{figure}
\newpage

\begin{table}[ht]
\centering
\renewrobustcmd{\bfseries}{\fontseries{b}\selectfont}
\renewrobustcmd{\boldmath}{}
\newrobustcmd{\B}{\bfseries}
\begin{tabular}{lrrrr}
  \hline
&1 Week Ahead & 2 Weeks Ahead & 3 Weeks Ahead & 4 Weeks Ahead \\ 
    \hline \multicolumn{1}{l}{RMSE} \\
 \hspace{1em} ARGO &  
36.85 & 41.81 & 50.74 & 66.80 \\ 
   \hspace{1em} ARGOX 2Step & 36.77 & 43.41 & 46.29 & 88.14 \\ 
   \hspace{1em} ARGOX NatConstraint & 46.54 & 50.64 & 72.46 & 102.88 \\ 
   \hspace{1em} Ensemble & 32.55 & 36.71 & 40.02 & 59.11 \\ 
   \hspace{1em} Naive& 36.06 & 45.21 & 49.81 & 59.77 \\ 
   \multicolumn{1}{l}{MAE} \\
  \hspace{1em} ARGO &24.48 & 29.54 & 37.16 & 45.38 \\ 
   \hspace{1em} ARGOX 2Step & 22.45 & 30.24 & 33.33 & 64.13 \\ 
   \hspace{1em} ARGOX NatConstraint & 30.78 & 37.15 & 55.19 & 80.26 \\ 
   \hspace{1em} Ensemble & 20.96 & 25.77 & 27.40 & 39.51 \\ 
   \hspace{1em} Naive& 22.63 & 31.08 & 35.18 & 43.58 \\ 
   \multicolumn{1}{l}{Correlation} \\
 \hspace{1em} ARGO &0.78 & 0.73 & 0.63 & 0.41 \\ 
   \hspace{1em} ARGOX 2Step & 0.80 & 0.78 & 0.79 & 0.60 \\ 
   \hspace{1em} ARGOX NatConstraint & 0.64 & 0.58 & 0.24 & 0.02 \\ 
   \hspace{1em} Ensemble & 0.84 & 0.80 & 0.78 & 0.58 \\ 
   \hspace{1em} Naive& 0.79 & 0.67 & 0.61 & 0.45 \\ 
   \hline
\end{tabular}
\caption{Comparison of different methods for state-level COVID-19 1 to 4 weeks ahead incremental death in West Virginia (WV). The MSE, MAE, and correlation are reported and best performed method is highlighted in boldface.} 
\end{table}

\begin{figure}[!h] 
  \centering 
\includegraphics[width=0.6\linewidth, page=50]{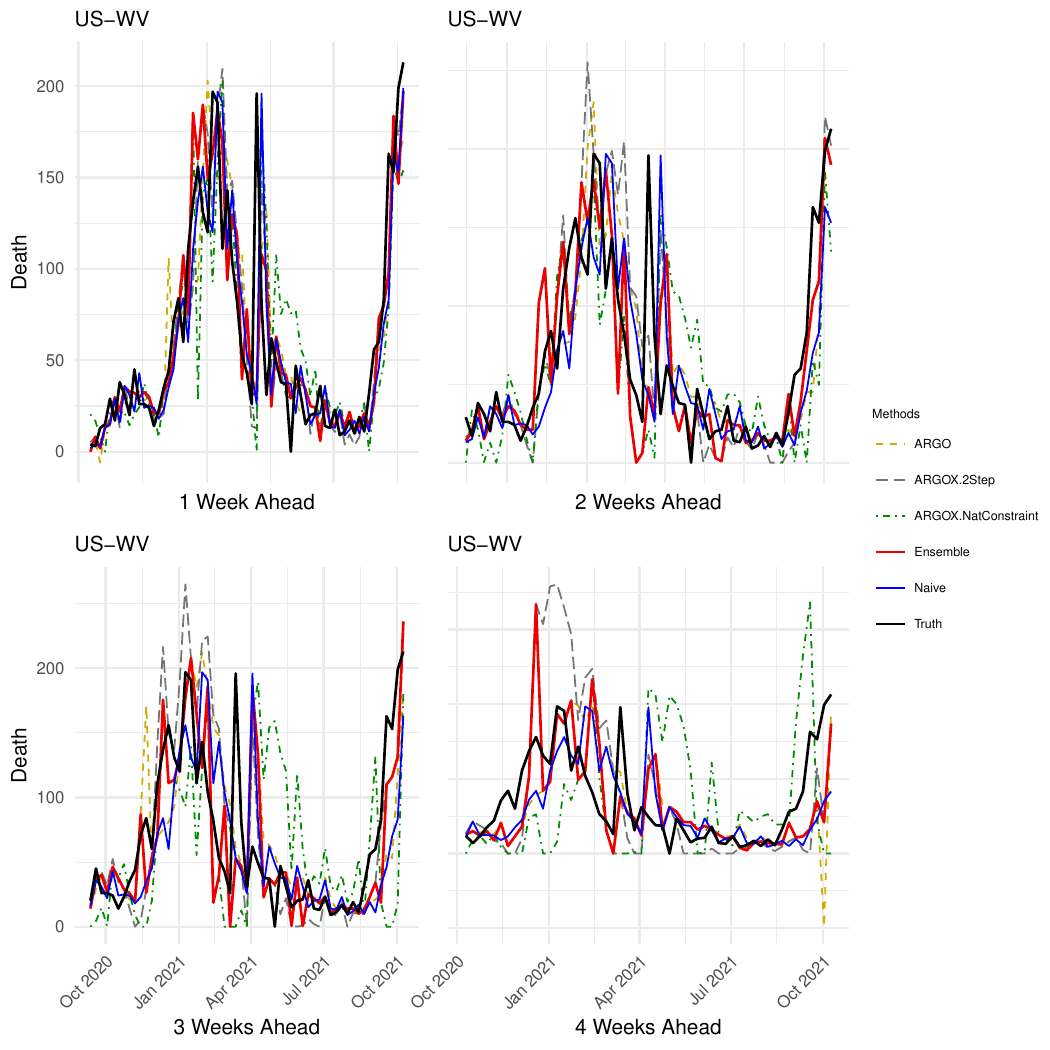} 
\caption{Plots of the COVID-19 1 week (top left), 2 weeks (top right), 3 weeks (bottom left), and 4 weeks (bottom right) ahead estimates for West Virginia (WV).}
\end{figure}
\newpage

\begin{table}[ht]
\centering
\renewrobustcmd{\bfseries}{\fontseries{b}\selectfont}
\renewrobustcmd{\boldmath}{}
\newrobustcmd{\B}{\bfseries}
\begin{tabular}{lrrrr}
  \hline
&1 Week Ahead & 2 Weeks Ahead & 3 Weeks Ahead & 4 Weeks Ahead \\ 
    \hline \multicolumn{1}{l}{RMSE} \\
 \hspace{1em} ARGO &  
12.47 & 13.55 & 14.24 & 16.60 \\ 
   \hspace{1em} ARGOX 2Step & 11.56 & 13.13 & 16.24 & 34.86 \\ 
   \hspace{1em} ARGOX NatConstraint & 22.08 & 26.09 & 44.07 & 80.49 \\ 
   \hspace{1em} Ensemble & 11.25 & 11.95 & 12.31 & 13.95 \\ 
   \hspace{1em} Naive& 9.87 & 12.03 & 11.67 & 15.42 \\ 
   \multicolumn{1}{l}{MAE} \\
  \hspace{1em} ARGO &8.45 & 9.46 & 9.72 & 12.36 \\ 
   \hspace{1em} ARGOX 2Step & 6.99 & 8.15 & 10.55 & 18.26 \\ 
   \hspace{1em} ARGOX NatConstraint & 16.11 & 19.93 & 31.02 & 48.69 \\ 
   \hspace{1em} Ensemble & 6.83 & 8.11 & 7.36 & 9.14 \\ 
   \hspace{1em} Naive& 6.37 & 7.69 & 8.89 & 11.64 \\ 
   \multicolumn{1}{l}{Correlation} \\
 \hspace{1em} ARGO &0.79 & 0.74 & 0.70 & 0.58 \\ 
   \hspace{1em} ARGOX 2Step & 0.81 & 0.80 & 0.83 & 0.64 \\ 
   \hspace{1em} ARGOX NatConstraint & 0.39 & 0.04 & 0.27 & 0.09 \\ 
   \hspace{1em} Ensemble & 0.81 & 0.77 & 0.76 & 0.76 \\ 
   \hspace{1em} Naive& 0.82 & 0.74 & 0.76 & 0.59 \\ 
   \hline
\end{tabular}
\caption{Comparison of different methods for state-level COVID-19 1 to 4 weeks ahead incremental death in Wyoming (WY). The RMSE, MAE, and correlation are reported and best performed method is highlighted in boldface.} 
\label{tab:State_Ours_WY}
\end{table}

\begin{figure}[!h] 
  \centering 
\includegraphics[width=0.6\linewidth, page=51]{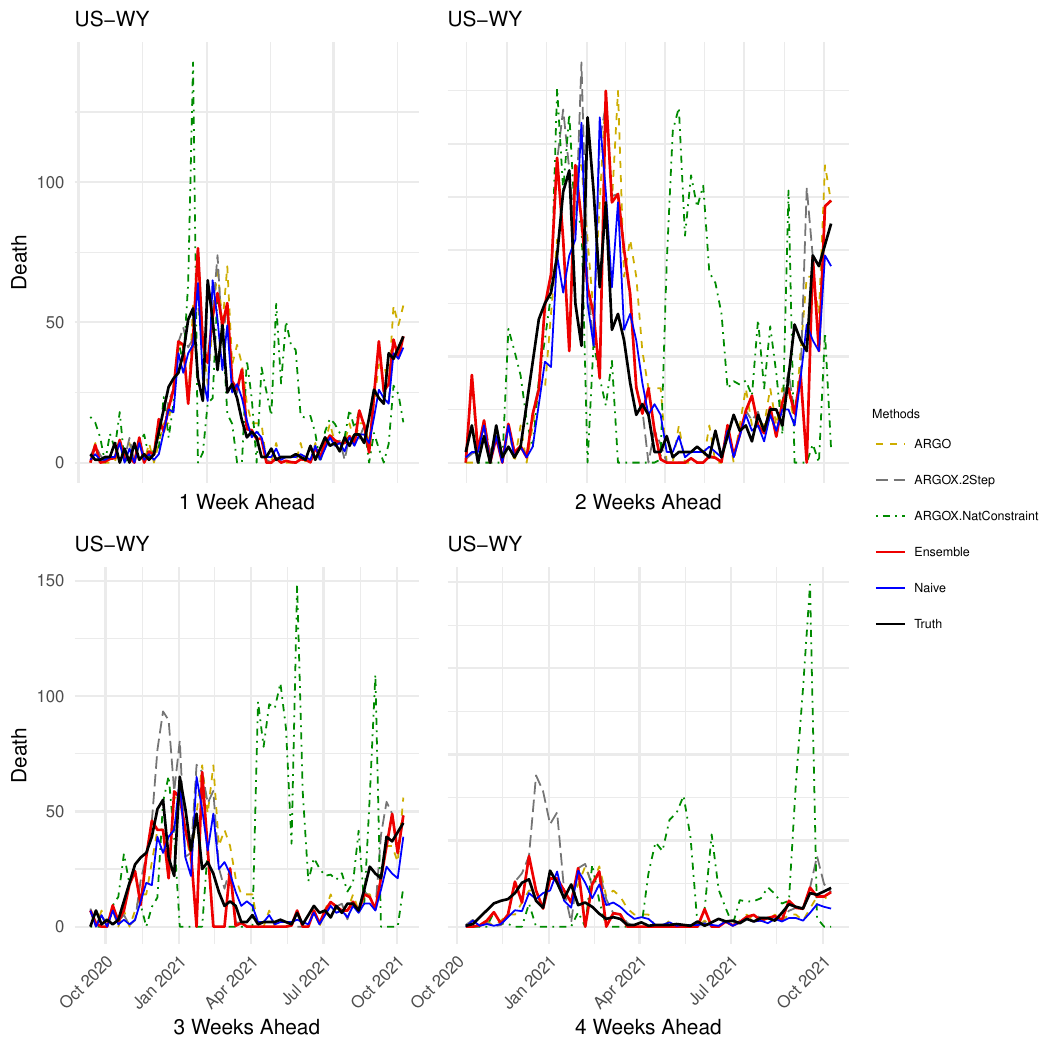} 
\caption{Plots of the COVID-19 1 week (top left), 2 weeks (top right), 3 weeks (bottom left), and 4 weeks (bottom right) ahead estimates for Wyoming (WY).}
\label{fig:State_Ours_WY}
\end{figure}
\newpage

\restoregeometry

\end{document}